\documentclass[aps, prx, two column, preprintnumbers, floatfix, nofootinbib]{revtex4-2}
\usepackage[utf8]{inputenc}
\usepackage[T1]{fontenc}
\usepackage{lmodern}
\usepackage{amsmath}
\usepackage{amsfonts}
\usepackage{amssymb}
\usepackage{amsthm}
\usepackage{thmtools}
\usepackage{dsfont}
\usepackage{graphicx}
\graphicspath{{figures/}{./figures/}{../figures/}}
\usepackage[dvipsnames]{xcolor}
\usepackage{nicefrac}
\usepackage{enumitem}
\usepackage{hyperref}
\usepackage{fullpage}
\usepackage{physics}
\usepackage{braket}
\usepackage{cleveref}
\usepackage{multirow}
\usepackage{float}
\usepackage{tikz}
\usetikzlibrary{arrows.meta,fit,backgrounds,calc}

\newtheorem{theorem}{Theorem}
\newtheorem{lemma}{Lemma}

\newtheorem{corollary}[lemma]{Corollary}

\makeatletter
\newcounter{od@anchor}
\let\od@orig@thm\@thm
\def\@thm{\stepcounter{od@anchor}\od@orig@thm}
\@for\od@name:={theorem,definition,corollary,lemma,prop,conjecture,condition,sublemma,remark}\do{%
  \expandafter\gdef\csname theH\od@name\endcsname{od\arabic{od@anchor}}%
}
\makeatother

\makeatletter
\newcommand{\od@env}{}
\newcommand{\od@key}{}
\newenvironment{informalthm}[2]{%
  \gdef\od@env{#1}\gdef\od@key{#2}%
  \expandafter\renewcommand\csname the#1\endcsname{\arabic{theorem}$'$}%
  \csname #1\endcsname
}{%
  \csname end\od@env\endcsname
  \expandafter\xdef\csname od@pair@\od@key\endcsname{\arabic{theorem}}%
}
\newenvironment{formalthm}[3][]{%
  \gdef\od@env{#2}%
  \@ifundefined{od@pair@#3}{\@latex@warning{No informal version for pair '#3'}}{}%
  \expandafter\renewcommand\csname the#2\endcsname{\csname od@pair@#3\endcsname #1}%
  \csname #2\endcsname
}{%
  \csname end\od@env\endcsname
  \addtocounter{theorem}{-1}%
}
\newenvironment{formalrestatable}[5][]{%
  \@ifundefined{od@pair@#2}{\@latex@warning{No informal version for pair '#2'}}{}%
  \expandafter\renewcommand\csname the#4\endcsname{\csname od@pair@#2\endcsname #1}%
  \restatable[#3]{#4}{#5}%
}{%
  \endrestatable
  \addtocounter{theorem}{-1}%
}
\newenvironment{corof}[2]{%
  \renewcommand{\thecorollary}{\ref*{#1}.#2}%
  \corollary
}{%
  \endcorollary
  \addtocounter{lemma}{-1}%
}
\makeatother

\newif\ifappendixtoc
\makeatletter
\AtBeginDocument{%
  \let\orig@contentsline\contentsline
  \def\contentsline{\ifappendixtoc\expandafter\orig@contentsline\else\expandafter\@gobblefour\fi}%
}
\makeatother

\begin{document}
\preprint{MIT-CTP/6118}

\title{Oracle Distillation}
\author{Ruohan Shen}
\email{rhshen@mit.edu}
\affiliation{MIT Center for Theoretical Physics - a Leinweber Institute, Massachusetts Institute of Technology, Cambridge, MA, 02139}
\author{Soonwon Choi}
\email{soonwon@mit.edu}
\affiliation{MIT Center for Theoretical Physics - a Leinweber Institute, Massachusetts Institute of Technology, Cambridge, MA, 02139}

%\date{\today}

\begin{abstract}
Learning and sensing are among the most promising applications of quantum technology, often with provable quantum advantages.
In any realistic experiment, however, noise threatens to erase these advantages as the problem size grows.
We introduce \textit{oracle distillation}, a procedure that distills many queries to a noisy oracle into a single high-fidelity oracle.
Here the oracle, an unknown unitary from a known family, models the interface through which a quantum device learns about an unknown system.
We construct an explicit protocol that distills any Boolean oracle, both under adversarial noise that corrupts a constant fraction of the input qubits and under i.i.d.\ depolarizing noise at a constant rate on every qubit.
At the heart of the protocol is the \textit{weak query}, in which the device deliberately queries the oracle only weakly, trading response strength for the ability to correct errors.
Our protocol is near-optimal among all protocols that distill the oracle while keeping its errors correctable, revealing a fundamental tension between these two requirements.
An important implication of the protocol is a threshold theorem: every Boolean oracle problem whose quantum advantage is polynomial in the domain size, including Grover search, $k$-forrelation, and Simon's problem, retains a quantum advantage whenever the per-qubit depolarizing rate is below a constant threshold.
In particular, the quadratic quantum advantage of Grover search, long believed fragile under noise, survives nearly intact at low error rates even when almost every query involves errors on one or more qubits.
We further extend the protocol to fractional and continuous-time Boolean oracles.
Quantum error correction can thus protect not only prescribed operations but also unknown dynamics.
Our results make a large family of quantum advantages in learning robust against noise and open a path toward robust computational sensing.

\end{abstract}

\maketitle

\begin{figure*}[t]
    \centering
    \includegraphics[width=\linewidth]{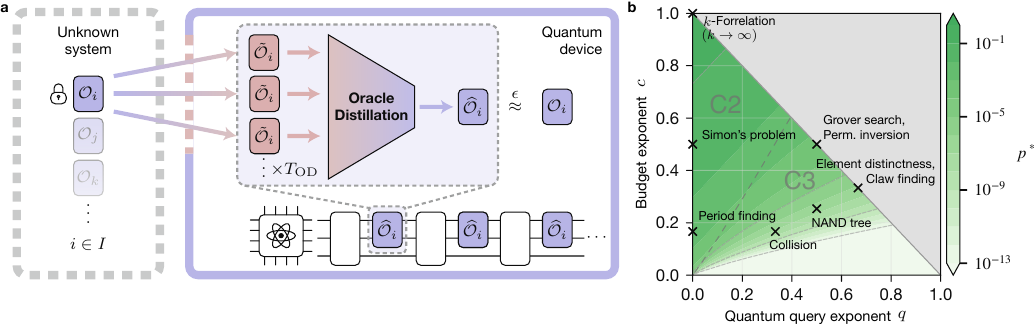}
    \caption{
        (a) The quantum device interacts with an unknown external system only through queries to an oracle.
        The oracle is promised to be a member $\mathcal{O}_i$ of the known family $\mathfrak{O}_I=\{\mathcal{O}_j\}_{j\in I}$, marked by the lock.
        The device does not know the index $i$, but $i$ remains fixed throughout, so every query calls the same oracle $\mathcal{O}_i$.
        Due to noise, each query applies a noisy version $\tilde{\mathcal{O}}_i$.
        Oracle distillation (OD) consumes $T_{\mathrm{OD}}$ noisy queries and outputs a distilled oracle $\widehat{\mathcal{O}}_i$ that is $\epsilon$-close to the ideal oracle $\mathcal{O}_i$ in diamond norm.
        Each oracle call in the downstream quantum algorithm (bottom circuit) uses a freshly distilled oracle, and each distillation consumes a new batch of noisy queries.
        (b) Advantage threshold $p^*$ for per-qubit depolarizing noise, as a function of the quantum query exponent $q$ (where $T_Q = \tilde{O}(N^q)$) and the budget exponent $c$ (where $T_C/T_Q = \tilde{\Omega}(N^c)$).
        The problem retains a quantum advantage whenever $p < p^*$.
        The labels C2 and C3 mark the regions, separated by the dark gray dashed line, where the threshold is attained by Construction 2 and by Construction 3 of Sec.~\ref{subsec:query_state_constructions}, respectively.
        Crosses mark representative oracle problems (Table~\ref{tab:oracle_separations}).
        No problem lies in the gray region $c > 1 - q$, because one classical query to each of the $N$ inputs solves any problem and hence $T_C \leq N$.
        See Theorem~\ref{thm:threshold}.
    }
    \label{fig:oracle_distillation}
\end{figure*}

\section{Introduction}

Quantum technologies promise advantages across a vast range of tasks, from speeding up computation to learning and sensing physical systems with otherwise unreachable precision~\cite{Huang2025}.
Among these tasks, the learning and sensing advantages are often practically useful and come with provable performance guarantees, from enhanced measurement precision~\cite{Giovannetti2004, Giovannetti2011} to reduced numbers of repeated experiments~\cite{Huang2022, Chen2022}.
To achieve quantum advantage at scale, however, one must confront noise.
In computation, every operation the device performs is known in advance.
Under this premise, a systematic theory for overcoming noise, known as fault-tolerant quantum computing (FTQC), has long been in place~\cite{Shor1996, Aharonov2008, Knill1998, Kitaev1997}, and recent experiments have demonstrated its key ingredients~\cite{Google2025, Bluvstein2024, Bluvstein2026}.
In learning and sensing, by contrast, the device interacts with an external system whose behavior is unknown, precisely because that behavior is what we want to learn.
Whether noise can be overcome in this setting is not clear, even in theory.

Making learning and sensing robust faces a fundamental obstacle: the very feature that defines these tasks is what makes them hard to protect against noise.
To learn about a system, the device must interact with it through an interface (Fig.~\ref{fig:oracle_distillation}a).
To the device, the interface is just another operation it performs, corrupted by noise like every other operation.
Yet standard FTQC cannot protect this one operation, because a fault-tolerant implementation presupposes knowing what the operation should be, and the action of the interface is determined by the very system being learned.

Consistent with this obstacle, research on overcoming noise in learning and sensing remains far more limited than in computation.
In the most celebrated case, overcoming noise is even provably impossible, since precision at the Heisenberg limit is unachievable under generic noise~\cite{Escher2011, Demkowicz2012, Zhou2018}.
This raises the question of whether FTQC has a counterpart for learning and sensing.

In this work, we establish a counterpart of FTQC for a broad class of learning problems.
We model the interface as an \emph{oracle}, an unknown unitary drawn from a known family, and every query applies this same dynamics (Fig.~\ref{fig:oracle_distillation}a).
Due to noise, however, each query applies its noisy version instead.
We then introduce \emph{oracle distillation} (OD), a procedure that consumes many queries to the noisy oracle and produces a single high-fidelity oracle (Fig.~\ref{fig:oracle_distillation}a).
We give an explicit protocol that distills any Boolean oracle under adversarial noise, even when the noise corrupts a constant fraction of the qubits in every query.
At the heart of the protocol is a new primitive we call the \emph{weak query}: a query that deliberately extracts only a little information from the noisy oracle, in exchange keeping the intended input hidden from the noise.
The protocol then aggregates many repetitions of the weak query into one clean query.
We further prove that this repetition is unavoidable: any protocol that can distill Boolean oracles while keeping their errors correctable must make a comparable number of queries, and our protocol is near-optimal in the low-noise regime.

Building on this protocol, we prove a threshold theorem for Boolean oracle problems under i.i.d.\ depolarizing noise, analogous to the threshold theorem of FTQC (Fig.~\ref{fig:oracle_distillation}b).
For any such problem whose quantum advantage is polynomial in the domain size, the quantum algorithm still outperforms its classical counterpart whenever the per-qubit error rate is below a constant threshold independent of the domain size, even though every query is noisy.
In particular, Grover search~\cite{Grover1996} retains an advantage below this threshold, approaching the full quadratic advantage as the error rate decreases.
This may appear to contradict several well-established no-go theorems showing that noisy oracles destroy the Grover advantage~\cite{Regev2012,Rosmanis2024,Rosmanis2023}, but those theorems concern noise correlated across all qubits, whereas the noise we consider acts on each qubit independently and only after each query.
Finally, we consider a more physical setting for both the noise and the control.
The same i.i.d.\ noise may strike before and during each query as well as after it, and the algorithm may in turn run the oracle for any fraction of its full duration and interleave its own operations between these fractions, as in the fractional and continuous-time query models~\cite{Farhi1998, Mochan2007, Cleve2009, Lee2011}.
We show that OD distills such oracles below the same error threshold, with a total evolution time that scales as the query complexity when the noise strikes only after each query.

Our results show that quantum error correction can be extended beyond FTQC to protect unknown dynamics, requiring only structure that can be exploited to distinguish the signal from the noise.
As in FTQC, our threshold theorem decouples algorithm design from error correction.
The algorithm can be designed as if the oracles were noiseless; OD delivers oracles accurate enough for as many queries as the algorithm makes; and their composition retains a quantum advantage whenever the advantage is sufficiently large.
These results also point to a route toward an end-to-end sensing advantage robust against noise.
Even though entanglement-enhanced sensitivity is provably fragile under generic noise~\cite{Escher2011, Demkowicz2012, Zhou2018, Huang2025}, some advantages in computational sensing arise instead from oracular speedups, enabled by synthesizing an oracle out of the continuous sensing dynamics~\cite{Allen2025}.
For these advantages to survive noise, the synthesized oracle need not be perfect, only distillable, due to the results presented in this work.

\section{Summary of Results}\label{sec:summary}
This section outlines the main conceptual and technical results of our work, leaving detailed discussion to later sections.

\emph{Problem setup.---}%
We assume a universal quantum computer with ideal gates and unlimited quantum memory, and study the effect of noise on the oracle alone.
The \emph{oracle} models the interface between the quantum device and the external system it probes (Fig.~\ref{fig:oracle_distillation}a), and is the sole source of noise.
Formally, an ideal oracle is a quantum channel $\mathcal{O}_i$, possibly unitary, drawn from a family of channels indexed by a label $i$.
Noise corrupts each ideal oracle $\mathcal{O}_i$ into a noisy counterpart $\tilde{\mathcal{O}}_i$, so each query, that is, each use of the oracle, applies the noisy channel instead of the ideal one.
Both the family of ideal oracles and the family of noisy oracles are known in advance.
What is unknown is the label $i$, which identifies the oracle instance actually being queried and stays the same across all queries.

\emph{Oracle distillation.---}%
We propose \textit{oracle distillation} (OD), a procedure that distills many queries to the noisy oracle $\tilde{\mathcal{O}}_i$ into a single high-fidelity \textit{distilled oracle} $\widehat{\mathcal{O}}_i$ (Fig.~\ref{fig:oracle_distillation}a).
We say the protocol distills with \textit{precision} $\epsilon$ if, for every label $i$, the distilled oracle is $\epsilon$-close to the ideal one in diamond norm,
\begin{equation}
    \left\| \widehat{\mathcal{O}}_i - \mathcal{O}_i \right\|_{\diamond} \leq \epsilon.
\end{equation}

An efficient OD protocol must satisfy three requirements.
First, it must avoid explicitly learning the label of the oracle (\textit{label-agnosticism}).
Knowing the label answers every question about the oracle without any further query, so the label carries far more information than distillation requires and learning it costs far more queries than necessary.
Therefore, explicitly learning the label is only a trivial baseline for the efficiency of OD, and an efficient protocol must do better (Sec.~\ref{sec:comparison}).
Second, the output must be $\epsilon$-close to the ideal oracle, so deviations must be detected and corrected (\textit{error correction}).
Third, the output must itself be an oracle, responding correctly to every input (\textit{functionality}).
These requirements pull against one another.
Label-agnosticism conflicts with error correction, because a protocol that does not know which oracle it is querying has no reference dynamics against which to define an error.
Functionality conflicts with error correction as well, because the states on which the protocol queries the noisy oracle must depend on the input, yet error correction requires that the states for different inputs be indistinguishable to the noise.
This work develops a protocol that satisfies all three requirements by exploiting structure in the oracles and the noise.
Table~\ref{tab:comparison} places OD among other related tasks.
Each related task involves at most two of the three requirements, and the absent one is what makes existing tools sufficient.
OD requires all three simultaneously.
\begin{table}[t!]
    \centering
    \begin{tabular}{lccc}
    \hline\hline
    Task & Label Agn. & Err.\ Corr. & Funct. \\
    \hline
    FTQC~\cite{Shor1996,Aharonov2008,Knill1998,Kitaev1997}                             & $\times$   & \checkmark & \checkmark \\
    MSD~\cite{Bravyi2005}                              & $\times$   & \checkmark & $\times$   \\
    Pur./Ent. Distill.~\cite{Bennett1996, Cirac1999, Li2025} & \checkmark & \checkmark & $\times$   \\
    QSP~\cite{Low2017, Gilyen2019, Martyn2021}                              & \checkmark & $\times$   & \checkmark \\
    \textbf{OD (this work)}          & \checkmark & \checkmark & \checkmark \\
    \hline\hline
    \end{tabular}
    \caption{Comparison of oracle distillation (OD) with related tasks.
    The three columns record label-agnosticism (Label Agn.), error correction (Err.\ Corr.), and functionality (Funct.).
    Fault-tolerant quantum computation (FTQC) corrects errors against a fully known target circuit, so label-agnosticism is absent.
    Magic state distillation (MSD) corrects errors toward a known target magic state, so functionality and label-agnosticism are absent.
    Purity and entanglement distillation are label-agnostic but output a clean state rather than a functioning dynamics, so functionality is absent.
    Quantum signal processing (QSP) achieves functionality and label-agnosticism but assumes noiseless oracle access, so error correction is absent.
    Each related task involves at most two of the three requirements, whereas OD requires all three simultaneously.}
    \label{tab:comparison}
\end{table}

\emph{Boolean oracle distillation.---}%
We now focus on a concrete family, the Boolean oracles.
The \textit{Boolean oracle} with label $f:\{0,1\}^n\to\{0,1\}^m$ is the unitary channel of the $(n+m)$-qubit unitary $O_f$, which acts as
\begin{equation}
    O_f \ket{x}\ket{y} = \ket{x}\, Z^{f(x)}\ket{y},
\end{equation}
where $x\in\{0,1\}^{n}$ is the index and $y\in\{0,1\}^{m}$ is the content of the register that receives the oracle's response.
Accordingly, we call the first $n$ qubits the \textit{index register} and the last $m$ qubits the \textit{response register}, and write $N=2^n$ for the domain size.
The oracle responds to the index $x$ by imprinting $f(x)$ onto the response register as the phase flip $Z^{f(x)} = Z^{f(x)_1}\otimes \cdots \otimes Z^{f(x)_m}$, where $f(x)_j$ denotes the $j$-th bit of $f(x)$.
This oracle differs from the standard bit-flip oracle $\ket{x}\ket{y} \mapsto \ket{x}\ket{y \oplus f(x)}$ only by a Hadamard on each response qubit.

Noise corrupts each Boolean oracle $\mathcal{O}_f$ into a noisy Boolean oracle $\tilde{\mathcal{O}}_f$.
Since $\mathcal{O}_f$ is a unitary channel and hence invertible, the noisy oracle always factors as the ideal oracle followed by a \textit{noise channel} $\mathcal{E}_f$,
\begin{equation}\label{eq:noisy-oracle-local}
    \tilde{\mathcal{O}}_f = \mathcal{E}_f \circ \mathcal{O}_f.
\end{equation}
A noise model is then imposed by assumptions on the structure of $\mathcal{E}_f$, a choice that is pivotal because it largely determines whether a quantum advantage survives noise.

Our first main result is that the noisy Boolean oracles can be distilled into the ideal ones, provided the noise has bounded weight.
Precisely, we call $\tilde{\mathcal{O}}_f$ a \textit{weight-$\lfloor \alpha n/2 \rfloor$ noisy Boolean oracle}, for a constant $\alpha$, if its noise channel $\mathcal{E}_f$ admits a Kraus representation in which every Kraus operator is a linear combination of Pauli strings, each of weight at most $\lfloor \alpha n/2 \rfloor$ and supported only on the index register.
Beyond the weight bound and the support restriction, the channel is arbitrary and may depend on the label $f$, so the noise is adversarial.

\begin{informalthm}{theorem}{main}[Boolean oracle distillation under adversarial noise, informal version of Corollary~\ref{cor:boolean_OD_encoding3}]\label{thm:main}
    For any constant $\alpha \in (0,0.16]$ and all sufficiently large $n$, our OD protocol distills every weight-$\lfloor \alpha n/2 \rfloor$ noisy Boolean oracle, with any number $m$ of response qubits, into the corresponding ideal Boolean oracle with precision $\epsilon$, using
    \begin{equation}\label{eq:T_OD_main}
        T_{\mathrm{OD}} = O\!\left(N^{H(\alpha)+2\alpha} \ln\frac{1}{\epsilon}\right)
    \end{equation}
    queries to the noisy oracle.
    Here $H(\alpha) = -\alpha \log_2 \alpha - (1-\alpha) \log_2 (1-\alpha)$ is the binary entropy.
\end{informalthm}
\noindent
We implement this protocol with an explicit circuit with bounded elementary gate and ancilla counts (Theorem~\ref{thm:gate_complexity_OD}).
The key idea of the protocol is the \textit{weak query}.
We query the oracle on codewords that are locally indistinguishable to the noise, so errors can be corrected, yet each retains large overlap with the corresponding bitstring state, so the oracle responds to the intended input, but only weakly.
We coherently aggregate the weak responses of many weak queries into the full oracle response, then uncompute the oracle's action to return the state to a fixed codespace independent of the label, where errors are corrected without knowledge of the label.

\emph{Optimality of Boolean oracle distillation.---}%
Our second main result is a lower bound on the number of queries that OD requires, showing that the rate $H(\alpha)$ in Theorem~\ref{thm:main} is not an artifact of our construction, but results from the fundamental tension between distilling the oracle and correcting its errors.
The bound makes this tension concrete through two conditions, one fixing which family of oracles the protocol must distill correctly and one fixing what type of error in the oracle the protocol intends to correct.
For the first, we consider the Grover oracles, the Boolean oracles whose label $f:\{0,1\}^n\to\{0,1\}$ takes the value $1$ on exactly one bitstring, and require only that the protocol can distill ideal Grover oracles into themselves.
For the second, we require that, for each of its $T_{\mathrm{OD}}$ queries, the states just before that query span a subspace satisfying the Knill--Laflamme conditions for all $Z$ errors of weight at most $\alpha n$ on the index register.
\begin{informalthm}{theorem}{opt}[Optimality of Boolean oracle distillation, informal version of Theorem~\ref{thm:OD_grover_lower_bound}]\label{thm:opt_grover}
    For any constant $\alpha \in (0, 0.061]$, any OD protocol that distills every ideal Grover oracle into itself with precision $\epsilon \leq 1/2$ must make
    \begin{equation}
        T_{\mathrm{OD}} = \tilde{\Omega}\left(N^{H(\alpha)}\right)
    \end{equation}
    queries, provided that, for each of these $T_{\mathrm{OD}}$ queries, its states just before that query span a subspace satisfying the Knill--Laflamme conditions for all $Z$ errors of weight at most $\alpha n$ on the index register.
\end{informalthm}
\noindent
Both conditions are mild.
The first is met by any more general protocol as long as it can distill the Grover oracles as part of a larger family.
Ours distills every Boolean oracle, so it qualifies.
The second is a natural requirement for any protocol that is able to correct these errors, as it only demands that the errors not irreversibly destroy the logical information.
When $\alpha \to 0$ and the precision is no smaller than any useful distillation needs, our upper bound exceeds this lower bound only by the factor $N^{2\alpha}$, which is negligibly small relative to $N^{H(\alpha)}$ (Sec.~\ref{sec:optimality}).
Our protocol is therefore near-optimal in this regime.

\emph{Quantum advantage using distilled oracles.---}%
Our third main result is a threshold theorem for Boolean oracle problems under i.i.d.\ depolarizing noise.
In this noise model, every qubit of the index and response registers independently depolarizes at rate $p$ after each query, which is more realistic than adversarial noise.
The theorem rests on an OD protocol that distills these noisy Boolean oracles into the ideal ones (Theorem~\ref{thm:iid_threshold_at_exponent}, Sec.~\ref{subsec:iid_noise}).
For any constant $\gamma \in (0,1)$, the protocol distills to precision $\epsilon$ using $T_{\mathrm{OD}} = O(N^{\gamma} \ln(m/\epsilon))$ queries to the noisy oracle, as long as $p$ is below a constant that depends only on $\gamma$ and how fast the precision $\epsilon$ decays with $N$.

We now define the problems the threshold theorem applies to and the two query counts it compares.
A \textit{Boolean oracle problem} asks, for each input size $n$, to compute a property of the label $f$, where $f$ is promised to lie in a given subset of the Boolean functions.
We fix a quantum algorithm that solves the problem with the ideal oracle $\mathcal{O}_f$, and write $T_Q$ for the number of queries it makes.
The classical query complexity $T_C$ is the minimum number of noiseless classical evaluations of $f$ that any algorithm needs to solve the problem.
Both counts are functions of $N$.
To solve the problem with noisy oracles, it suffices to distill them to precision $\epsilon = O(1/T_Q)$, so that the errors from all $T_Q$ queries add up to $O(1)$, and run the quantum algorithm on the distilled oracles.
This consumes $T_Q \cdot T_{\mathrm{OD}}$ noisy queries in total.
Whenever $T_{\mathrm{OD}} < T_C/T_Q$, the quantum algorithm makes fewer than $T_C$ queries even though each query is noisy, and we say that the problem retains a quantum advantage.
The ratio $T_C/T_Q$ is therefore the largest number of noisy queries the problem can afford per distillation, and we call it the \textit{advantage budget} of the problem.
\begin{informalthm}{theorem}{thr}[Threshold theorem for Boolean oracle problems, informal version of Theorem~\ref{thm:threshold_formal}]\label{thm:threshold}
    Every Boolean oracle problem with advantage budget $N^{\Omega(1)}$ and $\log_2 m = o(n)$ admits a constant threshold $p^* > 0$ with the following property.
    Whenever the per-qubit error rate is a constant $p < p^*$, the problem retains a quantum advantage in query complexity for all sufficiently large $n$ when the quantum algorithm queries the noisy oracle.
\end{informalthm}
\noindent
The theorem counts noisy quantum queries against noiseless classical queries.
This comparison is conservative, since noise can only increase the classical query complexity (Appendix~\ref{SM_sec:noisy_classical}).
Crucially, $p^*$ is a positive constant independent of $n$, even though the probability $(1-p)^{n+m}$ that a query is entirely error-free is exponentially small.

Many oracle problems of central interest have a large advantage budget and hence a constant threshold, with $p^*$ determined by how fast the advantage budget and $T_Q$ grow with $N$ (Table~\ref{tab:oracle_separations} and Fig.~\ref{fig:oracle_distillation}b).
This $p^*$ is the threshold our protocol achieves, not a limit of the problem itself, so the quantum advantage may survive error rates above it.
For Grover search, $T_Q = \Theta(\sqrt{N})$ and $T_C = \Theta(N)$ give the advantage budget $\Theta(N^{1/2})$, and the threshold evaluates to $p^* = 5.1 \times 10^{-4}$ (Sec.~\ref{subsec:grover_example}).
As $p \to 0$, the number of noisy queries needed to solve Grover search approaches the noiseless scaling $\tilde{O}(\sqrt{N})$.
That the quantum advantage of Grover search survives a constant per-qubit error rate may appear to contradict prior works~\cite{Regev2012,Rosmanis2023,Rosmanis2024}, which proved that certain noisy oracles destroy it at constant error rate.
The difference lies in the noise model.
In these works, the noise channel $\mathcal{E}_f$ is correlated across all qubits, whereas our noise strikes each qubit independently after the oracle, so typical errors touch only a small fraction of the qubits.

\emph{Distillation of fractional and continuous-time oracles.---}%
Our last result models both the noise and the control more physically.
The i.i.d.\ depolarizing noise may strike before, during, and after each query, and the algorithm may act during the implementation of the oracle rather than only after each query completes.
We model this control by the fractional query model~\cite{Farhi1998, Mochan2007, Cleve2009, Lee2011}, in which the algorithm may run the Hamiltonian $H_f := \sum_{x} \ket{x}\bra{x} \otimes \sum_{j=1}^m f(x)_j \ket{1}\bra{1}_j$ for any angle $\theta$ and act between such runs.
The \textit{fractional oracle} of angle $\theta$ is $O_f(\theta) := e^{-i\theta H_f}$, and $O_f(\pi)$ is the Boolean oracle $O_f$.
We consider two noise models, i.i.d.\ depolarizing noise of per-qubit rate proportional to $\theta$ before and after each fractional query, and continuous-time evolution under $H_f$ with i.i.d.\ depolarizing noise of rate $\Gamma$ added to the Lindbladian.
Since a fractional query runs $H_f$ for a shorter time than a Boolean oracle query, we measure the cost by the \textit{total evolution time}, the sum of the evolution times of all queries.
For the continuous-time Boolean oracle, we prove the following result (Sec.~\ref{sec:two_sided_noise}, Theorem~\ref{thm:continuous_OD_two_sided}).
\begin{informalthm}{theorem}{twosided}[Distillation of continuous-time Boolean oracles, informal version of Theorem~\ref{thm:continuous_OD_two_sided}]\label{thm:two_sided}
    For any constant $\gamma \in (0, 1)$, our protocol distills every continuous-time Boolean oracle into the corresponding ideal Boolean oracle with precision $\epsilon$, using $T_{\mathrm{OD}}$ queries of angle $\theta$ with total evolution time
    \begin{equation}\label{eq:T_ct_informal}
        \theta\, T_{\mathrm{OD}} = O\!\left( N^{\gamma} \ln \frac{m}{\epsilon} \right) .
    \end{equation}
    This guarantee holds for all sufficiently large $n$ whenever the noise rate is a constant $\Gamma$ below a positive constant threshold $\Gamma_{\mathrm{th}}$ and the angle $\theta$ is at most a positive threshold $\theta_{\mathrm{th}}$.
    Here $\epsilon = \Omega(1/N)$ and $\log_2 m = o(n)$.
\end{informalthm}
\noindent
A similar result holds for the fractional Boolean oracle under two-sided i.i.d.\ depolarizing noise (Theorem~\ref{thm:fractional_OD_two_sided}).

\emph{Outline.---}%
The rest of the paper develops the results summarized above.
Section~\ref{sec:oracle_distillation} formally defines oracle distillation and its query complexity.
Section~\ref{sec:grover_distillation} presents the two-stage protocol distilling Boolean oracles under adversarial noise and extends it to i.i.d.\ depolarizing noise.
Section~\ref{sec:optimality} states lower bounds on the query complexity of oracle distillation and compares them with the query complexity of the protocol.
Section~\ref{sec:threshold} states the threshold theorem formally and evaluates the threshold for Grover search, the $k$-forrelation problem, and Simon's problem.
Section~\ref{sec:two_sided_noise} extends distillation to fractional and continuous-time oracles.
Section~\ref{sec:comparison} compares oracle distillation with other approaches to correcting noisy operations.
Section~\ref{sec:discussion} discusses subtleties and implications of oracle distillation.
Section~\ref{sec:conclusion} concludes with open questions.
All proofs are given in the appendices.

\section{Oracle Distillation}\label{sec:oracle_distillation}
In this section, we present the precise definition of oracle distillation and establish technical notations that will be used in the later sections.

We define a \textit{family of oracles} to be a set of channels
\begin{equation}
    \mathfrak{O}_I = \{\mathcal{O}_i\}_{i\in I}
\end{equation}
indexed by a label set $I$, and call each channel $\mathcal{O}_i$ an \textit{oracle} with label $i$.
In this work, every oracle in $\mathfrak{O}_I$ acts on the same Hilbert space $\mathcal{H}_O$, serving as both its input and output space; the two may differ in the most general setting.
The Hilbert space an oracle acts on will always be clear from context, and oracles are implicitly tensored with the identity on registers they do not act on.

Now we define \textit{oracle distillation} (OD), which is specified by two families of oracles sharing the same label set $I$: $\mathfrak{O}_I^{\mathrm{raw}}$ and $\mathfrak{O}_I^{\mathrm{target}}$.
We say an OD protocol distills $\mathfrak{O}_I^{\mathrm{raw}}$ into $\mathfrak{O}_I^{\mathrm{target}}$ with precision $\epsilon$ if, for every $i\in I$, the protocol queries $\mathcal{O}_i^{\mathrm{raw}}$ multiple times and outputs a \textit{distilled oracle} $\widehat{\mathcal{O}}_i$ satisfying
\begin{equation}
    \left\|\widehat{\mathcal{O}}_i - \mathcal{O}_i^{\mathrm{target}} \right\|_{\diamond} \leq \epsilon,
\end{equation}
where $\|\cdot\|_{\diamond}$ is the diamond norm~\cite{Watrous2018}, which induces a distance between channels.
We call such an output an \textit{$\epsilon$-approximate oracle}.
This is the most stringent distance for channels: it quantifies the largest distinguishability achievable by applying the two channels to the same input state, which may be entangled with an auxiliary system.
This enables us to replace the target oracle by the distilled oracle in any algorithm that queries it, introducing only $\epsilon$ error per query.
The total error is upper bounded by $\epsilon$ times the number of queries, by the triangle inequality for the diamond norm.

Besides the queries to the raw oracle, the protocol has no dependence on the label $i$, yet it must work for all $i\in I$.
Formally, an OD protocol is specified by an ancilla state $\rho_A$ on an ancilla space $\mathcal{H}_A$ and unitaries $U_0, \dots, U_T$, none of which depends on the label $i$.
Running the protocol with the raw oracle $\mathcal{O}_i^{\mathrm{raw}}$ produces the distilled oracle
\begin{equation}\label{eq:od_protocol}
    \widehat{\mathcal{O}}_i := \Tr_A \circ\, \mathcal{U}_T \circ \mathcal{O}_i^{\mathrm{raw}} \circ \mathcal{U}_{T-1} \circ \dots \circ \mathcal{O}_i^{\mathrm{raw}} \circ \mathcal{U}_0 \circ \left(\,\cdot\, \otimes \rho_A\right).
\end{equation}
Each use of $\mathcal{O}_i^{\mathrm{raw}}$ in the composition is called a \textit{query} and acts on $\mathcal{H}_O$.
Each $\mathcal{U}_t$ is the unitary channel of $U_t$, acting on $\mathcal{H}_O \otimes \mathcal{H}_A$.
The input space of $\widehat{\mathcal{O}}_i$ is $\mathcal{H}_O \otimes \mathcal{H}'$, where $\mathcal{H}'$ is an arbitrary auxiliary space on which neither the queries nor the unitaries act.
Fixing every query to act on the same space $\mathcal{H}_O$ is only a formal convention: a concrete protocol can SWAP a different register into $\mathcal{H}_O$ before each query and SWAP it back afterwards, and these SWAPs can be absorbed into the neighboring $U_t$'s.

We define the \textit{query complexity} $T_{\mathrm{OD}}$ of an OD protocol as the number of queries it makes to the raw oracle, the $T$ in Eq.~\eqref{eq:od_protocol}.
We measure the cost of an OD protocol in queries rather than gates because queries are the more expensive resource.
Each query requires interacting with the external world, whereas the gates in between are internal operations that a fault-tolerant quantum computer can synthesize on its own.
Moreover, the downstream algorithms that consume the distilled oracles are usually themselves measured in queries.
Each call to a distilled oracle costs $T_{\mathrm{OD}}$ queries to the raw oracle, so the total number of raw-oracle queries is simply the product of the two counts.

\section{Boolean Oracle Distillation}\label{sec:grover_distillation}
In this section, we present an OD protocol that distills the family of noisy Boolean oracles into high-fidelity ones.
We consider two noise models: adversarial noise on the index register, and stochastic i.i.d.\ depolarizing noise on both index and response register.
We first present the protocol and show that it corrects adversarial noise, then show that a generalization of the same argument handles stochastic noise.

\subsection{Boolean oracles and noise models}
\begin{figure}[t]
    \centering
    \includegraphics[width=\linewidth]{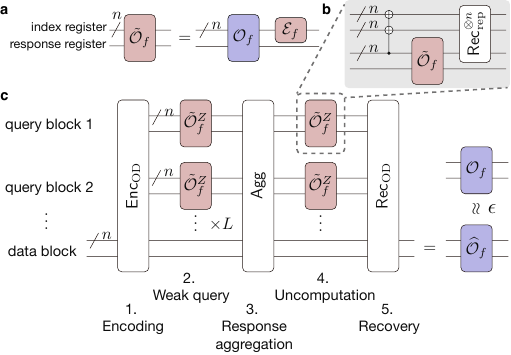}
    \caption{
    The two-stage protocol for distilling noisy Boolean oracles.
    \textbf{a} The noise model.
    The noisy Boolean oracle $\tilde{\mathcal{O}}_f^{\mathrm{adv}}$ is the ideal oracle $\mathcal{O}_f$ followed by an adversarial noise channel $\mathcal{E}_f$ on the index register.
    \textbf{b} The stage 1 gadget, which converts adversarial noise into phase noise with no query overhead.
    Each of the $n$ index qubits is encoded into a length-3 repetition code, and the noisy oracle $\tilde{\mathcal{O}}_f^{\mathrm{adv}}$ acts on the third qubit of each code block together with the response qubit.
    Recovery on each code block corrects all $X$-type errors, while $Z$-type errors remain unaffected and are corrected in stage 2.
    \textbf{c} The stage 2 circuit, which distills phase-noisy Boolean oracles into ideal Boolean oracles.
    The circuit acts on $L$ query blocks and one data block, and proceeds in five steps.
    Each red box is one call to the phase-noisy Boolean oracle $\tilde{\mathcal{O}}_f^{\mathrm{adv},Z}$ produced by the stage 1 gadget, written $\tilde{\mathcal{O}}_f^Z$ in the figure.
    The dashed box marks one such call, expanded in b.
    The full circuit implements the distilled oracle $\widehat{\mathcal{O}}_f$, which is $\epsilon$-close to the ideal oracle $\mathcal{O}_f$ in diamond norm.
    }
    \label{fig:architecture}
\end{figure}

The \textit{family of Boolean oracles} (or \textit{ideal Boolean oracles}, when contrasted with noisy ones) $\mathfrak{O}_{F} = \{\mathcal{O}_f\}_{f\in F}$ serves as the target oracle family of the OD protocol.
Its label set $F$ is the collection of all Boolean functions $f: \{0,1\}^n \rightarrow \{0,1\}^m$, where $n,m \ge 1$ are integers fixed by context.
The \textit{Boolean oracle} with label $f$ is the unitary channel on $n+m$ qubits $\mathcal{O}_f(\cdot) := O_f(\cdot)O_f^{\dagger}$, where
\begin{equation}
    O_f\ket{x}\ket{y} = \ket{x} Z^{f(x)}\ket{y}
\end{equation}
for all $x\in \{0,1\}^n$, $y\in \{0,1\}^m$, and $Z^{f(x)} := Z^{f(x)_1} \otimes \dots \otimes Z^{f(x)_m}$ is the $Z$-string defined by $f(x)$.
We call the first $n$ qubits the \textit{index register} and the last $m$ qubits the \textit{response register}.

The \textit{family of noisy Boolean oracles} $\tilde{\mathfrak{O}}_{F} = \{\tilde{\mathcal{O}}_f\}_{f\in F}$, a set of channels indexed by the same label set $F$, serves as the raw oracle family of the OD protocol.
Since each ideal Boolean oracle is a unitary channel and hence invertible, every noisy oracle factors as an ideal oracle followed by a \textit{noise channel}:
\begin{equation}
    \tilde{\mathcal{O}}_f = \mathcal{E}_f \circ \mathcal{O}_f, \qquad \mathcal{E}_f := \tilde{\mathcal{O}}_f \circ \mathcal{O}_f^{-1}.
\end{equation}

We focus on two noise models.

The first is adversarial noise on the index register (Fig.~\ref{fig:architecture}a).
Here $\mathcal{E}_f$ admits a Kraus representation in which every Kraus operator is a linear combination of Pauli strings, each of weight at most $w_P$ and supported only on the index register, where $w_P$ is a nonnegative integer.
We call such a channel a \textit{weight-$w_P$ noise channel}.
Beyond the weight bound and the support restriction, the channel is arbitrary and may depend on the label $f$, so the noise is adversarial.
We denote the family of noisy Boolean oracles under this model by $\tilde{\mathfrak{O}}_F^{\mathrm{adv}} := \{\tilde{\mathcal{O}}_f^{\mathrm{adv}}\}_{f\in F}$.
If moreover every Kraus operator is a linear combination of $Z$-strings only, we call the channel a \textit{weight-$w_P$ phase-noise channel}, denote the family of noisy Boolean oracles carrying such channels by $\tilde{\mathfrak{O}}_F^{\mathrm{adv},Z} := \{\tilde{\mathcal{O}}_f^{\mathrm{adv},Z}\}_{f\in F}$, and call each member a \textit{phase-noisy Boolean oracle}.
We restrict the noise to the index register because allowing it on the response register would make distillation impossible: a $Z$ error on a response qubit after $\mathcal{O}_f$ implements exactly another ideal oracle $\mathcal{O}_{f'}$, so the same channel is a valid raw oracle both for $f$ (with the $Z$ error) and for $f'$ (with no error).
Querying this channel, a protocol produces a single output, which cannot be close to both targets $\mathcal{O}_f$ and $\mathcal{O}_{f'}$ at once.

The second is stochastic noise acting on all qubits, modeled as the standard i.i.d.\ depolarizing channel
\begin{equation}\label{iidnoise}
    \mathcal{E}_f = \mathcal{D}^{\otimes (n+m)}_p
\end{equation}
where $\mathcal{D}_p(\cdot) := (1-p)(\cdot) + \frac{p}{3}\bigl(X(\cdot)X + Y(\cdot)Y + Z(\cdot)Z\bigr)$ and $p$ is the per-qubit error rate.
We denote the family of noisy Boolean oracles under this model by $\tilde{\mathfrak{O}}_F^{\mathrm{iid}} := \{\tilde{\mathcal{O}}_f^{\mathrm{iid}}\}_{f\in F}$, and call each member an \textit{i.i.d.-depolarizing Boolean oracle}.
Replacing the depolarizing channel by the single-qubit dephasing channel $\mathcal{D}^Z_p(\cdot) := (1-p)(\cdot) + p Z(\cdot)Z$ gives the noise channel $\mathcal{E}_f = (\mathcal{D}^Z_p)^{\otimes(n+m)}$.
We denote the corresponding family by $\tilde{\mathfrak{O}}_F^{\mathrm{iid},Z} := \{\tilde{\mathcal{O}}_f^{\mathrm{iid},Z}\}_{f\in F}$, and call each member an \textit{i.i.d.-dephasing Boolean oracle}.

\subsection{Overview of the protocol}\label{subsec:protocol_overview}
Our distillation protocol for adversarial noise achieves the following guarantee.
\begin{formalrestatable}[a]{main}{Distillation of Boolean oracles under adversarial noise}{theorem}{booleanodadv}\label{thm:boolean_OD_adv}
    The two-stage protocol distills the family of weight-$w_P$ noisy Boolean oracles $\tilde{\mathfrak{O}}_F^{\mathrm{adv}}$ into the family of ideal Boolean oracles $\mathfrak{O}_F$ with precision $\epsilon$ and query complexity
    \begin{equation}\label{eq:T_OD_adv}
        T_{\mathrm{OD}} = 2 \left\lceil \frac{2}{\eta} \ln\frac{4}{\epsilon} \right\rceil,
    \end{equation}
    provided the query states are generated from a seed state by Eq.~\eqref{eq:query-state-from-seedstate} and satisfy the error orthogonality conditions Eq.~\eqref{eq:error_orthogonality} for $r \geq 2 w_P$ errors with matched query power $\eta$.
\end{formalrestatable}
\noindent
The query states, the seed state, the error orthogonality conditions, and the matched query power $\eta$ are components of the protocol that we introduce shortly.
In the remainder of this section we outline the overall architecture of the protocol and the argument behind the theorem, and the full proof is given in Appendix~\ref{SM_sec:adv}.

The protocol is a concatenation of two stages.
Stage 1 converts the noisy oracle $\tilde{\mathcal{O}}_f^{\mathrm{adv}} = \mathcal{E}_f \circ \mathcal{O}_f$ into a phase-noisy Boolean oracle $\tilde{\mathcal{O}}_f^{\mathrm{adv},Z} = \mathcal{E}_f^Z \circ \mathcal{O}_f$ with the same label $f$, where $\mathcal{E}_f^Z$ is a weight-$w_P$ phase-noise channel on the index register.
The conversion is a repetition-code gadget (Fig.~\ref{fig:architecture}b) that implements $\tilde{\mathcal{O}}_f^{\mathrm{adv},Z}$ with a single query to $\tilde{\mathcal{O}}_f^{\mathrm{adv}}$, so it has no query overhead.
Because the Boolean oracle is diagonal in the computational basis, it acts as the logical Boolean oracle on the codespace, and the recovery corrects all $X$-type errors while $Z$-type errors remain unaffected and act as logical phase errors (Corollary~\ref{cor:stage1}).

Stage 2 is an OD protocol that distills $\tilde{\mathfrak{O}}_F^{\mathrm{adv},Z}$ into the family of ideal Boolean oracles $\mathfrak{O}_F$.
Whereas stage 1 is a standard application of the repetition code, stage 2 is where most of the new ideas and techniques of this work enter.
The stage 2 protocol consists of five steps, namely 1.~encoding, 2.~weak query, 3.~response aggregation, 4.~uncomputation, and 5.~recovery (Fig.~\ref{fig:architecture}c).
To illustrate how this works, we will demonstrate the action of the distilled oracle on an input basis state $\ket{x}\ket{y}$, where $x$ is a bitstring of length $n$ and $y$ is a single bit ($m=1$) for simplicity.
The analysis of superpositions of basis inputs, and of $m>1$, is given in Appendix~\ref{SM_sec:stage2_OD}.

Every query of the stage 2 protocol is a query to the phase-noisy Boolean oracle $\tilde{\mathcal{O}}_f^{\mathrm{adv},Z} = \mathcal{E}_f^Z \circ \mathcal{O}_f$, and the argument proceeds in two rounds.
In the first round, we analyze the circuit with every query replaced by the ideal oracle $\mathcal{O}_f$, and show that this ideal-oracle circuit acts on the input as the ideal oracle up to a small error.
In the second round, we return to the actual circuit with the noisy queries, and show that the noise adds no further error to the distilled oracle.

The first round of the argument begins with the encoding step, where we encode each basis state $\ket{x}\ket{y}$ into the logical state
\begin{equation}\label{eq:logical_state}
    \ket{\bar{x}}\ket{\bar{y}} := \underbrace{\left(\ket{\Theta(x)} \ket{+}\right)^{\otimes L}}_{L\ \text{query blocks}} \otimes \underbrace{\ket{x}\ket{y}}_{\text{data block}},
\end{equation}
and we call the span of these logical states over all inputs $x$ and $y$ the \textit{logical subspace}.
Here $\ket{\Theta(x)}$ is an $x$-dependent $n$-qubit pure state, which we call the \textit{query state}; the properties it must satisfy are specified in the following.
We call each group of $n+1$ qubits in the state $\ket{\Theta(x)}\ket{+}$ a \textit{query block}, where $\ket{+} = (\ket{0}+\ket{1})/\sqrt{2}$; there are $L$ query blocks in total.
In the upcoming oracle query, the first $n$ qubits of a query block serve as the index register and the last qubit serves as the response register.
We call the last $n+1$ qubits of the logical state the \textit{data block}.
The data block stores the input $\ket{x}\ket{y}$ intact, so the logical states of distinct inputs are orthogonal and the encoding is well defined on all possible input states.
The number of query blocks $L$ sets the query complexity of the protocol; its value will be fixed by the query state we use and the precision $\epsilon$ we demand of the distilled oracle.

Both the correctness and the query complexity of the protocol depend on the construction of the query states $\ket{\Theta(x)}$.
For correctness, we demand that local $Z$ errors map $\ket{\Theta(x)}$ to mutually orthogonal states.
Concretely, we require the \textit{error orthogonality conditions} (EOC) that for all $x$ and all $a, b \in \{0,1\}^n$ with $|a|, |b| \leq r$,
\begin{equation}\label{eq:error_orthogonality}
    \bra{\Theta(x)} Z^a Z^b \ket{\Theta(x)} = \delta_{a,b},
\end{equation}
where $r$ is the maximum weight of $Z$ errors on each query block that the logical subspace can correct.
We require $r \geq 2w_P$, for reasons that will become clear in the second round of argument.
The query complexity is set by the \textit{matched query power}
\begin{equation}
    \eta_x := \left| \braket{x | \Theta(x)}\right|^2,
\end{equation}
the population of $\ket{\Theta(x)}$ on the input $\ket{x}$ it stands for, and we write $\eta := \min_x \eta_x$ for the worst-case matched query power.
When the oracle acts on a query block, only this matched component acquires the response about $f(x)$.
Hence $\eta_x$ sets the strength of the response that each query extracts, and the query complexity Eq.~\eqref{eq:T_OD_adv} scales as $1/\eta$.
In Sec.~\ref{subsec:query_state_constructions} we present several concrete constructions of the query states, realizing different trade-offs between the matched query power $\eta$ and the maximum weight $r$ of correctable $Z$ errors.

In the weak query step, we query $\mathcal{O}_f$ once on each of the $L$ query blocks, and leave the data block untouched.
The query state $\ket{\Theta(x)}$ is a superposition of bitstring states $\ket{z}$, and the oracle answers every component in superposition.
Each component $\ket{z}$ has the response qubit flipped from $\ket{+}$ to $\ket{-}$ precisely when $f(z) = 1$.
In the post-query state, we will only focus on the population of the state $\ket{x}\ket{-}$.
Every component with index $z \neq x$ is orthogonal to $\ket{x}$ and contributes nothing to this population, so it records the oracle's response about $x$ and nothing else.
When $f(x)=0$, this population is exactly zero.
When $f(x)=1$, it is only the weight $\eta_x$ of the matched component, whereas a conventional query on the bare bitstring $\ket{x}$ would place the full population on $\ket{x}\ket{-}$.
This weakened response is what the name \textit{weak query} refers to.

In the response aggregation step, we use an \textit{aggregator} to collect the weak responses from the $L$ query blocks and write the full response into the data block, while leaving the query blocks nearly intact.
We say a query block \textit{responds} when its index matches the index in the data block and its response register is in $\ket{-}$.
For each query block $i$, we define the projector onto its response subspace,
\begin{equation}\label{eq:response_projector}
    \Pi^{(i)}_R := \sum_{z\in \{0,1\}^n} \left(\ket{z}\bra{z} \otimes \ket{-}\bra{-}\right) \otimes \ket{z}\bra{z} \otimes I,
\end{equation}
where the projector in the parenthesis acts on the $i$-th query block, the next projector acts on the index register in the data block, and $I$ is the identity on the rest of the Hilbert space, namely the other query blocks and the response register in the data block.
The aggregator takes the form
\begin{equation}\label{eq:aggregator}
    U_{w^*} = \Pi_{W \geq w^*} \otimes Z + (I-\Pi_{W \geq w^*}) \otimes I,
\end{equation}
where $\Pi_{W \geq w^*}$ projects onto the subspace where the \textit{response count} $W$, the number of responding query blocks, is at least $w^*$, and $Z$ acts on the response register in the data block.
$U_{w^*}$ is diagonal in the computational basis on the index register of each query block, a property we will use in the second round of the argument.
We will show that the post-query population concentrates on at least $w^*$ responding blocks when $f(x)=1$, and below $w^*$ when $f(x)=0$.
Hence, the aggregator implements an ideal oracle $\mathcal{O}_f$ on the data block without entangling it with the query blocks, up to an error of order $\sim e^{-\Theta(\eta L)}$ from the residual population on the wrong side of the threshold.
The precise construction of $\Pi_{W \geq w^*}$ and the error analysis are given in Sec.~\ref{subsec:response_aggregation}.

In the uncomputation step, we query $\mathcal{O}_f$ once more on each of the $L$ query blocks.
On each query block, the two oracle queries then act with nothing in between, since the aggregator, up to the aggregation error above, acts trivially there.
Because $O_f^2 = I$ for Boolean oracles, the two queries cancel, and the global state returns to the logical subspace, except that the response register in the data block now records the oracle's response.

Finally, in the recovery step, we apply a recovery channel that inverts the encoding, mapping each logical state $\ket{\bar{x}}\ket{\bar{y}}$ back to $\ket{x}\ket{y}$.
The EOC, Eq.~\eqref{eq:error_orthogonality}, together with the orthogonality of the data block, ensure that the logical subspace satisfies the Knill--Laflamme conditions~\cite{Knill1997} for a non-degenerate code with respect to the $Z$ errors whose weight on the index register of each query block is at most $r$ (Lemma~\ref{lem:global_KLC_OD} and the discussion following it).
The Knill--Laflamme conditions guarantee a channel that inverts the encoding even when a weight-$r$ phase-noise channel acts on the index register of each query block, and we fix one such channel as the recovery channel.
In the first round of the argument the queries are ideal, so the recovery channel only inverts the encoding, and its correction of phase noise enters in the second round.
The recovery channel is constructed from the logical subspace and the error weight $r$ alone, so it is independent of the label $f$.
The pre-recovery state deviates from the logical subspace only by the aggregation error, so the recovery is perfect up to this error.
The five steps together thus act on the input $\ket{x}\ket{y}$ as the ideal oracle, up to the aggregation error $\sim e^{-\Theta(\eta L)}$, which completes the first round of the argument.

We can now derive the query complexity of distilling $\epsilon$-approximate oracles.
The protocol queries the oracle twice on each query block, once in the weak query step and once in the uncomputation step, so $T_{\mathrm{OD}} = 2L$.
The only error of the protocol is the aggregation error $\sim e^{-\Theta(\eta L)}$, and demanding that it be at most $\epsilon$ yields the query complexity Eq.~\eqref{eq:T_OD_adv} of Theorem~\ref{thm:boolean_OD_adv}.

It remains to run the second round of the argument, where we return to the actual circuit and each query is to the noisy oracle $\tilde{\mathcal{O}}_f^{\mathrm{adv},Z} = \mathcal{E}_f^Z \circ \mathcal{O}_f$.
The Kraus operators of $\mathcal{E}_f^Z$ are $Z$-strings on the index register, while the ideal oracle $O_f$ and the aggregator are both diagonal in the computational basis on the index register of each query block, so $\mathcal{E}_f^Z$ commutes exactly with both.
In each query block, we can therefore push the phase-noise channel introduced by the weak query past the aggregator and the uncomputation query, where it merges with the phase-noise channel of the uncomputation query into $(\mathcal{E}_f^Z)^2$, a weight-$2w_P$ phase-noise channel acting just before the recovery step.
After this commutation, the circuit from encoding through uncomputation is exactly the ideal-oracle circuit analyzed in the first round.
Since we required $r \geq 2w_P$ for the query states, the merged channel is also a weight-$r$ phase-noise channel, and the recovery step removes it deterministically, up to the aggregation error.
Hence the protocol distills with precision $\epsilon$ using the same query complexity as in Eq.~\eqref{eq:T_OD_adv} even when the raw oracles are noisy.
The formal error bound and the resulting query complexity are derived in Appendix~\ref{SM_sec:error_bound_OD}.

The same protocol also distills i.i.d.-depolarizing Boolean oracles, where the index errors are no longer bounded in weight and the response register is noisy as well; Sec.~\ref{subsec:iid_noise} handles both by a specific choice of the recovery channel.
Section~\ref{subsec:gate_complexity} gives the gate and ancilla complexity of the protocol.

\subsection{Query state constructions}\label{subsec:query_state_constructions}
\begin{figure}[t]
    \centering
    \includegraphics[width=\linewidth]{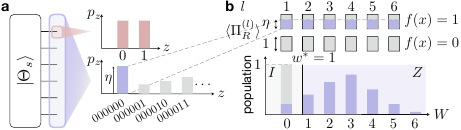}
    \caption{\textbf{a} The seed state $\ket{\Theta_s}$, which generates all query states via Eq.~\eqref{eq:query-state-from-seedstate}, shown here for $n=6$ and $r=1$.
    The seed state must generate query states that are indistinguishable to local $Z$ noise, so the probability distribution $p(z) := |\braket{z|\Theta_s}|^2$ has uniform marginals on every subset of at most $2r$ bits (top).
    It must also retain a large matched query power $\eta = p(0^n)$, which sets the strength of the response each query extracts (bottom).
    \textbf{b} Illustration of response aggregation.
    When $f(x)=1$, every query block responds with population $\eta$, so the population on the eigenspaces of the response count $W$, the number of responding query blocks, follows a binomial distribution peaked at $\eta L$.
    When $f(x)=0$, no block responds and the population concentrates at $W=0$.
    The aggregator thresholds $W$ at $w^* = 1$ and approximately produces the ideal response in the data block.}
    \label{fig:weakquery_aggregation}
\end{figure}
The query states are the only part of the protocol left to design.
Their choice determines the error correction capability of the protocol through the maximum weight $r$ of correctable $Z$ errors in the EOC Eq.~\eqref{eq:error_orthogonality}, and the query complexity through the matched query power $\eta$ via Eq.~\eqref{eq:T_OD_adv}.
We therefore seek query states that satisfy the EOC for a given $r$ while retaining a large matched query power.

We give four constructions.
Construction 1 is a pedagogical example with $r=1$.
Construction 2 gives a systematic construction for any $r$ from constant to linear in $n$, with closed-form expressions for all finite $n$.
Construction 3 handles $r = \lfloor \alpha n \rfloor$ for sufficiently large $n$ and has near-optimal matched query power as $\alpha \to 0$.
The explicit exponents of these two constructions, $H(2\alpha)$ for Construction 2 and $H(\alpha) + 2\alpha$ for Construction 3, are what we use to compute the noise threshold under i.i.d.\ noise in Sec.~\ref{subsec:iid_noise}.
Construction 4 formulates the construction problem as a linear program, which can be solved for any finite $n$ and $r$.
There is an analytic lower bound on the resulting $\eta$, but the constants are not explicit.

In all four constructions, we reduce the design of all the query states to that of a single \textit{seed state} $\ket{\Theta_s}$, which generates the query states by
\begin{equation}\label{eq:query-state-from-seedstate}
    \ket{\Theta(x)} := X^x \ket{\Theta_s}.
\end{equation}
Under this generation rule, the matched query power is uniform among all inputs,
$\eta_x = \left|\braket{x | \Theta(x)}\right|^2 = \left|\braket{0^n | \Theta_s}\right|^2 = \eta$ for every $x$, since $X^x$ maps $\ket{0^n}$ to $\ket{x}$.
Moreover, if the seed state satisfies the EOC, then so does every query state, since conjugation by $X_i$ sends $Z_i \mapsto -Z_i$ and zero expectation values remain zero.

Finding a seed state that satisfies the EOC while retaining a large matched query power reduces to designing a classical probability distribution (Fig.~\ref{fig:weakquery_aggregation}a).
Writing the seed state in the computational basis as $\ket{\Theta_s} = \sum_{z} c_{z} \ket{z}$, the squared amplitudes $p(z) := |c_{z}|^2$ form a probability distribution over bitstrings with $\eta = p(0^n)$.
Since $Z^a Z^b = Z^{a \oplus b}$, the EOC state that the expectation values $\bra{\Theta_s} Z^{a \oplus b} \ket{\Theta_s}$ vanish for all $a \neq b$.
The operator $Z^{a \oplus b}$ is diagonal in the computational basis, with eigenvalue $(-1)^{(a \oplus b) \cdot z}$ on $\ket{z}$, so its expectation value $\sum_z (-1)^{(a \oplus b) \cdot z}\, p(z)$ depends only on the distribution $p$, leaving the phases of the amplitudes free.
Since $a \oplus b$ ranges over all nonzero strings of weight at most $2r$, the EOC hold if and only if $p$ has uniform marginals on every subset of at most $2r$ bits (Lemma~\ref{lem:uniform_marginals}).
Constructing query states therefore amounts to finding a distribution that looks exactly uniform on every subset of $2r$ bits while placing as large a weight as possible on the single bitstring $0^n$.
This is the classical problem of maximizing a single pointwise probability over $2r$-wise independent distributions on $\{0,1\}^n$, studied in discrete probability and combinatorics~\cite{Benjamini2012, Peled2011, Berend2026}.

This classical problem further reduces to distributions over Hamming weights.
Averaging $p$ over all permutations of the $n$ bits preserves the uniform-marginal constraints and leaves $\eta = p(0^n)$ unchanged, since $0^n$ is a fixed point of every permutation, so we may take $p$ to be permutation-symmetric.
A permutation-symmetric distribution is determined by its weight distribution $p_w := \Pr[|z| = w]$, and the uniform-marginal conditions become the Krawtchouk moment conditions (Lemma~\ref{lem:krawtchouk_moments}),
\begin{equation}\label{eq:krawtchouk_moments}
    \sum_{w=0}^n p_w\, K_k(w) = \delta_{k,0}, \quad k = 0, 1, \ldots, 2r,
\end{equation}
where $K_k(w)$ is the binary Krawtchouk polynomial~\cite{Levenshtein1995, Coleman2011}, a degree-$k$ polynomial in $w$ whose explicit form we only need in the appendices.
All four seed states below take this permutation-symmetric form,
\begin{equation}\label{eq:seed_state_dicke}
    \ket{\Theta_s} = \sum_{w=0}^n \sqrt{p_w}\, \ket{D_w^n},
\end{equation}
where $\ket{D_w^n} = \binom{n}{w}^{-1/2} \sum_{|z| = w} \ket{z}$ are the Dicke states, and the matched query power is $\eta = p_0$.

We judge each construction by how close its matched query power comes to the following upper bound.
Any seed state satisfying the EOC obeys
\begin{equation}\label{eq:eta_upper_bound}
    \eta \leq M_r^{-1} =: \eta^*
\end{equation}
(Corollary~\ref{cor:seed_qp_bound}), where
\begin{equation}\label{eq:def_Mr}
    M_r := \sum_{j=0}^{r}\binom{n}{j}
\end{equation}
is the number of correctable $Z$ errors of one query block.
In the classical formulation, this is the known upper bound on the weight that a $2r$-wise independent distribution can place on a single bitstring~\cite{Benjamini2012}.

\begin{table}[t!]
\centering
\footnotesize
\setlength{\tabcolsep}{2.5pt}
\begin{tabular*}{\columnwidth}{@{\extracolsep{\fill}}lcccc}
\hline\hline
 & $r$ & $\eta$ & Optimal? & Explicit? \\
\hline
C1 & $1$ & $\Theta(M_1^{-1})$ & yes & yes \\
\hline
C2 & $\leq n/2$ & $M_{2r}^{-1}$ & no & yes \\
\hline
C3 & $\lfloor \alpha n \rfloor$ & $2^{o(n)-2r}\,M_r^{-1}$ & \begin{tabular}[c]{@{}c@{}}near-optimal\\as $\alpha\to 0$\end{tabular} & for large $n$ \\[4pt]
\hline
 & $O(1)$ & $\Theta(M_r^{-1})$ & yes & \\
\cline{2-4}
C4 & $o(n)$ & $\geq M_r^{-(1+o(1))}$ & \begin{tabular}[c]{@{}c@{}}yes,\\up to $M_r^{o(1)}$\end{tabular} & no \\[4pt]
\cline{2-4}
 & $\lfloor \alpha n \rfloor$ & $\geq e^{-O(r/V(1/\alpha))}M_r^{-1}$ & \begin{tabular}[c]{@{}c@{}}near-optimal\\as $\alpha\to 0$\end{tabular} & \\[4pt]
\hline\hline
\end{tabular*}
\caption{Summary of the four query state constructions.
The column $\eta$ gives the matched query power achieved by each construction, expressed through $M_r$ defined in Eq.~\eqref{eq:def_Mr}.
The entries for C2 and C3 are exact values, with the subexponential factor of C3 collected into $2^{o(n)}$, the $\Theta$ entries are two-sided asymptotic statements, and the entries marked $\geq$ are lower bounds.
The column Optimal? compares $\eta$ with the upper bound $\eta^* = M_r^{-1}$ of Eq.~\eqref{eq:eta_upper_bound}.
The column Explicit? records whether the seed state is given in closed form.
C4 computes the optimal $\eta$ by a linear program, and its three rows give the guarantee on the optimum in three regimes of $r$, where $V(x) = \exp(\sqrt{\log(x)\log\log(x)})$.}
\label{tab:constructions}
\end{table}

\emph{Construction 1 ($r=1$).}
This construction superposes $\ket{0^n}$ with a single Dicke state, and for odd $n$ it attains the upper bound $\eta^*$ exactly.
For $r = 1$, the EOC require the expectation value of every $Z$ string of weight one or two to vanish.
Since $Z$ strings are diagonal, they have no matrix elements between $\ket{0^n}$ and the Dicke state, so each expectation value is a weighted average of its values on the two states.
On $\ket{0^n}$ every $Z$ string has expectation value $+1$, while on a Dicke state the expectation value depends only on the weight of the string.
A single Dicke state must cancel the $+1$ for string weights one and two simultaneously, so its two expectation values must be equal and negative, which pins the Dicke weight to $w = \frac{n+1}{2}$, where both equal $-\frac{1}{n}$ (Appendix~\ref{SM_sec:construction1}).
Cancelling $+1$ against $-\frac{1}{n}$ requires $n$ times more mass on the Dicke state than on $\ket{0^n}$, so for odd $n \ge 3$ the seed state is
\begin{equation}\label{eq:Theta-0-odd}
    \ket{\Theta_s} = \frac{1}{\sqrt{n+1}} \ket{0^n} + \sqrt{\frac{n}{n+1}} \ket{D_{\frac{n+1}{2}}^n}.
\end{equation}
The matched query power is
\begin{equation}
    \eta = \frac{1}{n+1} = \eta^*.
\end{equation}
For even $n$, $\frac{n+1}{2}$ is not an integer, two Dicke states must be added, and $\eta = \Theta(n^{-1})$.
The even-$n$ state and the verification of both parities through the moment conditions are given in Appendix~\ref{SM_sec:construction1}.

\emph{Construction 2 (any $r \leq n/2$).}
This construction starts from $\ket{0^n}$ and expands its density matrix over Pauli-$Z$ strings rather than expanding the state over Dicke states,
\begin{equation}
    \ket{0}\!\bra{0}^{\otimes n} = \prod_{i=1}^n \left(\frac{I+Z_i}{2}\right) = \frac{1}{2^n}\sum_{a\in\{0,1\}^n} Z^a.
\end{equation}
The EOC demand vanishing expectation values for all $Z$ strings of weight between $1$ and $2r$, and the expectation value of each $Z$ string is $2^n$ times its coefficient in this expansion.
Removing these $Z$ strings from the density matrix $\ket{0}\!\bra{0}^{\otimes n}$ and mixing the resulting operator with the maximally mixed state to restore positivity gives a valid density matrix
\begin{equation}\label{eq:encoding2}
    \rho = \frac{1}{2^n}\left( I + \frac{1}{M_{2r}} \sum_{|a| > 2r} Z^a \right),
\end{equation}
where the coefficient $M_{2r}^{-1}$ is chosen so that every diagonal entry of $\rho$ is nonnegative.
As a linear combination of Pauli-$Z$ strings, $\rho$ is diagonal in the computational basis, and both the matched query power and the EOC depend only on its diagonal entries.
We therefore take the seed state to be the pure state with the same diagonal,
\begin{equation}
    \ket{\Theta_s} = \sum_{z \in \{0,1\}^n} \sqrt{\bra{z} \rho \ket{z}}\, \ket{z},
\end{equation}
which satisfies $\bra{\Theta_s} Z^a \ket{\Theta_s} = \Tr[Z^a \rho]$ for every $a$.
This seed state satisfies the EOC for every $r \leq n/2$, with matched query power exactly
\begin{equation}
    \eta = M_{2r}^{-1}.
\end{equation}
The verification is given in Appendix~\ref{SM_sec:construction2}.

\emph{Construction 3 ($r = \lfloor \alpha n \rfloor$).}
The full construction is involved and is deferred to Appendix~\ref{SM_sec:construction3}, so here we sketch the idea and state the result.
The starting point is $\ket{+}^{\otimes n}$, whose weight distribution $\mathrm{Bin}(n,1/2)$ satisfies the moment conditions Eq.~\eqref{eq:krawtchouk_moments} for every $r$.
However, its mass at $w = 0$, which is the matched query power, is only $2^{-n}$.
The construction raises this mass to $\eta$, and modifies the remaining weights in two steps in order to restore the moment conditions.
The first step truncates the remaining mass to a window of width $O(n)$ around $w = n/2$, so that the correction of the second step can stay small.
The second step multiplies the weights on the window by a polynomial of degree $2r$ in $w$, whose $2r+1$ coefficients are matched to the $2r+1$ moment conditions.
We show that this polynomial exists and stays uniformly small on the window, so the corrected weights remain nonnegative and form a valid distribution.
For every constant $0 < \alpha \leq 0.16$ and all sufficiently large $n$, the resulting matched query power is
\begin{equation}\label{eq:construction3_eta}
    \eta = \xi(n)\,2^{-n(H(\alpha)+2\alpha)},
\end{equation}
where $\xi(n)$ is any function with $\xi(n) \geq 1$ and $\log\xi(n) = o(n)$.
For $r = \lfloor \alpha n \rfloor$, Lemma~\ref{lem:Mr_asymptotics} gives $M_r = \Theta\!\left(2^{nH(\alpha)}/\sqrt{n}\right)$, so $\eta^{-1} = M_r^{1 + 2\alpha/H(\alpha) + o(1)}$.
The exponent exceeds the optimal value $1$ by $2\alpha/H(\alpha)$, which vanishes as $\alpha \to 0$.

\begin{figure}[t]
    \centering
    \includegraphics[width=\linewidth]{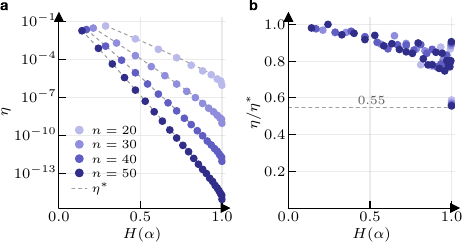}
    \caption{Matched query power of Construction 4, obtained by solving the LP \eqref{eq:construction4_LP} in exact rational arithmetic for $n = 20, 30, 40, 50$ and $1 \leq r \leq n/2$, plotted against $H(\alpha)$ with $\alpha = r/n$.
    (a) The LP optimum $\eta$ (dots), together with the optimal value $\eta^* = M_r^{-1}$ (dashed, one curve per $n$).
    (b) The ratio $\eta/\eta^*$, which stays above $0.55$ for every $(n, r)$ computed.}
    \label{fig:construction4_LP_numerics}
\end{figure}

\emph{Construction 4 (any $r\leq n/2$, numerical).}
This construction solves the moment conditions Eq.~\eqref{eq:krawtchouk_moments} directly as a linear program (LP):
\begin{equation}\label{eq:construction4_LP}
    \max\; p_0 \quad \text{s.t.} \quad
    \text{Eq.~\eqref{eq:krawtchouk_moments}},\;\;
    p_w \geq 0.
\end{equation}
The LP optimum is the matched query power $\eta$ of the resulting seed state, and it comes with an analytic guarantee.
Theorem 1.1 of Ref.~\cite{Peled2011} bounds how far the optimum can fall below the upper bound $\eta^*$.
The three rows for Construction 4 in Table~\ref{tab:constructions} record the resulting guarantee in three regimes of $r$, and Appendix~\ref{SM_sec:construction4} derives them.
The constants in the guarantee are implicit, however, so the noise threshold in Sec.~\ref{subsec:iid_noise} is computed from the explicit exponents of Constructions 2 and 3 instead.
We solve the LP numerically for $n = 20, 30, 40, 50$ and every $1 \leq r \leq n/2$, with the results shown in Fig.~\ref{fig:construction4_LP_numerics} as functions of $\alpha = r/n$.
For every $(n, r)$ computed, the ratio $\eta/\eta^*$ stays at least $0.55$, so in practice the LP saturates $\eta^*$ up to a constant factor, far better than the analytic guarantee indicates.

The four constructions are summarized in Table~\ref{tab:constructions}.

\subsection{Response aggregation}\label{subsec:response_aggregation}
Here we complete the construction of the aggregator in Eq.~\eqref{eq:aggregator} by defining the threshold projector $\Pi_{W \geq w^*}$, and then analyze the aggregation error.

The threshold projector is built from the response subspace projectors $\Pi^{(i)}_R$ of Eq.~\eqref{eq:response_projector}.
The $\Pi^{(i)}_R$ with different $i$ commute with each other, since their only overlap is on the data block and they are all diagonal there.
We define the response count $W := \sum_{i=1}^L \Pi_R^{(i)}$.
Since the projectors commute, they can be diagonalized simultaneously, so the eigenvalues of $W$ take values only in $\{0,1,\dots, L\}$.
We define $\Pi_{W \geq w^*}$ to be the spectral projector of $W$ associated with eigenvalues at least $w^*$.
For the adversarial noise model considered here, it suffices to set $w^* = 1$; the i.i.d.\ noise analysis in Sec.~\ref{subsec:iid_noise} requires a larger threshold.
Substituting this $\Pi_{W \geq w^*}$ into Eq.~\eqref{eq:aggregator} completes the definition of the aggregator.

Now we analyze the aggregation error.
We will show that the aggregator writes the response $Z^{f(x)}$ into the data block while leaving the query blocks intact, up to an error of size $e^{-\Theta(\eta L)}$.
Following the first round of the argument in the overview, we take the protocol input to be a basis state $\ket{x}\ket{y}$ and let the protocol query the ideal oracle $\mathcal{O}_f$.
Superposition inputs are handled in Appendix~\ref{SM_sec:stage2_OD}, and the noisy-oracle case reduces to the ideal one by the commutation argument in the second round.

The aggregator treats the threshold event $W \geq w^*$ as a detected response, and the error comes entirely from the population on which this detection is wrong, as the following decomposition shows.
Decompose the state before aggregation into the component with $W \geq w^*$ and the component with $W < w^*$.
On the former the aggregator applies $Z$ to the data block, and on the latter it acts as the identity.
The component receiving the wrong action is thus a false negative, the component with $W < w^*$ when $f(x)=1$, or a false positive, the component with $W \geq w^*$ when $f(x)=0$.
Hence it suffices to bound the false-negative and false-positive populations.

We now compute this population.
After the weak query, the state in each query block is $O_f \ket{\Theta(x)} \ket{+}$.
Since the index register of the data block holds the basis state $\ket{x}$, each $\Pi_R^{(i)}$ reduces on query block $i$ to the rank-one projector $\ket{x}\bra{x} \otimes \ket{-}\bra{-}$.
The per-block population in the response subspace is therefore
\begin{equation}
    \left|\bra{x}\bra{-}O_f \ket{\Theta(x)}\ket{+}\right|^2 =
    \begin{cases}
        \eta_x, & f(x)=1,\\
        0, & f(x)=0.
    \end{cases}
\end{equation}
Since the query blocks are in a product state and the reduced projectors act on separate blocks, the blocks contribute to $W$ independently.
The population on the eigenspaces of $W$ thus follows the binomial distribution $\mathrm{Bin}(L, \eta_x)$ when $f(x)=1$, and sits entirely on $W=0$ when $f(x)=0$ (Fig.~\ref{fig:weakquery_aggregation}b).
With $w^*=1$, the false-positive population is exactly zero, since the population sits entirely on $W=0$ when $f(x)=0$.
The false-negative population is the weight of $\mathrm{Bin}(L, \eta_x)$ at $W=0$, which is $(1-\eta_x)^L \leq e^{-\eta_x L} \leq e^{-\eta L}$.
The aggregation error, defined in full in Appendix~\ref{SM_sec:response_aggregation_OD}, is controlled by the false-negative population and is of the same order, $e^{-\Theta(\eta L)}$.

\subsection{Distillation of i.i.d.-depolarizing Boolean oracles}\label{subsec:iid_noise}
We now show that the same two-stage protocol distills the family of i.i.d.-depolarizing Boolean oracles $\tilde{\mathfrak{O}}_F^{\mathrm{iid}}$ into the family of ideal Boolean oracles $\mathfrak{O}_F$.
For every per-qubit error rate $p$ below a constant, which we call the \textit{distillation threshold}, the protocol with query states chosen to correct $r = \lfloor \alpha n \rfloor$ errors and with the number of query blocks set from $p$ distills with precision $\epsilon$.
The threshold depends only on the overhead exponent $\gamma$ and the decay rate $\nu := \limsup_{n\to\infty} \frac{1}{n}\log_2 \frac{1}{\epsilon}$ of the precision, which must be finite and is at most one in every application with constant success probability (Appendix~\ref{SM_sec:useful_precision}).

\begin{formalrestatable}[b]{main}{Distillation of Boolean oracles under i.i.d.\ depolarizing noise}{theorem}{iidthresholdexponent}\label{thm:iid_threshold_at_exponent}
    For any constant overhead exponent $\gamma \in (0, 1)$, any precision $\epsilon = \epsilon(n) \in (0,1)$ with finite decay rate $\nu := \limsup_{n\to\infty} \frac{1}{n} \log_2 \frac{1}{\epsilon}$, any number $m$ of response qubits with $\log_2 m = o(n)$, and any constant per-qubit error rate $p < p_{\mathrm{th}}\left( \gamma, \nu \right)$, the two-stage protocol, with its parameters set from $p$, distills the family of i.i.d.-depolarizing Boolean oracles $\tilde{\mathfrak{O}}_F^{\mathrm{iid}}$ with this error rate into the family of ideal Boolean oracles $\mathfrak{O}_F$ with precision $\epsilon$ for all sufficiently large $n$, with query complexity at most
    \begin{equation}\label{eq:T_OD_at_exponent}
        T_{\mathrm{OD}} = 2 \left\lceil \frac{48 \left( 9 - 4 p \right)}{\left( 3 - 4 p \right)^2} \, N^{\gamma} \, \ln \frac{12 m}{\epsilon} \right\rceil .
    \end{equation}
    The threshold is positive for every such $\gamma$ and $\nu$ (Lemma~\ref{lem:threshold_equation}) and equals
    \begin{equation}\label{eq:iid_pth_at_exponent}
    \begin{split}
        p_{\mathrm{th}}\left( \gamma, \nu \right) := \max\Big\{ &p_{\mathrm{th}}^{(3)}\big( \alpha^{(3)},\, \tfrac{1}{2} - \alpha^{(3)},\, \nu \big) , \\
        &p_{\mathrm{th}}^{(2)}\big( \alpha^{(2)},\, 1,\, \nu \big) \Big\} ,
    \end{split}
    \end{equation}
    where $p_{\mathrm{th}}^{(3)}$ and $p_{\mathrm{th}}^{(2)}$ are the thresholds Eq.~\eqref{eq:iid_pth_C3} and Eq.~\eqref{eq:iid_pth_C2} of Corollaries~\ref{cor:iid_boolean_query_complexity} and~\ref{cor:iid_boolean_query_complexity_C2}, $\alpha^{(3)}$ is the largest $\alpha \in (0, 0.16]$ with $H(\alpha) + 2 \alpha \leq \gamma$, and $\alpha^{(2)}$ is the largest $\alpha \in (0, 1/4)$ with $H(2\alpha) \leq \gamma$.
\end{formalrestatable}
\noindent The theorem is proved in Appendix~\ref{SM_sec:iid}.
As stated, the theorem fixes the scaling $N^{\gamma}$ of the query complexity and gives the largest error rate $p$ that can be tolerated at that scaling.
Conversely, for a given $p$ and $\epsilon$, one can choose the smallest $\gamma$ with $p < p_{\mathrm{th}}(\gamma, \nu)$, so that the distillation works while the query complexity scaling $N^{\gamma}$ is minimized.

The protocol and its analysis run as follows.
We first apply the stage 1 gadget to every qubit of the index and response registers, which converts each i.i.d.-depolarizing Boolean oracle with per-qubit error rate $p$ into an i.i.d.-dephasing Boolean oracle with per-qubit error rate $2p/3$ (Corollary~\ref{cor:stage1_depol}).
We then run the stage 2 protocol on the resulting i.i.d.-dephasing Boolean oracles, with the aggregator threshold raised to $w^* = \eta L/2$ as announced in Sec.~\ref{subsec:response_aggregation}.
Two assumptions of the adversarial analysis now fail: the errors on the index register are no longer bounded in weight, so in a query block they can have weight above $r$; and errors now also occur on the response register.
For the index register, we handle this by using the \textit{sequential recovery} (Appendix~\ref{SM_sec:seq_recovery}), which corrects the high-weight errors approximately.
For the response register, we show that the distribution of the weight of error pattern concentrates, and for every typical pattern the aggregation error remains $e^{-\Theta(\eta L)}$.

We first discuss the errors on the index register, with details in Appendix~\ref{SM_sec:iid_seq_recovery}.
In each query block, the oracle in the weak query step and the oracle in the uncomputation step each apply i.i.d.\ dephasing noise to the index register.
Since $Z$ errors commute with the ideal oracles and the aggregator, the two noise channels can be merged into a single effective channel acting before recovery.
Two $Z$ errors on the same qubit cancel, so in the effective channel each index qubit carries a $Z$ error at rate $p_{\mathrm{eff}} = \frac{4p}{3}\left(1-\frac{2p}{3}\right)$.
The error weight in each query block then follows $\mathrm{Bin}(n, p_{\mathrm{eff}})$ and concentrates around its mean $p_{\mathrm{eff}} n$.
EOC ensures every error pattern of weight at most $r$ can be identified and corrected exactly, because the corrupted seed states $Z^{e}\ket{\Theta_s}$ with $|e| \leq r$ are mutually orthogonal.
Hence, whenever $r > p_{\mathrm{eff}} n$, typical error patterns are corrected exactly.

The above argument, which uses only the exact orthogonality of the corrupted seed states with $|e| \leq r$, already yields a constant distillation threshold with query states of $r = \Theta(n)$.
A substantially larger threshold can be obtained by also correcting error patterns with $|e| > r$.
For Constructions 2 and 3, the corresponding corrupted seed states no longer satisfy the EOC exactly, but they remain \textit{approximately orthogonal} to each other, as stated in Lemmas~\ref{lem:construction2_high_weight} and~\ref{lem:construction3_high_weight}.
The sequential recovery (Appendix~\ref{SM_sec:seq_recovery}), inspired by Refs.~\cite{Lloyd2011, Giovannetti2012}, exploits this approximate orthogonality to correct error patterns of weight up to $r_{\mathrm{seq}} = \lfloor \alpha_{\mathrm{seq}} n \rfloor$, where $\alpha_{\mathrm{seq}} \in (\alpha, 1]$ is a tunable constant.
Theorem~\ref{thm:iid_threshold_at_exponent} takes $\alpha_{\mathrm{seq}}$ as large as each construction allows, and adopts whichever of Constructions 2 and 3 gives the larger distillation threshold, which is the maximum in Eq.~\eqref{eq:iid_pth_at_exponent}.

We now explain how the sequential recovery works.
The merged dephasing noise is a mixture over error patterns.
For illustration, suppose that the error patterns on the index registers of the $L$ query blocks are $e_1, \ldots, e_L \in \{0,1\}^n$, and that the data block holds the index $\ket{x}$.
The general case, where the noise is a mixture over such patterns and the input is a superposition over $x$, is treated in Appendix~\ref{SM_sec:iid_seq_recovery}.
As shown in Sec.~\ref{subsec:protocol_overview}, the circuit from encoding through uncomputation acts, up to the aggregation error, as if only an ideal oracle acted on the data block, which applies $Z^{f(x)}$ to the response register of that block and leaves the rest unchanged, so the index registers just before recovery are in the corrupted logical state
\begin{equation}
    \bigotimes_{l=1}^{L} Z^{e_l} \ket{\Theta(x)} \otimes \ket{x},
\end{equation}
where we omit the response registers, on which the recovery does not act.
The recovery first applies to each query block $n$ CNOT gates, controlled on the index register of the data block, which map $\ket{\Theta(x)} = X^{x}\ket{\Theta_s}$ (Eq.~\eqref{eq:query-state-from-seedstate}) back to $\ket{\Theta_s}$ and thereby undo the encoding.
Each $Z$ error on a query block propagates through the CNOT gates to a $Z$ on both the query block and the index register of the data block, so the state becomes
\begin{equation}
    \bigotimes_{l=1}^{L} Z^{e_l} \ket{\Theta_s} \otimes Z^{e_1 \oplus \cdots \oplus e_L} \ket{x}.
\end{equation}
Each query block is now in the corrupted seed state $Z^{e_l}\ket{\Theta_s}$, independent of the data block, while the data block carries the accumulated error $Z^{e_1 \oplus \cdots \oplus e_L}$.
It therefore suffices to identify $e_l$ on each query block and to correct only the data block.
For each query block, the recovery tests the candidate patterns $e$ with $|e| \leq r_{\mathrm{seq}}$ one at a time in order of nondecreasing weight, by a binary projective measurement onto $Z^{e}\ket{\Theta_s}$ and its complement, and stops at the first accepted candidate, which we denote by $\hat{e}_l$.
Finally, it applies $Z^{\hat{e}_1 \oplus \cdots \oplus \hat{e}_L}$ to the index register of the data block and discards the query blocks, which returns the data block to $\ket{x}$ whenever $\hat{e}_l = e_l$ for every $l$.
The recovery error is bounded by a union bound over the $L$ query blocks of the probability that a query block ends with $\hat{e}_l \neq e_l$.
When $|e_l| \leq r$, the EOC make $Z^{e_l}\ket{\Theta_s}$ orthogonal to every candidate tested before $e_l$, so each of these tests rejects without disturbing the state and the test of $e_l$ accepts, giving $\hat{e}_l = e_l$ with certainty.
The failure probability therefore has two sources.
First, when $|e_l| > r_{\mathrm{seq}}$, the pattern is never tested and hence never corrected.
Such patterns occur with probability exponentially small in $n$, by the Chernoff bound for the binomial distribution.
Second, when $r < |e_l| \leq r_{\mathrm{seq}}$, the candidates tested before $e_l$ are only approximately orthogonal to $Z^{e_l}\ket{\Theta_s}$, so each rejection disturbs the state slightly and may cause a later wrong candidate to be accepted.
The resulting misidentification probability is also exponentially small in $n$, guaranteed by the approximate orthogonality.

We then discuss the errors on the response register, with details in Appendix~\ref{SM_sec:iid_aggregation}.
Consider first the errors introduced by the oracles in the weak query step, deferring those from the uncomputation step to the end of this paragraph.
Across the $L$ query blocks, these errors form a $Z$-string $Z^{\tau}$ with $\tau \in \{0,1\}^L$, whose weight $|\tau|$ follows $\mathrm{Bin}(L, 2p/3)$.
The weight thus concentrates around its mean $2pL/3$ with fluctuations $O(\sqrt{pL}) \ll 2pL/3$.
We now show that for every typical $\tau$, the population on the eigenspaces of $W$ still concentrates above the threshold $w^*$ when $f(x) = 1$ and below it when $f(x) = 0$, so the response aggregation error remains small.
On a query block with $\tau_i = 1$, the $Z$ error swaps $\ket{+}$ and $\ket{-}$ on its response qubit, so a responding query block becomes non-responding and vice versa.
Since the query states of Constructions 2 and 3 are generated from seed states, the matched query power is uniform, $\eta_x = \eta$.
For $x$ with $f(x) = 1$, only the query blocks with $\tau_i = 0$ retain population in the response subspace, so the population on the eigenspaces of $W$ follows $\mathrm{Bin}(L - |\tau|, \eta)$.
For $x$ with $f(x) = 0$, the population instead follows $\mathrm{Bin}(|\tau|, \eta)$.
At the mean weight $|\tau| = 2pL/3$, the means of the two distributions, $\eta(1 - 2p/3)L$ and $2p\eta L/3$, lie on opposite sides of the threshold $w^* = \eta L/2$ whenever $p < 3/4$, each at distance $\eta L (1/2 - 2p/3)$.
A typical fluctuation of $|\tau|$ shifts both means by only $O(\eta\sqrt{pL})$, so the separation persists for every typical $\tau$.
By the Chernoff bound, the false-negative and false-positive populations are both $e^{-\Theta(\eta L)}$, so the response aggregation still applies an ideal oracle on the data block up to an error of this order, while leaving the query blocks nearly intact.
After aggregation, the residual $Z^{\tau}$ commutes through the oracles in uncomputation and merges with the response-register errors they introduce.
The sequential recovery channel traces out the response register of each query block, so these errors have no effect on the distilled oracle.

In conclusion, only three error sources contribute to the distillation error, namely the response aggregation error, the recovery error, and the probability of an atypical response error pattern.
The distilled oracles are $\epsilon$-close to the ideal ones once each source contributes $O(\epsilon)$.
Choosing $L = O(\eta^{-1}\ln(m/\epsilon))$ makes the aggregation error $O(\epsilon)$, and for every $p < 3/4$ the probability of an atypical pattern is then $O(\epsilon)$ as well.
This choice of $L$ gives the query complexity Eq.~\eqref{eq:T_OD_at_exponent}.
Hence, the distillation threshold is set by the recovery error alone.
As $p$ decreases, the distribution of the error weight $|e|$ shifts toward lower weights, where the patterns rarely exceed $r_{\mathrm{seq}}$ and the corrupted seed states tested by the recovery are closer to exactly orthogonal, so both sources of the recovery error shrink.
The recovery error is therefore $O(\epsilon)$ for every $p$ below a threshold, and this threshold is the $p_{\mathrm{th}}$ of Eq.~\eqref{eq:iid_pth_at_exponent}.

\subsection{Gate and ancilla complexity}\label{subsec:gate_complexity}
The two-stage protocol admits an exact circuit implementation with the sequential recovery as the recovery channel, and one distilled query uses
\begin{equation}\label{eq:main_gate_complexity}
    G = O\bigl((n^2 M_{r_{\mathrm{seq}}} + m) L\bigr), \qquad A = O\bigl((n+m)L\bigr)
\end{equation}
oracle-independent elementary gates and ancilla qubits (Theorem~\ref{thm:gate_complexity_OD}), where $M_{r_{\mathrm{seq}}}$ is the number of candidate patterns tested by the sequential recovery, given by Eq.~\eqref{eq:def_Mr}.
The dominant term comes from the sequential recovery, which tests each of the $M_{r_{\mathrm{seq}}}$ candidate patterns on each of the $L$ query blocks, at $O(n^2)$ gates per test using the circuits of Refs.~\cite{Bartschi2019, Nielsen_Chuang2010}.
Under adversarial noise, $r_{\mathrm{seq}} = r$ suffices.
Under i.i.d.\ noise, Theorem~\ref{thm:iid_threshold_at_exponent} takes $\alpha_{\mathrm{seq}}$ as large as each construction admits, which gives a larger distillation threshold than $r_{\mathrm{seq}} = r$ at the cost of more gates, while the query complexity is unchanged.
See details in Appendix~\ref{SM_sec:gate_complexity_OD}.

\section{Optimality of Boolean Oracle Distillation}\label{sec:optimality}
In this section we prove two lower bounds, Theorems~\ref{thm:OD_grover_lower_bound} and~\ref{thm:OD_label_indep}, on the query complexity of OD protocols.
Both bounds apply to the protocol of Sec.~\ref{sec:grover_distillation}, and its query complexity matches them up to subleading factors in a wide range of parameter regimes, so the protocol is near-optimal among all protocols that distill the family of ideal Boolean oracles $\mathfrak{O}_F$ into itself and retain the ability to correct $Z$ errors on the index register at every query.
The query overhead of oracle distillation is therefore not an artifact of our construction, but a consequence of the fundamental tension between distilling the oracle and keeping its errors correctable.

We state the lower bounds for protocols that distill a subset of the family of ideal Boolean oracles $\mathfrak{O}_F$ into itself and keep the $Z$ errors on the index register correctable at every query.
The first requirement, functionality, means that the protocol must work when the raw oracle is already ideal.
This rules out contrived protocols that succeed only because a label-dependent noise channel reveals the label.
Every protocol that works for adversarial noise on the index register satisfies this requirement, since the identity channel is a trivial example of weight-$w_P$ phase-noise channel.
The second requirement, an error correction condition on the internal states of the protocol, is therefore the only restrictive assumption.
Stating the requirements this way separates the tension between functionality and error correction from the noise channel of the noisy oracle, so the bounds constrain every protocol that keeps the $Z$ errors correctable, regardless of the noise channel the protocol is built to withstand.

We now make the error correction condition precise.
Consider the run of the protocol in Eq.~\eqref{eq:od_protocol} with the ideal oracle $\mathcal{O}_f$ in place of the raw oracle and the ancilla state purified to $\ket{A}$.
Every operation in this run is unitary, so the composition preceding the $t$-th query, $U_{t-1} O_f U_{t-2} \cdots O_f U_0 \left(\,\cdot\, \otimes \ket{A}\right)$, is an isometry from $\mathcal{H}_O$ into $\mathcal{H}_O \otimes \mathcal{H}_A$.
We define the $t$-th \textit{pre-query subspace} to be the image of this isometry, with projector $P_{t,f}$.
The pre-query subspace depends on the label $f$ in general, since the preceding queries all apply $\mathcal{O}_f$.
In the actual protocol, noise strikes at the queries, so we demand that each $P_{t,f}$ satisfy the Knill--Laflamme conditions (KLC) for all $Z$ errors of weight at most $r$,
\begin{equation}\label{eq:KLC_prequery}
    P_{t,f} Z^a Z^b P_{t,f} = C^{a,b}_{t,f} P_{t,f},\quad a,b\in \{0,1\}^n, |a|,|b|\leq r,
\end{equation}
where the $Z$ strings act on the index register of $\mathcal{H}_O$.
The numbers $C^{a,b}_{t,f}$ form a matrix $C_{t,f}$, which we call the \textit{KL matrix}.
The KLC mean that the $Z$ errors do not destroy the logical information in $P_{t,f}$ irreversibly, but do not imply active error correction to be performed at each step.
Together with functionality, this condition defines the class of protocols, parametrized by the weight $r$.

The two lower bounds below rest on two shared ingredients, which we outline here and prove in Appendix~\ref{SM_sec:od_optimality}.
The first ingredient is a bound on query power.
We call the population that a state places on a bitstring $\ket{z}$ of the index register its query power on $z$, generalizing the matched query power of Sec.~\ref{sec:grover_distillation}, which is the query power of the query state $\ket{\Theta(x)}$ on its own input $x$.
The KLC Eq.~\eqref{eq:KLC_prequery} limits the query power that any state in the pre-query subspace places on each bitstring, and the limit is at most $1/M_r$ on average over the $N$ bitstrings, where $M_r$, defined in Eq.~\eqref{eq:def_Mr}, is the number of $Z$ strings appearing in the KLC.
The second ingredient is a pair test.
Run the same protocol with two different oracles $\mathcal{O}_f$ and $\mathcal{O}_g$ from the family, on a common input state $\ket{z}\ket{+}^{\otimes m}$ with an index bitstring $z$ on which the two oracles disagree, $f(z) \neq g(z)$.
The two ideal oracles map this input to the orthogonal outputs $\ket{z}\, Z^{f(z)}\ket{+}^{\otimes m}$ and $\ket{z}\, Z^{g(z)}\ket{+}^{\otimes m}$, so distilling both into themselves with precision $\epsilon$ forces the final states of the two runs to be nearly orthogonal.
The unitaries between the queries are identical in the two runs, so the overlap of the two runs changes only at the queries, and each query changes the overlap only through the query power that the runs place on the bitstrings where the two oracles disagree.
For a single pair the disagreeing bitstrings may carry query power well above the average $1/M_r$, so we average the pair test over the pairs in the family, which lets the disagreeing bitstrings range over all bitstrings, and only the average query power enters the final bound.
Each query then changes the averaged overlap only slightly, so driving it from one to near zero requires many queries, in numbers quantified by the two theorems below.
The averaged pair test follows the spirit of the adversary method \cite{Ambainis2002, Hoyer2007}, with the difference explained in Appendix~\ref{SM_sec:od_optimality}.

Below, we present the two theorems.
They trade the generality of the KL matrices against the generality of the oracle family.
Theorem~\ref{thm:OD_grover_lower_bound} allows the KL matrices $C_{t,f}$ to depend arbitrarily on both the query round and the oracle label, and applies as long as the protocol can distill the family of ideal Grover oracles, defined below.
Theorem~\ref{thm:OD_label_indep} assumes KL matrices independent of the oracle label, $C_{t,f} = C_t$, and in exchange applies whenever the protocol distills any subset of the family of ideal Boolean oracles $\mathfrak{O}_F$ containing at least two oracles, even if it cannot distill the Grover family.

To state the first bound, we define the \textit{family of ideal Grover oracles}, $\mathfrak{O}_{F_G} := \{\mathcal{O}_f\}_{f \in F_G}$, where $F_G \subset F$ is the set of Boolean functions $f: \{0,1\}^n \rightarrow \{0,1\}$ for which exactly one bitstring $x$ satisfies $f(x) = 1$, called the marked element of $f$.
\begin{formalrestatable}[a]{opt}{Query complexity lower bound for distilling Grover oracles}{theorem}{odgroverlb}\label{thm:OD_grover_lower_bound}
    Any OD protocol that distills $\mathfrak{O}_{F_G}$ into itself with precision $\epsilon \leq 1/2$, and whose pre-query subspaces satisfy the KLC~\eqref{eq:KLC_prequery} for all $Z$ errors of weight at most $r$, has query complexity
    \begin{equation}\label{eq:OD_grover_lower_bound}
        T_{\mathrm{OD}} = \Omega\left(\min\left\{M_r,\ (M_r N)^{1/4}\right\}\right).
    \end{equation}
\end{formalrestatable}
\noindent Although each Grover oracle has a single response qubit, the bound also applies to protocols whose oracles have $m > 1$ response qubits, because a Grover oracle can be viewed as an oracle with $m$ response qubits that acts trivially on the last $m - 1$ of them.

The second bound carries a free parameter, the distribution that averages the pair tests.
The bound holds for every choice of the distribution, so for a given family one can pick the choice that makes it strongest.
Write $\mathfrak{O}_{F_0} = \{\mathcal{O}_f\}_{f\in F_0}$ with $F_0 \subseteq F$ for the family the protocol distills.
Let $\mu_{(f,g)} \geq 0$ be weights on the ordered pairs $(f,g)$ of distinct functions $f, g \in F_0$, normalized so that $\sum_{(f,g)} \mu_{(f,g)} = 1$, and for each $z \in \{0,1\}^n$ let
\begin{equation}\label{eq:lz_def}
    l_z := \sum_{(f,g)} \mu_{(f,g)} \mathbf{1}_{f(z) \neq g(z)} = \Pr_{(f,g) \sim \mu }[f(z) \neq g(z)]
\end{equation}
be the probability that a pair drawn from $\mu$ disagrees on the input $z$.

\begin{formalrestatable}[b]{opt}{Query complexity lower bound for distilling arbitrary families of Boolean oracles}{theorem}{odlabelindep}\label{thm:OD_label_indep}
    Any OD protocol that distills $\mathfrak{O}_{F_0}$ with $|F_0| \geq 2$ into itself with precision $\epsilon \leq 1/2$, and whose pre-query subspaces satisfy the KLC~\eqref{eq:KLC_prequery} for all $Z$ errors of weight at most $r$ with a label-independent KL matrix $C_t$ for each query round $t$, has query complexity
    \begin{equation}\label{eq:OD_label_indep_bound}
        T_{\mathrm{OD}} = \Omega\left(\frac{M_r}{N \max_z l_z}\right)
    \end{equation}
    for every choice of the weights $\mu$, with $l_z$ defined in Eq.~\eqref{eq:lz_def}.
\end{formalrestatable}
\noindent As a concrete instance, take $F_0 = F_G$ and let $\mu$ be uniform over the ordered pairs.
Two Grover oracles with marked elements $x \neq y$ disagree exactly on the two inputs $x$ and $y$, so under the uniform $\mu$ every input has $l_z = 2/N$, and Eq.~\eqref{eq:OD_label_indep_bound} gives $T_{\mathrm{OD}} = \Omega(M_r)$.
The bound is not tied to the Grover family.
Appendix~\ref{SM_sec:label_indep_lb} chooses the weights $\mu$ for two further families, the fixed-period Simon oracles and the balanced oracles, and obtains $T_{\mathrm{OD}} = \Omega(M_r)$ for both.

We now compare the lower bounds with our protocol of Sec.~\ref{sec:grover_distillation}, and first check that both theorems apply to it.
The protocol distills the family of all ideal Boolean oracles $\mathfrak{O}_F$ into itself, and since $F_G \subset F$, it in particular distills $\mathfrak{O}_{F_G}$ into itself.
In the protocol, $Z$ errors on the index qubits commute with each ideal oracle and with the aggregator, since both are diagonal in the computational basis on the index register.
Every pre-query subspace therefore inherits the KL matrix of the logical subspace, so it satisfies the KLC~\eqref{eq:KLC_prequery} with a KL matrix independent of both the query round and the oracle label.
Both theorems thus apply to the protocol.

On the lower bound side, Theorem~\ref{thm:OD_label_indep} applied to $\mathfrak{O}_{F_G}$ gives $T_{\mathrm{OD}} = \Omega(M_r)$ at every $r$, as computed below the theorem, for protocols whose KL matrices are independent of the oracle label.
Theorem~\ref{thm:OD_grover_lower_bound} extends the bound $\Omega(M_r)$ to arbitrary KL matrices when $M_r \lesssim N^{1/3}$, which covers all $r = o(n)$ and, since $M_r = \tilde{\Theta}(N^{H(\alpha)})$ for $r = \lfloor \alpha n \rfloor$, all $\alpha$ with $H(\alpha) \leq 1/3$, that is $\alpha \leq 0.061$.
For larger $\alpha$ it gives only the weaker bound $T_{\mathrm{OD}} = \tilde{\Omega}(N^{(1 + H(\alpha))/4})$.

On the upper bound side, the protocol has query complexity $T_{\mathrm{OD}} = O(\eta^{-1} \ln \epsilon^{-1})$ by Eq.~\eqref{eq:T_OD_adv}.
Its gap to the lower bound $\Omega(M_r)$ is therefore the product of two factors, the ratio $\eta^{-1}/M_r$ set by the query state construction and recorded in Table~\ref{tab:constructions}, and the factor $\ln \epsilon^{-1}$ set by the target precision.
The second factor is at most $O(n)$ in every application with constant success probability, for any number $m$ of response qubits, since below the precision $\epsilon = \Theta(1/N)$ it becomes cheaper to learn the label of the oracle and then synthesize the ideal oracle exactly (Appendix~\ref{SM_sec:useful_precision}).

Combining the two factors with the ratio achieved by each construction gives, in each regime of $r$, how close the protocol comes to optimal among all protocols that distill $\mathfrak{O}_{F_G}$ and satisfy the KLC~\eqref{eq:KLC_prequery}.
For constant $r$, Construction 4 achieves $\eta^{-1} = \Theta(M_r)$, so the protocol is optimal up to a constant factor when the application needs only constant precision, and up to a factor $O(n)$ in general.
For super-constant $r = o(n)$, Construction 4 achieves $\eta^{-1} = M_r^{1+o(1)}$, and the factor $\ln \epsilon^{-1} = O(n)$ is also $M_r^{o(1)}$ because $M_r$ is superpolynomial in $n$, so the protocol is optimal up to a factor $M_r^{o(1)}$.
For $r = \lfloor \alpha n \rfloor$ with $H(\alpha) \leq 1/3$, Construction 3 achieves $\eta^{-1} = O(4^{r} M_r^{1+o(1)})$, so the gap is $N^{2\alpha + o(1)}$, which becomes subleading in the exponent relative to $M_r = \tilde{\Theta}(N^{H(\alpha)})$ as $\alpha \to 0$, and Construction 4 improves this ratio to $e^{O(r/V(1/\alpha))}$ with constants that are not explicit.

Two questions about the lower bounds remain open.
The first is whether a protocol whose pre-query subspaces violate the KLC~\eqref{eq:KLC_prequery} can distill noisy oracles into ideal ones with query complexity below the bounds above.
The second is to prove an unconditional lower bound for a concrete distillation task, for example distilling the family of i.i.d.-depolarizing Boolean oracles $\tilde{\mathfrak{O}}_F^{\mathrm{iid}}$ into the family of ideal Boolean oracles $\mathfrak{O}_F$ with precision $\epsilon$.

\section{A Threshold Theorem for Boolean Oracle Problems}\label{sec:threshold}
In this section, we present the formal version of the threshold theorem for Boolean oracle problems, which turns a large enough noiseless quantum advantage into an advantage that survives qubit-level i.i.d.\ depolarizing noise on every query, and we work out the threshold it gives for concrete problems.

The theorem is stated in terms of two query complexities, both counted for noiseless oracles.
The classical query complexity $T_C$ is the minimum number of classical evaluations of $f$ that any algorithm needs to solve the problem with constant success probability, so solving the problem with fewer than $T_C$ queries beats every classical algorithm.
The quantum query complexity $T_Q$ is the number of queries to the ideal oracle $\mathcal{O}_f$ made by one fixed quantum algorithm that solves the problem with constant success probability.
The \textit{advantage budget} of the problem is the ratio $T_C/T_Q$, and its \textit{budget exponent} is $c := \liminf_{n\rightarrow \infty} \frac{1}{n} \log_2 \frac{T_C}{T_Q}$.
The \textit{quantum query exponent} is $q := \limsup_{n\rightarrow\infty} \frac{1}{n} \log_2 T_Q$.

\begin{formalrestatable}{thr}{Threshold theorem for Boolean oracle problems}{theorem}{odthresholdformal}\label{thm:threshold_formal}
    Every Boolean oracle problem with budget exponent $0 < c < 1$ and $\log_2 m = o(n)$ retains a quantum advantage in query complexity for all sufficiently large $n$ when the quantum algorithm queries the i.i.d.-depolarizing oracle with any constant per-qubit error rate $p$ below a positive constant threshold $p^*$ that depends only on $c$ and $q$,
    \begin{equation}\label{eq:pstar_def}
        p^* = p_{\mathrm{th}}(c, q) ,
    \end{equation}
    where $p_{\mathrm{th}}$ is the distillation threshold of Theorem~\ref{thm:iid_threshold_at_exponent}.
    The loss in success probability compared to the noiseless quantum algorithm is an arbitrarily small constant.
\end{formalrestatable}
\noindent The theorem is proved in Appendix~\ref{SM_sec:threshold}.

Many Boolean oracle problems of central interest have advantage budget $N^{\Omega(1)}$, so the theorem gives each of them a constant threshold, achieved by our explicit OD protocol of Theorem~\ref{thm:iid_threshold_at_exponent}, as summarized in Table~\ref{tab:oracle_separations}.
For any Boolean oracle problem, reading off the exponents $q$ and $c$ locates its threshold $p^*$ in Fig.~\ref{fig:oracle_distillation}b.
The two exponents always satisfy $c \leq 1 - q$, since $N$ classical evaluations of $f$ determine the label completely and hence $T_C \leq N$.
We work out three examples, each illustrating a different aspect of the theorem: Grover search (Sec.~\ref{subsec:grover_example}), the $k$-forrelation problem (Sec.~\ref{subsec:forrelation_example}), and Simon's problem (Sec.~\ref{subsec:simon_example}).

\begin{table*}[t]
\centering
\resizebox{\textwidth}{!}{%
\begin{tabular}{llcccc}
\hline
Problem & Task & $T_Q$ & $T_C$ & $c$ & $p^*$ \\
\hline
$k$-Forrelation ($k \geq 2$)~\cite{Aaronson2018, Bansal2021, Sherstov2023} & \begin{tabular}[c]{@{}l@{}}decide if $f_1,\dots,f_k$ interleaved with\\Fourier transforms are correlated\end{tabular} & $O_k(1)$ & $\tilde{\Omega}_k(N^{1-1/k})$ & $1-1/k$ & \begin{tabular}[c]{@{}c@{}}$3.4\times 10^{-2}$ at $k=2$\\$\to 3/4$ as $k \to \infty$\end{tabular} \\
\hline
Simon's problem~\cite{Simon1997} & find $s$: $f(x)=f(x\oplus s)$ for all $x$ & $O(\log N)$ & $\Omega(\sqrt{N})$ & $1/2$ & $3.4\times 10^{-2}$ \\
\hline
Period finding~\cite{Shor1997, Cleve2004}\textsuperscript{\ref{fn:period_finding}} & find $r$: $f(x)=f(x+r)$ for all $x$ & $O(1)$ & $\tilde{\Omega}(N^{1/6})$ & $1/6$ & $2.8\times 10^{-3}$ \\
\hline
Grover search~\cite{Grover1996} & find $x\in[N]$ with $f(x)=1$ & $O(\sqrt{N})$ & $\Omega(N)$ & \multirow{2}{*}{$1/2$} & \multirow{2}{*}{$5.1\times 10^{-4}$} \\
Permutation inversion~\cite{Nayak2011} & find $\pi^{-1}(y)$ for a permutation $\pi$ on $[N]$ & $O(\sqrt{N})$ & $\Omega(N)$ & & \\
\hline
Element distinctness~\cite{Ambainis2007} & decide if $f\colon[N]\to R$ has a collision & $O(N^{2/3})$ & $\Omega(N)$ & \multirow{2}{*}{$1/3$} & \multirow{2}{*}{$7.1\times 10^{-7}$} \\
Claw finding~\cite{Tani2009} & find $x,y$ with $f(x)=g(y)$ for injective $f,g$ & $O(N^{2/3})$ & $\Omega(N)$ & & \\
\hline
NAND tree~\cite{Saks1986, Farhi2008, Childs2009, Ambainis2010, Reichardt2012} & evaluate the balanced binary NAND tree with leaves $f(x)$ & $O(\sqrt{N})$ & $\Omega(N^{0.7537})$ & $0.2537$ & $3.1\times 10^{-7}$ \\
\hline
Collision~\cite{Brassard1998} & distinguish 1-to-1 from 2-to-1 function & $O(N^{1/3})$ & $\Omega(\sqrt{N})$ & $1/6$ & $5.6\times 10^{-8}$ \\
\hline
\end{tabular}%
}
\caption{Threshold error rate for Boolean oracle problems.
$N = 2^n$ is the domain size, $T_Q$ and $T_C$ are the quantum and classical query complexities, $q = \limsup_{n\to\infty} \frac{1}{n}\log_2 T_Q$ is the quantum query exponent, and $c = \liminf_{n\to\infty} \frac{1}{n}\log_2 (T_C/T_Q)$ is the budget exponent.
The column $p^*$ gives the threshold $p_{\mathrm{th}}(c, q)$ of our protocol, solved numerically from Eq.~\eqref{eq:iid_pth_at_exponent}.
For the $k$-forrelation problem the threshold depends on $k$ through $c = 1 - 1/k$, and grows from $3.4 \times 10^{-2}$ at $k = 2$ to $3/4$ as $k \to \infty$.}
\label{tab:oracle_separations}
\end{table*}

\subsection{Grover search}\label{subsec:grover_example}
Grover search~\cite{Grover1996} is the paradigmatic example of a quadratic quantum advantage, finding a single marked element of an unstructured database with $T_Q = O(\sqrt{N})$ quantum queries while any classical algorithm needs $T_C = \Omega(N)$ queries.
The advantage budget is therefore $\Omega(N^{1/2})$, with budget exponent $c=1/2$.
Together with $q = 1/2$, Theorem~\ref{thm:threshold_formal} guarantees a threshold $p^* = p_{\mathrm{th}}(1/2, 1/2) = 5.1\times 10^{-4}$, achieved by our explicit OD protocol of Theorem~\ref{thm:iid_threshold_at_exponent}.
Running the protocol at the smallest overhead exponent $\gamma$ with $p < p_{\mathrm{th}}(\gamma, 1/2)$ solves Grover search with $\tilde{O}(N^{1/2+\gamma})$ queries to the noisy oracle (Lemma~\ref{lem:distilled_simulation}).
This smallest $\gamma$ tends to zero as $p \to 0$ (Eq.~\eqref{eq:gamma_min_exp}), so the query complexity approaches the noiseless scaling $\tilde{O}(\sqrt{N})$, preserving nearly the full advantage.

That the quantum advantage of Grover search survives local noise may be surprising.
Each query is error-free only with probability $(1-p)^{n+1}$, which is exponentially small in $n$, so almost every query is corrupted.
Prior works proved that a constant error rate per query destroys the Grover advantage~\cite{Regev2012,Rosmanis2023,Rosmanis2024,Vrana2014}, and our result may appear to contradict them.
The difference lies in the noise model.
Most prior works model the noise as acting on the query register as a whole, treating the register as a single qudit of dimension $2^{n+m}$.
Common noise models on such a qudit correspond, at the qubit level, to a high-weight error correlated across all qubits.
In contrast, our noise affects each qubit independently after the oracle, so although almost every query is corrupted, a typical error touches only a $p$ fraction of the qubits.
Therefore, the robustness of the quantum advantage is not purely about the magnitude of the errors but rather about their structure.
The same locality is what makes conventional quantum error correction possible, since no code protects against arbitrary errors acting on all qubits at once, and it is what makes oracle distillation possible here.

\subsection{$k$-forrelation}\label{subsec:forrelation_example}

\refstepcounter{footnote}\label{fn:period_finding}\footnotetext[\value{footnote}]{Here $f(x) = \pi^{x}(y_0)$ for a permutation $\pi$ on $\{0,1\}^{n/2}$ and a given $y_0$.
The classical lower bound follows from Theorem~2 of Ref.~\cite{Cleve2004}, which holds for the oracle $(x,y) \mapsto \pi^{x}(y)$ and hence for its restriction to $y = y_0$.}

The $k$-forrelation problem~\cite{Bansal2021, Sherstov2023} gives oracle access to $k$ Boolean functions $f_1,\dots,f_k$ and asks whether they are correlated when interleaved with Fourier transforms, promised that this $k$-fold forrelation is either at least $2^{-5k}$ or at most half that value.
For constant $k$, a quantum algorithm solves it with $T_Q = O(1)$ queries, while any classical algorithm needs $T_C = \tilde{\Omega}(N^{1-1/k})$ queries~\cite{Aaronson2018, Bansal2021, Sherstov2023}.

We include this problem because it has $q = 0$ and its budget exponent $c = 1 - 1/k$ approaches the maximum possible value $c = 1$ as $k \to \infty$.
Theorem~\ref{thm:threshold_formal} guarantees a threshold $p^* = p_{\mathrm{th}}(1 - 1/k, 0)$, which by numerically solving Eq.~\eqref{eq:iid_pth_at_exponent} equals $3.4 \times 10^{-2}$ at $k = 2$ and increases to $3/4$ as $k \to \infty$ (Lemma~\ref{lem:pth_nu_zero}), the largest value any threshold can take since $p_{\mathrm{th}} < 3/4$ (Lemma~\ref{lem:threshold_equation}).
Hence, for every i.i.d.-depolarizing Boolean oracle that is not completely depolarizing, that is, for every $p < 3/4$, there is a $k$ for which $k$-forrelation retains a quantum advantage.

\subsection{Simon's problem}\label{subsec:simon_example}
In Simon's problem~\cite{Simon1997}, the oracle computes a function $f\colon \{0,1\}^n \rightarrow \{0,1\}^n$ promised to satisfy $f(x) = f(y)$ if and only if $x \oplus y \in \{0^n, s\}$ for a hidden nonzero string $s$, and the task is to find $s$.
Simon's algorithm solves it with $T_Q = O(\log N)$ quantum queries and $O(\mathrm{poly}(n))$ classical post-processing time for solving a linear system over $\{0,1\}^n$, while any classical algorithm needs $T_C = \Omega(\sqrt{N})$ queries, an exponential separation.

We now apply OD and run Simon's algorithm with the distilled oracle.
The advantage budget is $\tilde{\Theta}(N^{1/2})$, with budget exponent $c = 1/2$, and $T_Q = O(\log N)$ gives $q = 0$.
Theorem~\ref{thm:threshold_formal} guarantees a threshold $p^* = p_{\mathrm{th}}(1/2, 0)$, which by numerically solving Eq.~\eqref{eq:iid_pth_at_exponent} equals $3.4 \times 10^{-2}$.
For any $p < p^*$, OD incurs an overhead of $N^{\gamma}$ queries per distilled query for some positive constant $\gamma < 1/2$ that depends on $p$, so Simon's problem is solved with $\tilde{O}(N^{\gamma})$ queries to the i.i.d.-depolarizing oracle, strictly fewer than the $\Omega(\sqrt{N})$ queries any classical algorithm needs.
The quantum query complexity is now polynomial in $N$ rather than polylogarithmic, so the noise weakens the exponential separation to a polynomial one, $T_C = \tilde{\Omega}(T_Q^{1/(2\gamma)})$.
The degree $1/(2\gamma)$ of this separation grows without bound as $p \to 0$, since the smallest overhead exponent $\gamma$ then tends to zero (Eq.~\eqref{eq:gamma_min_const}).

Besides the classical algorithm, we compare OD against a second baseline, the naive approach where noisy oracle is directly plugged into Simon's algorithm.
The algorithm then samples bitstrings distributed exactly as in the learning Simon with noise (LSN) problem with parameter $\tau$ satisfying $1-2\tau = (1-4p/3)^{|s|}$ (Lemma~\ref{lem:noisy_simon_lsn}), and LSN is polynomial-time equivalent to learning parity with noise (LPN) with the same parameters $n$ and $\tau$~\cite{May2021}.
A random hidden string typically has weight $|s| = \Theta(n)$, so $1-2\tau = N^{-\Theta(p)}$, and recovering $s$ requires $N^{\Omega(p)}$ samples information-theoretically~\cite{Gentile2001}.
The classical post-processing reads every sample, so its time is at least $N^{\Omega(p)}$, exponential in $n$.
With the distilled oracle, in contrast, the classical post-processing solves the same linear system over $\{0,1\}^n$ as in the noiseless algorithm and takes $O(\mathrm{poly}(n))$ time, exponentially less than the naive approach.
This exponential reduction in post-processing time comes at a price in the other two measures, query complexity and gate complexity.
In query complexity, the naive approach can recover $s$ from $N^{\Theta(p)}$ samples, although the best known solver at this sample count takes $N^{1+o(1)}$ time~\cite{Levieil2006}.
The smallest overhead exponent of OD is instead $\Theta(\sqrt{p})$ as $p \to 0$ (Eq.~\eqref{eq:gamma_min_const}), so OD makes more queries than the naive approach at its fewest samples.
In gate complexity, the parameters of Theorem~\ref{thm:iid_threshold_at_exponent} that attain $p^*$ take $\alpha_{\mathrm{seq}} = 1$, so the sequential recovery tests all $N$ candidate patterns and solving Simon's problem takes $\tilde{O}(N^{3/2})$ elementary gates in total (Eq.~\eqref{eq:main_gate_complexity} with $M_{r_{\mathrm{seq}}} = N$).
Since $\alpha_{\mathrm{seq}}$ is a tunable parameter of the protocol, choosing a constant $\alpha_{\mathrm{seq}} \leq 1/2$ reduces the gate complexity to $\tilde{O}(N^{H(\alpha_{\mathrm{seq}}) + \gamma})$, at the price of a threshold below $p^*$ that remains positive (Corollaries~\ref{cor:iid_boolean_query_complexity} and~\ref{cor:iid_boolean_query_complexity_C2}).
For example, Construction 2 with $\alpha = 0.01$ and $\alpha_{\mathrm{seq}} = 0.05$ solves Simon's problem with $\tilde{O}(N^{0.15})$ queries and $\tilde{O}(N^{0.43})$ elementary gates for every $p$ below its threshold $1.9 \times 10^{-3}$, obtained by numerically solving Eq.~\eqref{eq:iid_pth_C2} at $\nu = 0$.

\section{Distillation of Fractional and Continuous-time Boolean Oracles}\label{sec:two_sided_noise}

In the previous sections, we considered noise models in which local noise acts after the ideal Boolean oracle.
In a physical implementation, however, errors may occur not only after the oracle but also before it, or even during its action.
Such an error is local at the moment it arises, but if it does not commute with the oracle then the remaining action of the oracle scrambles it into a non-local, correlated operator acting after the ideal oracle.
The absence of knowledge about the timing of an error, e.g. before versus after the oracle, forces any correction to handle both the local error and its scrambled non-local version at once, which can lead to a catastrophic effect.
For example, Ref.~\cite{Rosmanis2024} showed that when a single known qubit is depolarized at a constant rate before and after each query, quantum search over $N$ elements requires $\Omega(N)$ queries, so the Grover advantage is lost.

In this section, we show how to mitigate such noise models.
The key idea is to remedy the stronger noise model with a correspondingly finer control model.
In a physical implementation, every quantum operation, including the oracle, runs continuously in time, so the algorithm can act in the middle of a query rather than only after each whole query completes.
With this finer control, the algorithm corrects an error soon after it strikes, before the remaining action of the oracle scrambles it into a non-local operator.

To this end, we consider the fractional query model~\cite{Farhi1998, Mochan2007, Cleve2009, Lee2011}, which formalizes this finer control.
The unitary of an \textit{ideal fractional Boolean oracle} of angle $\theta$ is
\begin{align}
	  O_f(\theta) &:= e^{-i \theta H_f},\\
		H_f &:= \sum_{x\in\{0,1\}^n} \ket{x}\bra{x} \otimes \sum_{j=1}^m f(x)_j \ket{1}\bra{1}_j
\end{align}
which is the evolution under the Hamiltonian $H_f$ for time $\theta$ and recovers the ideal Boolean oracle $O_f$ at $\theta=\pi$.
We assume $\pi/\theta$ is an integer, so that $\pi/\theta$ repeated queries of the fractional oracle of angle $\theta$ compose one Boolean oracle query.

We show that fractional queries extend efficient distillation to noise acting not only after the oracle but also before it and during its action.
We consider two such noise models.
First, we consider the \textit{family of two-sided i.i.d.-depolarizing fractional oracles} $\tilde{\mathfrak{O}}_F^{\mathrm{frac}}(\theta) := \{\tilde{\mathcal{O}}_f(\theta)\}_{f\in F}$, where
\begin{equation}\label{eq:frac_noisy_oracle_main}
    \tilde{\mathcal{O}}_f(\theta) := \mathcal{D}^{\otimes (n+m)}_{p_\theta} \circ \mathcal{O}_f(\theta) \circ \mathcal{D}^{\otimes (n+m)}_{p_\theta}
\end{equation}
and $\mathcal{O}_f(\theta)(\cdot) := O_f(\theta)(\cdot)O_f(\theta)^{\dagger}$ is the unitary channel of the fractional oracle.
The per-qubit error rate $p_\theta := \frac{\theta}{2\pi} p$ of each depolarizing layer is proportional to the angle, so that the rates of the $2\pi/\theta$ layers accompanying the fractional queries that compose one Boolean oracle query sum to $p$.
Second, we consider the \textit{family of continuous-time i.i.d.-depolarizing oracles} $\tilde{\mathfrak{O}}_F^{\mathrm{ct}}(\theta) := \{e^{\theta \mathcal{L}_f}\}_{f\in F}$, the evolutions for time $\theta$ under the Lindbladian
\begin{equation}\label{eq:ct_lindbladian_main}
    \mathcal{L}_f(\rho) := -i[H_f,\rho]
    + \frac{\Gamma}{3}\sum_{j=1}^{n+m}\sum_{P\in\{X,Y,Z\}}
    \left(P_j\rho P_j-\rho\right),
\end{equation}
whose second term is i.i.d.\ depolarizing noise at rate $\Gamma$ on all $n+m$ oracle qubits.

We show that both families can be distilled into the family of ideal Boolean oracles $\mathfrak{O}_F$ with precision $\epsilon$.
A fractional query runs $H_f$ for time $\theta$ while a Boolean oracle query runs it for time $\pi$, so counting queries would treat these two unequal uses of $H_f$ as the same cost.
We therefore measure the cost by the \textit{total evolution time}, the sum of the evolution times of all queries.
Fix a constant overhead exponent $\gamma \in (0,1)$ and let $\log_2 m = o(n)$.
For any angle $\theta \leq \theta_{\mathrm{th}}$, any constant per-qubit error rate $p < p_{\mathrm{th}}(\gamma, \nu)$ of Eq.~\eqref{eq:iid_pth_at_exponent}, the same threshold as for the i.i.d.-depolarizing Boolean oracles, and all sufficiently large $n$, the family of noisy fractional oracles $\tilde{\mathfrak{O}}_F^{\mathrm{frac}}(\theta)$ can be distilled with precision $\epsilon$ and total evolution time $\pi\, T_{\mathrm{OD}}$, where $T_{\mathrm{OD}} = \tilde{O}(N^{\gamma})$ is the query complexity Eq.~\eqref{eq:T_OD_at_exponent} of the i.i.d.-depolarizing Boolean oracle at precision $\epsilon/2$ (Theorem~\ref{thm:fractional_OD_two_sided}).
The continuous-time family $\tilde{\mathfrak{O}}_F^{\mathrm{ct}}(\theta)$ admits the same guarantee whenever the noise rate satisfies $\Gamma < \frac{3}{4\pi} \ln \frac{3}{3 - 4 p_{\mathrm{th}}(\gamma, \nu)}$ (Theorem~\ref{thm:continuous_OD_two_sided}).
The angle threshold $\theta_{\mathrm{th}} = \Theta\big(\epsilon / (m (n+m)\, p\, T_{\mathrm{OD}})\big)$ for the fractional family, and the same expression with $p$ replaced by $\Gamma$ for the continuous-time family, shrinks as $n$ grows but is positive for every $n$, so the protocol needs to interrupt the oracle evolution only after a positive angle, never continuously.

We now describe the protocol behind the two theorems and explain why it works, leaving the full proofs to Appendix~\ref{SM_sec:frac_ct_distillation}.
The protocol runs the same two stages as the distillation of the i.i.d.-depolarizing Boolean oracles, with a composition step in between.
We apply the stage 1 gadget to each noisy query, compose $\pi/\theta$ of the gadget outputs into one Boolean oracle query, and run the stage 2 protocol on the composed queries.
The stage 1 gadget converts each noisy query into an i.i.d.-dephasing fractional oracle up to an error $O(\theta^2)$, as we explain in the next paragraph.
Composing $\pi/\theta$ gadget outputs multiplies the fractional oracles back into the Boolean oracle $O_f$, the dephasing layers merge into a single layer of i.i.d.\ dephasing noise, and the $\pi/\theta$ errors of $O(\theta^2)$ add up to $O(\theta)$, which vanishes as $\theta$ vanishes.
Each dephasing layer has a per-qubit rate of order $\theta$ and there are $\pi/\theta$ layers, so the rate of the merged layer never exceeds $2p/3$, the rate left by the stage 1 gadget in the Boolean case, no matter how small $\theta$ is.
The composed query is therefore an i.i.d.-dephasing Boolean oracle up to an error $O(\theta)$, which the stage 2 protocol distills below the same threshold $p_{\mathrm{th}}(\gamma, \nu)$.
Requiring the total error of the stage 2 protocol to stay below $\epsilon/2$ sets the angle threshold $\theta_{\mathrm{th}}$.

We now explain what the stage 1 gadget does to each noisy query.
In both families, the gadget corrects each error sooner after it occurs than in the Boolean case, after at most an angle $\theta$ of oracle evolution rather than a full Boolean oracle query, so the oracle changes the error by at most $O(\theta)$ before the gadget removes it.
We first consider the family $\tilde{\mathfrak{O}}_F^{\mathrm{frac}}(\theta)$.
An error after the oracle does not propagate, and the stage 1 gadget converts it into i.i.d.\ dephasing noise exactly as in the Boolean case.
An error before the oracle, in contrast, propagates through $O_f(\theta)$ before the gadget acts on it.
A $Z$-type error commutes with the diagonal unitary $O_f(\theta)$, so it passes through the oracle unchanged and the gadget converts it in the same way.
An $X$-type error does not commute, and moving it past the oracle conjugates the oracle,
\begin{equation}
	  O_f(\theta) X = X \left(X O_f(\theta) X\right) ,
\end{equation}
so the gadget removes the $X$-type error but leaves the conjugated oracle $X O_f(\theta) X$ in place of $O_f(\theta)$.
On each computational basis state, $O_f(\theta)$ and $X O_f(\theta) X$ have eigenvalues $e^{-i\theta k}$ and $e^{-i\theta k'}$ with $k, k' \in \{0, \ldots, m\}$.
For the Boolean oracle, $\theta = \pi$, these eigenvalues are $\pm 1$, so $X O_f X$ can differ from $O_f$ by a sign on some computational basis states, and indeed Ref.~\cite{Rosmanis2024} rules out efficient distillation of the Boolean oracle under two-sided noise.
For a fractional oracle, in contrast, the two eigenvalues differ by at most $\theta |k - k'| \leq m\theta$, so the two oracles differ by at most $O(\theta)$.
An $X$-type error before the oracle occurs with probability $O(\theta)$ in each query, since the error rate $p_\theta$ is proportional to $\theta$, and when it occurs it changes the oracle by the $O(\theta)$ difference above, so the output of the stage 1 gadget on one noisy query is within $O(\theta^2)$ of an i.i.d.-dephasing fractional oracle.
For the family $\tilde{\mathfrak{O}}_F^{\mathrm{ct}}(\theta)$, the first-order product formula~\cite{Wang2026} writes each query as i.i.d.\ depolarizing noise acting after an ideal fractional oracle, up to an error $O(\theta^2)$, and the stage 1 gadget converts that noise exactly, so its output on one noisy query is again within $O(\theta^2)$ of an i.i.d.-dephasing fractional oracle.

It is surprising that fractional queries help here, since prior work found that acting partway through a query brings no advantage, neither in the noiseless setting nor under noise correlated across all qubits.
In the noiseless setting, the fractional query model strictly contains the discrete query model, yet the query complexity of every problem is the same in the two models up to a constant factor~\cite{Cleve2009, Lee2011}.
Under Lindbladian noise correlated across all qubits, Grover search needs evolution time $\Omega(N)$ even with continuous-time queries~\cite{Temme2014}.
Two-sided i.i.d.\ depolarizing noise is different, because an error before the oracle is local at the moment it occurs and becomes correlated across all qubits only through the remaining action of the oracle, so acting partway through the query removes it immediately after it occurs.

\section{Comparison with Other Approaches to Correcting Noisy Operations}\label{sec:comparison}
Several lines of work also aim to turn a noisy quantum operation into an ideal one.
Here we explain how OD in the present work  differs from each of them.

\emph{Channel tomography and unitary purification.}
There are two existing approaches to turn a noisy operation into an ideal one without assuming any structure in its target.
Channel tomography~\cite{Chen2026, Mele2026} learns the noisy oracle channel to enough precision to infer its label, and then resynthesizes the ideal oracle on a fault-tolerant quantum computer.
A Boolean oracle acts on $n+m$ qubits, and tomography of a channel of dimension $d = 2^{n+m}$ to constant precision requires $\Omega(d^2) = \Omega(N^2)$ queries.
Unitary purification instead restores an arbitrary unknown unitary under global depolarizing noise without learning it. A method has been developed first for a single-qubit unitary~\cite{Zhao2026, Niwa2026} and then generalized for channels in any dimension~\cite{Niwa2026_2}.
In the asymptotic limit where the noise rate goes to zero, the optimal query cost for a unitary of dimension $d$ is quadratic in $d$~\cite{Niwa2026_2}, hence again $\Omega(N^2)$ for $d = 2^{n+m}$.

OD is distinguished from these results because those results cannot be directly used to retain any quantum advantage in Boolean oracle problems, while OD can.
Specifically, every Boolean oracle problem has $T_C \leq N$, so a distillation cost quadratic in $N$ already exceeds the classical query complexity.
OD instead exploits the structure of Boolean oracles, distilling the oracle without learning its label at a query cost sublinear in $N$ and logarithmic in $1/\epsilon$, and hence retains quantum advantage.

\emph{Error filtration.}
Error filtration~\cite{Lee2023} takes a noisy channel on any number of qubits that approximates an unknown ideal unitary with small infidelity $\epsilon$, and with $T$ uses of the channel and postselection reduces the infidelity to $\epsilon/T + O(\epsilon^2)$, without learning the unitary or accessing the inside of the channel.
It differs from OD in two respects.
First, the global infidelity $\epsilon$ must be small, since the output is never better than the $O(\epsilon^2)$ floor.
In contrast, the Boolean oracle under i.i.d.\ depolarizing noise of constant per-qubit rate, considered in the present work, has infidelity approaching one as $n$ grows; almost every query is affected by at least one error. Error filtration does not apply in this regime, whereas OD still distills the oracle to any precision.
Second, the output of error filtration is postselected, so an algorithm that queries it succeeds only when all $T_Q$ queries pass the postselection.
In contrast, our OD protocol outputs a distilled oracle that needs no postselection.

\emph{Robustification of faulty oracle.}
Ref.~\cite{Lolck2024} considered a specific noisy oracle model, called the faulty oracle: each query successfully applies the ideal Boolean oracle with probability $1-p$ and does nothing with probability $p$.
This noise model is special in that it can be viewed a classical error affecting only the response register.
More specifically, the faulty oracle can be interpreted as applying a faulty response, containing either bit-flip or phase-flip errors on the response registers, conditioned on the configuration of index registers without affecting them. 
Consequently, it suffices to extract the correct response from multiple faulty responses.
Indeed, Ref.~\cite{Lolck2024} constructs a gadget that coherently combines multiple \emph{weak responses} into a response with reduced error, enabling any quantum algorithm making $T_Q$ ideal queries to be simulated using $\tilde{O}(T_Q^3 m^2)$ faulty queries.
In contrast, our approach protects quantum information encoded in both index and response registers against generic quantum errors through the use of \emph{weak queries}.
We note that the approach in Ref.~\cite{Lolck2024} is complementary to ours. The faulty oracle can be also understood as a highly correlated nonlocal error where the error channels strongly depends on $f$.
This constitutes a setting for which the applicability of our locality-based OD protocol is limited.
Therefore, we find that there are multiple ways to distill oracles leveraging different structural properties.

\section{Discussion}\label{sec:discussion}
In this section, we discuss various aspects of the errors in our protocol as well as the implications of OD. 
Specifically, we analyze the origin of the remnant errors after distillation (Sec.~\ref{subsec:error_origin}), discuss the physical settings in which the oracle noise arises and in which OD is needed (Sec.~\ref{subsec:noise_origin}), and describe the implications of OD for quantum computational sensing (Sec.~\ref{subsec:sensing}) as well as for quantum random access memory (Sec.~\ref{subsec:qram}).

\subsection{Knill--Laflamme conditions and exact recoverability}\label{subsec:error_origin}
In conventional quantum error correction, a codespace and an error model that exactly satisfy the KLC admit a perfect recovery channel.
In contrast, the distillation error 
in our protocol remains nonzero even though every pre-query subspace satisfies the KLC~\eqref{eq:KLC_prequery} exactly.
Examining this discrepancy reveals a subtlety in the relation between the KLC and exact recoverability.
Specifically, the equivalence between perfect recovery and the exact KLC does not
hold in general for OD.
We illustrate this point with a simple example. 
Let us consider a quantum error-correcting code with codewords $\{ \ket{\bar{x}}\}_x$ and a noise model $\mathcal{N}[\cdot]=\sum_k E_k \cdot E_k^\dagger$ that exactly satisfy the simplest form of the KLC:
\begin{align}
    \bra{\bar{x}}E_\alpha^\dagger E_\beta \ket{\bar{y}} =\delta_{xy} \delta_{\alpha \beta}.
\end{align}
We further assume that the $E_k$ are unitary. 
If a particular error occurs, $\ket{\bar{x}}\mapsto E_\alpha \ket{\bar{x}}$, one can perform a measurement to identify $\alpha$ without revealing any information about the code state, and then apply the exact recovery $E_\alpha^\dagger$.
Now consider the analogous case in oracle distillation:
\begin{align}
    \bra{\bar{x}} O_f^\dagger E_\alpha^\dagger E_\beta O_f \ket{\bar{y}} = \delta_{xy} \delta_{\alpha \beta}
\end{align}
for every oracle label $f$.
If an error occurs after a query, $\ket{\bar{x}} \mapsto  E_\alpha O_f\ket{\bar{x}}$, the resulting states with different errors $\alpha$ are perfectly orthogonal for any (unknown) $f$.
Nevertheless, one cannot identify $\alpha$ without knowledge of the unknown label $f$, because the actions of $E_\alpha O_f$ and $E_\beta O_{g}$ on a logical state with $\alpha \neq \beta$ and $f\neq g$ may not be perfectly distinguishable.
This analysis implies that, in the most general setting, OD demands a technical challenge: a \emph{label-agnostic} recovery channel.

In the present work, we were able to resolve this challenge simply because one can uncompute the effect of $O_f$, leveraging the fact that $O_f^{-1} = O_f$ for Boolean oracles. 
Therefore, the only remaining error arises from the aggregator not perfectly recognizing ideal oracle responses.
In a general setting where one does not have the ability to apply the inverse of an unknown oracle~\cite{Tang2026}, designing an OD protocol might become substantially more challenging.

\subsection{Origin of the oracle noise}\label{subsec:noise_origin}
In this work, we model the interface between the quantum device and the unknown system as an oracle, whose noise model may have qualitatively different properties depending on how the oracle is constructed.
In computational settings, the oracle often represents a functional unit whose circuit decomposition is explicitly known.
In such cases, each gate in the circuit can be protected by conventional FTQC schemes, and oracle distillation is unnecessary.
In quantum learning settings, the specification of the noisy oracle is often part of the problem description. In fact, identifying the ``correct'' noise model is a nontrivial problem.
At the very fundamental level, the noise arises from physical origins, whose form may vary from one specific hardware device to another. Therefore, there is no single correct model. Nevertheless, many noise models share common features such as locality, which justifies the standard use of local i.i.d.\ depolarizing channels to model noise in quantum computation.
Unfortunately, the same reasoning cannot be directly applied to the noise models describing oracles. Even if the oracular dynamics is constructed from conventional physical interactions that obey locality, many such features may be lost at the level of the abstract oracle.
Consequently, the effective noise model varies across physical implementations of the oracle. 
Hence, whether the quantum advantage survives must be studied case by case, for the given noisy oracle family and the task.
No-go results show that several noise models destroy the quantum advantage in many cases~\cite{Regev2012, Vrana2014, Rosmanis2023, Rosmanis2024, Chen2023, Cotler2026}.
For other noise models, the quantum advantage is known to survive~\cite{Cross2015, Lee2023, Lolck2024}.
Our results show that quantum advantages based on Boolean oracles survive in a large class of tasks.
From the prior results and ours, a landscape of quantum learning advantage with noisy oracles emerges: the fate of the quantum advantage depends more on the structure of the noise than on its magnitude.

\subsection{Implication for quantum computational sensing}\label{subsec:sensing}
The most direct application of OD is computational sensing, in particular the sensing tasks whose quantum advantage comes from an oracular speedup.
In sensing, we probe an unknown object through the physical dynamics generated by its interaction with the sensor, and no such dynamics natively takes the form of a Boolean oracle.
The Boolean oracle is nevertheless a valid abstraction of the interface, as, for example,  Ref.~\cite{Allen2025} has shown that the continuous evolution of a sensor under an unknown AC signal can be \textit{digitized} into a Boolean oracle.
A sensing algorithm that queries the digitized oracle can draw on two sources of advantage, namely Heisenberg scaling in the sensing time needed to construct each oracle and the oracular speedup in the number of oracle calls. The advantage in total sensing time comes from combining the two.

The major obstacle to realizing this advantage at scale is noise, yet the digitization scheme in Ref.~\cite{Allen2025} assumes noiseless dynamics.
Under generic noise on the sensor, a high-fidelity oracle cannot be digitized with Heisenberg scaling, since a single query to it would then decide with Heisenberg scaling whether a signal at a known frequency is present, which the no-go theorems forbid~\cite{Escher2011, Demkowicz2012, Zhou2018, Huang2025}.
However, these no-go theorems do not apply to the oracular speedup, because the speedup resides in the number of oracle calls rather than in the time to construct each one.
Even if a high-fidelity oracle cannot be digitized with Heisenberg scaling, the oracular speedup may survive, such as the quadratic speedup in the bandwidth of the candidate signal in Ref.~\cite{Allen2025}.

The question thus becomes how to digitize a high-fidelity oracle out of noisy dynamics.
Such a task is much harder than in the noiseless case, because the digitization already requires a complex sequence of controls on the sensor, and one would like to correct any error before it propagates through such complicated dynamics.

Our results alleviate this challenge, because the digitized oracle need not be ideal, only distillable.
The physical interaction involves its own noise, and different digitization circuits shape this noise into different noise channels.
Therefore, a no-go result for one oracle noise model does not show that the sensing task is fragile under noise, since a different digitization could avoid that noise model.
Conversely, our positive result alone does not show that the advantage can be made robust, since the guarantee holds only when there exists a digitization that produces a distillable noisy oracle.
We call the design of a digitization circuit that yields a distillable noisy oracle \textit{noise shaping}.
Identifying the controls and the conditions on the signal that permit noise shaping remains an open problem.

\subsection{Implication for quantum random access memory}\label{subsec:qram}
Although OD is designed for oracles whose identity is unknown, we expect it to also be useful for implementing quantum random access memory (QRAM), where the main difficulty is the cost of making it fault-tolerant.
The baseline approach is to decompose a QRAM into elementary gates, and implement each gate fault-tolerantly.
However, such a decomposition requires $\tilde{\Omega}(N)$ elementary gates per access~\cite{Jaques2025}, so making each gate fault-tolerant destroys the quantum advantage in many cases, including Grover search.
Ref.~\cite{Cesa2026} addresses this challenge by moving all non-Clifford operations into offline resource-state preparation, so that online queries use only Clifford operations and Pauli measurements, but the number of required fault-tolerant operations is still $\Theta(N)$ when both the online and offline stages are taken into account.
Another recent method, proposed in Ref.~\cite{Dalzell2025}, approaches this challenge from a different angle.
It assumes that a noisy QRAM can be implemented efficiently at the physical level and then ``distilled'' into a fault-tolerant one.
It prepares many copies of a noisy resource state, each by applying the noisy physical QRAM once to a fixed reference state, distills these copies into one high-fidelity resource state, and then fault-tolerantly teleports this state into a QRAM gate.
This method only requires $\mathrm{poly}(n)/\epsilon$ queries to the physical QRAM and $\mathrm{poly}(n)/\epsilon$ fault-tolerant gates.
However, it requires $\Omega(N)$ classical processing time, and assumes that the noise in the physical QRAM is independent of the dataset.
OD provides a potential alternative approach to implementing QRAM efficiently.
The key idea is again noise shaping, now applied to the physical QRAM instead of a digitization circuit.
Since the dataset is known, the physical QRAM can be designed with the dataset in hand, and its noise may depend on the dataset.
If we can implement a noisy QRAM that has the form of a distillable oracle, then we can apply OD and distill one high-fidelity QRAM using $\tilde{O}(N^{\gamma})$ queries to the physical QRAM and $\tilde{O}(N^{H(\alpha_{\mathrm{seq}}) + \gamma})$ elementary gates and classical processing time.
For suitably small $\gamma< 1$ and $\alpha_{\mathrm{seq}}$, all these costs are $o(N)$, so in particular the classical processing time of OD is asymptotically smaller than the $\Omega(N)$ required by Ref.~\cite{Dalzell2025}.

\section{Summary and Outlook}\label{sec:conclusion}

We have shown that error correction does not require knowing the dynamics it protects: an oracle can be distilled to high fidelity even when its identity is unknown.
This extends quantum error correction and fault-tolerant quantum computation beyond their traditional scope of protecting prescribed operations.
Concretely, we presented a universal oracle distillation protocol that works for all Boolean oracles under bounded-weight noise or i.i.d.\ depolarizing noise.
We proved asymptotic upper bounds on both its query and gate complexity and established a threshold theorem, and for finite sizes we gave a numerical optimization procedure for explicit constructions.
We also proved a nearly matching lower bound for any protocol whose pre-query subspaces satisfy the Knill--Laflamme conditions, showing that the tension between error correction and functionality is fundamental and that our protocol is near-optimal.
It therefore remains open whether a better construction of query states can close the $2\alpha$ gap in the exponent in the $r=\lfloor \alpha n \rfloor$ regime.
It would also be interesting to tailor OD protocols to specific families of oracles, trading the generality of our protocol, which works for every Boolean oracle, for better performance.

Conceptually, our protocol works because it exploits a property that all Boolean oracles share but the noise lacks, namely that each Boolean oracle applies a phase flip that depends on the entire input bitstring, whereas the noise acts locally.
This difference is what allows the protocol to distinguish the oracle response from the noise.

A broader direction is to understand the general principle behind efficient distillation.
We expect that any structural property that systematically distinguishes the oracle response from the noise can be harnessed for distillation.
Such structure may exist in other settings, for example in state-preparation oracles or block-encoding oracles~\cite{Gilyen2019} under physically relevant noise models.
A further goal is a criterion that decides when efficient OD is possible, in the spirit of the Hamiltonian-not-in-Lindblad-span (HNLS) condition~\cite{Zhou2018} in noisy metrology.
The HNLS condition states that the Heisenberg limit is achievable if and only if the signal Hamiltonian lies outside the span of the Lindblad operators, and Ref.~\cite{Zhou2018} gives an explicit protocol that reaches the Heisenberg limit whenever the condition holds.
Similarly, identifying such algebraic conditions associated with the distillation of oracular operations would provide both a deeper understanding of, and practical recipes for, robust quantum learning and computational sensing protocols.

Another important future direction is the fault-tolerant construction of a robust oracle, taking into account the noise of physical interactions as well as of all gates in both the digitization circuit and the distillation protocol.
In such cases, all gates are known and can therefore be performed fault-tolerantly at the logical level.
The physical interaction, in contrast, is unknown and acts directly on physical qubits.
Then, the difficulty lies at the interface between the physical interaction and the logical gates around it.
One route is to design an error-correcting code on which the physical interaction itself implements a desired logical operation.
Another route is to apply a gadget: (i) decode logical information into physical qubits, (ii) apply physical interactions, and (iii) finally encode back to logical subspace.
By performing the encoding and decoding steps based on the fault-tolerant interfaces of Ref.~\cite{Christandl2025}, such a sequence enables constructing a \emph{logical noisy oracle}, wherein ideal physical interactions and noise are all applied at the logical level.
As long as the resulting logical noisy oracle is distillable, one obtains an end-to-end robust oracle.
Therefore, the challenge reduces, again, to noise shaping in the construction of fault-tolerant interfaces.
An analogous question has recently been studied in noisy metrology, with the positive result that Heisenberg scaling can be preserved under biased noise when the physical error rate is below a constant threshold~\cite{Sahu2026, Conlon2026}.

Our ultimate goal is to develop an end-to-end quantum process---from physical signals to final answers---that achieves quantum advantage under realistic conditions. Reaching this goal will likely require concerted, vertically integrated efforts spanning hardware design, oracle digitization, noise shaping, oracle distillation, and quantum algorithm design. 
This is an area that remains relatively underexplored with many challenges ahead.
However, successful outcomes would establish practically useful and robust applications of scalable quantum devices with provable advantages. We hope this work stimulates further progress toward realizing this vision.

\begin{acknowledgments}
We thank Byungmin Kang for insightful discussions at the early stage of this project. 
In particular, we would like to acknowledge his detailed comments and suggestions on the manuscript, especially on Appendix E.
We also thank Richard Allen, Fernando Brand\~ao, Alexander Dalzell, Weiyuan Gong, Yu-Jie Liu, John Preskill, Xinyu Tan, Bingtian Ye, and Haimeng Zhao for insightful discussions.
We used Claude (Opus 4.6-5.5, Fable 5-5.1) and ChatGPT (GPT 5.5-6) to improve the presentation of our work and to assist with technical calculations.
All calculations were independently verified by the authors, who take full responsibility for the content.
\end{acknowledgments}

\clearpage
\appendix
\providecommand{\theHequation}{}
\renewcommand{\theHequation}{\theequation}
\onecolumngrid
\addtocontents{toc}{\protect\appendixtoctrue}
\begin{center}
    {\Large\bfseries Appendices}
\end{center}
\vspace{0.5em}
\tableofcontents

\vspace{1.5em}
The diagram below shows the main results of this paper and the dependencies among them.

\begin{center}
\hyphenpenalty=10000\exhyphenpenalty=10000
\definecolor{generalgate}{HTML}{BAB9EB}
\definecolor{almostblack}{HTML}{0D0D0D}
\begin{tikzpicture}[
  box/.style={draw=almostblack, rounded corners=2pt, align=center, text width=2.5cm,
              minimum height=1.9cm, inner sep=4pt, font=\footnotesize, anchor=north, text=almostblack},
  thmbox/.style={box, fill=generalgate!45},
  lembox/.style={box, fill=black!6},
  secbox/.style={box, fill=white, draw=black!55, dashed},
  arr/.style={-{Latex[length=2mm]}, semithick, draw=almostblack},
  wire/.style={semithick, draw=almostblack},
  grplab/.style={font=\normalsize\itshape, text=almostblack},
  x=3.5cm, y=2.8cm]
\node[lembox] (oderr)  at (0.5,0)   {Lemma~\ref{lem:OD_error}\\Error of the distilled oracle};
\node[lembox] (rec)    at (1.5,0)   {Lemma~\ref{lem:agg_error_classical_OD}\\Aggregation error as a classical probability};
\node[thmbox] (gate)   at (-1,-1)   {Theorem~\ref{thm:gate_complexity_OD}\\Gate and ancilla complexity};
\node[thmbox] (adv)    at (0,-1)    {Theorem~\ref{thm:boolean_OD_adv}\\Distillation under adversarial noise};
\node[thmbox] (c3)     at (1,-1)    {Theorem~\ref{thm:near_optimal_construction}\\Construction 3};
\node[lembox] (iidlem) at (2,-1)    {Lemma~\ref{lem:boolean_OD_iid}\\Distillation under i.i.d.\ depolarizing noise at finite $n$};
\node[secbox] (c2)     at (3,-1)    {Sec.~\ref{SM_sec:construction2}\\Construction 2};
\node[thmbox] (corenc) at (0.5,-2)  {Corollary~\ref{cor:boolean_OD_encoding3}\\Distillation under adversarial noise via Construction 3};
\node[thmbox] (iidthm) at (2,-2)    {Theorem~\ref{thm:iid_threshold_at_exponent}\\Distillation under i.i.d.\ depolarizing noise};
\node[thmbox] (thr)    at (1,-3)    {Theorem~\ref{thm:threshold_formal}\\Threshold theorem};
\node[thmbox] (frac)   at (2,-3)    {Theorem~\ref{thm:fractional_OD_two_sided}\\Distillation of fractional Boolean oracles};
\node[thmbox] (ct)     at (3,-3)    {Theorem~\ref{thm:continuous_OD_two_sided}\\Distillation of continuous-time Boolean oracles};
\node[lembox] (tqp)    at (-0.5,-2) {Lemma~\ref{lem:total_qp}\\Total bound on query power};
\node[thmbox] (glb)    at (-1,-3)   {Theorem~\ref{thm:OD_grover_lower_bound}\\Lower bound for Grover oracles};
\node[thmbox] (llb)    at (0,-3)    {Theorem~\ref{thm:OD_label_indep}\\Lower bound for arbitrary families of Boolean oracles};
\coordinate (bus) at ($(oderr.south)!0.5!(adv.north)$);
\draw[wire] (oderr.south) -- (oderr.south |- bus);
\draw[wire] (rec.south)   -- (rec.south |- bus);
\draw[wire] (adv.north |- bus) -- (iidlem.north |- bus);
\draw[arr]  (adv.north |- bus)    -- (adv.north);
\draw[arr]  (iidlem.north |- bus) -- (iidlem.north);
\coordinate (mid1) at ($(adv.south)!0.5!(corenc.north)$);
\draw[wire] (adv.south) -- (adv.south |- mid1) -- (c3.south |- mid1) -- (c3.south);
\draw[arr]  (corenc.north |- mid1) -- (corenc.north);
\draw[arr] (c3.south) |- (iidthm.west);
\draw[arr] (c2.south) |- (iidthm.east);
\draw[arr] (iidlem) -- (iidthm);
\coordinate (mid2) at ($(iidthm.south)!0.5!(frac.north)$);
\draw[wire] (iidthm.south) -- (iidthm.south |- mid2);
\draw[wire] (thr.north |- mid2) -- (ct.north |- mid2);
\draw[arr]  (thr.north |- mid2)  -- (thr.north);
\draw[arr]  (frac.north |- mid2) -- (frac.north);
\draw[arr]  (ct.north |- mid2)   -- (ct.north);
\coordinate (mid3) at ($(tqp.south)!0.5!(glb.north)$);
\draw[arr] (tqp.south) |- (llb.west);
\draw[arr] (tqp.south |- mid3) -| (glb.north);
% dashed frame around the lower-bound group, stepping around the corollary
\coordinate (fTL) at ($(glb.west |- tqp.north)+(-6pt,6pt)$);
\coordinate (fMX) at ($(tqp.east)!0.5!(corenc.west)$);
\coordinate (fSY) at ($(corenc.south)!0.5!(llb.north)$);
\coordinate (fBR) at ($(llb.east |- llb.south)+(6pt,-6pt)$);
\begin{scope}[on background layer]
\draw[black!55, dashed, thin, rounded corners=4pt]
  (fTL) -- (fMX |- fTL) -- (fMX |- fSY) -- (fBR |- fSY) -- (fBR) -- (fTL |- fBR) -- cycle;
\node[grplab, fill=white, inner xsep=3pt, anchor=west] at ($(fTL)+(4pt,0)$) {Lower bounds};
\end{scope}
\node[grplab, anchor=south west] at ($(gate.west |- oderr.north)+(0,3pt)$) {Upper bounds};
\end{tikzpicture}
\end{center}

\clearpage
\section{Preliminaries}\label{SM_sec:preliminaries}
This section collects the notation and the standard facts used throughout the appendices, so that later sections can invoke them by reference.
Section~\ref{SM_sec:prelim_notation} summarizes the notation in Table~\ref{SM_tab:notation}.
Section~\ref{SM_sec:prelim_klc} states the Knill--Laflamme and error orthogonality conditions.
Section~\ref{SM_sec:prelim_channels} states properties of the trace and diamond norms.
Section~\ref{SM_sec:prelim_tails} states tail bounds for binomial distributions and entropy bounds on binomial sums.

\subsection{Notation}\label{SM_sec:prelim_notation}

Table~\ref{SM_tab:notation} collects the symbols that recur throughout the appendices.

\begin{table}[H]
\centering
\renewcommand{\arraystretch}{1.25}
\begin{tabular}{@{}p{0.22\textwidth} p{0.31\textwidth} p{0.41\textwidth}@{}}
\hline\hline
\textbf{Name} & \textbf{Notation} & \textbf{Meaning} \\
\hline
\multicolumn{3}{@{}l}{\textit{Norms and channel distances}}\\
trace norm & $\|\cdot\|_1$ & sum of singular values of a matrix \\
operator norm & $\|\cdot\|_{\mathrm{op}}$ & largest singular value of a matrix\\
vector 2-norm & $\|\ket{\psi}\|_2 = \sqrt{\braket{\psi|\psi}}$ & Euclidean norm of a vector\\
diamond norm & $\|\Lambda_1-\Lambda_2\|_{\diamond}$ & largest trace-norm distance between the two channel outputs over all inputs, possibly entangled with a reference system~\cite{Watrous2018} \\
\hline
\multicolumn{3}{@{}l}{\textit{Oracles, noise, and distillation}}\\
oracle family & $\mathfrak{O}_I = \{\mathcal{O}_i\}_{i\in I}$ & channels on a common space $\mathcal{H}_O$, indexed by a label set $I$ \\
Boolean oracle & $O_f\ket{x}\ket{y} = \ket{x} Z^{f(x)}\ket{y}$ & $f:\{0,1\}^n\to\{0,1\}^m$; the first $n$ qubits are the \textit{index register}, the last $m$ the \textit{response register} (Sec.~\ref{sec:grover_distillation}) \\
Boolean oracle family & $\mathfrak{O}_F = \{\mathcal{O}_f\}_{f\in F}$ & ideal Boolean oracles for all $f$; the target family of every distillation task in this work (Sec.~\ref{sec:grover_distillation}) \\
noise channel & $\mathcal{E}_f = \tilde{\mathcal{O}}_f \circ \mathcal{O}_f^{-1}$ & every noisy Boolean oracle factors as $\tilde{\mathcal{O}}_f = \mathcal{E}_f\circ\mathcal{O}_f$, Eq.~\eqref{eq:noisy-oracle-local} \\
{\raggedright noisy Boolean oracle families\par} & $\tilde{\mathfrak{O}}_F^{\mathrm{adv}}$, $\tilde{\mathfrak{O}}_F^{\mathrm{adv},Z}$, $\tilde{\mathfrak{O}}_F^{\mathrm{iid}}$, $\tilde{\mathfrak{O}}_F^{\mathrm{iid},Z}$ & families of noisy Boolean oracles under adversarial, adversarial phase-only, i.i.d.\ depolarizing, and i.i.d.\ dephasing noise, respectively (Secs.~\ref{SM_sec:adv} and~\ref{SM_sec:iid}) \\
noise weight & $w_P$ & maximal Pauli weight of the adversarial noise on the index register (Sec.~\ref{sec:grover_distillation}) \\
per-qubit error rate & $p$ & parameter of $\mathcal{D}_p$ in the i.i.d.\ models, Eq.~\eqref{iidnoise} \\
{\raggedright effective dephasing rate\par} & $p_{\mathrm{eff}} = \frac{4p}{3}\left(1-\frac{2p}{3}\right)$ & per-index-qubit $Z$-error rate of the merged weak-query and uncomputation noise, Eq.~\eqref{eq:iid_peff} \\
distilled oracle & $\widehat{\mathcal{O}}_i$ & output of the protocol, Eq.~\eqref{eq:od_protocol} \\
precision & $\epsilon$ & $\|\widehat{\mathcal{O}}_i - \mathcal{O}_i^{\mathrm{target}}\|_{\diamond} \leq \epsilon$ for every label $i$ \\
query complexity & $T_{\mathrm{OD}}$ & the number $T$ of raw-oracle queries in Eq.~\eqref{eq:od_protocol}; $T_{\mathrm{OD}} = 2L$ for the two-stage protocol (Sec.~\ref{SM_sec:error_bound_OD}) \\
\hline\hline
\end{tabular}
\caption{Notation used throughout the appendices.
Each entry gives a short description; the referenced equations and sections contain the full definitions.}
\label{SM_tab:notation}
\end{table}

\begin{table}[H]
\centering
\renewcommand{\arraystretch}{1.25}
\begin{tabular}{@{}p{0.22\textwidth} p{0.31\textwidth} p{0.41\textwidth}@{}}
\hline\hline
\textbf{Name} & \textbf{Notation} & \textbf{Meaning} \\
\hline
\multicolumn{3}{@{}l}{\textit{Encoding and query states}}\\
logical state & $\ket{\bar{x}}\ket{\bar{y}} = \left(\ket{\Theta(x)}\ket{+}\right)^{\otimes L}\otimes\ket{x}\ket{y}$ & $L$ \textit{query blocks} of $n+1$ qubits each, followed by the \textit{data block}; their span is the \textit{logical subspace} \\
bitstring power & $P^{a} = P^{a_1} \otimes \cdots \otimes P^{a_n}$ & for a single-qubit operator $P$ and a bitstring $a$: acts as $P$ on the qubits with $a_i = 1$ and trivially elsewhere, e.g.\ $X^x$, $Z^a$ \\
query state & $\ket{\Theta(x)} = X^x\ket{\Theta_s}$ & $n$-qubit state carrying the index $x$ into the weak query \\
seed state & $\ket{\Theta_s} = \sum_{w} \sqrt{p_w}\,\ket{D_w^n}$ & generates every query state; $p_w$ is its weight distribution \\
{\raggedright maximum weight of correctable $Z$ errors\par} & $r$ & the logical subspace can correct $Z$ errors of weight up to $r$ on each query block \\
matched query power & $\eta = \min_x \left|\braket{x|\Theta(x)}\right|^2$ & population of $\ket{\Theta(x)}$ on the index $\ket{x}$ it stands for; bounded by $\eta \leq \eta^* = M_r^{-1}$ \\
aggregator & $U_{w^*}$, $\mathsf{Agg}(\cdot) = U_{w^*}\,(\cdot)\,U_{w^*}^{\dagger}$ & writes the response into the data block when the \textit{response count} $W$, the number of responding query blocks, reaches the threshold $w^*$ \\
\hline
\multicolumn{3}{@{}l}{\textit{Combinatorics and entropies}}\\
Hamming weight & $|a|$ & number of nonzero entries of a bitstring \\
Dicke state & $\ket{D_w^n} = \binom{n}{w}^{-1/2}\sum_{|z|=w}\ket{z}$ & uniform superposition over weight-$w$ bitstrings \\
Hamming ball size & $M_r = \sum_{j=0}^{r}\binom{n}{j}$ & number of correctable $Z$ errors of one query block \\
binary entropy & $H(\alpha)$ & $-\alpha\log_2\alpha - (1-\alpha)\log_2(1-\alpha)$, in bits \\
binary relative entropy & $D(a\|b)$ & $a\log_2\frac{a}{b} + (1-a)\log_2\frac{1-a}{1-b}$, in bits \\
\hline\hline
\end{tabular}
\addtocounter{table}{-1}
\caption{Notation used throughout the appendices (continued).}
\end{table}

\subsection{Knill--Laflamme and error orthogonality conditions}\label{SM_sec:prelim_klc}

The Knill--Laflamme conditions and the error orthogonality conditions recur throughout the appendices, and we state both here for reference.
A projector $P$ satisfies the \textit{Knill--Laflamme conditions} (KLC)~\cite{Knill1997} for an error set $\{E_a\}$ if
\begin{equation}\label{SM_eq:KLC_general}
    P E_a^{\dagger} E_b P = C^{a,b} P
\end{equation}
for all pairs of errors, where the numbers $C^{a,b}$ form the \textit{KL matrix} $C$.
The KLC guarantee the existence of a recovery channel that undoes any noise channel whose Kraus operators are linear combinations of the errors $E_a$~\cite{Knill1997, Nielsen_Chuang2010}.
When the KL matrix has full rank, the code is \textit{non-degenerate}.

The conditions invoked most often are the \textit{error orthogonality conditions} (EOC) on the query states, which require that for all $x \in \{0,1\}^n$ and all $a, b \in \{0,1\}^n$ with $|a|, |b| \leq r$,
\begin{equation}\label{SM_eq:error_orthogonality}
    \bra{\Theta(x)} Z^a Z^b \ket{\Theta(x)} = \delta_{a,b},
\end{equation}
where $r$ is the maximum weight of $Z$ errors on each query block that the logical subspace can correct.
We call Eq.~\eqref{SM_eq:error_orthogonality} the EOC for $r$ errors.
The EOC imply that the logical subspace satisfies the KLC for a non-degenerate code, as we prove in Lemma~\ref{lem:global_KLC_OD}.

\subsection{Channel distances}\label{SM_sec:prelim_channels}

\begin{lemma}[Trace and diamond norm facts~\cite{Watrous2018}]\label{lem:channel_norm_facts}
    Let $\Lambda$, $\Lambda_1$, $\Lambda_2$, $\Lambda_3$ be quantum channels with the same input space, where $\Lambda_1$, $\Lambda_2$, and $\Lambda_3$ also share their output space, and let $\rho$, $\sigma$ be states on the input space, possibly entangled with a reference system on which all channels act as the identity.
    Then
    \begin{enumerate}[label=(\roman*)]
        \item $\left\| \Lambda_1(\rho) - \Lambda_2(\rho) \right\|_1 \leq \left\| \Lambda_1 - \Lambda_2 \right\|_{\diamond}$,
        \item $\left\| \Lambda(\rho) - \Lambda(\sigma) \right\|_1 \leq \left\| \rho - \sigma \right\|_1$,
        \item $\left\| \Lambda_1 - \Lambda_2 \right\|_{\diamond} \leq 2$,
        \item $\left\| \sum_i p_i \Lambda_i - \Lambda \right\|_{\diamond} \leq \sum_i p_i \left\| \Lambda_i - \Lambda \right\|_{\diamond}$ for any probability distribution $p_i$ and channels $\Lambda_i$,
        \item $\left\| \Lambda_1 - \Lambda_3 \right\|_{\diamond} \leq \left\| \Lambda_1 - \Lambda_2 \right\|_{\diamond} + \left\| \Lambda_2 - \Lambda_3 \right\|_{\diamond}$,
        \item $\left\| \Theta \circ \Lambda_1 \circ \Psi - \Theta \circ \Lambda_2 \circ \Psi \right\|_{\diamond} \leq \left\| \Lambda_1 - \Lambda_2 \right\|_{\diamond}$ for any channels $\Theta$ and $\Psi$ composable with $\Lambda_1$ and $\Lambda_2$,
        \item $\left\| \Lambda_1 \otimes \mathcal{I}_B - \Lambda_2 \otimes \mathcal{I}_B \right\|_{\diamond} = \left\| \Lambda_1 - \Lambda_2 \right\|_{\diamond}$ for any register $B$.
    \end{enumerate}
\end{lemma}
\begin{proof}
    Item (i) holds because the diamond norm is the largest trace distance between the outputs of the two channels over all inputs, including entangled ones.
    Item (ii) is the data processing inequality for the trace norm, and taking $\Lambda$ to be a partial trace covers the special case used most often below.
    Item (iii) follows from item (i), since the outputs are states and the trace norm of a difference of two states is at most $2$.
    Item (iv) applies the triangle inequality to $\sum_i p_i \left( \Lambda_i - \Lambda \right)$ and uses $\left\| p_i \left( \Lambda_i - \Lambda \right) \right\|_{\diamond} = p_i \left\| \Lambda_i - \Lambda \right\|_{\diamond}$.
    Item (v) is the triangle inequality for the diamond norm.
    Item (vi) holds because the supremum defining the diamond norm runs over all input states entangled with a reference system, so the states prepared by $\Psi$ form a subset of these inputs, and the postprocessing $\Theta$ does not increase the trace distance of the two outputs by item (ii).
    Item (vii) is the stability of the diamond norm under tensoring with an identity channel, which holds because the reference system in its definition can absorb $B$.
\end{proof}

\begin{lemma}[Trace distance between pure states~\cite{Nielsen_Chuang2010}]\label{lem:pure_state_trace_dist}
    For any pure states $\ket{\psi}$ and $\ket{\phi}$,
    \begin{equation}
        \left\| \ket{\psi}\bra{\psi} - \ket{\phi}\bra{\phi} \right\|_1
        = 2\sqrt{1 - \left|\braket{\psi|\phi}\right|^2}
        \leq 2 \left\| \ket{\psi} - \ket{\phi} \right\|_2.
    \end{equation}
\end{lemma}
\begin{proof}
    The equality is standard, so we prove only the inequality.
    It follows from $1 - \left|\braket{\psi|\phi}\right|^2 \leq 2\left(1 - \left|\braket{\psi|\phi}\right|\right) \leq 2\left(1 - \mathrm{Re}\braket{\psi|\phi}\right) = \left\| \ket{\psi} - \ket{\phi} \right\|_2^2$.
\end{proof}

\begin{lemma}[Hybrid argument~\cite{Bennett1997}]\label{lem:hybrid_argument}
    Consider a process that prepares a state and then applies channels $V_0, \Lambda_1, V_1, \ldots, \Lambda_T, V_T$ in sequence.
    If each $\Lambda_t$ is replaced by a channel $\Lambda_t'$ with $\left\| \Lambda_t - \Lambda_t' \right\|_{\diamond} \leq \epsilon_t$, then the final state changes by at most $\sum_{t=1}^{T} \epsilon_t$ in trace norm, and the outcome distribution of any measurement of the final state changes by at most $\sum_{t=1}^{T} \epsilon_t$ in total variation distance.
\end{lemma}
\begin{proof}
    Let $\rho$ denote the prepared state, and for $0 \leq t \leq T$ let
    \begin{equation*}
        \rho_t = V_T \circ \Lambda_T \circ \cdots \circ V_t \circ \Lambda_t' \circ V_{t-1} \circ \cdots \circ \Lambda_1' \circ V_0 (\rho)
    \end{equation*}
    be the final state of the process in which $\Lambda_1, \ldots, \Lambda_t$ are replaced by $\Lambda_1', \ldots, \Lambda_t'$ and all other channels are unchanged, so $\rho_0$ is the final state of the original process and $\rho_T$ is the final state of the fully replaced process.
    The processes producing $\rho_{t-1}$ and $\rho_t$ agree before the $t$-th replaced channel, so the same state enters $\Lambda_t$ in one and $\Lambda_t'$ in the other.
    The outputs of these two channels differ by at most $\left\| \Lambda_t - \Lambda_t' \right\|_{\diamond} \leq \epsilon_t$ in trace norm by Lemma~\ref{lem:channel_norm_facts}(i), and the subsequent channels $V_T \circ \Lambda_T \circ \cdots \circ V_t$, common to both processes, do not increase this distance by Lemma~\ref{lem:channel_norm_facts}(ii), so $\left\| \rho_t - \rho_{t-1} \right\|_1 \leq \epsilon_t$.
    The triangle inequality then gives
    \begin{equation*}
        \left\| \rho_T - \rho_0 \right\|_1 \leq \sum_{t=1}^{T} \left\| \rho_t - \rho_{t-1} \right\|_1 \leq \sum_{t=1}^{T} \epsilon_t,
    \end{equation*}
    and the total variation distance is at most half the trace distance.
\end{proof}

\subsection{Tail bounds and binomial sums}\label{SM_sec:prelim_tails}

\begin{lemma}[Entropy bounds on binomial sums]\label{lem:entropy_binomial}
    Let $n \geq 1$.
    For every integer $0 \leq w \leq n$,
    \begin{equation}\label{eq:binom_entropy_lower}
        \binom{n}{w} \geq \frac{1}{n+1}\, 2^{nH(w/n)},
    \end{equation}
    and the partial binomial sums $M_r = \sum_{k=0}^{r} \binom{n}{k}$ satisfy
    \begin{equation}\label{eq:entropy_binomial}
        M_r \leq 2^{nH(\lambda)} \quad \text{for any } 0 < \lambda \leq \tfrac{1}{2} \text{ and integer } r \leq \lambda n.
    \end{equation}
    In particular, taking $w = r$ and $\lambda = r/n$ gives $\frac{1}{n+1}\,2^{nH(r/n)} \leq \binom{n}{r} \leq M_r \leq 2^{nH(r/n)}$ for every integer $1 \leq r \leq n/2$.
    With the saturated binary entropy $\bar{H}(x) := H\left( \min\{ x, 1/2 \} \right)$, the bound extends to every integer $0 \leq r \leq n$,
    \begin{equation}\label{eq:entropy_binomial_saturated}
        M_r \leq 2^{n \bar{H}(r/n)}.
    \end{equation}
\end{lemma}

\begin{proof}
    For Eq.~\eqref{eq:binom_entropy_lower}, the cases $w \in \{0, n\}$ hold since $H(0) = H(1) = 0$, so assume $0 < w < n$ and set $p = w/n$.
    The binomial distribution $B_p(k) = \binom{n}{k} p^k (1-p)^{n-k}$ is maximized at $k = w$, since the ratio $B_p(k+1)/B_p(k) = \frac{n-k}{k+1}\cdot\frac{p}{1-p}$ exceeds $1$ for $k < w$ and is less than $1$ for $k \geq w$.
    Therefore $B_p(w)$ is the largest of the $n+1$ non-negative terms summing to $1$, so $B_p(w) \geq 1/(n+1)$, and dividing by $p^{w}(1-p)^{n-w} = 2^{-nH(w/n)}$ gives Eq.~\eqref{eq:binom_entropy_lower}.

    For Eq.~\eqref{eq:entropy_binomial}, the normalization of $B_\lambda$ gives
    \begin{equation}
        1 \geq \sum_{k=0}^{r} \binom{n}{k}\, \lambda^k (1-\lambda)^{n-k} \geq 2^{-nH(\lambda)} \sum_{k=0}^{r} \binom{n}{k},
    \end{equation}
    where the second inequality holds because $\lambda^k (1-\lambda)^{n-k} = (1-\lambda)^n \left(\frac{\lambda}{1-\lambda}\right)^{k}$ is non-increasing in $k$ for $\lambda \leq 1/2$, so every term with $k \leq \lambda n$ is at least its value at $k = \lambda n$, which is $(1-\lambda)^n \left(\frac{\lambda}{1-\lambda}\right)^{\lambda n} = \lambda^{\lambda n} (1-\lambda)^{(1-\lambda) n} = 2^{-nH(\lambda)}$.

    For Eq.~\eqref{eq:entropy_binomial_saturated}, the case $1 \leq r \leq n/2$ is Eq.~\eqref{eq:entropy_binomial} with $\lambda = r/n$, the case $r = 0$ holds with equality since $M_0 = 1$, and the case $r > n/2$ follows from $M_r \leq 2^n$ and $\bar{H}(r/n) = H(1/2) = 1$.
\end{proof}

\begin{lemma}[Entropy cost of the floor in $r = \lfloor \alpha n \rfloor$]\label{lem:entropy_floor}
    For all $0 < x \leq y \leq 1/2$,
    \begin{equation}\label{eq:entropy_increment}
        0 \leq H(y) - H(x) \leq (y - x) \log_2 \frac{1-x}{x}.
    \end{equation}
    In particular, for $0 < \alpha \leq 1/4$ and $n \geq 2/\alpha$, the integer $r = \lfloor \alpha n \rfloor$ satisfies
    \begin{equation}\label{eq:entropy_floor}
        2^{n \left( H(2\alpha) - H(2r/n) \right)} \leq \left( \frac{1-\alpha}{\alpha} \right)^{2}.
    \end{equation}
\end{lemma}
\begin{proof}
    The derivative $H'(t) = \log_2 \frac{1-t}{t}$ is non-negative and decreasing on $(0, 1/2]$.
    Non-negativity gives the first inequality of Eq.~\eqref{eq:entropy_increment}, and integrating $H'(t) \leq H'(x)$ over $t \in [x, y]$ gives the second.

    For Eq.~\eqref{eq:entropy_floor}, the floor satisfies $r > \alpha n - 1$, so $n \geq 2/\alpha$ gives $2r/n > 2\alpha - 2/n \geq \alpha$, and $2\alpha - 2r/n \leq 2/n$.
    Applying Eq.~\eqref{eq:entropy_increment} with $x = 2r/n$ and $y = 2\alpha$, which satisfy $0 < x \leq y \leq 1/2$, and then $\log_2 \frac{1-x}{x} \leq \log_2 \frac{1-\alpha}{\alpha}$ by $x \geq \alpha$, gives $H(2\alpha) - H(2r/n) \leq \frac{2}{n} \log_2 \frac{1-\alpha}{\alpha}$.
\end{proof}

\begin{lemma}[Chernoff bound, relative entropy form~\cite{CoverThomas2006}]\label{lem:chernoff_KL}
    Let $W \sim \mathrm{Bin}(n, q)$.
    For every $a$ with $q \leq a \leq 1$,
    \begin{equation}
        \Pr[W \geq a n] \leq 2^{-n D(a \| q)},
    \end{equation}
    and for every $a$ with $0 \leq a \leq q$,
    \begin{equation}
        \Pr[W \leq a n] \leq 2^{-n D(a \| q)}.
    \end{equation}
\end{lemma}

\begin{lemma}[Monotonicity of the binary relative entropy]\label{lem:divergence_monotone}
For any $a \in (0,1]$, the following hold.
\begin{enumerate}[label=(\roman*)]
        \item $D(a \| b)$ is continuous and strictly decreasing in $b$ on $(0, a]$,
        \item $D(a \| a) = 0$ and $D(a \| b) \to \infty$ as $b \to 0^{+}$,
        \item for every $t > 0$ there is a unique $b^* \in (0, a)$ with $D(a \| b^*) = t$,
        \item for this $b^*$, a given $b \in (0, a)$ satisfies $D(a \| b) > t$ if and only if $b < b^*$.
    \end{enumerate}

\end{lemma}
\begin{proof}
For item (i) with $a < 1$, differentiating in $b$ gives $\partial_b D(a \| b) = \frac{b - a}{b(1-b)} \log_2 e$, which is negative for $0 < b < a$, so $D(a \| b)$ is continuous and strictly decreasing in $b$ on $(0, a]$.
For $a = 1$ the divergence is $D(1 \| b) = \log_2 \frac{1}{b}$, which is continuous and strictly decreasing on $(0, 1]$.
For item (ii), both logarithms in $D(a \| a)$ vanish, so $D(a \| a) = 0$, and as $b \to 0^{+}$ the term $a \log_2 \frac{a}{b}$ diverges to $\infty$ while $(1-a)\log_2 \frac{1-a}{1-b}$ stays bounded, under the convention $0 \log_2 0 = 0$ when $a = 1$, so $D(a \| b) \to \infty$.
For item (iii), items (i) and (ii) show that $D(a \| b)$ decreases continuously and strictly from $\infty$ to $0$ as $b$ increases from $0$ to $a$, so the intermediate value theorem gives a solution $b^* \in (0, a)$ of $D(a \| b^*) = t$, and strict monotonicity makes this solution unique.
For item (iv), since $D(a \| b)$ is strictly decreasing in $b$ on $(0, a)$ by item (i), the inequality $D(a \| b) > t = D(a \| b^*)$ holds if and only if $b < b^*$.
\end{proof}

\clearpage
\section{Distillation of Boolean Oracles under Adversarial Noise}\label{SM_sec:adv}
This section gives the full proof of the oracle distillation protocol for Boolean oracles under adversarial noise.
Section~\ref{sec:grover_distillation} presented a high-level description of this protocol.

We first recall the oracle families involved.
The target oracle family is the family of \textit{ideal Boolean oracles} $\mathfrak{O}_F = \{\mathcal{O}_f\}_{f \in F}$, where $F$ is the set of all Boolean functions $f \colon \{0,1\}^n \to \{0,1\}^m$.
The Boolean oracle $\mathcal{O}_f$ is the unitary channel of the unitary
\begin{equation}
    O_f \ket{x}\ket{y} = \ket{x} Z^{f(x)} \ket{y},
\end{equation}
acting on the $n$-qubit \textit{index register} and the $m$-qubit \textit{response register}.
The raw oracle family is the family of \textit{weight-$w_P$ noisy Boolean oracles} $\tilde{\mathfrak{O}}_F^{\mathrm{adv}} = \{\tilde{\mathcal{O}}_f^{\mathrm{adv}}\}_{f \in F}$.
Each noisy oracle factors as
\begin{equation}
    \tilde{\mathcal{O}}_f^{\mathrm{adv}} = \mathcal{E}_f \circ \mathcal{O}_f,
\end{equation}
where the noise channel $\mathcal{E}_f$ is a \textit{weight-$w_P$ noise channel}.
A weight-$w_P$ noise channel admits a Kraus representation in which each Kraus operator is a linear combination of Pauli strings of weight at most $w_P$ supported on the index register.
If each Kraus operator is a linear combination of $Z$ strings only, we call the noise channel a \textit{weight-$w_P$ phase-noise channel} and write it as $\mathcal{E}_f^Z$.
We denote the family of noisy Boolean oracles carrying such channels by $\tilde{\mathfrak{O}}_F^{\mathrm{adv},Z} = \{\tilde{\mathcal{O}}_f^{\mathrm{adv},Z}\}_{f \in F}$, where $\tilde{\mathcal{O}}_f^{\mathrm{adv},Z} = \mathcal{E}_f^Z \circ \mathcal{O}_f$.
The main result of this section is the following theorem.
\booleanodadv*

The rest of this section is organized as follows.
Section~\ref{SM_sec:stage1_new} proves that stage 1 shapes any weight-$w_P$ noise channel into a weight-$w_P$ phase-noise channel via the repetition code.
Section~\ref{SM_sec:stage2_OD} describes the stage 2 protocol and proves that the index logical subspace satisfies the KLC.
We then replace the noise channels of the $2L$ noisy oracle calls by a single equivalent channel acting before recovery (Lemma~\ref{lem:reduction_OD}), and bound the response aggregation error as a classical probability (Sec.~\ref{SM_sec:response_aggregation_OD}).
Section~\ref{SM_sec:error_bound_OD} assembles these pieces into the proof of Theorem~\ref{thm:boolean_OD_adv}.
Finally, Sec.~\ref{SM_sec:gate_complexity_OD} analyzes the gate and ancilla complexity of the protocol.

\subsection{Stage 1 protocol}\label{SM_sec:stage1_new}

The repetition code converts any weight-$w_P$ noise channel into a weight-$w_P$ phase-noise channel.
We prove this noise-conversion lemma as Lemma~\ref{lem:stage1_rewrite} and then derive the stage 1 protocol as Corollaries~\ref{cor:stage1} and \ref{cor:stage1_depol}.

The encoding channel $\mathsf{Enc}_{\mathrm{rep}}$ maps one qubit into a 3-qubit repetition code block,
\begin{equation}
    \mathsf{Enc}_{\mathrm{rep}}(\rho) = V \rho\, V^{\dagger}, \quad V = \ket{000}\bra{0} + \ket{111}\bra{1},
\end{equation}
and encoding $k$ qubits into $k$ blocks is the tensor product channel $\mathsf{Enc}_{\mathrm{rep}}^{\otimes k}$.
The codespace projector for one block is $\Pi = \ket{000}\bra{000} + \ket{111}\bra{111}$.
The recovery channel $\mathsf{Rec}_{\mathrm{rep}}$ acts on one block with four Kraus operators
\begin{equation}
    D_s = V^{\dagger} \Pi\, X_s, \quad s \in \{0,1,2,3\}
\end{equation}
where $X_0 = I$ and $X_s$ ($s=1,2,3$) is a Pauli $X$ on the $s$-th qubit of the block.
One can verify $\sum_{s=0}^{3} D_s^{\dagger} D_s = I$, so this defines a valid quantum channel.
Recovery on $k$ blocks is the tensor product channel $\mathsf{Rec}_{\mathrm{rep}}^{\otimes k}$.

\begin{lemma}[Noise conversion by repetition code]\label{lem:stage1_rewrite}
    Let $\mathcal{E}$ be a $k$-qubit weight-$w_P$ noise channel. Then
    \begin{equation}
        \mathsf{Rec}_{\mathrm{rep}}^{\otimes k} \circ (\mathcal{E} \otimes \mathcal{I}) \circ \mathsf{Enc}_{\mathrm{rep}}^{\otimes k} = \mathcal{E}_Z
    \end{equation}
    where $\mathcal{E} \otimes \mathcal{I}$ acts as $\mathcal{E}$ on the third qubit of each code block and identity on the rest, and $\mathcal{E}_Z$ is a weight-$w_P$ phase-noise channel.
\end{lemma}
\begin{proof}
    Each Kraus operator of $\mathcal{E}$ is a linear combination of $k$-qubit Pauli strings,
    \begin{equation}
        K_i = \sum_{a,b \in \{0,1\}^k} c_{i,a,b}\, X^a Z^b
    \end{equation}
    with $c_{i,a,b} = 0$ unless $|\mathrm{supp}(a) \cup \mathrm{supp}(b)| \leq w_P$. Under $\mathcal{E} \otimes \mathcal{I}$, each $X^a Z^b$ acts on the third qubit of each code block. Since the noise creates $X$ errors only on the third qubit of each block, only the $s \in \{0,3\}$ branches of the recovery channel contribute. Writing $D_a = (V^{\dagger})^{\otimes k} \Pi^{\otimes k} X^a$ for these branches, we compute
    \begin{equation}\begin{split}
        D_a K_i \Pi^{\otimes k}
        &= (V^{\dagger})^{\otimes k} \Pi^{\otimes k} X^a \sum_{a'} X^{a'} \left(\sum_{b} c_{i,a',b}\,Z^b\right) \Pi^{\otimes k}\\
        &= (V^{\dagger})^{\otimes k} \sum_{a'} \Pi^{\otimes k}\, X^{a\oplus a'}\, \Pi^{\otimes k} \left(\sum_{b} c_{i,a',b}\,Z^b\right)
    \end{split}\end{equation}
    where we used that $Z$ on any qubit of a repetition code block acts as a logical operator and therefore preserves the codespace, so $Z^b$ commutes with $\Pi^{\otimes k}$. Since $X$ on the third qubit of any block takes the codespace orthogonal to itself, $\Pi^{\otimes k}\, X^{a\oplus a'}\, \Pi^{\otimes k} = 0$ whenever $a \neq a'$, and the sum collapses to
    \begin{equation}
        D_a K_i \Pi^{\otimes k} = (V^{\dagger})^{\otimes k} \Pi^{\otimes k} \left(\sum_{b} c_{i,a,b}\,Z^b\right).
    \end{equation}

    Assembling the full channel and using $V^{\otimes k}(V^{\dagger})^{\otimes k} = \Pi^{\otimes k}$, we get
    \begin{equation}\begin{split}
        \mathsf{Rec}_{\mathrm{rep}}^{\otimes k} \circ (\mathcal{E} \otimes \mathcal{I}) \circ \mathsf{Enc}_{\mathrm{rep}}^{\otimes k}(\rho)
        = &\sum_{a,i} (V^{\dagger})^{\otimes k} \Pi^{\otimes k} \left(\sum_{b} c_{i,a,b}\,Z^b\right) V^{\otimes k}\, \rho\, (V^{\dagger})^{\otimes k} \left(\sum_{b} c_{i,a,b}^*\,Z^b\right) \Pi^{\otimes k} V^{\otimes k}.
    \end{split}\end{equation}
    Since $Z$ on any qubit of a code block acts as logical $Z$, we have per block $V^{\dagger} \Pi\, Z\, V = Z$, and tensoring gives $(V^{\dagger})^{\otimes k} \Pi^{\otimes k}\, Z^b\, V^{\otimes k} = Z^b$ on the logical qubits. The Kraus operators of the resulting channel are
    \begin{equation}
        K'_{i,a} = \sum_{b} c_{i,a,b}\,Z^b
    \end{equation}
    Each $K'_{i,a}$ is a linear combination of $Z$ strings with weight at most $w_P$, so the output is a weight-$w_P$ phase-noise channel $\mathcal{E}_Z$.
\end{proof}

\begin{corof}{lem:stage1_rewrite}{1}[Stage 1 for noisy Boolean oracle]\label{cor:stage1}
    Let $\tilde{\mathcal{O}}_f$ be a weight-$w_P$ noisy Boolean oracle. Encode the $n$ index qubits with the repetition code and leave the response register unencoded. Then
    \begin{equation}
        \mathsf{Rec}_{\mathrm{rep}}^{\otimes n} \circ (\tilde{\mathcal{O}}_f \otimes \mathcal{I}) \circ \mathsf{Enc}_{\mathrm{rep}}^{\otimes n} = \tilde{\mathcal{O}}_f^Z
    \end{equation}
    where $\tilde{\mathcal{O}}_f \otimes \mathcal{I}$ acts as $\tilde{\mathcal{O}}_f$ on the third qubit of each index code block and on the response register, and identity on the rest, and $\tilde{\mathcal{O}}_f^Z$ is a weight-$w_P$ phase-noisy Boolean oracle.
\end{corof}
\begin{proof}
    The Boolean oracle is diagonal in the computational basis, so it commutes with the repetition-code encoding: $(\mathcal{O}_f \otimes \mathcal{I}) \circ \mathsf{Enc}_{\mathrm{rep}}^{\otimes n} = \mathsf{Enc}_{\mathrm{rep}}^{\otimes n} \circ \mathcal{O}_f$.
    The result follows from Lemma~\ref{lem:stage1_rewrite}.
\end{proof}
\begin{corof}{lem:stage1_rewrite}{2}[Stage 1 for i.i.d.\ depolarizing noise after a diagonal unitary]\label{cor:stage1_depol}
    Let $U$ be a unitary on the $n+m$ oracle qubits that is diagonal in the computational basis, write $\mathcal{U}(\cdot) := U(\cdot)U^{\dagger}$ for its channel, and let $p \in [0,1]$.
    Encode all $n+m$ qubits, including the response register, with the repetition code.
    Then
    \begin{equation}\label{eq:stage1_depol}
        \mathsf{Rec}_{\mathrm{rep}}^{\otimes (n+m)} \circ \left( \left( \mathcal{D}_p^{\otimes (n+m)} \circ \mathcal{U} \right) \otimes \mathcal{I} \right) \circ \mathsf{Enc}_{\mathrm{rep}}^{\otimes (n+m)}
        = \left( \mathcal{D}^Z_{2p/3} \right)^{\otimes (n+m)} \circ \mathcal{U} ,
    \end{equation}
    where $\left( \mathcal{D}_p^{\otimes (n+m)} \circ \mathcal{U} \right) \otimes \mathcal{I}$ acts on the third qubit of each code block and identity on the rest.
\end{corof}
\begin{proof}
    Since $U$ is diagonal with entries $u_y$, the encoding isometry satisfies $(U \otimes I)\, V^{\otimes (n+m)} \ket{y} = u_y V^{\otimes (n+m)} \ket{y} = V^{\otimes (n+m)} U \ket{y}$ for every computational basis state $\ket{y}$, so $\left( \mathcal{U} \otimes \mathcal{I} \right) \circ \mathsf{Enc}_{\mathrm{rep}}^{\otimes (n+m)} = \mathsf{Enc}_{\mathrm{rep}}^{\otimes (n+m)} \circ \mathcal{U}$ and it remains to show that $\mathsf{Rec}_{\mathrm{rep}}^{\otimes (n+m)} \circ \left( \mathcal{D}_p^{\otimes (n+m)} \otimes \mathcal{I} \right) \circ \mathsf{Enc}_{\mathrm{rep}}^{\otimes (n+m)} = \left( \mathcal{D}^Z_{2p/3} \right)^{\otimes (n+m)}$.
    On the third qubit of each code block, $\mathcal{D}_p$ applies $I$, $X$, $Y$ or $Z$ with probabilities $1-p$, $p/3$, $p/3$, $p/3$, independently across blocks.
    By Lemma~\ref{lem:stage1_rewrite}, the recovery returns the logical qubit unchanged for $I$ and $X$, and applies a logical $Z$ for $Y = iXZ$ and $Z$, so each logical qubit receives a $Z$ with probability $2p/3$ independently, which is $\left( \mathcal{D}^Z_{2p/3} \right)^{\otimes (n+m)}$.
\end{proof}

\subsection{Stage 2 protocol}\label{SM_sec:stage2_OD}
The stage 2 protocol distills the family of phase-noisy Boolean oracles $\tilde{\mathfrak{O}}_F^{\mathrm{adv},Z}$, the output of the stage 1 protocol, into the family of ideal Boolean oracles $\mathfrak{O}_F$ with precision $\epsilon$.
Throughout, we consider general Boolean functions $f \colon \{0,1\}^n \to \{0,1\}^m$.
We now give a concrete channel-level description, specifying each of the five steps as a map on the data block $D$ and the $L$ query blocks $Q_1, \dots, Q_L$.

\paragraph{Setup.}
The protocol acts on a data block and $L$ query blocks.
The data block consists of an index register $D^{\mathrm{I}}$ of $n$ qubits and a response register $D^{\mathrm{R}}$ of $m$ qubits, so its Hilbert space is $\mathcal{H}_D = \mathcal{H}_{D^{\mathrm{I}}} \otimes \mathcal{H}_{D^{\mathrm{R}}}$ with $\mathcal{H}_{D^{\mathrm{I}}} = (\mathbb{C}^2)^{\otimes n}$ and $\mathcal{H}_{D^{\mathrm{R}}} = (\mathbb{C}^2)^{\otimes m}$.
Each query block has the same structure, an index register $Q_l^{\mathrm{I}}$ of $n$ qubits and a response register $Q_l^{\mathrm{R}}$ of $m$ qubits, whose $j$-th qubit we denote $Q_{l,j}^{\mathrm{R}}$.
Its Hilbert space is $\mathcal{H}_{Q_l} = \mathcal{H}_{Q_l^{\mathrm{I}}} \otimes \mathcal{H}_{Q_l^{\mathrm{R}}}$, and we write $\mathcal{H}_Q = \bigotimes_{l=1}^{L} \mathcal{H}_{Q_l}$ for all query blocks together.
We reserve the index letters $i \in [n]$ for index qubits, $j \in [m]$ for response bits, and $l \in [L]$ for query blocks.
For a unitary $U$ we write $\mathcal{U}(\cdot) := U(\cdot)U^{\dagger}$ for the corresponding channel.
Since the query states are generated from a seed state by Eq.~\eqref{eq:query-state-from-seedstate}, the matched query power is uniform, $\eta_x = \eta$ for all $x$.
We decompose each query state into its matched and mismatched components,
\begin{equation}\label{eq:matched_decomposition_OD}
    \ket{\Theta(x)} = \sqrt{\eta} \ket{x} + \ket{\Theta_{\perp}(x)},  \qquad \braket{x|\Theta_{\perp}(x)} = 0.
\end{equation}
Finally, the correctable error set is the set of bitstrings $\mathcal{K}_r := \{ e \in \{0,1\}^n : |e| \leq r \}$, where the query states satisfy the EOC Eq.~\eqref{SM_eq:error_orthogonality} for $r$ errors, and the error operator associated with a pattern $e$ is the $Z$ string $Z^{e}$ on the index register of a single query block.

\paragraph{Encoding.}
The nontrivial part of the encoding acts only on the index registers $Q_1^{\mathrm{I}}, \dots, Q_L^{\mathrm{I}}$.
We define the index-register encoding isometry $V_{\mathrm{Enc}}^{\mathrm{I}} \colon \mathcal{H}_{D^{\mathrm{I}}} \to \mathcal{H}_{Q^{\mathrm{I}}} \otimes \mathcal{H}_{D^{\mathrm{I}}}$, with $\mathcal{H}_{Q^{\mathrm{I}}} := \bigotimes_{l=1}^{L} \mathcal{H}_{Q_l^{\mathrm{I}}}$, which prepares on each $Q_l^{\mathrm{I}}$ a query state controlled on the index stored in $D^{\mathrm{I}}$, together with the corresponding channel $\mathsf{Enc}^{\mathrm{I}}$,
\begin{equation}\label{eq:V_enc_OD}
    V_{\mathrm{Enc}}^{\mathrm{I}} = \sum_{x} \bigotimes_{l=1}^{L} \ket{\Theta(x)}_{Q_l^{\mathrm{I}}} \otimes \ket{x}\bra{x}_{D^{\mathrm{I}}},
    \qquad
    \mathsf{Enc}^{\mathrm{I}}(\cdot) := V_{\mathrm{Enc}}^{\mathrm{I}}\, (\cdot)\, V_{\mathrm{Enc}}^{\mathrm{I}\dagger}.
\end{equation}
The full encoding isometry $V_{\mathrm{Enc}} \colon \mathcal{H}_{D} \to \mathcal{H}_{Q} \otimes \mathcal{H}_{D}$ additionally initializes every response qubit of the query blocks in $\ket{+}$,
\begin{equation}\label{eq:Enc_OD_decomposition}
    V_{\mathrm{Enc}} = V_{\mathrm{Enc}}^{\mathrm{I}} \otimes \bigotimes_{l=1}^{L} \ket{+}^{\otimes m}_{Q_l^{\mathrm{R}}} \otimes I_{D^{\mathrm{R}}},
    \qquad
    \mathsf{Enc}_{\mathrm{OD}} = \mathsf{Enc}^{\mathrm{I}} \otimes \mathsf{Enc}^{\mathrm{R}} \otimes \mathcal{I}_{D^{\mathrm{R}}},
\end{equation}
where the encoding channel is $\mathsf{Enc}_{\mathrm{OD}}(\rho) = V_{\mathrm{Enc}}\, \rho\, V_{\mathrm{Enc}}^{\dagger}$ and $\mathsf{Enc}^{\mathrm{R}}$ is the state preparation channel that initializes the query-block response registers in $\ket{+}^{\otimes Lm}$.

An efficient circuit implementation of this isometry is given in Sec.~\ref{SM_sec:gate_complexity_OD}.

The encoded index registers form a quantum error-correcting code.
We call its code space the \textit{index logical subspace} $\mathcal{C}^{\mathrm{I}}$, the index-register part of the logical subspace spanned by the logical states of Eq.~\eqref{eq:logical_state}.
It is the image of the index-register encoding isometry,
\begin{equation}\label{eq:code_space_OD}
    \mathcal{C}^{\mathrm{I}} := V_{\mathrm{Enc}}^{\mathrm{I}}\, \mathcal{H}_{D^{\mathrm{I}}}
    = \mathrm{span} \left\{ \ket{\bar{x}^{\mathrm{I}}} : x \in \{0,1\}^n \right\},
    \qquad
    \ket{\bar{x}^{\mathrm{I}}} := V_{\mathrm{Enc}}^{\mathrm{I}} \ket{x} = \bigotimes_{l=1}^{L} \ket{\Theta(x)}_{Q_l^{\mathrm{I}}} \otimes \ket{x}_{D^{\mathrm{I}}} .
\end{equation}
The global correctable error set is $\mathcal{K}_r^{L}$, whose elements $\vec{e} = (e_1, \dots, e_L)$ label the error operators $Z^{\vec{e}} := \bigotimes_{l=1}^{L} Z^{e_l}$ acting on the query-block index registers.
The following lemma shows that $\mathcal{C}^{\mathrm{I}}$ is a non-degenerate code correcting this error set.

\begin{lemma}[Index logical subspace satisfies the KLC]\label{lem:global_KLC_OD}
    Suppose each query state satisfies the EOC Eq.~\eqref{SM_eq:error_orthogonality} for $r$ errors.
    Then the index logical states $\{ \ket{\bar{x}^{\mathrm{I}}} \}_{x \in \{0,1\}^n}$ satisfy the KLC for a non-degenerate code with respect to the error set $\{ Z^{\vec{e}} : \vec{e} \in \mathcal{K}_r^{L} \}$,
    \begin{equation}
        \bra{\bar{x}'^{\mathrm{I}}} Z^{\vec{e}} Z^{\vec{e}\,'} \ket{\bar{x}^{\mathrm{I}}} = \delta_{x,x'}\, \delta_{\vec{e},\vec{e}\,'} .
    \end{equation}
\end{lemma}
\begin{proof}
    The factor on $D^{\mathrm{I}}$ gives $\delta_{x,x'}$, since no error acts there.
    For $x = x'$ the matrix element factors over the blocks as $\prod_{l=1}^{L} \bra{\Theta(x)} Z^{e_l \oplus e'_l} \ket{\Theta(x)}$, and each factor equals $\delta_{e_l, e'_l}$ by Eq.~\eqref{SM_eq:error_orthogonality}.
\end{proof}

The full logical subspace inherits the KLC from the index logical subspace.
Indeed, each logical state of Eq.~\eqref{eq:logical_state} is the index logical state $\ket{\bar{x}^{\mathrm{I}}}$ tensored with the fixed state $\ket{+}^{\otimes Lm}$ on the query-block response registers and the intact input $\ket{y}$ on $D^{\mathrm{R}}$.
Since the errors $Z^{\vec{e}}$ act only on the index registers, every KLC matrix element between logical states factors into the matrix element of Lemma~\ref{lem:global_KLC_OD} times the overlaps of these error-free factors, which contribute $\delta_{y,y'}$.

\paragraph{Weak query.}
$L$ independent phase-noisy Boolean oracles $\left( \tilde{\mathcal{O}}_f^{\mathrm{adv},Z} \right)^{\otimes L}$ are applied, one call per query block, performing a weak query on each.
When the oracle is ideal and the index register of the data block is in $\ket{x}$, the encoding prepares each query block in $\ket{\Theta(x)} \otimes \ket{+}^{\otimes m}$, and the oracle call maps it to the post-query state
\begin{equation}\label{eq:post_query_state_OD}
    \ket{Q_x^f} := O_f \left( \ket{\Theta(x)} \otimes \ket{+}^{\otimes m} \right).
\end{equation}

\paragraph{Response aggregation.}
We say query block $l$ \textit{responds} on bit $j$ when its index agrees with the index in the data block and its $j$-th response qubit is in $\ket{-}$.
The corresponding projector is
\begin{equation}\label{eq:response_projector_OD}
    \Pi_R^{(l,j)} := \sum_{z \in \{0,1\}^n} \ket{z}\bra{z}_{Q_l^{\mathrm{I}}} \otimes \ket{-}\bra{-}_{Q_{l,j}^{\mathrm{R}}} \otimes \ket{z}\bra{z}_{D^{\mathrm{I}}} .
\end{equation}
Any two of these projectors commute.
If they belong to different query blocks, they share only the register $D^{\mathrm{I}}$, on which both are diagonal in the computational basis.
If they belong to the same query block $l$, they share in addition the index register $Q_l^{\mathrm{I}}$, on which both are again diagonal, while their $\ket{-}\bra{-}$ factors act on distinct response qubits.
For each response bit $j$ we define the response count
\begin{equation}\label{eq:response_count_OD}
    W_j := \sum_{l=1}^{L} \Pi_R^{(l,j)},
\end{equation}
whose eigenvalues take values in $\{0, 1, \dots, L\}$, and let $\Pi_{W_j \geq w^*}$ be the spectral projector of $W_j$ onto eigenvalues at least $w^*$.
The aggregator applies a threshold-controlled $Z$ for every response bit,
\begin{equation}\label{eq:aggregator_OD}
    U_{w^*} := \prod_{j=1}^{m} \left( \Pi_{W_j \geq w^*} \otimes Z_{D_j^{\mathrm{R}}} + \left( I - \Pi_{W_j \geq w^*} \right) \otimes I \right)
    = \sum_{v \in \{0,1\}^m} \Pi_v \otimes Z^{v},
\end{equation}
where $Z_{D_j^{\mathrm{R}}}$ acts on the $j$-th response qubit of the data block.
Since the $\Pi_R^{(l,j)}$ all commute, the counts $W_j$ commute with one another, and hence so do the projectors $\Pi_{W_j \geq w^*}$.
Expanding the product therefore gives the second equality, a sum over orthogonal sectors labeled by strings $v \in \{0,1\}^m$, where $v_j$ records whether the count $W_j$ reaches the threshold and the sector projector is $\Pi_v := \prod_{j=1}^{m} \Pi_{W_j \geq w^*}^{\,v_j} \left( I - \Pi_{W_j \geq w^*} \right)^{1-v_j}$. On each sector $\Pi_v$, the aggregator applies exactly the phase flip $Z^{v}$ to the response register of the data block.

The ideal oracle applies $Z^{f(x)}$ regardless of the sector, so the aggregator reproduces its action exactly on the single sector $v = f(x)$.
For $m=1$ the aggregator reduces to Eq.~\eqref{eq:aggregator} in the main text.
The threshold $w^*$ is a parameter of the protocol, chosen by the response aggregation analysis of Sec.~\ref{SM_sec:response_aggregation_OD}.
For the adversarial noise model it suffices to set $w^* = 1$, which makes the aggregation error decay exponentially in $\eta L$ (Corollary~\ref{cor:agg_error_tau0_OD}). The i.i.d.\ noise analysis instead sets $w^* = \eta L/2$ to tolerate response-register errors (Corollary~\ref{cor:agg_error_lowweight_OD}).
The aggregation channel is $\mathsf{Agg}(\cdot) := U_{w^*} (\cdot) U_{w^*}^{\dagger}$.

\paragraph{Uncomputation.}
A second round of phase-noisy Boolean oracles $\left( \tilde{\mathcal{O}}_f^{\mathrm{adv},Z} \right)^{\otimes L}$ is applied, one call per query block, to uncompute the oracle action of the weak query.

\paragraph{Recovery.}
The recovery channel restores the logical subspace exactly under any correctable index noise.
By Lemma~\ref{lem:global_KLC_OD}, the index logical subspace $\mathcal{C}^{\mathrm{I}}$ is a non-degenerate code on the index registers $Q^{\mathrm{I}}$ and $D^{\mathrm{I}}$, so the standard recovery constructed in the sufficiency proof of the KLC~\cite{Knill1997, Nielsen_Chuang2010}, followed by the decoding $V_{\mathrm{Enc}}^{\mathrm{I}\dagger}$, yields an index-register recovery channel $\mathsf{Rec}^{\mathrm{I}} \colon \mathcal{H}_{Q^{\mathrm{I}}} \otimes \mathcal{H}_{D^{\mathrm{I}}} \to \mathcal{H}_{D^{\mathrm{I}}}$ satisfying the recovery identity
\begin{equation}\label{eq:recovery_OD}
    \mathsf{Rec}^{\mathrm{I}} \circ \mathcal{E}^{\mathrm{I}} \circ \mathsf{Enc}^{\mathrm{I}} = \mathcal{I}_{D^{\mathrm{I}}}
\end{equation}
for every channel $\mathcal{E}^{\mathrm{I}}$ whose Kraus operators are linear combinations of the error operators $\{ Z^{\vec{e}} : \vec{e} \in \mathcal{K}_r^{L} \}$ on the query-block index registers.
An explicit instance, the sequential recovery, is constructed in Sec.~\ref{SM_sec:seq_recovery}.
Since $\mathsf{Rec}^{\mathrm{I}}$ acts only on index registers, the full recovery channel $\mathsf{Rec}_{\mathrm{OD}} \colon \mathcal{H}_Q \otimes \mathcal{H}_D \to \mathcal{H}_D$, which additionally discards the query-block response registers, decomposes as
\begin{equation}\label{eq:Rec_OD_decomposition}
    \mathsf{Rec}_{\mathrm{OD}} = \mathsf{Rec}^{\mathrm{I}} \otimes \Tr_{Q^{\mathrm{R}}} \otimes\, \mathcal{I}_{D^{\mathrm{R}}} .
\end{equation}
Correctable index noise is thus undone by $\mathsf{Rec}^{\mathrm{I}}$, and noise on the query-block response registers is removed by the partial trace $\Tr_{Q^{\mathrm{R}}}$.

\paragraph{Distilled oracle.}
Composing the five steps gives the distilled oracle, a channel on the data block built from $2L$ noisy oracle calls.
For generality we insert a fixed $Z$-error pattern on the query-block response registers after each of the two oracle rounds.
A response error pattern is a binary matrix $\tau \in \{0,1\}^{L \times m}$ with rows $\tau_l \in \{0,1\}^m$, and its error operator is $Z^{\tau} := \bigotimes_{l=1}^{L} Z^{\tau_l}$, where $Z^{\tau_l}$ acts on $Q_l^{\mathrm{R}}$.
With a pattern $\tau$ after the weak query and a pattern $\sigma$ after the uncomputation, the distilled oracle is
\begin{equation}\label{eq:distilled_oracle_OD}
    \widehat{\mathcal{O}}{}_f^{\,\tau,\sigma}
    := \mathsf{Rec}_{\mathrm{OD}}
    \circ \mathcal{Z}^{\sigma}
    \circ \left( \tilde{\mathcal{O}}_f^{\mathrm{adv},Z} \right)^{\otimes L}
    \circ \mathsf{Agg}
    \circ \mathcal{Z}^{\tau}
    \circ \left( \tilde{\mathcal{O}}_f^{\mathrm{adv},Z} \right)^{\otimes L}
    \circ \mathsf{Enc}_{\mathrm{OD}} .
\end{equation}
In the noise model of this section the response registers are noiseless, so the distilled oracle is $\widehat{\mathcal{O}}_f := \widehat{\mathcal{O}}{}_f^{\,0,0}$.
The general patterns are used in the i.i.d.\ analysis of Sec.~\ref{SM_sec:iid}.

\subsection{Equivalent noise channel before recovery}

Here we show that the distilled oracle equals one with ideal oracle queries followed by a single noise channel acting just before recovery.
This lets us analyze the bulk of the OD protocol assuming ideal oracle queries.
The mechanism is that the index phase errors from the $2L$ noisy oracle calls commute past every step separating them from the recovery, and merge into one channel on the query-block index registers.

\begin{lemma}[Equivalent noise channel before recovery]\label{lem:reduction_OD}
    Suppose each noisy oracle factorizes as $\tilde{\mathcal{O}}_f^{\mathrm{adv},Z} = \mathcal{E}_f^Z \circ \mathcal{O}_f$, where $\mathcal{E}_f^Z$ is a channel on the index register admitting a Kraus representation in which each Kraus operator is a linear combination of $Z$ strings.
    Then for all response error patterns $\tau$ and $\sigma$, the distilled oracle Eq.~\eqref{eq:distilled_oracle_OD} equals
    \begin{equation}\label{eq:distilled_oracle_reduced_OD}
        \widehat{\mathcal{O}}{}_f^{\,\tau,\sigma}
        = \mathsf{Rec}_{\mathrm{OD}}
        \circ \left(\bigl(\mathcal{E}_f^Z\bigr)^{2}\right)^{\otimes L}
        \circ \mathcal{Z}^{\sigma}
        \circ \mathcal{O}_f^{\otimes L}
        \circ \mathsf{Agg}
        \circ \mathcal{Z}^{\tau}
        \circ \mathcal{O}_f^{\otimes L}
        \circ \mathsf{Enc}_{\mathrm{OD}},
    \end{equation}
    where $\bigl(\mathcal{E}_f^Z\bigr)^{2} := \mathcal{E}_f^Z \circ \mathcal{E}_f^Z$ acts on the index register of each query block.
\end{lemma}
\begin{proof}
    Substituting the factorization into Eq.~\eqref{eq:distilled_oracle_OD} places one error layer $\bigl(\mathcal{E}_f^Z\bigr)^{\otimes L}$ after the weak query and another after the uncomputation.
    Fix a Kraus representation of $\mathcal{E}_f^Z$ in which each Kraus operator is a linear combination of $Z$ strings on the index register.
    It suffices to show that these $Z$ strings commute with every operation separating the two layers.
    They commute with $\mathcal{Z}^{\tau}$, which acts only on the query-block response registers.
    They commute with $\mathsf{Agg}$, because the aggregator Eq.~\eqref{eq:aggregator_OD} is diagonal in the computational basis of the index registers.
    They commute with the uncomputation oracles, because $O_f$ commutes with every operator diagonal in the computational basis of the index register.
    The first error layer therefore moves past these three operations and merges with the second into $\bigl(\mathcal{E}_f^Z\bigr)^{2}$ on each query block.
    Finally, $\left(\bigl(\mathcal{E}_f^Z\bigr)^{2}\right)^{\otimes L}$ commutes with $\mathcal{Z}^{\sigma}$ for the same reason as with $\mathcal{Z}^{\tau}$, and arrives immediately before recovery.
\end{proof}

When $\mathcal{E}_f^Z$ is a weight-$w_P$ phase-noise channel, composing two of its Kraus representations gives Kraus operators that are linear combinations of $Z$ strings of weight at most $2w_P$, so the merged channel $\bigl(\mathcal{E}_f^Z\bigr)^{2}$ is a weight-$2w_P$ phase-noise channel.

The reduced form Eq.~\eqref{eq:distilled_oracle_reduced_OD} depends on the noise only through the channel $\left(\bigl(\mathcal{E}_f^Z\bigr)^{2}\right)^{\otimes L}$ acting just before recovery.
This motivates a more general object.
For any channel $\mathcal{E}_f^{\mathrm{tot}}$ acting on the query-block index registers and any index-register recovery channel $\mathsf{Rec}^{\mathrm{I}}$ satisfying the recovery identity Eq.~\eqref{eq:recovery_OD}, we define the general distilled oracle
\begin{equation}\label{eq:general_distilled_oracle_OD}
    \widehat{\mathcal{O}}{}_f^{\,\tau,\sigma}\bigl[\mathcal{E}_f^{\mathrm{tot}}, \mathsf{Rec}^{\mathrm{I}}\bigr]
    := \mathsf{Rec}_{\mathrm{OD}}
    \circ \mathcal{E}_f^{\mathrm{tot}}
    \circ \mathcal{Z}^{\sigma}
    \circ \mathcal{O}_f^{\otimes L}
    \circ \mathsf{Agg}
    \circ \mathcal{Z}^{\tau}
    \circ \mathcal{O}_f^{\otimes L}
    \circ \mathsf{Enc}_{\mathrm{OD}},
\end{equation}
in which every oracle query is ideal, the index registers carry the noise $\mathcal{E}_f^{\mathrm{tot}}$, the response registers carry the patterns $\tau$ and $\sigma$, and the recovery $\mathsf{Rec}_{\mathrm{OD}}$ is built from the chosen $\mathsf{Rec}^{\mathrm{I}}$ by Eq.~\eqref{eq:Rec_OD_decomposition}.
Lemma~\ref{lem:reduction_OD} states that the distilled oracle of this section is the instance with $\mathcal{E}_f^{\mathrm{tot}} = \left(\bigl(\mathcal{E}_f^Z\bigr)^{2}\right)^{\otimes L}$ and the same recovery channel $\mathsf{Rec}^{\mathrm{I}}$.
Here $\mathcal{E}_f^{\mathrm{tot}}$ need not factorize across query blocks, which is the form needed in the i.i.d.\ analysis of Sec.~\ref{SM_sec:iid}.

\subsection{Response aggregation}\label{SM_sec:response_aggregation_OD}
In this subsection, we define the response aggregation error, relate it to a classical probability, and show that this error is exponentially small in $\eta L$.

In the OD protocol, the aggregator is the only operation between encoding and recovery that acts on the response register of the data block, so it alone is responsible for implementing the correct oracle response $Z^{f(x)}$ there.
To quantify how well it succeeds, we define the response aggregation error for each input basis state $\ket{x}\ket{y}$ with $x\in \{0,1\}^n$ and $y\in \{0,1\}^m$, oracle label $f$, and response error pattern $\tau \in \{0,1\}^{L \times m}$ as
\begin{equation}\label{eq:agg_error_def_OD}
    \epsilon_{\mathrm{agg}}(x,y;f,\tau) := \left\|U_{w^*} Z^{\tau} \ket{Q^f_x}^{\otimes L} \ket{x}\ket{y} -  Z^{\tau} \ket{Q^f_x}^{\otimes L} \ket{x} Z^{f(x)}\ket{y}\right\|_{2} .
\end{equation}
The first term is the state of the protocol just after the aggregator on input $\ket{x}\ket{y}$.
The second term is the reference state, obtained from the same input by replacing the aggregator with the correct response $Z^{f(x)}$ on the data-block response register.
The error pattern is included so that the same quantity applies in the i.i.d.\ analysis of Sec.~\ref{SM_sec:iid}, and in this section only $\tau = 0$ is needed.
To simplify the notation, we fix $x$, $y$, $f$, and $\tau$ for the rest of this subsection, and every statement below holds for each such choice.

We decompose the state of each query block right before the aggregator into a matched and a mismatched component, according to whether its index register matches the data-block index $\ket{x}$.
For the $l$-th query block, combining Eq.~\eqref{eq:post_query_state_OD} with the matched decomposition Eq.~\eqref{eq:matched_decomposition_OD} gives
\begin{equation}
    Z^{\tau_l}\ket{Q^f_x} = \ket{\beta^{(l)}_1} +  \ket{\beta^{(l)}_0}
\end{equation}
where
\begin{equation}
    \ket{\beta^{(l)}_1} := \sqrt{\eta} \ket{x} Z^{f(x) \oplus \tau_l} \ket{+}^{\otimes m}, \qquad
    \ket{\beta^{(l)}_0} := Z^{\tau_l} O_f \left(\ket{\Theta_\perp(x)}\ket{+}^{\otimes m}\right)
\end{equation}
Here $\ket{\beta^{(l)}_1}$ is the matched component and $\ket{\beta^{(l)}_0}$ the mismatched component.
We further define the branch decomposition
\begin{equation}\label{eq:branch_decomposition_OD}
    Z^{\tau}\ket{Q^f_x}^{\otimes L} \ket{x}\ket{y} = \sum_{h \in \{0,1\}^{L}} \ket{B_h},\qquad \ket{B_h} := \bigotimes_{l=1}^L \ket{\beta^{(l)}_{h_l}} \otimes \ket{x}\ket{y}
\end{equation}
where the string $h$ records which query blocks contribute their matched component, $h_l = 1$ when block $l$ does and $h_l = 0$ otherwise.
We call $h$ the \textit{hit pattern} and $\ket{B_h}$ the \textit{branch vector} associated with $h$.

\begin{lemma}[Branch decomposition]\label{lem:branch_decomposition_OD}
    The branch vectors $\{\ket{B_h}\}_{h \in \{0,1\}^L}$ defined in Eq.~\eqref{eq:branch_decomposition_OD} satisfy the following properties.
    \begin{enumerate}
        \item The branch vectors are mutually orthogonal.
        \item Their squared norms are $\left\|\ket{B_h} \right\|_2^2 = \eta^{|h|} (1-\eta)^{L-|h|}$, where $|h|$ denotes the Hamming weight of $h$.
        \item For every response bit $j$, each branch vector is an eigenvector of the response count $W_j$ of Eq.~\eqref{eq:response_count_OD},
        \begin{equation}\label{eq:branch_eigenvalue_OD}
            W_j \ket{B_h} = W_j(h) \ket{B_h},\qquad W_j(h):= \left|\{l \in [L]: h_l=1,\ \tau_{lj} \oplus f(x)_j = 1\}\right| .
        \end{equation}
        On the branch $\ket{B_h}$, the count $W_j(h)$ is the number of matched blocks that respond on bit $j$.
    \end{enumerate}
\end{lemma}
\begin{proof}
    All three properties follow from the index-register structure of the two components.
    The matched component $\ket{\beta^{(l)}_1}$ has its index register in $\ket{x}$.
    The mismatched component $\ket{\beta^{(l)}_0}$ has no support on $\ket{x}$, because $\braket{x|\Theta_{\perp}(x)} = 0$ and both $O_f$ and $Z^{\tau_l}$ are diagonal in the computational basis of the index register.

    For the first property, the inner product $\braket{B_{h'}|B_h}$ factors over the blocks.
    Two distinct hit patterns $h \neq h'$ differ at some block $l$, where the factor $\braket{\beta^{(l)}_{h'_l} | \beta^{(l)}_{h_l}}$ vanishes because the two components have orthogonal index-register supports.

    For the second property, the state $Z^{f(x) \oplus \tau_l} \ket{+}^{\otimes m}$ is normalized, so $\|\ket{\beta^{(l)}_1}\|_2^2 = \eta$.
    The operator $Z^{\tau_l} O_f$ is unitary, so $\|\ket{\beta^{(l)}_0}\|_2^2 = \|\ket{\Theta_{\perp}(x)}\|_2^2 = 1-\eta$ by Eq.~\eqref{eq:matched_decomposition_OD}.
    Multiplying over the $L$ blocks proves the claim.

    For the third property, the data-block index register of $\ket{B_h}$ is in $\ket{x}$, so only the $z = x$ term of the projector Eq.~\eqref{eq:response_projector_OD} survives,
    \begin{equation}
        \Pi_R^{(l,j)} \ket{B_h} = \left( \ket{x}\bra{x}_{Q_l^{\mathrm{I}}} \otimes \ket{-}\bra{-}_{Q_{l,j}^{\mathrm{R}}} \right) \ket{B_h} .
    \end{equation}
    When $h_l = 0$, the projector $\ket{x}\bra{x}_{Q_l^{\mathrm{I}}}$ annihilates the mismatched component, so $\Pi_R^{(l,j)} \ket{B_h} = 0$.
    When $h_l = 1$, the projector $\ket{x}\bra{x}_{Q_l^{\mathrm{I}}}$ acts trivially, and the $j$-th response qubit of block $l$ is in the state $Z^{\tau_{lj} \oplus f(x)_j} \ket{+}$, which equals $\ket{-}$ if $\tau_{lj} \oplus f(x)_j = 1$ and $\ket{+}$ otherwise.
    Hence $\Pi_R^{(l,j)} \ket{B_h}$ equals $\ket{B_h}$ if $h_l = 1$ and $\tau_{lj} \oplus f(x)_j = 1$, and vanishes otherwise.
    Summing over $l \in [L]$ gives Eq.~\eqref{eq:branch_eigenvalue_OD}.
\end{proof}

Now we define the \textit{response vector} $v(h)$ of a hit pattern $h$, whose entries record whether each response count reaches the threshold $w^*$,
\begin{equation}\label{eq:response_vector_OD}
    v(h)_j := \begin{cases}
        1, & W_j(h) \geq w^*\\
        0, & W_j(h) < w^*
    \end{cases}.
\end{equation}
On the branch $\ket{B_h}$, the response vector determines the phase flip that the aggregator applies to the data-block response register.
Specifically, by Lemma~\ref{lem:branch_decomposition_OD}, $\ket{B_h}$ is an eigenvector of every $W_j$ with eigenvalue $W_j(h)$, so it lies in the range of the sector projector $\Pi_{v(h)}$ defined below Eq.~\eqref{eq:aggregator_OD}.
Hence, using Eq.~\eqref{eq:aggregator_OD},
\begin{equation}\label{eq:aggregator_on_branch_OD}
    U_{w^*} \ket{B_h} = \left(\sum_{v \in \{0,1\}^{m}} \Pi_{v} \otimes Z^v_{D^\mathrm{R}} \right)\ket{B_h} = Z^{v(h)}_{D^\mathrm{R}} \ket{B_h} = (-1)^{v(h) \cdot y} \ket{B_h},
\end{equation}
where the last equality holds because the data-block response register is in $\ket{y}$.
In the reference state, the correct response $Z^{f(x)}$ acts on the same register in $\ket{y}$, contributing the phase $(-1)^{f(x) \cdot y}$ instead.
Comparing the two phases branch by branch bounds the aggregation error by a classical probability.
\begin{lemma}[Aggregation error as a classical probability]\label{lem:agg_error_classical_OD}
    Let $H \in \{0,1\}^L$ be a random hit pattern whose entries $H_1, \dots, H_L$ are i.i.d.\ Bernoulli random variables with $\Pr[H_l = 1] = \eta$.
    Then for all $x$, $y$, $f$, and $\tau$, the aggregation error satisfies
    \begin{equation}\label{eq:agg_error_classical_OD}
        \epsilon_{\mathrm{agg}}(x,y;f,\tau) \leq 2\sqrt{ \Pr\left[v(H) \neq f(x)\right]} .
    \end{equation}
\end{lemma}
\begin{proof}
    Substituting the branch decomposition Eq.~\eqref{eq:branch_decomposition_OD} and the aggregator action Eq.~\eqref{eq:aggregator_on_branch_OD} into the definition Eq.~\eqref{eq:agg_error_def_OD} gives
    \begin{equation}
        \epsilon_{\mathrm{agg}}(x,y;f,\tau)
        = \left\| \sum_{h \in \{0,1\}^L} \left((-1)^{v(h)\cdot y} - (-1)^{f(x) \cdot y} \right) \ket{B_h}\right\|_2 ,
    \end{equation}
    where the reference state contributes the second phase.
    By Lemma~\ref{lem:branch_decomposition_OD}, the branch vectors are mutually orthogonal with squared norms $\left\|\ket{B_h}\right\|_2^2 = \eta^{|h|}(1-\eta)^{L-|h|}$, so
    \begin{align}
        \epsilon_{\mathrm{agg}}(x,y;f,\tau)^2
        &= \sum_{h \in \{0,1\}^L}  \left( (-1)^{v(h)\cdot y} - (-1)^{f(x) \cdot y} \right)^2 \left\| \ket{B_h}\right\|_2^2\\
        &\leq 4 \sum_{h: v(h)\neq f(x)} \eta^{|h|} (1-\eta)^{L-|h|}\\
        &= 4 \Pr\left[v(H) \neq f(x)\right] .
    \end{align}
    The inequality holds because each squared coefficient vanishes when $v(h) = f(x)$ and is at most $4$ otherwise.
    The last equality holds because $\eta^{|h|}(1-\eta)^{L-|h|} = \Pr[H = h]$.
    Taking square roots proves the claim.
\end{proof}

\begin{corof}{lem:agg_error_classical_OD}{1}[Aggregation error without response errors]\label{cor:agg_error_tau0_OD}
    Set the threshold to $w^* = 1$.
    Then for all $x$, $y$, and $f$, the aggregation error with $\tau = 0$ satisfies
    \begin{equation}\label{eq:agg_error_tau0_OD}
        \epsilon_{\mathrm{agg}}(x,y;f,0) \leq 2 (1-\eta)^{L/2} \leq 2 e^{-\eta L/2} .
    \end{equation}
\end{corof}
\begin{proof}
    For $\tau = 0$, the eigenvalue in Eq.~\eqref{eq:branch_eigenvalue_OD} reduces to $W_j(h) = |h|$ if $f(x)_j = 1$ and $W_j(h) = 0$ if $f(x)_j = 0$.
    Hence $v(h)_j = f(x)_j$ holds for every $j$ with $f(x)_j = 0$, since the count $W_j(h) = 0$ never reaches the threshold, while for every $j$ with $f(x)_j = 1$ it holds exactly when $|h| \geq 1$.
    Therefore $v(H) \neq f(x)$ occurs only if $|H| = 0$, so
    \begin{equation}
        \Pr\left[v(H) \neq f(x)\right] \leq \Pr\left[|H| = 0\right] = (1-\eta)^L \leq e^{-\eta L} .
    \end{equation}
    Substituting this bound into Lemma~\ref{lem:agg_error_classical_OD} proves the claim.
\end{proof}

\subsection{Precision and query complexity}\label{SM_sec:error_bound_OD}
We now assemble the previous subsections into a bound on the precision $\epsilon$ of the whole protocol.
Two errors add up to $\epsilon$.
The first is the response aggregation error defined in Sec.~\ref{SM_sec:response_aggregation_OD}.
The second is the \textit{recovery error}, $\left\| \mathsf{Rec}^{\mathrm{I}} \circ \mathcal{E}_f^{\mathrm{tot}} \circ \mathsf{Enc}^{\mathrm{I}} - \mathcal{I}_{D^{\mathrm{I}}} \right\|_{\diamond}$, which measures how well the recovery channel $\mathsf{Rec}^{\mathrm{I}}$ corrects the noise $\mathcal{E}_f^{\mathrm{tot}}$ on states encoded in the index logical subspace.
The following lemma states the resulting bound.
\begin{lemma}[Error of the distilled oracle]\label{lem:OD_error}
    For all response error patterns $\tau$ and $\sigma$, every channel $\mathcal{E}_f^{\mathrm{tot}}$ on the query-block index registers, and every recovery channel $\mathsf{Rec}^{\mathrm{I}}$, the general distilled oracle Eq.~\eqref{eq:general_distilled_oracle_OD} satisfies
    \begin{equation}\label{eq:OD_error}
        \left\| \widehat{\mathcal{O}}{}_f^{\,\tau,\sigma}\bigl[\mathcal{E}_f^{\mathrm{tot}}, \mathsf{Rec}^{\mathrm{I}}\bigr] - \mathcal{O}_f \right\|_{\diamond}
        \leq 2 \max_{x,y}\, \epsilon_{\mathrm{agg}}(x,y;f,\tau)
        + \left\| \mathsf{Rec}^{\mathrm{I}} \circ \mathcal{E}_f^{\mathrm{tot}} \circ \mathsf{Enc}^{\mathrm{I}} - \mathcal{I}_{D^{\mathrm{I}}} \right\|_{\diamond} .
    \end{equation}
\end{lemma}
\begin{proof}
    We compare the distilled oracle with a reference evolution in which the aggregator is replaced by an ideal oracle query, in three steps.
    First, we cut the protocol at the aggregator and define the reference evolution.
    Second, we show that the reference evolution, continued by the rest of the protocol, equals the ideal oracle followed by $\mathsf{Rec}^{\mathrm{I}} \circ \mathcal{E}_f^{\mathrm{tot}} \circ \mathsf{Enc}^{\mathrm{I}}$ on the index register of the data block.
    Third, we show that the actual and the reference evolution differ on each input basis state by exactly the response aggregation error, and combine the two steps into the diamond norm bound Eq.~\eqref{eq:OD_error}.

    \paragraph{Actual and reference factorizations.}
    We cut the general distilled oracle just after the aggregator.
    The part of the protocol before the cut is the isometry
    \begin{equation}\label{eq:act_isometry_OD}
        V^{\tau}_{\mathrm{act}} := U_{w^*} Z^{\tau} O_f^{\otimes L} V_{\mathrm{Enc}}
    \end{equation}
    from $\mathcal{H}_D$ to $\mathcal{H}_Q \otimes \mathcal{H}_D$, and the remaining steps form the channel
    \begin{equation}\label{eq:post_rec_channel_OD}
        \Lambda := \mathsf{Rec}_{\mathrm{OD}}
        \circ \mathcal{E}_f^{\mathrm{tot}}
        \circ \mathcal{Z}^{\sigma}
        \circ \mathcal{O}_f^{\otimes L} ,
    \end{equation}
    so that the general distilled oracle Eq.~\eqref{eq:general_distilled_oracle_OD} factorizes as
    $\widehat{\mathcal{O}}{}_f^{\,\tau,\sigma}\bigl[\mathcal{E}_f^{\mathrm{tot}}, \mathsf{Rec}^{\mathrm{I}}\bigr] = \Lambda \circ \mathcal{V}^{\tau}_{\mathrm{act}}$,
    with $\mathcal{V}^{\tau}_{\mathrm{act}}(\cdot) := V^{\tau}_{\mathrm{act}}\, (\cdot)\, V^{\tau\dagger}_{\mathrm{act}}$.
    The reference evolution removes the aggregator and instead applies one ideal oracle query to the data block before the encoding,
    \begin{equation}\label{eq:ref_isometry_OD}
        V^{\tau}_{\mathrm{ref}} := Z^{\tau} O_f^{\otimes L} V_{\mathrm{Enc}} O_f,
    \end{equation}
    where the rightmost $O_f$ acts on the data block, and $\mathcal{V}^{\tau}_{\mathrm{ref}}$ denotes the corresponding channel.
    On an input basis state $\ket{x}\ket{y}$, the two isometries produce exactly the two states whose distance defines the response aggregation error Eq.~\eqref{eq:agg_error_def_OD}, so
    \begin{equation}\label{eq:isometry_difference_OD}
        \left\| \left( V^{\tau}_{\mathrm{act}} - V^{\tau}_{\mathrm{ref}} \right) \ket{x}\ket{y} \right\|_2
        = \epsilon_{\mathrm{agg}}(x,y;f,\tau) .
    \end{equation}

    \paragraph{Reference evolution up to the recovery error.}
    We show that the reference evolution followed by the remaining steps equals the ideal oracle query followed by $\mathsf{Rec}^{\mathrm{I}} \circ \mathcal{E}_f^{\mathrm{tot}} \circ \mathsf{Enc}^{\mathrm{I}}$ on the index register of the data block.
    Substituting the definitions of $\Lambda$ Eq.~\eqref{eq:post_rec_channel_OD} and of $V^{\tau}_{\mathrm{ref}}$ Eq.~\eqref{eq:ref_isometry_OD} and cancelling the two ideal query layers,
    \begin{align}
        \Lambda \circ \mathcal{V}^{\tau}_{\mathrm{ref}}
        &= \mathsf{Rec}_{\mathrm{OD}}
        \circ \mathcal{E}_f^{\mathrm{tot}}
        \circ \mathcal{Z}^{\sigma}
        \circ \mathcal{O}_f^{\otimes L}
        \circ \mathcal{Z}^{\tau}
        \circ \mathcal{O}_f^{\otimes L}
        \circ \mathsf{Enc}_{\mathrm{OD}}
        \circ \mathcal{O}_f \\
        &= \mathsf{Rec}_{\mathrm{OD}}
        \circ \left( \mathcal{E}_f^{\mathrm{tot}} \otimes \mathcal{Z}^{\sigma \oplus \tau} \right)
        \circ \mathsf{Enc}_{\mathrm{OD}}
        \circ \mathcal{O}_f .
    \end{align}
    The cancellation $\mathcal{O}_f^{\otimes L} \circ \mathcal{Z}^{\tau} \circ \mathcal{O}_f^{\otimes L} = \mathcal{Z}^{\tau}$ holds because $O_f$ and $Z^{\tau}$ are both diagonal in the computational basis, so they commute, and $O_f^2 = I$.
    The two response error patterns then merge as $\mathcal{Z}^{\sigma} \circ \mathcal{Z}^{\tau} = \mathcal{Z}^{\sigma \oplus \tau}$, and the tensor product separates $\mathcal{E}_f^{\mathrm{tot}}$ on the query-block index registers from $\mathcal{Z}^{\sigma \oplus \tau}$ on the query-block response registers.
    The encoding Eq.~\eqref{eq:Enc_OD_decomposition}, the noise, and the recovery Eq.~\eqref{eq:Rec_OD_decomposition} all factor over the query-block index registers, the query-block response registers, and $D^{\mathrm{R}}$, so their composition factors over these three groups as well,
    \begin{equation}
        \mathsf{Rec}_{\mathrm{OD}} \circ \left( \mathcal{E}_f^{\mathrm{tot}} \otimes \mathcal{Z}^{\sigma \oplus \tau} \right) \circ \mathsf{Enc}_{\mathrm{OD}}
        = \left( \mathsf{Rec}^{\mathrm{I}} \circ \mathcal{E}_f^{\mathrm{tot}} \circ \mathsf{Enc}^{\mathrm{I}} \right)
        \otimes \left( \Tr_{Q^{\mathrm{R}}} \circ\, \mathcal{Z}^{\sigma \oplus \tau} \circ \mathsf{Enc}^{\mathrm{R}} \right)
        \otimes \mathcal{I}_{D^{\mathrm{R}}} .
    \end{equation}
    In the response factor, $\mathsf{Enc}^{\mathrm{R}}$ prepares the query-block response registers in $\ket{+}^{\otimes Lm}$, $\mathcal{Z}^{\sigma \oplus \tau}$ adds phases to them, and $\Tr_{Q^{\mathrm{R}}}$ discards them, so this factor drops out, leaving
    \begin{equation}\label{eq:ref_evolution_OD}
        \Lambda \circ \mathcal{V}^{\tau}_{\mathrm{ref}}
        = \left[ \left( \mathsf{Rec}^{\mathrm{I}} \circ \mathcal{E}_f^{\mathrm{tot}} \circ \mathsf{Enc}^{\mathrm{I}} \right) \otimes \mathcal{I}_{D^{\mathrm{R}}} \right] \circ \mathcal{O}_f .
    \end{equation}

    \paragraph{Diamond-norm bound.}
    The triangle inequality (Lemma~\ref{lem:channel_norm_facts}(v)) splits the distance at the reference evolution,
    \begin{equation}\label{eq:OD_error_split}
        \left\| \widehat{\mathcal{O}}{}_f^{\,\tau,\sigma}\bigl[\mathcal{E}_f^{\mathrm{tot}}, \mathsf{Rec}^{\mathrm{I}}\bigr] - \mathcal{O}_f \right\|_{\diamond}
        \leq
        \left\| \Lambda \circ \mathcal{V}^{\tau}_{\mathrm{act}} - \Lambda \circ \mathcal{V}^{\tau}_{\mathrm{ref}} \right\|_{\diamond}
        +
        \left\| \Lambda \circ \mathcal{V}^{\tau}_{\mathrm{ref}} - \mathcal{O}_f \right\|_{\diamond} ,
    \end{equation}
    using the factorization $\widehat{\mathcal{O}}{}_f^{\,\tau,\sigma}\bigl[\mathcal{E}_f^{\mathrm{tot}}, \mathsf{Rec}^{\mathrm{I}}\bigr] = \Lambda \circ \mathcal{V}^{\tau}_{\mathrm{act}}$ in the first term.
    We bound the second term first.
    By Eq.~\eqref{eq:ref_evolution_OD}, its two channels are $\mathsf{Rec}^{\mathrm{I}} \circ \mathcal{E}_f^{\mathrm{tot}} \circ \mathsf{Enc}^{\mathrm{I}}$ and $\mathcal{I}_{D^{\mathrm{I}}}$, each tensored with $\mathcal{I}_{D^{\mathrm{R}}}$ and precomposed with $\mathcal{O}_f$, so dropping the common $\mathcal{O}_f$ (Lemma~\ref{lem:channel_norm_facts}(vi)) and the common $\mathcal{I}_{D^{\mathrm{R}}}$ (Lemma~\ref{lem:channel_norm_facts}(vii)) gives
    \begin{equation}\label{eq:ref_to_ideal_OD}
        \left\| \Lambda \circ \mathcal{V}^{\tau}_{\mathrm{ref}} - \mathcal{O}_f \right\|_{\diamond}
        \leq \left\| \mathsf{Rec}^{\mathrm{I}} \circ \mathcal{E}_f^{\mathrm{tot}} \circ \mathsf{Enc}^{\mathrm{I}} - \mathcal{I}_{D^{\mathrm{I}}} \right\|_{\diamond} ,
    \end{equation}
    which is the second term of Eq.~\eqref{eq:OD_error}.
    In the first term the two channels share the final channel $\Lambda$, which the data processing inequality (Lemma~\ref{lem:channel_norm_facts}(vi)) removes,
    \begin{equation}\label{eq:data_processing_OD}
        \left\| \Lambda \circ \mathcal{V}^{\tau}_{\mathrm{act}} - \Lambda \circ \mathcal{V}^{\tau}_{\mathrm{ref}} \right\|_{\diamond}
        \leq \left\| \mathcal{V}^{\tau}_{\mathrm{act}} - \mathcal{V}^{\tau}_{\mathrm{ref}} \right\|_{\diamond} .
    \end{equation}
    The remaining diamond norm is the worst-case trace distance over inputs entangled with an auxiliary register~\cite{Watrous2018},
    \begin{equation}\label{eq:diamond_sup_OD}
        \left\| \mathcal{V}^{\tau}_{\mathrm{act}} - \mathcal{V}^{\tau}_{\mathrm{ref}} \right\|_{\diamond}
        = \sup_{\ket{\psi}} \left\| \left( \mathcal{V}^{\tau}_{\mathrm{act}} \otimes \mathcal{I}_{\mathrm{aux}} \right)(\psi) - \left( \mathcal{V}^{\tau}_{\mathrm{ref}} \otimes \mathcal{I}_{\mathrm{aux}} \right)(\psi) \right\|_1 ,
    \end{equation}
    where the supremum runs over normalized states $\ket{\psi} \in \mathcal{H}_D \otimes \mathcal{H}_{\mathrm{aux}}$ with an auxiliary space $\mathcal{H}_{\mathrm{aux}}$ of the same dimension as $\mathcal{H}_D$, and $\psi := \ket{\psi}\bra{\psi}$.
    Fix such a state $\ket{\psi}$.
    The two states in Eq.~\eqref{eq:diamond_sup_OD} are the pure states $\ket{\psi'} := \left( V^{\tau}_{\mathrm{act}} \otimes I_{\mathrm{aux}} \right) \ket{\psi}$ and $\ket{\psi'_{\mathrm{ref}}} := \left( V^{\tau}_{\mathrm{ref}} \otimes I_{\mathrm{aux}} \right) \ket{\psi}$, and the trace distance between pure states is bounded by the Euclidean distance, $\left\| \psi' - \psi'_{\mathrm{ref}} \right\|_1 \leq 2 \left\| \ket{\psi'} - \ket{\psi'_{\mathrm{ref}}} \right\|_2$ (Lemma~\ref{lem:pure_state_trace_dist}).
    It remains to bound this Euclidean distance by the basis-state bound Eq.~\eqref{eq:isometry_difference_OD}.
    Decompose $\ket{\psi} = \sum_{x,y} \ket{x}\ket{y} \otimes \ket{\psi_{x,y}}$ in the computational basis of the data block, with $\sum_{x,y} \left\| \ket{\psi_{x,y}} \right\|_2^2 = 1$.
    On the input $\ket{x}\ket{y}$, both isometries output a state carrying the factor $\ket{x}\ket{y}$ on the data block.
    For $V^{\tau}_{\mathrm{ref}}$ this is visible in the reference state of Eq.~\eqref{eq:agg_error_def_OD}, since $Z^{f(x)}\ket{y} = (-1)^{f(x)\cdot y}\ket{y}$.
    For $V^{\tau}_{\mathrm{act}}$ it holds because the aggregator Eq.~\eqref{eq:aggregator_OD} is diagonal in the computational basis of $D^{\mathrm{I}}$ and $D^{\mathrm{R}}$, and no other operation in $V^{\tau}_{\mathrm{act}}$ acts on the data block.
    The difference vectors $\left( V^{\tau}_{\mathrm{act}} - V^{\tau}_{\mathrm{ref}} \right) \ket{x}\ket{y}$ for distinct pairs $(x,y)$ are therefore mutually orthogonal, so the cross terms vanish and the basis-state bound Eq.~\eqref{eq:isometry_difference_OD} gives
    \begin{equation}\label{eq:lifted_bound_OD}
        \left\| \ket{\psi'} - \ket{\psi'_{\mathrm{ref}}} \right\|_2^2
        = \sum_{x,y} \epsilon_{\mathrm{agg}}(x,y;f,\tau)^2 \left\| \ket{\psi_{x,y}} \right\|_2^2
        \leq \max_{x,y}\, \epsilon_{\mathrm{agg}}(x,y;f,\tau)^2 .
    \end{equation}
    Combining Eqs.~\eqref{eq:data_processing_OD} and~\eqref{eq:lifted_bound_OD} with the pure-state bound gives the claimed $2 \max_{x,y} \epsilon_{\mathrm{agg}}(x,y;f,\tau)$ and completes the proof.
\end{proof}

Combining stage 1 and stage 2, we can now prove Theorem~\ref{thm:boolean_OD_adv}.
\begin{proof}[Proof of Theorem~\ref{thm:boolean_OD_adv}]
    Set $L = \left\lceil \frac{2}{\eta} \ln\frac{4}{\epsilon} \right\rceil$ and run the stage 2 protocol with threshold $w^* = 1$.
    By Corollary~\ref{cor:stage1}, each query to the noisy oracle $\tilde{\mathcal{O}}_f^{\mathrm{adv}}$ implements one call to a phase-noisy Boolean oracle $\tilde{\mathcal{O}}_f^{\mathrm{adv},Z} = \mathcal{E}_f^Z \circ \mathcal{O}_f$, where $\mathcal{E}_f^Z$ is a weight-$w_P$ phase-noise channel on the index register.
    The stage 2 protocol consumes $2L$ such calls and outputs the distilled oracle $\widehat{\mathcal{O}}_f = \widehat{\mathcal{O}}{}_f^{\,0,0}$ of Eq.~\eqref{eq:distilled_oracle_OD}, so the full protocol uses $T_{\mathrm{OD}} = 2L$ queries, which is Eq.~\eqref{eq:T_OD_adv}.
    It remains to verify $\| \widehat{\mathcal{O}}_f - \mathcal{O}_f \|_{\diamond} \leq \epsilon$.

    We bound this error by Lemma~\ref{lem:OD_error} with $\tau = \sigma = 0$, and show that its recovery error vanishes here.
    By Lemma~\ref{lem:reduction_OD}, $\widehat{\mathcal{O}}_f$ equals the general distilled oracle Eq.~\eqref{eq:general_distilled_oracle_OD} with $\mathcal{E}_f^{\mathrm{tot}} = \left(\bigl(\mathcal{E}_f^Z\bigr)^{2}\right)^{\otimes L}$.
    The recovery identity Eq.~\eqref{eq:recovery_OD} applies to $\mathcal{E}_f^{\mathrm{tot}}$ for two reasons.
    First, the query states satisfy the EOC Eq.~\eqref{SM_eq:error_orthogonality} for $r$ errors by the assumption of the theorem, so Lemma~\ref{lem:global_KLC_OD} makes the index logical subspace a non-degenerate code with respect to the error set $\{ Z^{\vec{e}} : \vec{e} \in \mathcal{K}_r^{L} \}$.
    Second, the merged channel $\bigl(\mathcal{E}_f^Z\bigr)^{2}$ is a weight-$2w_P$ phase-noise channel, and $2 w_P \leq r$, so each Kraus operator of $\mathcal{E}_f^{\mathrm{tot}}$ is a linear combination of these error operators.
    The recovery error $\left\| \mathsf{Rec}^{\mathrm{I}} \circ \mathcal{E}_f^{\mathrm{tot}} \circ \mathsf{Enc}^{\mathrm{I}} - \mathcal{I}_{D^{\mathrm{I}}} \right\|_{\diamond}$ therefore vanishes, and Lemma~\ref{lem:OD_error} gives
    \begin{equation}
        \left\| \widehat{\mathcal{O}}_f - \mathcal{O}_f \right\|_{\diamond} \leq 2 \max_{x,y}\, \epsilon_{\mathrm{agg}}(x,y;f,0) .
    \end{equation}
    Finally, $L \geq \frac{2}{\eta} \ln\frac{4}{\epsilon}$ gives $e^{-\eta L/2} \leq \epsilon/4$, so Corollary~\ref{cor:agg_error_tau0_OD} gives $\epsilon_{\mathrm{agg}}(x,y;f,0) \leq 2 e^{-\eta L/2} \leq \epsilon/2$ for all $x$ and $y$, hence $\left\| \widehat{\mathcal{O}}_f - \mathcal{O}_f \right\|_{\diamond} \leq \epsilon$.
\end{proof}

\begin{corof}{thm:boolean_OD_adv}{1}[Distillation of Boolean oracles under adversarial noise via Construction 3]\label{cor:boolean_OD_encoding3}
    For any $\alpha \in (0, 0.16]$ and sufficiently large $n$, the two-stage protocol with the query states of Construction 3 distills the family of weight-$\lfloor \alpha n/2 \rfloor$ noisy Boolean oracles $\tilde{\mathfrak{O}}_F^{\mathrm{adv}}$ into the family of ideal Boolean oracles $\mathfrak{O}_F$ with precision $\epsilon$ and query complexity
    \begin{equation}\label{eq:T_OD_adv_C3}
        T_{\mathrm{OD}} = 2 \left\lceil \frac{2}{\xi(n)} \cdot 2^{n(H(\alpha) + 2\alpha)} \ln\frac{4}{\epsilon} \right\rceil,
    \end{equation}
    where $\xi(n) \geq 1$ is any function with $\log\xi(n) = o(n)$.
\end{corof}
\begin{proof}
    By Theorem~\ref{thm:near_optimal_construction}, the query states of Construction 3 satisfy the EOC Eq.~\eqref{SM_eq:error_orthogonality} for $r = \lfloor \alpha n \rfloor$ errors with matched query power $\eta = \xi(n)\, 2^{-n(H(\alpha)+2\alpha)}$.
    This maximum weight of correctable $Z$ errors covers the merged noise, $r = \lfloor \alpha n \rfloor \geq 2 \lfloor \alpha n/2 \rfloor = 2 w_P$.
    Substituting this $\eta$ into Theorem~\ref{thm:boolean_OD_adv} gives Eq.~\eqref{eq:T_OD_adv_C3}.
\end{proof}

\subsection{Explicit circuit implementation and its gate and ancilla complexity}\label{SM_sec:gate_complexity_OD}
Here we give explicit circuits for the two-stage protocol and count the gates and ancilla qubits of one distilled query.
An elementary gate is a single-qubit or two-qubit gate~\cite{Barenco1995}.
The \textit{gate complexity} $G$ of a circuit is the number of elementary gates it uses, and its \textit{ancilla complexity} $A$ is the number of ancilla qubits.
A Toffoli gate decomposes exactly into a constant number of elementary gates~\cite{Barenco1995}, so we count each Toffoli gate as $O(1)$ elementary gates.
All ancilla qubits are initialized in $\ket{0}$.
We state the total complexity first, and the subsections below construct the circuits and count their gates and ancilla qubits.

\begin{theorem}[Gate and ancilla complexity of the two-stage protocol]\label{thm:gate_complexity_OD}
    One distilled query of the two-stage protocol (Secs.~\ref{SM_sec:stage1_new} and~\ref{SM_sec:stage2_OD}), with the sequential recovery of maximal candidate weight $r_{\mathrm{seq}}$ (Sec.~\ref{SM_sec:seq_recovery}) as the index-register recovery channel, admits an exact circuit implementation with
    \begin{equation}\label{eq:total_gate_complexity_OD}
        G = O\bigl( (n^2 M_{r_{\mathrm{seq}}} + m) L\bigr), \qquad A = O\bigl((n+m)L\bigr),
    \end{equation}
    where $M_{r_{\mathrm{seq}}} = |\mathcal{K}_{r_{\mathrm{seq}}}|$ is the number of candidate patterns of the sequential recovery.
\end{theorem}
\begin{proof}
    One distilled query consumes $2L$ noisy queries, and the stage 1 gadget wraps each noisy query in repetition encoding and recovery circuits (Sec.~\ref{SM_sec:stage1_new}), which cost two $\mathsf{CNOT}$ gates per encoded qubit for encoding, two $\mathsf{CNOT}$ gates and one Toffoli-controlled correction per block for syndrome extraction and recovery, and $2(n+m)$ reusable ancilla qubits, contributing $O\bigl((n+m)L\bigr)$ to both complexities.
    The remaining steps of the stage 2 protocol are the encoding $V_{\mathrm{Enc}}$, the aggregator $U_{w^*}$, and the recovery $\mathsf{Rec}_{\mathrm{OD}}$, which applies the index-register recovery channel $\mathsf{Rec}^{\mathrm{I}}$ and discards the query-block response registers (Eq.~\eqref{eq:Rec_OD_decomposition}), so only $\mathsf{Rec}^{\mathrm{I}}$ needs a circuit.
    The rest of this section constructs their circuits, with complexities given by Eqs.~\eqref{eq:enc_gate_complexity}, \eqref{eq:agg_gate_complexity}, and~\eqref{eq:seq_rec_gate_complexity}.
    Summing all contributions gives Eq.~\eqref{eq:total_gate_complexity_OD}.
\end{proof}

\subsubsection{Encoding}
The encoding isometry $V_{\mathrm{Enc}}$ of Eq.~\eqref{eq:V_enc_OD} admits an exact implementation with
\begin{equation}\label{eq:enc_gate_complexity}
    G_{\mathrm{enc}} = O\bigl((n^2+m) L\bigr), \qquad A_{\mathrm{enc}} = (n+m)L .
\end{equation}
We introduce $(n+m)$ ancilla qubits for each of the $L$ query blocks, giving $A_{\mathrm{enc}} = (n+m)L$.
In each query block, we prepare the index register in $\ket{\Theta_s}$ and the response register in $\ket{+}^{\otimes m}$.
Since $\ket{\Theta_s}$ is a permutation-symmetric pure state, preparing it takes $O(n^2)$ elementary gates and no ancilla qubits~\cite[Thm.~2]{Bartschi2019}, and preparing $\ket{+}^{\otimes m}$ takes $m$ Hadamard gates, so preparing all $L$ query blocks takes $O\bigl((n^2+m) L\bigr)$ gates.
Next, the unitary
\begin{equation}\label{eq:def_U_enc_OD}
    U_{\mathrm{Enc}}^{(l)} = \prod_{i=1}^{n} \mathsf{CNOT}_{D_i^{\mathrm{I}} \to Q_{l,i}^{\mathrm{I}}}
\end{equation}
applies a CNOT to the $i$-th qubit of $Q_l^{\mathrm{I}}$, controlled on the $i$-th qubit of $D^{\mathrm{I}}$, for each $i = 1, \ldots, n$.
We apply this unitary to every query block $l = 1, \ldots, L$.
Since $\ket{\Theta(x)} = X^{x} \ket{\Theta_s}$ (Eq.~\eqref{eq:query-state-from-seedstate}), this prepares $\ket{\Theta(x)}$ in each query block conditioned on the index register of the data block, using $nL$ CNOT gates in total.
The total gate count is dominated by the state preparations, giving $G_{\mathrm{enc}} = O\bigl((n^2+m) L\bigr)$.

\subsubsection{Aggregator}
The aggregator $U_{w^*}$ of Eq.~\eqref{eq:aggregator_OD} admits an exact implementation with
\begin{equation}\label{eq:agg_gate_complexity}
    G_{\mathrm{agg}} = O\bigl( (n+m) L \bigr), \qquad A_{\mathrm{agg}} = O(L+n).
\end{equation}
Let $k := \lceil w^* \rceil$.
Since the eigenvalues of $W_j$ are integers, $\Pi_{W_j \geq w^*} = \Pi_{W_j \geq k}$, so the circuit compares each response count against the integer $k$.

We first describe the logic of the circuit.
For each query block $l$, a match ancilla $M_l$ initialized to $\ket{0}$ records whether the index of the query block equals the index of the data block,
\begin{equation}
    \ket{z}_{Q_l^{\mathrm{I}}}\ket{x}_{D^{\mathrm{I}}}\ket{0}_{M_l}
    \longmapsto
    \ket{z}_{Q_l^{\mathrm{I}}}\ket{x}_{D^{\mathrm{I}}}\ket{\delta_{z,x}}_{M_l}.
\end{equation}
A Hadamard gate on every response qubit of the query blocks converts the factor $\ket{-}\bra{-}_{Q_{l,j}^{\mathrm{R}}}$ in Eq.~\eqref{eq:response_projector_OD} into $\ket{1}\bra{1}_{Q_{l,j}^{\mathrm{R}}}$, so query block $l$ responds on bit $j$ exactly when $M_l$ and $Q_{l,j}^{\mathrm{R}}$ are both in $\ket{1}$.
For each response bit $j$, the circuit computes a flag bit $F_l = M_l \wedge Q_{l,j}^{\mathrm{R}}$ for every query block, so that the sum of the flag bits is the eigenvalue of the response count $W_j$.
It then computes this sum, compares it against $k$, applies $Z$ to $D_j^{\mathrm{R}}$ controlled on the outcome $W_j \geq k$, and uncomputes the sum and the flag bits.
The threshold-controlled $Z$ factors for different response bits commute, so processing the response bits one at a time realizes their product $U_{w^*}$ in Eq.~\eqref{eq:aggregator_OD}.
After the last response bit, we reverse the preparation of $K$, apply Hadamard gates again to all query-block response qubits, and reverse the match circuit.
This restores the query registers and returns all ancilla qubits to $\ket{0}$.

We now implement each step and count its gates and ancilla qubits.
To compute the match ancilla of block $l$, we apply a $\mathsf{CNOT}$ from $D_i^{\mathrm{I}}$ to $Q_{l,i}^{\mathrm{I}}$ for each $i = 1, \ldots, n$, mapping $\ket{z}_{Q_l^{\mathrm{I}}}$ to $\ket{z \oplus x}_{Q_l^{\mathrm{I}}}$, so that $z = x$ exactly when $Q_l^{\mathrm{I}}$ is in $\ket{0^n}$.
We apply $X$ to all $n$ qubits of $Q_l^{\mathrm{I}}$, converting this all-zero condition into an all-one condition, apply an $n$-controlled $\mathsf{NOT}$ with these qubits as controls and $M_l$ as the target, and then undo the $X$ and $\mathsf{CNOT}$ gates, restoring $Q_l^{\mathrm{I}}$ to $\ket{z}$.
The clean-ancilla decomposition of Ref.~\cite[Sec.~4.3]{Nielsen_Chuang2010} implements the $n$-controlled $\mathsf{NOT}$ with $2(n-1)$ Toffoli gates and $n-1$ ancilla qubits that start and end in $\ket{0}$ and are reused across blocks.
The match circuit and its inverse therefore use $O(nL)$ gates, the $L$ match ancillas, and $n-1$ shared ancillas.

For the comparisons, we introduce a $q$-qubit register $K$ with $q := \lceil \log_2(L+1) \rceil$ and prepare it in $\ket{k}$ by applying $X$ to the qubits at which the binary expansion of $k$ is one, which fits since $0 \leq k \leq L$.
Preparing and restoring $K$ uses $O(\log L)$ gates, and $K$ is reused for every response bit.

Fix a response bit $j$.
A Toffoli gate with controls $M_l$ and $Q_{l,j}^{\mathrm{R}}$ and target a clean ancilla $F_l$ computes each flag bit, using $2L$ Toffoli gates together with the uncomputation.
To sum the flag bits, we apply the in-place ripple-carry adder of Ref.~\cite[Secs.~2--3]{Cuccaro2004} in a binary tree, storing each sum in one of the two input registers and using one additional qubit for its high output bit.
At level $t$ of the tree there are $O(L/2^t)$ additions on $O(t)$-bit registers at $O(t)$ gates each, so the tree and its inverse use $O\bigl(\sum_{t \geq 1} t L / 2^t\bigr) = O(L)$ gates, the $L-1$ high output bits use $O(L)$ additional qubits, and the one ancilla qubit of the adder is reused between additions.
The final $q$-qubit partial-sum register $S$ contains $\sum_{l=1}^{L} F_l$.
The comparator built from reversible subtraction in Ref.~\cite[Sec.~4.3 and Table~1]{Cuccaro2004} compares $S$ with $K$ using $O(q)$ gates and two ancilla qubits, writing $\mathbf{1}\{s < k\}$ into one of them, denoted by $C$.
An $X$ gate on $C$ puts $C$ in $\ket{1}$ exactly when $s \geq k$, and a controlled-$Z$ from $C$ to $D_j^{\mathrm{R}}$, implemented as a Hadamard gate, a $\mathsf{CNOT}$, and a second Hadamard gate on $D_j^{\mathrm{R}}$, applies $Z$ exactly on the subspace where $W_j \geq k$.
Reversing the $X$ gate, the comparator, the adder tree, and the Toffoli gates returns all workspace of this response bit to $\ket{0}$, and the workspace is reused for the next response bit.
Each response bit therefore uses $O(L)$ gates.

Processing all $m$ response bits, together with the $2mL$ Hadamard gates on the response qubits, uses $O(mL)$ gates.
Combining with the match circuit gives $G_{\mathrm{agg}} = O(nL) + O(mL) = O\bigl((n+m)L\bigr)$.
At any time the circuit uses the $L$ match ancillas, the $L$ flag qubits, $O(L)$ additional qubits in the adder tree, the $q$ qubits of $K$, the two comparator ancillas, and the $n-1$ shared ancillas of the $n$-controlled $\mathsf{NOT}$, so $A_{\mathrm{agg}} = O(L+n)$.

\subsubsection{Recovery: sequential recovery channel}\label{SM_sec:seq_recovery}
We construct an explicit index-register recovery channel, the \textit{sequential recovery}, which detects the index error pattern of each query block by a fixed sequence of projective tests and then applies a single correction.
The \textit{candidate patterns} of this recovery channel $\mathsf{Rec}_{r_{\mathrm{seq}}}$, the error patterns it tests for, are the elements of $\mathcal{K}_{r_{\mathrm{seq}}} = \{ e \in \{0,1\}^n : |e| \leq r_{\mathrm{seq}} \}$, and the channel is parametrized by the \textit{maximal candidate weight} $r_{\mathrm{seq}}$ with $r_{\mathrm{seq}} \leq n$.
For every $r_{\mathrm{seq}} \geq r$, the sequential recovery satisfies the recovery identity Eq.~\eqref{eq:recovery_OD} (Lemma~\ref{lem:seq_recovery_identity}), so it is a valid instance of the recovery channel $\mathsf{Rec}^{\mathrm{I}}$, and taking $r_{\mathrm{seq}} = r$ would suffice for the adversarial protocol.
The freedom to take $r_{\mathrm{seq}} > r$ is used in the i.i.d.\ analysis of Sec.~\ref{SM_sec:iid_seq_recovery}.
It admits an exact circuit implementation with
\begin{equation}\label{eq:seq_rec_gate_complexity}
    G_{\mathrm{rec}} = O\left(n^2 M_{r_{\mathrm{seq}}} L\right), \qquad A_{\mathrm{rec}} = O(n),
\end{equation}
where $M_{r_{\mathrm{seq}}} = |\mathcal{K}_{r_{\mathrm{seq}}}|$ is the number of candidate patterns.

The sequential recovery is a composition of recovery channels, one per query block, applied in order,
\begin{equation}
    \mathsf{Rec}_{r_{\mathrm{seq}}} := \mathsf{Rec}_{r_{\mathrm{seq}}}^{(L)} \circ \mathsf{Rec}_{r_{\mathrm{seq}}}^{(L-1)} \circ \dots \circ \mathsf{Rec}_{r_{\mathrm{seq}}}^{(1)},
\end{equation}
where $\mathsf{Rec}_{r_{\mathrm{seq}}}^{(l)}$ acts on query block $l$ and the index register of the data block.
Each $\mathsf{Rec}_{r_{\mathrm{seq}}}^{(l)}$ performs three steps, first undoing the encoding of query block $l$, then identifying its index error pattern by a sequence of binary projective tests, and finally applying a single correction to the data block.

The first step applies $U_{\mathrm{Enc}}^{(l)\dagger}$, the inverse of the encoding unitary Eq.~\eqref{eq:def_U_enc_OD}.
Writing the encoding unitary as $U_{\mathrm{Enc}}^{(l)} = \sum_{z} X^{z}_{Q_l^{\mathrm{I}}} \otimes \ket{z}\bra{z}_{D^{\mathrm{I}}}$ and using $X^{z} Z^{e} X^{z} = (-1)^{e\cdot z} Z^{e}$ and $\sum_{z} (-1)^{e\cdot z} \ket{z}\bra{z}_{D^{\mathrm{I}}} = \left(Z^{e}\right)_{D^{\mathrm{I}}}$ gives the conjugation identity
\begin{equation}\label{eq:enc_conjugation_OD}
    U_{\mathrm{Enc}}^{(l)\dagger} \left( Z^{e}_{Q_l^{\mathrm{I}}} \otimes I_{D^{\mathrm{I}}} \right) U_{\mathrm{Enc}}^{(l)}
    = Z^{e}_{Q_l^{\mathrm{I}}} \otimes Z^{e}_{D^{\mathrm{I}}}.
\end{equation}
When query block $l$ is corrupted by $Z^{e_l}$ on its index register, the conjugation identity shows that the first step leaves $Q_l^{\mathrm{I}}$ in the corrupted seed state $Z^{e_l}\ket{\Theta_s}$, independently of the data block, and applies the same error $Z^{e_l}$ to $D^{\mathrm{I}}$.
It therefore suffices to detect the error only on $Q_l^{\mathrm{I}}$ and to correct it only on $D^{\mathrm{I}}$.

The second step identifies this state by testing the candidate patterns one at a time.
For each pattern $e \in \{0,1\}^{n}$, we define the projector onto the corrupted seed state,
\begin{equation}\label{eq:def_Pi}
    \Pi_{e} := Z^{e} \ket{\Theta_s}\bra{\Theta_s} Z^{e}.
\end{equation}
Order the candidate patterns in $\mathcal{K}_{r_{\mathrm{seq}}}$ by nondecreasing Hamming weight, with an arbitrary but fixed order within each weight, and write $e' \prec e''$ when $e'$ precedes $e''$.
For each candidate $e' \in \mathcal{K}_{r_{\mathrm{seq}}}$ in this order, the recovery performs the binary projective measurement $\left\{ \Pi_{e'},\, I - \Pi_{e'} \right\}$ on $Q_l^{\mathrm{I}}$.
Once a measurement accepts, the recovery stops testing and performs no further measurement.
Let $\hat{e}_l$ be the accepted candidate, with $\hat{e}_l := 0^n$ when every candidate is rejected, so that the correction below defaults to the identity.
The third step applies the single correction $Z^{\hat{e}_l}$ to $D^{\mathrm{I}}$ and discards $Q_l^{\mathrm{I}}$.
The sequential recovery depends only on the seed state $\ket{\Theta_s}$, the candidate set $\mathcal{K}_{r_{\mathrm{seq}}}$, and the order $\prec$, so it is independent of the label $f$.

We now verify that the sequential recovery satisfies the recovery identity.
\begin{lemma}[Sequential recovery satisfies the recovery identity]\label{lem:seq_recovery_identity}
    For every maximal candidate weight $r_{\mathrm{seq}}$ with $r \leq r_{\mathrm{seq}} \leq n$, the sequential recovery satisfies the recovery identity Eq.~\eqref{eq:recovery_OD},
    \begin{equation}
        \mathsf{Rec}_{r_{\mathrm{seq}}} \circ \mathcal{E}^{\mathrm{I}} \circ \mathsf{Enc}^{\mathrm{I}} = \mathcal{I}_{D^{\mathrm{I}}},
    \end{equation}
    for every channel $\mathcal{E}^{\mathrm{I}}$ whose Kraus operators are linear combinations of the error operators $\{ Z^{\vec{e}} : \vec{e} \in \mathcal{K}_r^{L} \}$.
\end{lemma}
\begin{proof}
    Write $\Lambda := \mathsf{Rec}_{r_{\mathrm{seq}}} \circ \mathcal{E}^{\mathrm{I}} \circ \mathsf{Enc}^{\mathrm{I}}$, a channel on $D^{\mathrm{I}}$.
    Pure-state projectors span the space of operators on $D^{\mathrm{I}}$ and the superoperator $\Lambda$ is linear, so it suffices to verify $\Lambda\left( \ket{\psi}\bra{\psi} \right) = \ket{\psi}\bra{\psi}$ for every pure state $\ket{\psi}$ on $D^{\mathrm{I}}$.

    Fix a Kraus operator $K = \sum_{\vec{e} \in \mathcal{K}_r^{L}} c^{K}_{\vec{e}}\, Z^{\vec{e}}$ of $\mathcal{E}^{\mathrm{I}}$ and a pure state $\ket{\psi}$ on $D^{\mathrm{I}}$.
    We bring the state before the tests into product form.
    The unitary $U_{\mathrm{Enc}}^{(l)\dagger}$ acts on $Q_l^{\mathrm{I}}$ and diagonally on $D^{\mathrm{I}}$, so it commutes with the tests and corrections of the other query blocks, and we may apply the inverse encoding of every block before any test.
    Using $\ket{\Theta(x)} = X^{x} \ket{\Theta_s}$ (Eq.~\eqref{eq:query-state-from-seedstate}), the encoding isometry Eq.~\eqref{eq:V_enc_OD} acts as $V_{\mathrm{Enc}}^{\mathrm{I}} \ket{\psi} = \prod_{l} U_{\mathrm{Enc}}^{(l)} \bigl( \bigotimes_{l} \ket{\Theta_s}_{Q_l^{\mathrm{I}}} \otimes \ket{\psi} \bigr)$, so undoing the encodings and applying the conjugation identity Eq.~\eqref{eq:enc_conjugation_OD} to each factor $Z^{e_l}$ of each error operator gives
    \begin{equation}\label{eq:seq_rec_branch_state}
        \left(\prod_{l} U_{\mathrm{Enc}}^{(l)\dagger} \right) K V_{\mathrm{Enc}}^{\mathrm{I}} \ket{\psi}
        = \sum_{\vec{e} \in \mathcal{K}_r^{L}} c^{K}_{\vec{e}} \bigotimes_{l=1}^{L} \left( Z^{e_l}\ket{\Theta_s} \right)_{Q_l^{\mathrm{I}}} \otimes Z^{e_1 \oplus \cdots \oplus e_L}_{D^{\mathrm{I}}} \ket{\psi}.
    \end{equation}
    We next run the tests on this superposition and track the component with error pattern $\vec{e}$, whose query block $l$ carries $Z^{e_l}\ket{\Theta_s}$ with $e_l \in \mathcal{K}_r$.
    In block $l$, every candidate $e'$ tested before $e_l$ has weight $|e'| \leq |e_l| \leq r$ by the weight ordering, so $e'$ and $e_l$ are distinct elements of $\mathcal{K}_r$, hence $1 \leq |e' \oplus e_l| \leq 2r$, and the EOC Eq.~\eqref{SM_eq:error_orthogonality} give $\Pi_{e'} Z^{e_l} \ket{\Theta_s} = 0$.
    The component therefore falls in the rejection branch of each of these tests and is left unchanged by the rejection projector $I - \Pi_{e'}$, until the test of $e_l$ itself accepts it, since $\Pi_{e_l} Z^{e_l}\ket{\Theta_s} = Z^{e_l}\ket{\Theta_s}$, records $\hat{e}_l = e_l$, and stops the testing of the block.
    Every component thus survives the tests unchanged and imprints its error pattern on the measurement record, so the tests of the $L$ query blocks produce the outcomes $(\hat{e}_1, \ldots, \hat{e}_L) = \vec{e}$ with probability $|c^{K}_{\vec{e}}|^2$ and leave the branch $\vec{e}$ of Eq.~\eqref{eq:seq_rec_branch_state}.
    In this branch, the corrections $Z^{\hat{e}_l}_{D^{\mathrm{I}}}$ multiply to $Z^{e_1 \oplus \cdots \oplus e_L}_{D^{\mathrm{I}}}$ and cancel the accumulated error operator on $D^{\mathrm{I}}$, the registers $Q_l^{\mathrm{I}}$ are discarded, and the branch outputs $\ket{\psi}\bra{\psi}$.
    Summing over the branches and the Kraus operators of $\mathcal{E}^{\mathrm{I}}$ gives $\Lambda\left( \ket{\psi}\bra{\psi} \right) = \sum_{K} \sum_{\vec{e}} |c^{K}_{\vec{e}}|^2\, \ket{\psi}\bra{\psi}$.
    Since $\Lambda$ is a composition of channels, it preserves the trace, so $\sum_{K} \sum_{\vec{e}} |c^{K}_{\vec{e}}|^2 = 1$ and $\Lambda$ fixes every pure state on $D^{\mathrm{I}}$, completing the proof.
\end{proof}

We now give an explicit circuit implementation of the sequential recovery channel, and analyze its gate and ancilla complexity.
Among the three steps of each $\mathsf{Rec}_{r_{\mathrm{seq}}}^{(l)}$, the inverse encoding $U_{\mathrm{Enc}}^{(l)\dagger}$ is a circuit of $n$ $\mathsf{CNOT}$ gates and the correction $Z^{\hat{e}_l}$ is a circuit of at most $n$ $Z$ gates, so the only step that needs a construction is the binary projective measurement $\left\{ \Pi_{e},\, I - \Pi_{e} \right\}$ of a candidate $e \in \mathcal{K}_{r_{\mathrm{seq}}}$.
Let $U_s$ be the preparation unitary of the seed state, $U_s \ket{0^n} = \ket{\Theta_s}$.
We realize this measurement with a single flag ancilla $F$ initialized in $\ket{0}$, by the unitary that applies $X$ to $F$ controlled on $Q_l^{\mathrm{I}}$ being in the corrupted seed state $Z^{e} \ket{\Theta_s}$,
\begin{equation}\label{eq:def_test_unitary}
    \left(\Pi_{e}\right)_{Q_l^{\mathrm{I}}} \otimes X_F + \left(I - \Pi_{e}\right)_{Q_l^{\mathrm{I}}} \otimes I_F
    = Z^{e} U_s \left( \ket{0^n}\bra{0^n} \otimes X_F + \left(I - \ket{0^n}\bra{0^n}\right) \otimes I_F \right) U_s^{\dagger} Z^{e},
\end{equation}
followed by a measurement of $F$, whose outcome $1$ accepts the candidate.
Here $Z^{e}$ and $U_s$ act on $Q_l^{\mathrm{I}}$, and the equality uses $\Pi_{e} = Z^{e} U_s \ket{0^n}\bra{0^n} U_s^{\dagger} Z^{e}$.
The measured flag is reset to $\ket{0}$ and reused for the next test.
Once a flag measurement returns $1$, the remaining tests of the query block are skipped by classical control, realizing the stop rule of the second step, so each query block performs at most $M_{r_{\mathrm{seq}}}$ tests.
Since $\ket{\Theta_s}$ is a permutation-symmetric pure state, $U_s$ takes $O(n^2)$ elementary gates and no ancilla qubits~\cite[Thm.~2]{Bartschi2019}.
The middle factor of the right-hand side applies $X$ to $F$ controlled on $Q_l^{\mathrm{I}}$ being in $\ket{0^n}$.
As in the match circuit of the aggregator, we apply $X$ to all $n$ qubits of $Q_l^{\mathrm{I}}$, apply an $n$-controlled $\mathsf{NOT}$ onto $F$, and then uncompute the $X$ gates, where the $n$-controlled $\mathsf{NOT}$ uses the decomposition of Ref.~\cite[Sec.~4.3]{Nielsen_Chuang2010} with $2(n-1)$ Toffoli gates and the $n-1$ shared ancilla qubits.
Each test is therefore dominated by the two applications of $U_s$ and takes $O(n^2)$ gates.

We finally count the whole circuit.
Each query block applies the $n$ $\mathsf{CNOT}$ gates of $U_{\mathrm{Enc}}^{(l)\dagger}$, performs at most $M_{r_{\mathrm{seq}}}$ tests at $O(n^2)$ gates each, and applies the correction $Z^{\hat{e}_l}$ to $D^{\mathrm{I}}$ with at most $n$ $Z$ gates conditioned on the measured flags.
Summing over the $L$ query blocks gives $G_{\mathrm{rec}} = O\left(n^2 M_{r_{\mathrm{seq}}} L\right)$.
For the ancilla count, the flag ancilla and the $n-1$ shared ancillas are reused across the tests and the query blocks, giving $A_{\mathrm{rec}} = n = O(n)$ and establishing Eq.~\eqref{eq:seq_rec_gate_complexity}.

\clearpage
\section{Distillation of Boolean Oracles under i.i.d.\ Depolarizing Noise}\label{SM_sec:iid}

In this section, we distill noisy Boolean oracles under i.i.d.\ depolarizing noise into the ideal ones with precision $\epsilon$.
Specifically, we consider the family of noisy oracles $\tilde{\mathfrak{O}}_F^{\mathrm{iid}} = \left\{\tilde{\mathcal{O}}_f^{\mathrm{iid}}\right\}_{f\in F}$, where each $f :\{0,1\}^n \rightarrow \{0,1\}^m$ and
\begin{equation}
    \tilde{\mathcal{O}}_f^{\mathrm{iid}} = \mathcal{D}_p^{\otimes (n+m)} \circ \mathcal{O}_f, \qquad
    \mathcal{D}_p(\cdot) := (1-p)(\cdot) + \frac{p}{3}\left(X(\cdot)X + Y(\cdot)Y + Z(\cdot)Z\right).
\end{equation}
We call each member an \textit{i.i.d.-depolarizing Boolean oracle} with per-qubit error rate $p$.
We distill this family with the same two-stage protocol as in the adversarial setting, but the analysis must be refined, since the index error on a query block can now fall outside the correctable error set $\mathcal{K}_r$ and the response register is no longer noiseless.
By the stage 1 protocol (Corollary~\ref{cor:stage1_depol}), each call to $\tilde{\mathcal{O}}_f^{\mathrm{iid}}$ can be converted into a call to the \textit{i.i.d.-dephasing Boolean oracle} $\tilde{\mathcal{O}}_f^{\mathrm{iid},Z}$, whose noise factorizes over the index and response registers,
\begin{equation}\label{eq:iid_noise_factorization}
    \tilde{\mathcal{O}}_f^{\mathrm{iid},Z}
    = \left( \mathcal{E}^{Z,\mathrm{I}} \otimes \mathcal{E}^{Z,\mathrm{R}} \right) \circ \mathcal{O}_f,
    \qquad
    \mathcal{E}^{Z,\mathrm{I}} := \left( \mathcal{D}^Z_{p_t} \right)^{\otimes n},
    \qquad
    \mathcal{E}^{Z,\mathrm{R}} := \left( \mathcal{D}^Z_{p_t} \right)^{\otimes m},
\end{equation}
where $\mathcal{D}^Z_{p_t}(\cdot) := (1-p_t)(\cdot) + p_t\,Z(\cdot)Z$ is the single-qubit dephasing channel with error rate $p_t := 2p/3$, $\mathcal{E}^{Z,\mathrm{I}}$ acts on the index register, and $\mathcal{E}^{Z,\mathrm{R}}$ acts on the response register.
The main task of this section is therefore to analyze stage 2, which distills these i.i.d.-dephasing oracle calls into the ideal oracle.

Throughout this section, $r = \lfloor \alpha n \rfloor$ for a constant $\alpha \in (0, 1/2]$, the maximal candidate weight of the sequential recovery is $r_{\mathrm{seq}} := \lfloor \alpha_{\mathrm{seq}} n \rfloor$ for a constant $\alpha_{\mathrm{seq}} \in (\alpha, 1]$, and $n \geq 1/\left( \alpha_{\mathrm{seq}} - \alpha \right)$, so that $r_{\mathrm{seq}} \geq r + 1$.

Section~\ref{SM_sec:iid_from_bounded} describes the distilled oracle as a mixture of the general distilled oracles of Sec.~\ref{SM_sec:stage2_OD} over random response error patterns.
Section~\ref{SM_sec:iid_aggregation} analyzes the response errors, bounding the response aggregation error of the components under a random response error pattern together with the probability of the patterns on which the bound fails.
Section~\ref{SM_sec:iid_seq_recovery} analyzes the index errors, including the high-weight patterns outside the correctable set $\mathcal{K}_r$, by running the sequential recovery with a maximal candidate weight above $r$ and bounding the recovery error that remains.
Section~\ref{SM_sec:iid_error_bound} combines these bounds into the distillation error and query complexity, which gives the main result of this section, and then specializes it to $r = \lfloor \alpha n \rfloor$ with the query states of Constructions 3 and 2, where its condition becomes a threshold on the per-qubit error rate.
Section~\ref{SM_sec:small_gamma} determines how the resulting threshold vanishes as the overhead exponent tends to zero, and inverts this into the smallest overhead exponent that admits a given error rate.

\subsection{Distilled oracle as a mixture over response error patterns}\label{SM_sec:iid_from_bounded}
Here we show that the distilled oracle can be written as a mixture of the general distilled oracles Eq.~\eqref{eq:general_distilled_oracle_OD} over random response error patterns.
The distilled oracle of this section, $\widehat{\mathcal{O}}_f^{\mathrm{iid}}$, is the stage 2 circuit Eq.~\eqref{eq:distilled_oracle_OD} with each of its $2L$ oracle calls made to $\tilde{\mathcal{O}}_f^{\mathrm{iid},Z}$ in place of $\tilde{\mathcal{O}}_f^{\mathrm{adv},Z}$.
The two response error patterns inserted on the query-block response registers in Eq.~\eqref{eq:distilled_oracle_OD} are set to zero, since here those registers are hit by the noise factor $\mathcal{E}^{Z,\mathrm{R}}$ of each call instead.
Recall from Eq.~\eqref{eq:distilled_oracle_OD} the response error patterns, a pattern $\tau \in \{0,1\}^{L \times m}$ acting after the weak query and a pattern $\sigma \in \{0,1\}^{L\times m}$ acting after the uncomputation.
The noise channel $\mathcal{E}^{Z,\mathrm{R}}$ on the response register in Eq.~\eqref{eq:iid_noise_factorization} applies $Z$ independently to each of the $m$ response qubits with probability $p_t = 2p/3$, the per-qubit dephasing rate left by stage 1 (Corollary~\ref{cor:stage1_depol}).
It is therefore a probabilistic mixture of $Z$ strings, acting exactly at the locations of the two inserted patterns.
Collecting the $Z$ strings of the $L$ calls in the weak query into $\tau$ and those of the $L$ calls in the uncomputation into $\sigma$ thus expands the distilled oracle into a mixture over the two patterns, whose entries are i.i.d.\ Bernoulli$(p_t)$.
The component with fixed $\tau$ and $\sigma$ is the distilled oracle Eq.~\eqref{eq:distilled_oracle_OD} built from the noisy oracle $\mathcal{E}^{Z,\mathrm{I}} \circ \mathcal{O}_f$.
Since $\mathcal{E}^{Z,\mathrm{I}}$ is itself a probabilistic mixture of $Z$ strings, this noisy oracle satisfies the conditions of Lemma~\ref{lem:reduction_OD}, so the component equals the general distilled oracle Eq.~\eqref{eq:general_distilled_oracle_OD} with $\mathcal{E}_f^{\mathrm{tot}} = \left( \left( \mathcal{E}^{Z,\mathrm{I}} \right)^2 \right)^{\otimes L}$.
Two dephasing layers on the same qubit leave a net $Z$ error exactly when one layer applies $Z$ and the other does not, which happens with probability $2p_t(1-p_t)$, so the merged channel is again an i.i.d.\ dephasing channel,
\begin{equation}\label{eq:iid_peff}
    \left( \left( \mathcal{E}^{Z,\mathrm{I}} \right)^2 \right)^{\otimes L} = \left( \mathcal{D}^Z_{p_{\mathrm{eff}}} \right)^{\otimes nL},
    \qquad
    p_{\mathrm{eff}} := 2 p_t \left( 1 - p_t \right),
\end{equation}
acting on the $nL$ qubits of the query-block index registers.
The distilled oracle is therefore the mixture
\begin{equation}\label{eq:iid_pattern_mixture}
    \widehat{\mathcal{O}}_f^{\mathrm{iid}}
    = \sum_{\tau, \sigma} p(\tau)\, p(\sigma)\;
    \widehat{\mathcal{O}}{}_f^{\,\tau,\sigma}\left[ \left( \mathcal{D}^Z_{p_{\mathrm{eff}}} \right)^{\otimes nL},\, \mathsf{Rec}^{\mathrm{I}} \right],
\end{equation}
with probabilities
\begin{equation}\label{eq:iid_pattern_weights}
    p(\tau) := p_t^{|\tau|} \left( 1 - p_t \right)^{Lm - |\tau|},
    \qquad
    p(\sigma) := p_t^{|\sigma|} \left( 1 - p_t \right)^{Lm - |\sigma|},
\end{equation}
where $|\tau|$ and $|\sigma|$ denote the total weight of the pattern.

\subsection{Response aggregation under random response errors}\label{SM_sec:iid_aggregation}
Here we bound the response aggregation error of the components of the mixture Eq.~\eqref{eq:iid_pattern_mixture}, by collecting the patterns on which the aggregator is guaranteed to succeed into a typical set and bounding the probability of the rest.
Write $\tau^{(j)} \in \{0,1\}^L$ for the $j$-th column of $\tau$, which records the blocks in which the response error pattern acts on the $j$-th response qubit.
We collect the patterns with low column weights into the typical set
\begin{equation}\label{eq:iid_typical_set}
    \mathcal{T} := \left\{ \tau \in \{0,1\}^{L \times m} \,\middle|\, |\tau^{(j)}| < \left( \frac{1}{4} + \frac{p_t}{2} \right) L \text{ for all } j \in [m] \right\}.
\end{equation}
The threshold fraction $1/4 + p_t/2$ is the midpoint of $p_t$ and $1/2$, so $p_t < 1/4 + p_t/2 < 1/2$ whenever $p_t < 1/2$.
Each column weight $|\tau^{(j)}|$ follows $\mathrm{Bin}(L, p_t)$, and a pattern is atypical exactly when at least one of its $m$ columns reaches the threshold.
Applying the union bound over the columns and then the Chernoff bound (Lemma~\ref{lem:chernoff_KL}) to each column, whose condition $1/4 + p_t/2 \geq p_t$ holds for every $p_t < 1/2$, gives
\begin{equation}\label{eq:iid_target_atyp}
    p_{\mathrm{atyp}} := \Pr\left[ \tau \notin \mathcal{T} \right]
    \leq \sum_{j=1}^{m} \Pr\left[ |\tau^{(j)}| \geq \left( \frac{1}{4} + \frac{p_t}{2} \right) L \right]
    \leq m \cdot 2^{-L D\left( \tfrac{1}{4} + \tfrac{p_t}{2} \,\big\|\, p_t \right)}.
\end{equation}
For typical response error patterns, applying Lemma~\ref{lem:agg_error_classical_OD} gives the following bound on the aggregation error.
\begin{corof}{lem:agg_error_classical_OD}{2}[Aggregation error under low-weight response errors]\label{cor:agg_error_lowweight_OD}
    Set the threshold to $w^* = \eta L/2$.
    If $p_t < 1/2$ and the response error pattern is typical, $\tau \in \mathcal{T}$, then for all $x$, $y$, and $f$,
    \begin{equation}\label{eq:agg_error_lowweight_OD}
        \epsilon_{\mathrm{agg}}(x,y;f,\tau)
        \leq 2 \sqrt{m}\, \exp\left( -\frac{\left( 1 - 2 p_t \right)^2}{16 \left( 3 - 2 p_t \right)}\, \eta L \right) .
    \end{equation}
\end{corof}
\begin{proof}
    The union bound over the $m$ response bits gives
    \begin{equation}
        \Pr\left[ v(H) \neq f(x) \right] \leq \sum_{j=1}^{m} \Pr\left[ v(H)_j \neq f(x)_j \right] .
    \end{equation}
    Fix a response bit $j$, and say it fails when $v(H)_j \neq f(x)_j$.
    By Eq.~\eqref{eq:branch_eigenvalue_OD}, $W_j(H) = \sum_{l :\, \tau_{lj} \oplus f(x)_j = 1} H_l$ is a sum of i.i.d.\ Bernoulli variables with mean $\eta$.
    When $f(x)_j = 1$, the sum runs over the $L - |\tau^{(j)}|$ blocks with $\tau_{lj} = 0$, so $W_j(H) \sim \mathrm{Bin}(L - |\tau^{(j)}|, \eta)$, and the bit fails exactly when $W_j(H) < w^*$.
    The mean $\mu_1 = \eta (L - |\tau^{(j)}|)$ exceeds $\frac{3 - 2p_t}{4}\, \eta L$ since $|\tau^{(j)}| < \frac{1 + 2p_t}{4} L$, so the deviation $\delta := \frac{1 - 2p_t}{3 - 2p_t} \in (0,1)$ satisfies
    \begin{equation}
        (1 - \delta)\, \mu_1 = \frac{2 \mu_1}{3 - 2p_t} > \frac{\eta L}{2} = w^* ,
    \end{equation}
    and the multiplicative Chernoff bound~\cite{MitzenmacherUpfal2005} gives
    \begin{equation}
        \Pr\left[ v(H)_j \neq f(x)_j \right]
        = \Pr\left[ W_j(H) < w^* \right]
        \leq \exp\left( -\frac{\mu_1 \delta^2}{2} \right)
        \leq \exp\left( -\frac{\left( 1 - 2p_t \right)^2}{8 \left( 3 - 2p_t \right)}\, \eta L \right) ,
    \end{equation}
    where the last step uses $\mu_1 > \frac{3 - 2p_t}{4} \eta L$ again.
    When $f(x)_j = 0$, the sum runs over the $|\tau^{(j)}|$ blocks with $\tau_{lj} = 1$, so $W_j(H) \sim \mathrm{Bin}(|\tau^{(j)}|, \eta)$, and the bit fails exactly when $W_j(H) \geq w^*$.
    The mean $\mu_2 = \eta\, |\tau^{(j)}|$ is below $\frac{1 + 2p_t}{4} \eta L$, which is itself below $w^*$ since $p_t < 1/2$, and the right-hand side of the upper-tail Chernoff bound $\Pr\left[ W_j(H) \geq w^* \right] \leq \exp\left( -\frac{(w^* - \mu_2)^2}{w^* + \mu_2} \right)$ is increasing in $\mu_2$, so
    \begin{equation}
        \Pr\left[ v(H)_j \neq f(x)_j \right]
        = \Pr\left[ W_j(H) \geq w^* \right]
        \leq \exp\left( - \frac{\left( \frac{\eta L}{2} - \frac{1 + 2p_t}{4} \eta L \right)^2}{\frac{\eta L}{2} + \frac{1 + 2p_t}{4} \eta L} \right)
        = \exp\left( -\frac{\left( 1 - 2p_t \right)^2}{4 \left( 3 + 2p_t \right)}\, \eta L \right) .
    \end{equation}
    Since $p_t < 1/2$ gives $4(3 + 2p_t) < 8(3 - 2p_t)$, the exponent of the $f(x)_j = 1$ case is the smaller one, so every bit fails with probability at most $\exp\left( -\frac{\left( 1 - 2p_t \right)^2}{8 \left( 3 - 2p_t \right)} \eta L \right)$.
    Therefore $\Pr\left[ v(H) \neq f(x) \right] \leq m \exp\left( -\frac{\left( 1 - 2p_t \right)^2}{8 \left( 3 - 2p_t \right)} \eta L \right)$, and Lemma~\ref{lem:agg_error_classical_OD} proves the claim.
\end{proof}

\subsection{Sequential recovery and error}\label{SM_sec:iid_seq_recovery}
In this subsection, we run the sequential recovery of Sec.~\ref{SM_sec:seq_recovery} with a maximal candidate weight larger than $r$, so that it also corrects index error patterns of weight above $r$.
The tests identify the patterns in $\mathcal{K}_r$ with certainty because the EOC Eq.~\eqref{SM_eq:error_orthogonality} make their corrupted seed states $Z^{e} \ket{\Theta_s}$ pairwise orthogonal.
For the seed states of Constructions 2 and 3, the corrupted seed states remain approximately orthogonal even when the error weight exceeds $r$ (Lemmas~\ref{lem:construction2_high_weight} and~\ref{lem:construction3_high_weight}), so the tests still identify candidate patterns of weight above $r$, no longer with certainty but with a failure probability, which we bound in this subsection through the expectations of high-weight $Z$ strings.
The cost is the circuit complexity Eq.~\eqref{eq:seq_rec_gate_complexity}, which grows with the number of candidate patterns $M_{r_{\mathrm{seq}}}$.

We first reduce the recovery error to the failure probability of the error pattern identification in a single query block.
Let $e \in \{0,1\}^n$ be the index error pattern of a query block, with i.i.d.\ Bernoulli($p_{\mathrm{eff}}$) entries as in the merged index noise Eq.~\eqref{eq:iid_peff}, and hence $p(e) = p_{\mathrm{eff}}^{|e|} (1-p_{\mathrm{eff}})^{n-|e|}$.
Let $\hat{e} \in \mathcal{K}_{r_{\mathrm{seq}}}$ be the accepted candidate of the sequential recovery in this block.
Then we define the failure probability
\begin{equation}\label{eq:def_p_fail}
    p_{\mathrm{fail}} := \sum_{e \in \{0,1\}^n} p(e) \Pr\left[ \hat{e} \neq e \,\middle|\, e \right]
\end{equation}
and it is the same for every query block, since the block patterns are identically distributed.
\begin{lemma}[Recovery error bound for the sequential recovery]\label{lem:seq_recovery_reduction}
    For every maximal candidate weight $r_{\mathrm{seq}} \leq n$, the recovery error of the sequential recovery satisfies
    \begin{equation}\label{eq:seq_rec_reduction}
        \left\| \mathsf{Rec}_{r_{\mathrm{seq}}} \circ \left( \mathcal{D}^Z_{p_{\mathrm{eff}}} \right)^{\otimes nL} \circ \mathsf{Enc}^{\mathrm{I}} - \mathcal{I}_{D^{\mathrm{I}}} \right\|_{\diamond} \leq 2 L\, p_{\mathrm{fail}} .
    \end{equation}
\end{lemma}
\begin{proof}
    We expand the noise over index error patterns, bound the term of each fixed pattern by the probability that some query block is misidentified, and finally average over the patterns.
    Each single-qubit dephasing channel is a mixture of the conjugations by $I$ and $Z$, so $\left( \mathcal{D}^Z_{p_{\mathrm{eff}}} \right)^{\otimes nL}$ is the mixture of the error channels $\mathcal{Z}^{\vec{e}}$ over all patterns $\vec{e} \in \left( \{0,1\}^n \right)^L$ with the probabilities $p(\vec{e}) = p_{\mathrm{eff}}^{|\vec{e}|} \left( 1 - p_{\mathrm{eff}} \right)^{Ln - |\vec{e}|}$, and the triangle inequality for the diamond norm (Lemma~\ref{lem:channel_norm_facts}) gives
    \begin{equation}\label{eq:seq_rec_pattern_split}
        \left\| \mathsf{Rec}_{r_{\mathrm{seq}}} \circ \left( \mathcal{D}^Z_{p_{\mathrm{eff}}} \right)^{\otimes nL} \circ \mathsf{Enc}^{\mathrm{I}} - \mathcal{I}_{D^{\mathrm{I}}} \right\|_{\diamond}
        \leq \sum_{\vec{e}} p(\vec{e}) \left\| \mathsf{Rec}_{r_{\mathrm{seq}}} \circ \mathcal{Z}^{\vec{e}} \circ \mathsf{Enc}^{\mathrm{I}} - \mathcal{I}_{D^{\mathrm{I}}} \right\|_{\diamond} .
    \end{equation}

    We bound the term of a fixed pattern $\vec{e}$.
    Fix a pure input $\ket{\psi}$ on $\mathcal{H}_{D^{\mathrm{I}}} \otimes \mathcal{H}_{\mathrm{aux}}$, where $\mathcal{H}_{\mathrm{aux}}$ is an auxiliary register that the recovery never touches.
    The channel $\mathcal{Z}^{\vec{e}}$ has the single Kraus operator $Z^{\vec{e}}$.
    The unitary $U_{\mathrm{Enc}}^{(l)\dagger}$ acts on $Q_l^{\mathrm{I}}$ and diagonally on $D^{\mathrm{I}}$, so it commutes with the tests and corrections of the other query blocks, and we may apply the inverse encoding of every block before any test.
    Applying the conjugation identity Eq.~\eqref{eq:enc_conjugation_OD} to each factor $Z^{e_l}$ then brings the state before the tests to the product form
    \begin{equation}\label{eq:seq_rec_single_pattern_state}
        \bigotimes_{l=1}^{L} \left( Z^{e_l}\ket{\Theta_s} \right)_{Q_l^{\mathrm{I}}} \otimes \left( Z^{e_1 \oplus \cdots \oplus e_L}_{D^{\mathrm{I}}} \otimes I_{\mathrm{aux}} \right) \ket{\psi}.
    \end{equation}
    The tests act only on the registers $Q_l^{\mathrm{I}}$, which are in product with the rest of the state, so the joint distribution of the accepted candidates $(\hat{e}_1, \ldots, \hat{e}_L)$ does not depend on $\ket{\psi}$, and the distribution of $\hat{e}_l$ depends only on $e_l$.
    On the outcome $\hat{e}_l = e_l$ for every $l$, the corrections $Z^{\hat{e}_l}_{D^{\mathrm{I}}}$ multiply to $Z^{e_1 \oplus \cdots \oplus e_L}_{D^{\mathrm{I}}}$ and cancel the error operator in Eq.~\eqref{eq:seq_rec_single_pattern_state}, so after the registers $Q_l^{\mathrm{I}}$ are discarded this outcome returns exactly $\ket{\psi}\bra{\psi}$.
    The channel $\mathsf{Rec}_{r_{\mathrm{seq}}} \circ \mathcal{Z}^{\vec{e}} \circ \mathsf{Enc}^{\mathrm{I}}$ is thus a mixture of channels weighted by the distribution of the accepted candidates, in which the outcome $\hat{e}_l = e_l$ for every $l$ contributes the identity and every other outcome contributes a CPTP channel.
    The triangle inequality for the diamond norm and the diamond distance of at most $2$ between two channels (Lemma~\ref{lem:channel_norm_facts}) then give
    \begin{equation}\label{eq:seq_rec_fixed_pattern_bound}
        \left\| \mathsf{Rec}_{r_{\mathrm{seq}}} \circ \mathcal{Z}^{\vec{e}} \circ \mathsf{Enc}^{\mathrm{I}} - \mathcal{I}_{D^{\mathrm{I}}} \right\|_{\diamond}
        \leq 2 \Pr\left[ \hat{e}_l \neq e_l \text{ for some } l \mid \vec{e}\, \right]
        \leq 2 \sum_{l=1}^{L} \Pr\left[ \hat{e}_l \neq e_l \mid e_l \right],
    \end{equation}
    where the last inequality is the union bound over the blocks.

    Finally, the probability factorizes over the blocks as $p(\vec{e}) = \prod_{l=1}^{L} p(e_l)$, so in the average of Eq.~\eqref{eq:seq_rec_fixed_pattern_bound} over $p(\vec{e})$ the sum over the patterns of the other blocks evaluates to one in each summand,
    \begin{equation}
        \sum_{\vec{e}} p(\vec{e}) \sum_{l=1}^{L} \Pr\left[ \hat{e}_l \neq e_l \mid e_l \right]
        = \sum_{l=1}^{L} \sum_{e_l \in \{0,1\}^n} p(e_l) \Pr\left[ \hat{e}_l \neq e_l \mid e_l \right]
        = L\, p_{\mathrm{fail}},
    \end{equation}
    where the last equality holds by Eq.~\eqref{eq:def_p_fail} since each block pattern $e_l$ has the distribution of $e$.
    Substituting into Eq.~\eqref{eq:seq_rec_pattern_split} proves the lemma.
\end{proof}

We now bound $p_{\mathrm{fail}}$, beginning with the conditional probability in Eq.~\eqref{eq:def_p_fail} for each fixed pattern $e$.
\begin{lemma}[Failure probability of the error pattern identification]\label{lem:seq_recovery_error}
    Suppose the seed state satisfies the EOC Eq.~\eqref{SM_eq:error_orthogonality} for $r$ errors.
    Let $e \in \{0,1\}^n$ be the index error pattern of a query block, and let $\hat{e} \in \mathcal{K}_{r_{\mathrm{seq}}}$ be the accepted candidate of the sequential recovery $\mathsf{Rec}_{r_{\mathrm{seq}}}$ in this block.
    Then
    \begin{equation}\label{eq:seq_rec_failure_prob}
        \Pr[\hat{e} \neq e | e] \leq
        \begin{cases}
            0, & e \in \mathcal{K}_r\\
            4\sum\limits_{\substack{e'\in \mathcal{K}_{r_{\mathrm{seq}}},\; e' \prec e \\ |e \oplus e'| > 2r}} \left|\bra{\Theta_s} Z^{e\oplus e'} \ket{\Theta_s}\right|^2, & e \in  \mathcal{K}_{r_{\mathrm{seq}}}, e \notin \mathcal{K}_r\\
            1, & e \notin  \mathcal{K}_{r_{\mathrm{seq}}} .
        \end{cases}
    \end{equation}
\end{lemma}
\begin{proof}
    Conditioned on the index error pattern $e$, the noise acts on the query block as the single error operator $Z^{e}$, so after the inverse encoding the query register $Q_l^{\mathrm{I}}$ holds the corrupted seed state $Z^{e}\ket{\Theta_s}$, in product with the remaining registers, by the conjugation identity Eq.~\eqref{eq:enc_conjugation_OD}.
    The accepted candidate $\hat{e}$ depends only on the second step of the recovery, the tests of the candidates in $\mathcal{K}_{r_{\mathrm{seq}}}$ on $Q_l^{\mathrm{I}}$ in the order $\prec$ until one accepts, so the proof reduces to analyzing this measurement sequence on the corrupted seed state.

    If $e \notin \mathcal{K}_{r_{\mathrm{seq}}}$, the accepted candidate $\hat{e}$ lies in $\mathcal{K}_{r_{\mathrm{seq}}}$ and can never equal $e$, so $\Pr[\hat{e} \neq e | e] = 1$.

    Now let $e \in \mathcal{K}_{r_{\mathrm{seq}}}$.
    The event $\hat{e} = e$ occurs whenever the tests of all candidates $e' \prec e$ reject and the test of $e$ then accepts, and for $e = 0^n$ also when every test rejects, through the default $\hat{e} = 0^n$.
    Keeping only the first event, the quantum union bound~\cite{Gao2015, ODonnell2022}, applied to the pure state $Z^{e}\ket{\Theta_s}$ and the projectors $I - \Pi_{e'}$ for $e' \prec e$ followed by $\Pi_{e}$, gives
    \begin{equation}\label{eq:quantum_union_bound}
        \Pr[\hat{e} \neq e | e]
        \leq 1 - \left\| \Pi_{e} \prod_{e' \prec e} \left( I - \Pi_{e'} \right) Z^{e} \ket{\Theta_s} \right\|_2^2
        \leq 4 \left( \bra{\Theta_s} Z^{e} \left( I - \Pi_{e} \right) Z^{e} \ket{\Theta_s} + \sum_{e' \prec e} \bra{\Theta_s} Z^{e}\, \Pi_{e'}\, Z^{e} \ket{\Theta_s} \right),
    \end{equation}
    where the product applies the rejection projectors in the order $\prec$, the earliest rightmost.
    The definition of the projector Eq.~\eqref{eq:def_Pi} evaluates every term on the right-hand side.
    The first term vanishes since $\Pi_{e} Z^{e}\ket{\Theta_s} = Z^{e}\ket{\Theta_s}$, and the term of each candidate $e'$ equals $\bra{\Theta_s} Z^{e}\, \Pi_{e'}\, Z^{e} \ket{\Theta_s} = \left|\bra{\Theta_s} Z^{e'} Z^{e} \ket{\Theta_s}\right|^2 = \left|\bra{\Theta_s} Z^{e \oplus e'} \ket{\Theta_s}\right|^2$.
    Every candidate $e' \prec e$ is distinct from $e$, so $1 \leq |e \oplus e'|$, and the EOC Eq.~\eqref{SM_eq:error_orthogonality} make the terms with $|e \oplus e'| \leq 2r$ vanish, which leaves the restricted sum of the second case and proves it for every $e \in \mathcal{K}_{r_{\mathrm{seq}}}$.

    Finally, if $e \in \mathcal{K}_r$, every candidate $e' \prec e$ satisfies $|e'| \leq |e| \leq r$ by the weight ordering, so $|e \oplus e'| \leq 2r$ and the restricted sum of the second case is empty, giving $\Pr[\hat{e} \neq e | e] = 0$.
\end{proof}

Combining Lemma~\ref{lem:seq_recovery_error} with the minimization of Sec.~\ref{SM_sec:exponent_minimization} now gives a bound on $p_{\mathrm{fail}}$ for any seed state.
The EOC Eq.~\eqref{SM_eq:error_orthogonality} make the corrupted seed states $Z^{e} \ket{\Theta_s}$ exactly orthogonal for error patterns of weight at most $r$, and above that weight they remain only approximately orthogonal.
The seed state enters the bound only through how nearly orthogonal they stay, measured by the decay rate
\begin{equation}\label{eq:def_b_high_weight}
    b(n) := - \frac{1}{n} \log_2 \max_{2r < |a| \leq \min\left\{ 2 r_{\mathrm{seq}},\, n \right\}} \left| \bra{\Theta_s} Z^{a} \ket{\Theta_s} \right| ,
\end{equation}
whose range of weights is the one that the sequential recovery can confuse.
Here $b(n)$ takes values in $\left[ 0, \infty \right]$, with $b(n) = \infty$ when the maximum vanishes or when $2r = n$ leaves no such weight, and the bounds below are read with $2^{-\infty} = 0$.
This rate depends on $\alpha_{\mathrm{seq}}$ through the upper end of the range, but the two constructions used below give lower bounds on it that hold for every $\alpha_{\mathrm{seq}}$ in their range.
How the exponent of the resulting bound moves with the effective index error rate is recorded in Lemma~\ref{lem:exponent_monotone}, which is what fixes the distillation threshold in Sec.~\ref{SM_sec:iid_error_bound}.

\begin{lemma}[Bound on the failure probability from the high-weight $Z$ expectations]\label{lem:p_fail_high_weight}
    Suppose the seed state satisfies the EOC Eq.~\eqref{SM_eq:error_orthogonality} for $r = \lfloor \alpha n \rfloor$ errors with a constant $\alpha \in (0, 1/2]$, and let $0 < p_{\mathrm{eff}} \leq \min\left\{ \alpha_{\mathrm{seq}},\, 1/2 \right\}$.
    Then
    \begin{equation}\label{eq:p_fail_high_weight}
        p_{\mathrm{fail}} \leq
        4 n\, 2^{-n \left( 2 b(n) + E_0 \right)}
        + 2^{-n D\left( \alpha_{\mathrm{seq}} \,\big\|\, p_{\mathrm{eff}} \right)} ,
    \end{equation}
    where the exponent $E_0$ is the minimum of Lemma~\ref{lem:exponent_minimization} over the interval $\left[ \alpha,\, \alpha_{\mathrm{seq}} \right]$,
    \begin{equation}\label{eq:p_fail_E0}
        E_0 := \min_{\lambda \in \left[ \alpha,\, \alpha_{\mathrm{seq}} \right]} \left( D\left( \lambda \,\big\|\, p_{\mathrm{eff}} \right) - \bar{H}(\lambda) \right) =
        \begin{cases}
            D\left( \alpha \,\big\|\, p_{\mathrm{eff}} \right) - H\left( \alpha \right), & p_{\mathrm{eff}} \leq \dfrac{\alpha^2}{\alpha^2 + \left( 1 - \alpha \right)^2} \\[10pt]
            D\left( \alpha_{\mathrm{seq}} \,\big\|\, p_{\mathrm{eff}} \right) - H\left( \alpha_{\mathrm{seq}} \right), & p_{\mathrm{eff}} \geq \dfrac{\alpha_{\mathrm{seq}}^2}{\alpha_{\mathrm{seq}}^2 + \left( 1 - \alpha_{\mathrm{seq}} \right)^2} \\[10pt]
            - \log_2\left( 1 + 2 \sqrt{p_{\mathrm{eff}} \left( 1 - p_{\mathrm{eff}} \right)} \right), & \text{otherwise} .
        \end{cases}
    \end{equation}
\end{lemma}
\begin{proof}
    Splitting the average Eq.~\eqref{eq:def_p_fail} according to the three cases of Lemma~\ref{lem:seq_recovery_error} gives
    \begin{equation}\label{eq:p_fail_split}
        p_{\mathrm{fail}}
        = \sum_{e \in \mathcal{K}_r} p(e) \Pr[\hat{e} \neq e | e]
        + \sum_{\substack{e \in \mathcal{K}_{r_{\mathrm{seq}}} \\ e \notin \mathcal{K}_r}} p(e) \Pr[\hat{e} \neq e | e]
        + \sum_{e \notin \mathcal{K}_{r_{\mathrm{seq}}}} p(e) \Pr[\hat{e} \neq e | e] .
    \end{equation}
    The first sum vanishes term by term by the first case of Lemma~\ref{lem:seq_recovery_error}.
    In the third sum the third case of Lemma~\ref{lem:seq_recovery_error} bounds every conditional probability by $1$, so this sum is at most $\Pr\left[ |e| > r_{\mathrm{seq}} \right]$.
    The weight $|e|$ follows $\mathrm{Bin}(n, p_{\mathrm{eff}})$, and an integer weight above $r_{\mathrm{seq}} = \lfloor \alpha_{\mathrm{seq}} n \rfloor$ is at least $\alpha_{\mathrm{seq}} n$, so the Chernoff bound (Lemma~\ref{lem:chernoff_KL}), whose condition $p_{\mathrm{eff}} \leq \alpha_{\mathrm{seq}}$ holds by assumption, gives
    \begin{equation}\label{eq:p_fail_tail}
        \Pr\left[ |e| > r_{\mathrm{seq}} \right]
        \leq \Pr\left[ |e| \geq \alpha_{\mathrm{seq}} n \right]
        \leq 2^{-n D\left( \alpha_{\mathrm{seq}} \,\big\|\, p_{\mathrm{eff}} \right)} ,
    \end{equation}
    which is the second term of Eq.~\eqref{eq:p_fail_high_weight}.
    The rest of the proof bounds the second sum of Eq.~\eqref{eq:p_fail_split}, which is where the decay rate Eq.~\eqref{eq:def_b_high_weight} enters.

    We first bound the conditional failure probability of a single pattern $e$ with $r < |e| \leq r_{\mathrm{seq}}$ through the second case of Lemma~\ref{lem:seq_recovery_error}, whose sum runs over the candidates $e' \prec e$ with $|e \oplus e'| > 2r$.
    Every such candidate has $|e'| \leq |e|$ by the weight ordering, so the sum has at most $M_{|e|}$ summands.
    Every summand also obeys $|e \oplus e'| \leq \min\left\{ 2 r_{\mathrm{seq}},\, n \right\}$, since $|e'|, |e| \leq r_{\mathrm{seq}}$ and the weight of an XOR is at most the sum of the two weights as well as at most $n$.
    Together with $|e \oplus e'| > 2r$ from the restriction of the sum, the definition Eq.~\eqref{eq:def_b_high_weight} applies to every summand and gives $\left|\bra{\Theta_s} Z^{e \oplus e'} \ket{\Theta_s}\right|^2 \leq 2^{-2 n b(n)}$.
    With the candidate count bounded by $M_{|e|} \leq 2^{n \bar{H}\left( |e|/n \right)}$ by Lemma~\ref{lem:entropy_binomial}, that case gives
    \begin{equation}\label{eq:p_fail_per_pattern}
        \Pr[\hat{e} \neq e | e]
        \leq 4 M_{|e|}\, 2^{-2 n b(n)}
        \leq 4 \cdot 2^{n \left( \bar{H}\left( \frac{|e|}{n} \right) - 2 b(n) \right)} ,
    \end{equation}
    a bound that depends on $e$ only through its weight.

    We next bound the probability that the pattern has a given weight $w$.
    Expanding the definitions of the binary entropy and the KL divergence gives $H\left( \frac{w}{n} \right) + D\left( \frac{w}{n} \,\big\|\, p_{\mathrm{eff}} \right) = - \frac{w}{n} \log_2 p_{\mathrm{eff}} - \left( 1 - \frac{w}{n} \right) \log_2 \left( 1 - p_{\mathrm{eff}} \right)$, so each pattern of weight $w$ has probability $p_{\mathrm{eff}}^{w} \left( 1 - p_{\mathrm{eff}} \right)^{n-w} = 2^{-n \left( H\left( \frac{w}{n} \right) + D\left( \frac{w}{n} \,\big\|\, p_{\mathrm{eff}} \right) \right)}$, and summing over the $\binom{n}{w}$ patterns of weight $w$ gives
    \begin{equation}\label{eq:p_fail_weight_prob}
        \Pr\left[ |e| = w \right]
        = \binom{n}{w}\, 2^{-n \left( H\left( \frac{w}{n} \right) + D\left( \frac{w}{n} \,\big\|\, p_{\mathrm{eff}} \right) \right)}
        \leq 2^{-n D\left( \frac{w}{n} \,\big\|\, p_{\mathrm{eff}} \right)} ,
    \end{equation}
    where the last inequality uses $\binom{n}{w} \leq 2^{n H\left( \frac{w}{n} \right)}$, which holds for $w \leq n/2$ by $\binom{n}{w} \leq M_w$ and Lemma~\ref{lem:entropy_binomial} with $\lambda = w/n$, and for $w > n/2$ by $\binom{n}{w} = \binom{n}{n-w}$ together with $H(1-x) = H(x)$.

    Finally, we bound the second sum of Eq.~\eqref{eq:p_fail_split}, grouping its summands by the weight $w = |e|$, which runs over the integers with $r < w \leq r_{\mathrm{seq}}$.
    Within each group, Eq.~\eqref{eq:p_fail_per_pattern} bounds the conditional failure probability uniformly, so the group of weight $w$ contributes at most $\Pr\left[ |e| = w \right]$ times this bound.
    Inserting Eq.~\eqref{eq:p_fail_weight_prob} and keeping the largest of the $r_{\mathrm{seq}} - r \leq n$ terms gives
    \begin{equation}\label{eq:p_fail_middle_min}\begin{split}
        \sum_{\substack{e \in \mathcal{K}_{r_{\mathrm{seq}}} \\ e \notin \mathcal{K}_r}} p(e) \Pr[\hat{e} \neq e | e]
        &\leq \sum_{w = r+1}^{r_{\mathrm{seq}}} \Pr\left[ |e| = w \right] \cdot 4 \cdot 2^{n \left( \bar{H}\left( \frac{w}{n} \right) - 2 b(n) \right)} \\
        &\leq 4 \sum_{w = r+1}^{r_{\mathrm{seq}}} 2^{-n \left( 2 b(n) + D\left( \frac{w}{n} \,\big\|\, p_{\mathrm{eff}} \right) - \bar{H}\left( \frac{w}{n} \right) \right)} \\
        &\leq 4 n\, 2^{-n \left( 2 b(n) + \min\limits_{r < w \leq r_{\mathrm{seq}}} \left( D\left( \frac{w}{n} \,\big\|\, p_{\mathrm{eff}} \right) - \bar{H}\left( \frac{w}{n} \right) \right) \right)} .
    \end{split}\end{equation}
    It remains to show that the minimum in Eq.~\eqref{eq:p_fail_middle_min} is bounded from below by the exponent $E_0$ of Eq.~\eqref{eq:p_fail_E0}.
    Since $r = \lfloor \alpha n \rfloor > \alpha n - 1$, every integer weight $w$ with $r < w \leq r_{\mathrm{seq}}$ satisfies $w \geq r + 1 > \alpha n$ and $w \leq r_{\mathrm{seq}} \leq \alpha_{\mathrm{seq}} n$.
    Hence every such $w/n$ lies in $\left[ \alpha,\, \alpha_{\mathrm{seq}} \right]$, so relaxing $w/n$ to a real variable in that interval can only decrease the minimum.
    We can therefore apply Lemma~\ref{lem:exponent_minimization} of Sec.~\ref{SM_sec:exponent_minimization} with $p = p_{\mathrm{eff}}$, $\lambda_1 = \alpha$ and $\lambda_2 = \alpha_{\mathrm{seq}}$, whose conditions hold since $0 < p_{\mathrm{eff}} \leq 1/2$ by assumption and $\alpha < \alpha_{\mathrm{seq}} \leq 1$ by the convention on $\alpha_{\mathrm{seq}}$, and Eq.~\eqref{eq:exponent_minimization} evaluates the relaxed minimum in the three cases of Eq.~\eqref{eq:p_fail_E0}.
    The last case of Eq.~\eqref{eq:exponent_minimization} involves no $\bar{H}$ and carries over directly, while the two other cases need $\bar{H} = H$ at their endpoints.
    This holds at $\alpha$ because $\alpha \leq 1/2$, and at $\alpha_{\mathrm{seq}}$ because in that case $p_{\mathrm{eff}} \geq \alpha_{\mathrm{seq}}^2 \big/ \left( \alpha_{\mathrm{seq}}^2 + \left( 1 - \alpha_{\mathrm{seq}} \right)^2 \right)$ together with $p_{\mathrm{eff}} \leq 1/2$ gives $\alpha_{\mathrm{seq}}^2 \leq \left( 1 - \alpha_{\mathrm{seq}} \right)^2$, hence $\alpha_{\mathrm{seq}} \leq 1/2$.
    The relaxed minimum is therefore $E_0$, so Eq.~\eqref{eq:p_fail_middle_min} is at most the first term of Eq.~\eqref{eq:p_fail_high_weight}, and adding the bounds on the three sums proves Eq.~\eqref{eq:p_fail_high_weight}.
\end{proof}

\subsection{Precision and query complexity}\label{SM_sec:iid_error_bound}

Here we combine the response aggregation error, the recovery error, and the probability of an atypical response error pattern into the error of the whole protocol, and then read off the precision and the query complexity.
\begin{lemma}[Distillation error under i.i.d.\ dephasing]\label{lem:stage2_error_iid_new}
    If the recovery channel is the sequential recovery, $\mathsf{Rec}^{\mathrm{I}} = \mathsf{Rec}_{r_{\mathrm{seq}}}$ with maximal candidate weight $r_{\mathrm{seq}} \leq n$, then the distilled oracle $\widehat{\mathcal{O}}_f^{\mathrm{iid}}$ built from the i.i.d.-dephasing Boolean oracle $\tilde{\mathcal{O}}_f^{\mathrm{iid},Z}$ satisfies
    \begin{equation}\label{eq:stage2_error_iid_new}
        \left\| \widehat{\mathcal{O}}_f^{\mathrm{iid}} - \mathcal{O}_f \right\|_{\diamond}
        \leq 2 \max_{x,y,\,\tau \in \mathcal{T}}\, \epsilon_{\mathrm{agg}}(x,y;f,\tau)
        + 2 L\, p_{\mathrm{fail}}
        + 2 p_{\mathrm{atyp}} ,
    \end{equation}
    where $\epsilon_{\mathrm{agg}}$ is the response aggregation error Eq.~\eqref{eq:agg_error_def_OD}, $p_{\mathrm{fail}}$ the failure probability Eq.~\eqref{eq:def_p_fail}, and $p_{\mathrm{atyp}}$ the probability Eq.~\eqref{eq:iid_target_atyp} that the response error pattern is atypical.
\end{lemma}
\begin{proof}
    We bound each component of the mixture Eq.~\eqref{eq:iid_pattern_mixture} separately for typical and for atypical response error patterns, and then average over the patterns.
    Throughout the proof we write
    \begin{equation}
        \widehat{\mathcal{O}}{}_f^{\,\tau,\sigma}
        := \widehat{\mathcal{O}}{}_f^{\,\tau,\sigma}\left[ \left( \mathcal{D}^Z_{p_{\mathrm{eff}}} \right)^{\otimes nL},\, \mathsf{Rec}_{r_{\mathrm{seq}}} \right]
    \end{equation}
    for the component of the mixture with response error patterns $\tau$ and $\sigma$.
    Applying the triangle inequality for the diamond norm (Lemma~\ref{lem:channel_norm_facts}(iv)) to the mixture Eq.~\eqref{eq:iid_pattern_mixture} gives
    \begin{equation}\label{eq:iid_mixture_split}
        \left\| \widehat{\mathcal{O}}_f^{\mathrm{iid}} - \mathcal{O}_f \right\|_{\diamond}
        \leq \sum_{\tau,\sigma} p(\tau)\, p(\sigma)
        \left\| \widehat{\mathcal{O}}{}_f^{\,\tau,\sigma} - \mathcal{O}_f \right\|_{\diamond} .
    \end{equation}

    We first bound the summands with a typical response error pattern $\tau \in \mathcal{T}$.
    Lemma~\ref{lem:OD_error} with $\mathcal{E}_f^{\mathrm{tot}} = \left( \mathcal{D}^Z_{p_{\mathrm{eff}}} \right)^{\otimes nL}$ and $\mathsf{Rec}^{\mathrm{I}} = \mathsf{Rec}_{r_{\mathrm{seq}}}$ bounds such a summand by a sum of the response aggregation error and the recovery error.
    Lemma~\ref{lem:seq_recovery_reduction}, whose condition $r_{\mathrm{seq}} \leq n$ holds by assumption, bounds the recovery error by $2 L\, p_{\mathrm{fail}}$, so for every $\tau \in \mathcal{T}$ and every $\sigma$,
    \begin{equation}\label{eq:iid_typical_component}
        \left\| \widehat{\mathcal{O}}{}_f^{\,\tau,\sigma} - \mathcal{O}_f \right\|_{\diamond}
        \leq 2 \max_{x,y}\, \epsilon_{\mathrm{agg}}(x,y;f,\tau) + 2 L\, p_{\mathrm{fail}}
        \leq 2 \max_{x,y,\,\tau \in \mathcal{T}}\, \epsilon_{\mathrm{agg}}(x,y;f,\tau) + 2 L\, p_{\mathrm{fail}} .
    \end{equation}
    For an atypical response error pattern $\tau \notin \mathcal{T}$, both $\widehat{\mathcal{O}}{}_f^{\,\tau,\sigma}$ and $\mathcal{O}_f$ are CPTP maps, so their diamond distance is at most $2$ (Lemma~\ref{lem:channel_norm_facts}(iii)).

    Splitting the sum in Eq.~\eqref{eq:iid_mixture_split} according to whether $\tau$ is typical, and using $\sum_{\sigma} p(\sigma) = 1$, gives
    \begin{equation}\begin{split}
        \left\| \widehat{\mathcal{O}}_f^{\mathrm{iid}} - \mathcal{O}_f \right\|_{\diamond}
        &\leq \left( 2 \max_{x,y,\,\tau \in \mathcal{T}}\, \epsilon_{\mathrm{agg}}(x,y;f,\tau) + 2 L\, p_{\mathrm{fail}} \right) \Pr\left[ \tau \in \mathcal{T} \right]
        + 2 \Pr\left[ \tau \notin \mathcal{T} \right] \\
        &\leq 2 \max_{x,y,\,\tau \in \mathcal{T}}\, \epsilon_{\mathrm{agg}}(x,y;f,\tau) + 2 L\, p_{\mathrm{fail}} + 2 p_{\mathrm{atyp}} ,
    \end{split}\end{equation}
    where the last inequality uses $\Pr\left[ \tau \in \mathcal{T} \right] \leq 1$ and $\Pr\left[ \tau \notin \mathcal{T} \right] = p_{\mathrm{atyp}}$ by Eq.~\eqref{eq:iid_target_atyp}.
\end{proof}

These bounds now combine into a distillation guarantee whose conditions on the error rate depend on $n$.

\begin{lemma}[Distillation under i.i.d.\ depolarizing noise at finite $n$]\label{lem:boolean_OD_iid}
    The two-stage protocol distills the family of i.i.d.-depolarizing Boolean oracles $\tilde{\mathfrak{O}}_F^{\mathrm{iid}}$ with per-qubit error rate $p$ into the family of ideal Boolean oracles $\mathfrak{O}_F$ with precision $\epsilon$ and query complexity
    \begin{equation}\label{eq:T_OD_iid}
        T_{\mathrm{OD}} = 2 L ,
        \qquad
        L = \left\lceil L_0 \right\rceil ,
        \qquad
        L_0 := \frac{16 \left( 3 - 2 p_t \right)}{\left( 1 - 2 p_t \right)^2} \, \frac{1}{\eta} \, \ln \frac{12 m}{\epsilon} ,
    \end{equation}
    where the number of query blocks $L$ is a parameter of the protocol set from $p$, $\eta$, $m$ and $\epsilon$, provided $p < 3/4$, $p_{\mathrm{eff}} \leq \min\left\{ \alpha_{\mathrm{seq}},\, 1/2 \right\}$, and
    \begin{equation}\label{eq:iid_index_condition}
        \min\left\{ 2 b(n) + E_0,\; D\left( \alpha_{\mathrm{seq}} \,\big\|\, p_{\mathrm{eff}} \right) \right\}
        > \frac{1}{n} \log_2 \frac{48\, n \left( L_0 + 1 \right)}{\epsilon} .
    \end{equation}
    Here $p_{\mathrm{eff}} := 2 p_t \left( 1 - p_t \right)$ is the effective index error rate Eq.~\eqref{eq:iid_peff}, $b(n)$ the decay rate Eq.~\eqref{eq:def_b_high_weight} of the high-weight $Z$ expectations, and $E_0$ the exponent Eq.~\eqref{eq:p_fail_E0}.
    The query states must be generated from a seed state by Eq.~\eqref{eq:query-state-from-seedstate} whose EOC Eq.~\eqref{SM_eq:error_orthogonality} hold for $r$ errors, with matched query power $\eta$.
    The protocol uses the sequential recovery $\mathsf{Rec}_{r_{\mathrm{seq}}}$ as the index recovery, and its response aggregator uses the threshold $w^* = \eta L / 2$, which depends on $p$ through $L$.
\end{lemma}

\begin{proof}[Proof of Lemma~\ref{lem:boolean_OD_iid}]
    The hypothesis $p < 3/4$ gives $p_t = 2p/3 < 1/2$, and $p_{\mathrm{eff}} = 0$ exactly when $p = 0$.
    The protocol runs the $L = \left\lceil L_0 \right\rceil$ query blocks of Eq.~\eqref{eq:T_OD_iid}, each making one oracle call in the weak query and one in the uncomputation, so $T_{\mathrm{OD}} = 2L$.
    Lemma~\ref{lem:stage2_error_iid_new}, whose condition $r_{\mathrm{seq}} \leq n$ holds because $\alpha_{\mathrm{seq}} \leq 1$, bounds the diamond distance by the three terms of Eq.~\eqref{eq:stage2_error_iid_new}, and we bound each of them by $\epsilon/3$.

    \textit{Response aggregation error.}
    Since $L \geq L_0$, the definition of $L_0$ in Eq.~\eqref{eq:T_OD_iid} rearranges to $\frac{\left( 1 - 2 p_t \right)^2}{16 \left( 3 - 2 p_t \right)}\, \eta L \geq \ln\frac{12m}{\epsilon}$.
    Since the protocol uses the threshold $w^* = \eta L / 2$ and $p_t < 1/2$, Corollary~\ref{cor:agg_error_lowweight_OD} applies to every typical response error pattern and gives
    \begin{equation}\label{eq:iid_sa_term}
        \max_{x,y,\,\tau \in \mathcal{T}}\, \epsilon_{\mathrm{agg}}(x,y;f,\tau)
        \leq 2 \sqrt{m}\, \exp\left( -\frac{\left( 1 - 2 p_t \right)^2}{16 \left( 3 - 2 p_t \right)}\, \eta L \right)
        \leq 2 \sqrt{m}\, \frac{\epsilon}{12 m}
        = \frac{\epsilon}{6 \sqrt{m}}
        \leq \frac{\epsilon}{6} ,
    \end{equation}
    where the last step uses $m \geq 1$, so the first term of Eq.~\eqref{eq:stage2_error_iid_new} is at most $\epsilon/3$.

    \textit{Recovery error.}
    If $p = 0$, the index error pattern of every query block is $e = 0$ with probability $1$, and $0 \in \mathcal{K}_r$, so $\Pr[\hat{e} \neq e \,|\, e] = 0$ by Lemma~\ref{lem:seq_recovery_error} and $p_{\mathrm{fail}} = 0$ by Eq.~\eqref{eq:def_p_fail}, so the second term of Eq.~\eqref{eq:stage2_error_iid_new} vanishes.
    If $p > 0$, the hypotheses of Lemma~\ref{lem:p_fail_high_weight} hold, namely the EOC Eq.~\eqref{SM_eq:error_orthogonality} for $r = \lfloor \alpha n \rfloor$ errors and $0 < p_{\mathrm{eff}} \leq \min\left\{ \alpha_{\mathrm{seq}},\, 1/2 \right\}$, so Eq.~\eqref{eq:p_fail_high_weight} bounds $p_{\mathrm{fail}}$.
    Both of its exponents exceed $\frac{1}{n}\log_2\frac{48 n \left( L_0 + 1 \right)}{\epsilon}$ by Eq.~\eqref{eq:iid_index_condition}, which is at least $\frac{1}{n}\log_2\frac{48 n L}{\epsilon}$ because $L \leq L_0 + 1$, so each of the two powers of two is smaller than $\frac{\epsilon}{48 n L}$ and
    \begin{equation}\label{eq:iid_index_term}
        p_{\mathrm{fail}}
        \leq 4 n\, 2^{-n \left( 2 b(n) + E_0 \right)}
        + 2^{-n D\left( \alpha_{\mathrm{seq}} \,\big\|\, p_{\mathrm{eff}} \right)}
        < 4 n \cdot \frac{\epsilon}{48\, n\, L} + \frac{\epsilon}{48\, n\, L}
        \leq \frac{5 \epsilon}{48 L} ,
    \end{equation}
    where the last step uses $n \geq 1$.
    Hence the second term of Eq.~\eqref{eq:stage2_error_iid_new} is $2 L\, p_{\mathrm{fail}} < \frac{5\epsilon}{24} < \frac{\epsilon}{3}$.

    \textit{Atypical response error patterns.}
    If $p = 0$, the response error pattern is $\tau = 0$ with probability $1$ by Eq.~\eqref{eq:iid_pattern_weights}, and $0 \in \mathcal{T}$, so $p_{\mathrm{atyp}} = 0$.
    If $p > 0$, we bound $p_{\mathrm{atyp}}$ from $L \geq L_0$ and $p_t < 1/2$ alone.
    The threshold fraction $1/4 + p_t/2$ of the typical set Eq.~\eqref{eq:iid_typical_set} lies strictly between $p_t$ and $1/2$, and Eq.~\eqref{eq:iid_target_atyp} gives $p_{\mathrm{atyp}} \leq m\, 2^{-L\, D\left( \frac{1}{4} + \frac{p_t}{2} \| p_t \right)}$, so it remains to bound that exponent from below.
    The two Bernoulli distributions have total variation distance $\left( \frac{1}{4} + \frac{p_t}{2} \right) - p_t = \frac{1 - 2 p_t}{4}$, so Pinsker's inequality~\cite{CoverThomas2006} in bits gives $D\left( \frac{1}{4} + \frac{p_t}{2} \,\big\|\, p_t \right) \geq \frac{2}{\ln 2} \left( \frac{1 - 2 p_t}{4} \right)^2 = \frac{\left( 1 - 2 p_t \right)^2}{8 \ln 2}$.
    Together with $L \geq L_0 \geq \frac{32}{\left( 1 - 2 p_t \right)^2}\, \frac{1}{\eta} \ln\frac{12m}{\epsilon}$, which uses $3 - 2 p_t > 2$, this gives
    \begin{equation}\label{eq:iid_target_exponent}
        L\, D\left( \frac{1}{4} + \frac{p_t}{2} \,\bigg\|\, p_t \right)
        \geq \frac{32}{\left( 1 - 2 p_t \right)^2}\, \frac{1}{\eta} \ln\frac{12m}{\epsilon} \cdot \frac{\left( 1 - 2 p_t \right)^2}{8 \ln 2}
        = \frac{4}{\eta} \log_2 \frac{12m}{\epsilon}
        > \log_2 \frac{6m}{\epsilon} ,
    \end{equation}
    where the last step uses $\eta \leq 1$.
    Hence $p_{\mathrm{atyp}} < m \cdot \frac{\epsilon}{6m} = \frac{\epsilon}{6}$, the third term of Eq.~\eqref{eq:stage2_error_iid_new} is at most $\epsilon/3$, and adding the three terms bounds the diamond distance by $\epsilon$.
\end{proof}

Both sides of the condition Eq.~\eqref{eq:iid_index_condition} of Lemma~\ref{lem:boolean_OD_iid} depend on $n$, so the lemma on its own does not single out a range of per-qubit error rates that works at every $n$.
The first lemma below defines a constant threshold through an equation in the per-qubit error rate and characterizes the error rates below it.
The second lemma shows that in the parameter regime we consider, the conditions of Lemma~\ref{lem:boolean_OD_iid} hold at all sufficiently large $n$ whenever the per-qubit error rate is below this threshold.

\begin{lemma}[Threshold equation]\label{lem:threshold_equation}
    Let constants $\alpha$, $\alpha_{\mathrm{seq}}$, $h > 0$, $B \geq 0$ and $\nu \geq 0$ satisfy one of the following two conditions.
    \begin{enumerate}[label=(\roman*)]
        \item $0 < \alpha < \alpha_{\mathrm{seq}} \leq 1/2$.
        \item $0 < \alpha \leq 1/2$, $\alpha_{\mathrm{seq}} = 1$, and $h + \nu > B - 1$.
    \end{enumerate}
    Then there is a unique $p_{\mathrm{th}} \in (0, 3/4)$ solving
    \begin{equation}\label{eq:iid_pth_asymptotic}
        \min\left\{ B + E_0,\; D\left( \alpha_{\mathrm{seq}} \,\big\|\, p_{\mathrm{eff}} \right) \right\}
        = h + \nu
        \qquad\text{with}\quad
        p_{\mathrm{eff}} < \alpha_{\mathrm{seq}} ,
    \end{equation}
    where $p_{\mathrm{eff}}$ is the effective index error rate Eq.~\eqref{eq:iid_peff} at $p_{\mathrm{th}}$ and $E_0$ the exponent Eq.~\eqref{eq:p_fail_E0} at this $p_{\mathrm{eff}}$ over the interval $\left[ \alpha, \alpha_{\mathrm{seq}} \right]$.
    Moreover, a constant error rate $p \in (0, 3/4)$ lies below $p_{\mathrm{th}}$ if and only if its effective index error rate satisfies $p_{\mathrm{eff}} \leq \alpha_{\mathrm{seq}}$ and
    \begin{equation}\label{eq:iid_pth_characterization}
        \min\left\{ B + E_0,\; D\left( \alpha_{\mathrm{seq}} \,\big\|\, p_{\mathrm{eff}} \right) \right\} > h + \nu ,
    \end{equation}
    where now $p_{\mathrm{eff}}$ and $E_0$ are evaluated at $p$.
\end{lemma}
\begin{proof}
    We first show that Eq.~\eqref{eq:iid_pth_asymptotic} has a unique solution.
    Its right-hand side $h + \nu$ is a positive constant, because $h > 0$ and $\nu \geq 0$.
    Lemma~\ref{lem:exponent_monotone}(ii) with $\lambda_1 = \alpha$, $\lambda_2 = \alpha_{\mathrm{seq}}$, $c = B$, and $t = h + \nu$, whose hypothesis is condition (i) or condition (ii), gives a unique $p_{\mathrm{eff}}^* \in \left( 0, \min\left\{ \alpha_{\mathrm{seq}},\, 1/2 \right\} \right)$ solving Eq.~\eqref{eq:iid_pth_asymptotic}.
    The effective index error rate $p_{\mathrm{eff}} = 2 p_t \left( 1 - p_t \right)$ with $p_t = 2p/3$ is continuous and strictly increasing in $p$ on $\left( 0, 3/4 \right)$ with values $\left( 0, 1/2 \right)$, so exactly one $p_{\mathrm{th}} \in \left( 0, 3/4 \right)$ attains $p_{\mathrm{eff}}(p_{\mathrm{th}}) = p_{\mathrm{eff}}^*$, and this $p_{\mathrm{th}}$ is the unique solution of Eq.~\eqref{eq:iid_pth_asymptotic} with $p_{\mathrm{eff}} < \alpha_{\mathrm{seq}}$.

    We next prove the characterization of $p < p_{\mathrm{th}}$, with $p_{\mathrm{eff}}$ and $E_0$ evaluated at $p$ here and below.
    If $p = 0$, both sides hold, because $p_{\mathrm{th}} > 0$, $p_{\mathrm{eff}} = 0 \leq \alpha_{\mathrm{seq}}$, and both entries of the minimum in Eq.~\eqref{eq:iid_pth_characterization} are infinite.
    If $p \in (0, 3/4)$, then $p_{\mathrm{eff}} > 0$, and $p < p_{\mathrm{th}}$ is equivalent to $p_{\mathrm{eff}} < p_{\mathrm{eff}}^*$ because $p_{\mathrm{eff}}$ is strictly increasing in $p$ on $(0, 3/4)$.
    Since $p_{\mathrm{eff}}^* < \alpha_{\mathrm{seq}}$, this forces $p_{\mathrm{eff}} \leq \alpha_{\mathrm{seq}}$, and for $p_{\mathrm{eff}} \in \left( 0, \min\left\{ \alpha_{\mathrm{seq}},\, 1/2 \right\} \right]$, which contains every $p_{\mathrm{eff}} \leq \alpha_{\mathrm{seq}}$ because $p_{\mathrm{eff}} = 2 p_t \left( 1 - p_t \right) \leq 1/2$, Lemma~\ref{lem:exponent_monotone}(iii) at the same $\lambda_1$, $\lambda_2$, $c$, and $t$ makes $p_{\mathrm{eff}} < p_{\mathrm{eff}}^*$ equivalent to Eq.~\eqref{eq:iid_pth_characterization}.
\end{proof}

\begin{lemma}[Constant threshold for the conditions of Lemma~\ref{lem:boolean_OD_iid}]\label{lem:iid_asymptotic_threshold}
    For any constants $\alpha$ and $\alpha_{\mathrm{seq}}$, any precision $\epsilon = \epsilon(n) \in (0,1)$, and any $m$ with $\log_2 m = o(n)$, let the query states satisfy the EOC Eq.~\eqref{SM_eq:error_orthogonality} for $r = \lfloor \alpha n \rfloor$ errors with matched query power $\eta$.
    Suppose that the precision, the matched query power, and the decay rate Eq.~\eqref{eq:def_b_high_weight} of the seed state satisfy
    \begin{equation}\label{eq:iid_asymptotic_constants}
        \nu := \limsup_{n \to \infty} \frac{1}{n} \log_2 \frac{1}{\epsilon} < \infty ,
        \qquad
        h \geq \limsup_{n \to \infty} \frac{1}{n} \log_2 \frac{1}{\eta} ,
        \qquad
        B \leq \liminf_{n \to \infty} 2 b(n) ,
    \end{equation}
    for constants $h > 0$ and $B \geq 0$ such that $\alpha$, $\alpha_{\mathrm{seq}}$, $h$, $B$ and $\nu$ satisfy condition (i) or (ii) of Lemma~\ref{lem:threshold_equation}, and let $p_{\mathrm{th}}$ be the threshold of that lemma at these constants.
    Then for every constant per-qubit error rate $p < p_{\mathrm{th}}$ and all sufficiently large $n$, the three conditions of Lemma~\ref{lem:boolean_OD_iid} on the error rate hold, namely $p < 3/4$, $p_{\mathrm{eff}} \leq \min\left\{ \alpha_{\mathrm{seq}},\, 1/2 \right\}$, and Eq.~\eqref{eq:iid_index_condition}.
\end{lemma}
\begin{proof}
    We show that for any constant $p < p_{\mathrm{th}}$, the three conditions hold for all sufficiently large $n$.
    If $p = 0$, then $p = 0 < 3/4$ and $p_{\mathrm{eff}} = 0 \leq \min\left\{ \alpha_{\mathrm{seq}},\, 1/2 \right\}$, and both entries of the minimum on the left-hand side of Eq.~\eqref{eq:iid_index_condition} are infinite while the right-hand side is finite, so all three conditions hold at every $n$.
    Now let $p > 0$.
    Then $p < 3/4$ because $p_{\mathrm{th}} < 3/4$ by Lemma~\ref{lem:threshold_equation}, and the characterization Eq.~\eqref{eq:iid_pth_characterization} of that lemma gives $p_{\mathrm{eff}} \leq \alpha_{\mathrm{seq}}$, which together with $p_{\mathrm{eff}} = 2 p_t \left( 1 - p_t \right) \leq 1/2$ gives $p_{\mathrm{eff}} \leq \min\left\{ \alpha_{\mathrm{seq}},\, 1/2 \right\}$, so the first two conditions hold, together with the constant gap
    \begin{equation}\label{eq:iid_asymptotic_gap}
        \delta := \min\left\{ B + E_0,\; D\left( \alpha_{\mathrm{seq}} \,\big\|\, p_{\mathrm{eff}} \right) \right\} - \left( h + \nu \right) > 0 .
    \end{equation}
    It remains to verify Eq.~\eqref{eq:iid_index_condition}.
    The left-hand side of Eq.~\eqref{eq:iid_index_condition} differs from the minimum in Eq.~\eqref{eq:iid_asymptotic_gap} only by $2 b(n)$ in place of $B$ in its first entry.
    The hypothesis on the decay rate gives $2 b(n) \geq B - o(1)$, and lowering one entry of a minimum by at most $o(1)$ lowers the minimum by at most $o(1)$, so the left-hand side is at least $h + \nu + \delta - o(1)$.
    For the right-hand side of Eq.~\eqref{eq:iid_index_condition}, the prefactor $\frac{16 \left( 3 - 2 p_t \right)}{\left( 1 - 2 p_t \right)^2}$ of $L_0$ in Eq.~\eqref{eq:T_OD_iid} is a finite constant because $p_t = 2p/3 < 1/2$, and $\frac{1}{\eta} \ln \frac{12m}{\epsilon} \geq 1$ because $\eta \leq 1$ and $\ln \frac{12m}{\epsilon} \geq 1$, so $L_0 + 1$ is at most a constant times $\frac{1}{\eta} \ln \frac{12m}{\epsilon}$.
    Here $\ln \frac{12m}{\epsilon} \leq \log_2 \frac{12m}{\epsilon} \leq 4 + \log_2 m + \log_2 \frac{1}{\epsilon} = O(n)$ by $\log_2 m = o(n)$ and the limit superior defining $\nu$, so $48\, n \left( L_0 + 1 \right) / \epsilon$ is at most $\frac{1}{\eta}\, \frac{1}{\epsilon}$ times a polynomial in $n$ and
    \begin{equation}
        \frac{1}{n} \log_2 \frac{48\, n \left( L_0 + 1 \right)}{\epsilon}
        \leq \frac{1}{n} \log_2 \frac{1}{\eta}
        + \frac{1}{n} \log_2 \frac{1}{\epsilon}
        + o(1)
        \leq h + \nu + o(1)
    \end{equation}
    by the two limits superior in Eq.~\eqref{eq:iid_asymptotic_constants}.
    The two sides of Eq.~\eqref{eq:iid_index_condition} are therefore separated by at least $\delta - o(1)$, which is positive for all sufficiently large $n$, so the third condition holds as well.
\end{proof}

With these two lemmas in place, applying Lemma~\ref{lem:boolean_OD_iid} to a concrete construction only requires reading off the constants $h$ and $B$ of Eq.~\eqref{eq:iid_asymptotic_constants} from its matched query power and its seed state.
The following two corollaries do this for Constructions 3 and 2 at $r = \lfloor \alpha n \rfloor$, and state the resulting protocols with their constant thresholds and query complexities.

\begin{corof}{lem:boolean_OD_iid}{1}[Distillation of Boolean oracles under i.i.d.\ depolarizing noise via Construction 3]\label{cor:iid_boolean_query_complexity}
    For any constants $\alpha$ and $\alpha_{\mathrm{seq}}$ with $0 < \alpha \leq 0.16$ and $\alpha < \alpha_{\mathrm{seq}} \leq 1/2 - \alpha$, any precision $\epsilon = \epsilon(n) \in (0,1)$ with finite decay rate $\nu := \limsup_{n\to\infty} \frac{1}{n} \log_2 \frac{1}{\epsilon}$, any $m$ with $\log_2 m = o(n)$, any constant per-qubit error rate $p < p_{\mathrm{th}}^{(3)}\left( \alpha, \alpha_{\mathrm{seq}}, \nu \right)$, and all sufficiently large $n$, the two-stage protocol with the query states of Construction 3, the sequential recovery with maximal candidate weight $r_{\mathrm{seq}} = \lfloor \alpha_{\mathrm{seq}} n \rfloor$, and the number of query blocks $L$ of Eq.~\eqref{eq:T_OD_iid}, which depends on $p$, distills the family of i.i.d.-depolarizing Boolean oracles $\tilde{\mathfrak{O}}_F^{\mathrm{iid}}$ with this error rate into the family of ideal Boolean oracles $\mathfrak{O}_F$ with precision $\epsilon$ and query complexity
    \begin{equation}\label{eq:T_OD_C3}
        T_{\mathrm{OD}} = 2 \left\lceil \frac{16 \left( 3 - 2 p_t \right)}{\left( 1 - 2 p_t \right)^2} \, \frac{1}{\xi(n)} \, N^{H(\alpha) + 2\alpha} \ln \frac{12m}{\epsilon} \right\rceil ,
    \end{equation}
    where $p_t = 2p/3$ and $\xi(n)$ is the function in the matched query power Eq.~\eqref{eq:construction3_eta} of Construction 3, which the construction leaves free subject to $\xi(n) \geq 1$ and $\log \xi(n) = o(n)$.
    The threshold $p_{\mathrm{th}}^{(3)}\left( \alpha, \alpha_{\mathrm{seq}}, \nu \right)$ is the unique solution of
    \begin{equation}\label{eq:iid_pth_C3}
        \min\left\{ 6 \alpha + \alpha^2 + H(2\alpha) + E_0,\; D\left( \alpha_{\mathrm{seq}} \,\big\|\, p_{\mathrm{eff}} \right) \right\}
        = H(\alpha) + 2 \alpha + \nu
        \qquad\text{with}\quad
        p_{\mathrm{eff}} < \alpha_{\mathrm{seq}} ,
    \end{equation}
    where $p_{\mathrm{eff}}$ is the effective index error rate Eq.~\eqref{eq:iid_peff} at $p_{\mathrm{th}}^{(3)}$ and $E_0$ the exponent Eq.~\eqref{eq:p_fail_E0} at this $p_{\mathrm{eff}}$.
\end{corof}
\begin{proof}
    We derive the corollary from Lemma~\ref{lem:boolean_OD_iid}, whose three conditions on the error rate we obtain from Lemma~\ref{lem:iid_asymptotic_threshold} with the constants $h := H(\alpha) + 2 \alpha$ and $B := 6 \alpha + \alpha^2 + H(2\alpha)$.
    The threshold $p_{\mathrm{th}}^{(3)}$ is the threshold $p_{\mathrm{th}}$ of Lemma~\ref{lem:threshold_equation} at these constants.

    We first verify the assumptions of Lemma~\ref{lem:iid_asymptotic_threshold}.
    Condition (i) of Lemma~\ref{lem:threshold_equation} holds because $\alpha < \alpha_{\mathrm{seq}} \leq 1/2 - \alpha \leq 1/2$, and $\nu < \infty$ and $\log_2 m = o(n)$ are assumptions of the corollary.
    By Theorem~\ref{thm:near_optimal_construction}, the query states of Construction 3 satisfy the EOC for $r = \lfloor \alpha n \rfloor$ errors with matched query power $\eta = \xi(n)\, 2^{-n \left( H(\alpha) + 2 \alpha \right)}$, so $\xi(n) \geq 1$ gives $\frac{1}{n} \log_2 \frac{1}{\eta} \leq H(\alpha) + 2 \alpha = h$ at every $n$.

    It remains to bound the decay rate of the high weight expectations, which we do by Lemma~\ref{lem:construction3_high_weight}.
    The lemma covers only weights $|a| \leq n - 2r$, so we first check that every weight in the maximum of Eq.~\eqref{eq:def_b_high_weight} is at most $n - 2r$.
    Its upper end satisfies $\min\left\{ 2 r_{\mathrm{seq}},\, n \right\} \leq 2 \alpha_{\mathrm{seq}} n \leq \left( 1 - 2\alpha \right) n \leq n - 2r$, by $\alpha_{\mathrm{seq}} \leq 1/2 - \alpha$ and $r \leq \alpha n$.
    Applying the bound of Lemma~\ref{lem:construction3_high_weight} at every weight of the maximum and taking $-\frac{1}{n} \log_2$ gives, for all sufficiently large $n$,
    \begin{equation}\label{eq:C3_b_lower}
        b(n) \geq 3 \alpha + \frac{\alpha^2}{2} + \frac{H(2\alpha)}{2} - \frac{1}{n} \log_2 \left[ \xi(n) + \frac{1-\alpha}{\alpha} \sqrt{n+1} \left( 2\sqrt{2} + 8 \sqrt{\alpha n} + 8 \xi(n) \right) \right] .
    \end{equation}
    Since $\xi(n) \geq 1$ and $\log \xi(n) = o(n)$, the argument of the logarithm is $2^{o(n)}$, so the last term is $o(1)$.
    Hence $\liminf_{n \to \infty} 2 b(n) \geq 6 \alpha + \alpha^2 + H(2\alpha) = B$.

    Lemma~\ref{lem:iid_asymptotic_threshold} therefore applies, and gives, for every constant $p < p_{\mathrm{th}}^{(3)}$ and all sufficiently large $n$, the three conditions of Lemma~\ref{lem:boolean_OD_iid} on the error rate.
    Lemma~\ref{lem:boolean_OD_iid} then gives the claimed distillation, and substituting $\eta = \xi(n)\, 2^{-n \left( H(\alpha) + 2 \alpha \right)}$ with $N = 2^n$ into its query complexity gives Eq.~\eqref{eq:T_OD_C3}.
\end{proof}

\begin{corof}{lem:boolean_OD_iid}{2}[Distillation of Boolean oracles under i.i.d.\ depolarizing noise via Construction 2]\label{cor:iid_boolean_query_complexity_C2}
    For any constants $\alpha$ and $\alpha_{\mathrm{seq}}$ with $0 < \alpha \leq 1/4$ and either $\alpha < \alpha_{\mathrm{seq}} \leq 1/2$, or $\alpha_{\mathrm{seq}} = 1$ and $\alpha < 1/4$, any precision $\epsilon = \epsilon(n) \in (0,1)$ with finite decay rate $\nu := \limsup_{n\to\infty} \frac{1}{n} \log_2 \frac{1}{\epsilon}$, any $m$ with $\log_2 m = o(n)$, any constant per-qubit error rate $p < p_{\mathrm{th}}^{(2)}\left( \alpha, \alpha_{\mathrm{seq}}, \nu \right)$, and all sufficiently large $n$, the two-stage protocol with the query states of Construction 2, the sequential recovery with maximal candidate weight $r_{\mathrm{seq}} = \lfloor \alpha_{\mathrm{seq}} n \rfloor$, and the number of query blocks $L$ of Eq.~\eqref{eq:T_OD_iid}, which depends on $p$, distills the family of i.i.d.-depolarizing Boolean oracles $\tilde{\mathfrak{O}}_F^{\mathrm{iid}}$ with this error rate into the family of ideal Boolean oracles $\mathfrak{O}_F$ with precision $\epsilon$ and query complexity
    \begin{equation}\label{eq:T_OD_C2}
        T_{\mathrm{OD}} = 2 \left\lceil \frac{16 \left( 3 - 2 p_t \right)}{\left( 1 - 2 p_t \right)^2} \, N^{H(2\alpha)} \ln \frac{12m}{\epsilon} \right\rceil ,
    \end{equation}
    where $p_t = 2p/3$.
    The threshold $p_{\mathrm{th}}^{(2)}\left( \alpha, \alpha_{\mathrm{seq}}, \nu \right)$ is the unique solution of
    \begin{equation}\label{eq:iid_pth_C2}
        \min\left\{ 2 H(2\alpha) + E_0,\; D\left( \alpha_{\mathrm{seq}} \,\big\|\, p_{\mathrm{eff}} \right) \right\}
        = H(2\alpha) + \nu
        \qquad\text{with}\quad
        p_{\mathrm{eff}} < \alpha_{\mathrm{seq}} ,
    \end{equation}
    where $p_{\mathrm{eff}}$ is the effective index error rate Eq.~\eqref{eq:iid_peff} at $p_{\mathrm{th}}^{(2)}$ and $E_0$ the exponent Eq.~\eqref{eq:p_fail_E0} at this $p_{\mathrm{eff}}$.
\end{corof}
\begin{proof}
    We derive the corollary from Lemma~\ref{lem:boolean_OD_iid}, whose three conditions on the error rate we obtain from Lemma~\ref{lem:iid_asymptotic_threshold} with the constants $h := H(2\alpha)$ and $B := 2 H(2\alpha)$.
    The threshold $p_{\mathrm{th}}^{(2)}$ is the threshold $p_{\mathrm{th}}$ of Lemma~\ref{lem:threshold_equation} at these constants.

    We first verify the assumptions of Lemma~\ref{lem:iid_asymptotic_threshold}.
    The conditions $\nu < \infty$ and $\log_2 m = o(n)$ are assumptions of the corollary.
    If $\alpha < \alpha_{\mathrm{seq}} \leq 1/2$, condition (i) of Lemma~\ref{lem:threshold_equation} holds.
    If $\alpha_{\mathrm{seq}} = 1$, condition (ii) holds, because $\alpha < 1/4 \leq 1/2$ and $h + \nu - (B - 1) = 1 - H(2\alpha) + \nu > 0$ for $\alpha < 1/4$.
    By Sec.~\ref{SM_sec:construction2}, the query states of Construction 2 satisfy the EOC for $r = \lfloor \alpha n \rfloor$ errors with matched query power exactly $\eta = M_{2r}^{-1}$, and Lemma~\ref{lem:entropy_binomial} applied with the integer $2r = 2 \lfloor \alpha n \rfloor \leq 2 \alpha n$ and with $2 \alpha \leq 1/2$ gives $M_{2r} \leq 2^{n H(2\alpha)}$, so $\frac{1}{n} \log_2 \frac{1}{\eta} \leq H(2\alpha) = h$ at every $n$.

    It remains to bound the decay rate of the high weight expectations, which we do by Lemma~\ref{lem:construction2_high_weight}.
    The lemma gives $\left| \bra{\Theta_s} Z^{a} \ket{\Theta_s} \right| = M_{2r}^{-1}$ at every weight $|a| > 2r$, so the maximum in Eq.~\eqref{eq:def_b_high_weight} gives $b(n) = \frac{1}{n} \log_2 M_{2r}$ whatever the maximal candidate weight is.
    Lemma~\ref{lem:entropy_binomial} gives $M_{2r} \geq \binom{n}{2r} \geq \frac{1}{n+1}\, 2^{n H\left( 2r/n \right)}$, so $b(n) \geq H\left( 2r/n \right) - \frac{1}{n} \log_2 \left( n + 1 \right)$.
    Here $2r/n = 2 \lfloor \alpha n \rfloor / n$ tends to $2\alpha$ and $H$ is continuous, so $\liminf_{n \to \infty} 2 b(n) \geq 2 H(2\alpha) = B$.

    Lemma~\ref{lem:iid_asymptotic_threshold} therefore applies, and gives, for every constant $p < p_{\mathrm{th}}^{(2)}$ and all sufficiently large $n$, the three conditions of Lemma~\ref{lem:boolean_OD_iid} on the error rate.
    Lemma~\ref{lem:boolean_OD_iid} then gives the claimed distillation, and substituting $\eta = M_{2r}^{-1} \geq 2^{-n H(2\alpha)}$ with $N = 2^n$ into its query complexity gives Eq.~\eqref{eq:T_OD_C2}.
\end{proof}

We now combine the two corollaries above into a single statement that takes the exponent $\gamma$ of the query complexity as given and returns the distillation threshold.
For each construction we take the largest $\alpha$ whose query complexity stays within $N^{\gamma}$ and the largest $\alpha_{\mathrm{seq}}$ that the protocol admits, and the threshold is the larger of the two.

\iidthresholdexponent*
\begin{proof}
    We run the two-stage protocol with Construction 3 as in Corollary~\ref{cor:iid_boolean_query_complexity} and with Construction 2 as in Corollary~\ref{cor:iid_boolean_query_complexity_C2}, in each case at the largest $\alpha$ at which the exponent of $N$ in the query complexity is at most $\gamma$ and at the largest $\alpha_{\mathrm{seq}}$ that the protocol admits, and we use whichever construction has the larger threshold.

    The two values $\alpha^{(3)}$ and $\alpha^{(2)}$ exist, because $H(\alpha) + 2 \alpha$ is continuous and strictly increasing on $(0, 0.16]$ with limit zero as $\alpha \to 0^{+}$, and $H(2\alpha)$ is continuous and strictly increasing on $(0, 1/4)$ with limit zero as $\alpha \to 0^{+}$ and limit $1 > \gamma$ as $\alpha \to 1/4$.
    With Construction 3 we take $\alpha_{\mathrm{seq}} = \frac{1}{2} - \alpha^{(3)}$, which Corollary~\ref{cor:iid_boolean_query_complexity} admits because $\alpha^{(3)} \leq 0.16 < 1/4$ gives $\alpha^{(3)} < \frac{1}{2} - \alpha^{(3)}$.
    With Construction 2 we take $\alpha_{\mathrm{seq}} = 1$, which Corollary~\ref{cor:iid_boolean_query_complexity_C2} admits because $\alpha^{(2)} < 1/4$.
    The present theorem assumes exactly the conditions on $\epsilon$ and $m$ that both corollaries require, so Corollary~\ref{cor:iid_boolean_query_complexity} applies at $\left( \alpha^{(3)}, \tfrac{1}{2} - \alpha^{(3)}, \nu \right)$ with $\xi(n) = 1$ and Corollary~\ref{cor:iid_boolean_query_complexity_C2} at $\left( \alpha^{(2)}, 1, \nu \right)$, which are the arguments in Eq.~\eqref{eq:iid_pth_at_exponent}.

    A constant $p < p_{\mathrm{th}}\left( \gamma, \nu \right)$ lies below at least one of the two thresholds in Eq.~\eqref{eq:iid_pth_at_exponent}, and the corresponding corollary then gives distillation with precision $\epsilon$ at query complexity Eq.~\eqref{eq:T_OD_C3} with $\xi(n) = 1$ or Eq.~\eqref{eq:T_OD_C2}.
    Both are at most Eq.~\eqref{eq:T_OD_at_exponent}, since their exponents $H\left( \alpha^{(3)} \right) + 2 \alpha^{(3)}$ and $H\left( 2 \alpha^{(2)} \right)$ are at most $\gamma$ by the definitions of $\alpha^{(3)}$ and $\alpha^{(2)}$.
\end{proof}

\subsection{Distillation threshold at small overhead exponent}\label{SM_sec:small_gamma}
This section determines how the distillation threshold $p_{\mathrm{th}}\left( \gamma, \nu \right)$ of Eq.~\eqref{eq:iid_pth_at_exponent} vanishes as the overhead exponent $\gamma$ tends to zero, and then inverts the function $\gamma \mapsto p_{\mathrm{th}}\left( \gamma, \nu \right)$ to obtain the smallest overhead exponent $\gamma_{\min}(p)$ at which $p_{\mathrm{th}}\left( \gamma, \nu \right) > p$ for a given error rate $p$, as $p \to 0$.
The computation has three steps.
First we expand the thresholds $p_{\mathrm{th}}^{(3)}$ and $p_{\mathrm{th}}^{(2)}$ of the two constructions as $\alpha \to 0$.
Second we substitute $\alpha^{(3)}$ and $\alpha^{(2)}$ of Eq.~\eqref{eq:iid_pth_at_exponent} as functions of $\gamma$, which expresses $p_{\mathrm{th}}\left( \gamma, \nu \right)$ in terms of $\gamma$ as $\gamma \to 0$.
Third we invert this function, which gives $\gamma_{\min}(p)$ as $p \to 0$.
The regimes $\nu = 0$ and $\nu > 0$ behave differently and are treated separately in each step.
No quantity in this section depends on $n$, since $\gamma$, $\nu$, $\alpha$, $p$ and every threshold are constants, and every $o(\cdot)$ and $O(\cdot)$ is taken in the variable of the current step.

The thresholds $p_{\mathrm{th}}^{(3)}\left( \alpha, \frac{1}{2} - \alpha, \nu \right)$ and $p_{\mathrm{th}}^{(2)}\left( \alpha, 1, \nu \right)$ that enter Eq.~\eqref{eq:iid_pth_at_exponent} are defined by the threshold equation~\eqref{eq:iid_pth_asymptotic} with the constants of Eq.~\eqref{eq:iid_asymptotic_constants} supplied by the proofs of Corollaries~\ref{cor:iid_boolean_query_complexity} and~\ref{cor:iid_boolean_query_complexity_C2}, namely $h_3(\alpha) := H(\alpha) + 2 \alpha$ and $B_3(\alpha) := 6 \alpha + \alpha^2 + H(2\alpha)$ for Construction 3 at $\alpha_{\mathrm{seq}} = \frac{1}{2} - \alpha$, and $h_2(\alpha) := H(2\alpha)$ and $B_2(\alpha) := 2 H(2\alpha)$ for Construction 2 at $\alpha_{\mathrm{seq}} = 1$.
Expanding $\log_2 \left( 1 - x \right) = - x \log_2 e + O(x^2)$ in the binary entropy gives $H(\alpha) = \alpha \log_2 \frac{1}{\alpha} + \alpha \log_2 e + O(\alpha^2)$ and hence $H(2\alpha) = 2 H(\alpha) - 2 \alpha + O(\alpha^2)$, so the only two facts about these constants that we need are
\begin{equation}\label{eq:small_alpha_h}
    h_3 = \alpha \log_2 \frac{1}{\alpha} \left( 1 + o(1) \right) ,
    \qquad
    h_2 = 2 \alpha \log_2 \frac{1}{\alpha} \left( 1 + o(1) \right) ,
    \qquad
    B_3 = 2 h_3 + O(\alpha^2) ,
    \qquad
    B_2 = 2 h_2 .
\end{equation}
In particular $h \to 0$ and $B = 2 h + o(h)$ as $\alpha \to 0$ for either construction, where from here on $h$, $B$, $\alpha_{\mathrm{seq}}$ and $p_{\mathrm{th}}$ denote the constants and the threshold of whichever construction is under consideration.

The first step reduces the threshold equation~\eqref{eq:iid_pth_asymptotic} of either construction to an equation for the exponent $E_0$ of Eq.~\eqref{eq:p_fail_E0} alone, evaluated at $u := p_{\mathrm{eff}}\left( p_{\mathrm{th}} \right)$, the effective index error rate Eq.~\eqref{eq:iid_peff} at the threshold.
Like $p_{\mathrm{th}}$, the rate $u$ is a function of $\alpha$, and we first show $u \to 0$ as $\alpha \to 0$.
The minimum in Eq.~\eqref{eq:iid_pth_asymptotic} equals $h + \nu$, so its first entry satisfies $B + E_0 \geq h + \nu \geq 0$.
If $u \geq \alpha$, then $\lambda = u$ is admissible in the minimum Eq.~\eqref{eq:p_fail_E0} because $u < \alpha_{\mathrm{seq}}$, and $D\left( u \,\big\|\, u \right) = 0$ with $u \leq 1/2$ gives $E_0 \leq - \bar{H}(u) = - H(u)$, so $H(u) \leq B - h - \nu \leq B \to 0$ and hence $u \to 0$.
If $u < \alpha$, then $u \to 0$ as well.
Given $u \to 0$, the second entry of the minimum diverges, $D\left( \alpha_{\mathrm{seq}} \,\big\|\, u \right) \geq \alpha_{\mathrm{seq}} \log_2 \frac{\alpha_{\mathrm{seq}}}{u} - 1 \to \infty$, since its second term $\left( 1 - \alpha_{\mathrm{seq}} \right) \log_2 \frac{1 - \alpha_{\mathrm{seq}}}{1 - u}$ is at least $\left( 1 - \alpha_{\mathrm{seq}} \right) \log_2 \left( 1 - \alpha_{\mathrm{seq}} \right) \geq -1$ and $\alpha_{\mathrm{seq}} \geq 1/4$ for both constructions.
The right-hand side $h + \nu$ of the threshold equation stays bounded, so the minimum is attained by its first entry and the threshold equation reads $B + E_0 = h + \nu$, which by Eq.~\eqref{eq:small_alpha_h} is
\begin{equation}\label{eq:small_alpha_E0}
    E_0 = h + \nu - B = \nu - h \left( 1 + o(1) \right) .
\end{equation}
It remains to solve Eq.~\eqref{eq:small_alpha_E0} for $u$, which requires knowing which case of Eq.~\eqref{eq:p_fail_E0} gives $E_0$.
The second case never applies, because it requires $u \geq \alpha_{\mathrm{seq}}^2 \big/ \left( \alpha_{\mathrm{seq}}^2 + \left( 1 - \alpha_{\mathrm{seq}} \right)^2 \right) \geq \frac{1}{10}$ while $u \to 0$.
Which of the other two cases applies depends on $\nu$, and we treat $\nu = 0$ and $\nu > 0$ separately.

At $\nu = 0$, Eq.~\eqref{eq:small_alpha_E0} reads $E_0 = - h \left( 1 + o(1) \right)$, and we show that the first case of Eq.~\eqref{eq:p_fail_E0} cannot produce this value.
In the first case, $u \leq u_\alpha := \alpha^2 \big/ \left( \alpha^2 + \left( 1 - \alpha \right)^2 \right) \leq \alpha$, and since $D\left( \alpha \,\big\|\, u \right)$ decreases in $u$ on $u \leq \alpha$, $E_0 = D\left( \alpha \,\big\|\, u \right) - H(\alpha) \geq D\left( \alpha \,\big\|\, u_\alpha \right) - H(\alpha)$.
Writing out $D\left( \alpha \,\big\|\, u_\alpha \right) = \alpha \log_2 \frac{1}{u_\alpha} - H(\alpha) - \left( 1 - \alpha \right) \log_2 \left( 1 - u_\alpha \right)$ with $\log_2 \frac{1}{u_\alpha} = 2 \log_2 \frac{1}{\alpha} + \log_2 \left( \alpha^2 + \left( 1 - \alpha \right)^2 \right) = 2 \log_2 \frac{1}{\alpha} + O(\alpha)$, $H(\alpha) = \alpha \log_2 \frac{1}{\alpha} + \alpha \log_2 e + O(\alpha^2)$ and $u_\alpha = O(\alpha^2)$ gives $E_0 \geq - 2 \alpha \log_2 e + O(\alpha^2)$.
This contradicts $E_0 = - h \left( 1 + o(1) \right)$, because Eq.~\eqref{eq:small_alpha_h} gives $h / \alpha \geq \log_2 \frac{1}{\alpha} \left( 1 + o(1) \right) \to \infty$ for either construction.
Hence the third case applies, which reads $- E_0 = \log_2 \left( 1 + 2 \sqrt{u \left( 1 - u \right)} \right) = 2 \sqrt{u} \log_2 e + O(u)$ by $\sqrt{1 - u} = 1 + O(u)$ and $\ln \left( 1 + x \right) = x + O(x^2)$.
Equating this with $- E_0 = h \left( 1 + o(1) \right)$ gives $u = \frac{\ln^2 2}{4}\, h^2 \left( 1 + o(1) \right)$, and inverting $u = \frac{4 p_{\mathrm{th}}}{3} \left( 1 - \frac{2 p_{\mathrm{th}}}{3} \right)$ from Eq.~\eqref{eq:iid_peff} as $p_{\mathrm{th}} = \frac{3 u}{4} \left( 1 + O(u) \right)$ gives
\begin{equation}\label{eq:small_alpha_pth_const}
    p_{\mathrm{th}} = \frac{3 \ln^2 2}{16}\, h^2 \left( 1 + o(1) \right)
\end{equation}
for either construction, which depends on $\alpha$ only through its exponent $h$.

At constant $\nu > 0$, Eq.~\eqref{eq:small_alpha_E0} makes $E_0 \to \nu > 0$, while the third case of Eq.~\eqref{eq:p_fail_E0} gives $E_0 = - \log_2 \left( 1 + 2 \sqrt{u \left( 1 - u \right)} \right) \leq 0$, so the first case applies.
It reads $E_0 = D\left( \alpha \,\big\|\, u \right) - H(\alpha) = \alpha \log_2 \frac{1}{u} - 2 H(\alpha) - \left( 1 - \alpha \right) \log_2 \left( 1 - u \right)$, so Eq.~\eqref{eq:small_alpha_E0} becomes
\begin{equation}\label{eq:small_alpha_nu_equation}
    \alpha \log_2 \frac{1}{u} = \nu - h \left( 1 + o(1) \right) + 2 H(\alpha) + \left( 1 - \alpha \right) \log_2 \left( 1 - u \right) = \nu + o(1) ,
\end{equation}
where the last equality uses $h \to 0$, $H(\alpha) \to 0$ and $u \to 0$.
Hence $\log_2 \frac{1}{u} = \frac{\nu}{\alpha} \left( 1 + o(1) \right)$, and $p_{\mathrm{th}} = \frac{3 u}{4} \left( 1 + O(u) \right)$ gives
\begin{equation}\label{eq:small_alpha_pth_exp}
    p_{\mathrm{th}} = 2^{- \frac{\nu}{\alpha} \left( 1 + o(1) \right)}
\end{equation}
for either construction, where the factor $\frac{3}{4}$ is absorbed into the $o(1)$ in the exponent because $\frac{\nu}{\alpha}$ diverges.
Unlike Eq.~\eqref{eq:small_alpha_pth_const}, this threshold depends on $\alpha$ itself and not only on $h$.

The second step expresses $p_{\mathrm{th}}\left( \gamma, \nu \right)$ of Eq.~\eqref{eq:iid_pth_at_exponent} as a function of $\gamma$, by substituting $\alpha^{(3)}$ and $\alpha^{(2)}$ as functions of $\gamma$ into Eqs.~\eqref{eq:small_alpha_pth_const} and~\eqref{eq:small_alpha_pth_exp}, and from here on every $o(\cdot)$ is taken as $\gamma \to 0$.
In Eq.~\eqref{eq:iid_pth_at_exponent}, $\alpha^{(3)}$ is the largest $\alpha \in (0, 0.16]$ with $h_3(\alpha) \leq \gamma$ and $\alpha^{(2)}$ is the largest $\alpha \in (0, 1/4)$ with $h_2(\alpha) \leq \gamma$.
Both $h_3 = H(\alpha) + 2 \alpha$ and $h_2 = H(2\alpha)$ are continuous and strictly increasing on these intervals with limit zero as $\alpha \to 0$, since $H$ is so on $\left[ 0, 1/2 \right]$, so at small $\gamma$ both maxima are attained with equality, $h_3\left( \alpha^{(3)} \right) = h_2\left( \alpha^{(2)} \right) = \gamma$, and both $\alpha^{(3)}$ and $\alpha^{(2)}$ tend to zero as $\gamma \to 0$.
Solving $\alpha \log_2 \frac{1}{\alpha} \left( 1 + o(1) \right) = \gamma$ and $2 \alpha \log_2 \frac{1}{\alpha} \left( 1 + o(1) \right) = \gamma$ from Eq.~\eqref{eq:small_alpha_h} gives $\log_2 \frac{1}{\alpha} = \log_2 \frac{1}{\gamma} \left( 1 + o(1) \right)$ in both cases and hence
\begin{equation}\label{eq:small_gamma_alpha}
    \alpha^{(3)} = \frac{\gamma}{\log_2 (1/\gamma)} \left( 1 + o(1) \right) ,
    \qquad
    \alpha^{(2)} = \frac{\gamma}{2 \log_2 (1/\gamma)} \left( 1 + o(1) \right) .
\end{equation}
The $o(1)$ terms in Eqs.~\eqref{eq:small_alpha_pth_const} and~\eqref{eq:small_alpha_pth_exp} were taken as $\alpha \to 0$, and since $\alpha^{(3)}$ and $\alpha^{(2)}$ tend to zero as $\gamma \to 0$, they remain $o(1)$ as $\gamma \to 0$ after substituting $\alpha = \alpha^{(3)}$ or $\alpha = \alpha^{(2)}$.
At $\nu = 0$, Eq.~\eqref{eq:small_alpha_pth_const} depends on $\alpha$ only through $h$, and both entries of the maximum in Eq.~\eqref{eq:iid_pth_at_exponent} have $h = \gamma$, so
\begin{equation}\label{eq:small_gamma_pth_const}
    p_{\mathrm{th}}\left( \gamma, 0 \right) = \frac{3 \ln^2 2}{16}\, \gamma^2 \left( 1 + o(1) \right) .
\end{equation}
At $\nu > 0$, substituting Eq.~\eqref{eq:small_gamma_alpha} into Eq.~\eqref{eq:small_alpha_pth_exp} gives $2^{- \frac{\nu}{\gamma} \log_2 \frac{1}{\gamma} \left( 1 + o(1) \right)}$ for the Construction 3 entry and $2^{- \frac{2 \nu}{\gamma} \log_2 \frac{1}{\gamma} \left( 1 + o(1) \right)}$ for the Construction 2 entry, so the maximum is the former and
\begin{equation}\label{eq:small_gamma_pth_exp}
    p_{\mathrm{th}}\left( \gamma, \nu \right) = 2^{- \frac{\nu}{\gamma} \log_2 \frac{1}{\gamma} \left( 1 + o(1) \right)} .
\end{equation}

The third step inverts the function $\gamma \mapsto p_{\mathrm{th}}\left( \gamma, \nu \right)$ of Eqs.~\eqref{eq:small_gamma_pth_const} and~\eqref{eq:small_gamma_pth_exp}, and from here on every $o(\cdot)$ is taken as $p \to 0$.
Let $\gamma_{\min}(p) := \inf\left\{ \gamma \in (0, 1) : p_{\mathrm{th}}\left( \gamma, \nu \right) > p \right\}$.
The right-hand sides of Eqs.~\eqref{eq:small_gamma_pth_const} and~\eqref{eq:small_gamma_pth_exp} are increasing in $\gamma$ at small $\gamma$, so $\gamma_{\min}(p)$ is obtained up to a factor $1 + o(1)$ by solving Eq.~\eqref{eq:small_gamma_pth_const} or Eq.~\eqref{eq:small_gamma_pth_exp} for $\gamma$ at $p_{\mathrm{th}}\left( \gamma, \nu \right) = p$.
At $\nu = 0$, Eq.~\eqref{eq:small_gamma_pth_const} gives
\begin{equation}\label{eq:gamma_min_const}
    \gamma_{\min}(p) = \frac{4}{\sqrt{3}\, \ln 2}\, \sqrt{p} \left( 1 + o(1) \right) = \Theta\!\left( \sqrt{p} \right) .
\end{equation}
At $\nu > 0$, Eq.~\eqref{eq:small_gamma_pth_exp} gives $\log_2 \frac{1}{p} = \frac{\nu}{\gamma} \log_2 \frac{1}{\gamma} \left( 1 + o(1) \right)$, whose logarithm gives $\log_2 \frac{1}{\gamma} = \log_2 \log_2 \frac{1}{p} \left( 1 + o(1) \right)$, and substituting back,
\begin{equation}\label{eq:gamma_min_exp}
    \gamma_{\min}(p) = \nu\, \frac{\log_2 \log_2 (1/p)}{\log_2 (1/p)} \left( 1 + o(1) \right) .
\end{equation}
The smallest overhead exponent therefore vanishes as $p \to 0$ in both regimes, and faster at $\nu = 0$ than at any constant $\nu > 0$.

\subsection{Supporting lemmas}\label{SM_sec:exponent_minimization}
The exponent $E_0$ of the failure probability bound Eq.~\eqref{eq:p_fail_high_weight} is the minimum of $D\left( \lambda \,\big\|\, p_{\mathrm{eff}} \right) - \bar{H}(\lambda)$ over the weight fractions $\lambda$ of the index error patterns that the sequential recovery can misidentify.
Here we evaluate this minimum over an arbitrary interval of weight fractions and then determine how it moves with $p$.

\begin{lemma}[Minimizer and minimum of $D(\,\cdot \,\|\, p) - \bar{H}(\cdot)$ over an interval]\label{lem:exponent_minimization}
    Let $0 < p \leq 1/2$ and $0 \leq \lambda_1 \leq \lambda_2 \leq 1$.
    The function $D\left( \lambda \,\big\|\, p \right) - \bar{H}(\lambda)$ attains its minimum over $\lambda \in \left[ \lambda_1, \lambda_2 \right]$ at
    \begin{equation}\label{eq:exponent_minimizer}
        \lambda_{\min} := \min\left\{ \max\left\{ \lambda^*,\, \lambda_1 \right\},\, \lambda_2 \right\} ,
        \qquad \lambda^* := \frac{\sqrt{p}}{\sqrt{p} + \sqrt{1 - p}} ,
    \end{equation}
    and
    \begin{equation}\label{eq:exponent_minimization}
        \min_{\lambda \in [\lambda_1, \lambda_2]} \left( D\left( \lambda \,\big\|\, p \right) - \bar{H}(\lambda) \right) =
        \begin{cases}
            D\left( \lambda_1 \,\big\|\, p \right) - \bar{H}(\lambda_1), & p \leq \dfrac{\lambda_1^2}{\lambda_1^2 + \left( 1 - \lambda_1 \right)^2} \\[10pt]
            D\left( \lambda_2 \,\big\|\, p \right) - \bar{H}(\lambda_2), & p \geq \dfrac{\lambda_2^2}{\lambda_2^2 + \left( 1 - \lambda_2 \right)^2} \\[10pt]
            - \log_2\left( 1 + 2 \sqrt{p \left( 1 - p \right)} \right), & \text{otherwise} .
        \end{cases}
    \end{equation}
\end{lemma}
\begin{proof}
    Write $G(\lambda) := D\left( \lambda \,\big\|\, p \right) - \bar{H}(\lambda)$ for the function to be minimized.
    Since $\bar{H}(\lambda) = H(\lambda)$ for $\lambda \leq 1/2$ and $\bar{H}(\lambda) = 1$ for $\lambda \geq 1/2$ (Lemma~\ref{lem:entropy_binomial}), $G$ has the two branches
    \begin{equation}\label{eq:exponent_two_branches}
        G(\lambda) =
        \begin{cases}
            D\left( \lambda \,\big\|\, p \right) - H(\lambda), & \lambda \leq 1/2 \\[4pt]
            D\left( \lambda \,\big\|\, p \right) - 1, & \lambda \geq 1/2 ,
        \end{cases}
    \end{equation}
    which agree at $\lambda = 1/2$.

    The branch for $\lambda \leq 1/2$ is itself a KL divergence.
    Expanding the definitions of the KL divergence and the binary entropy, and then writing $\sqrt{p} = c\, \lambda^*$ and $\sqrt{1 - p} = c \left( 1 - \lambda^* \right)$ with $c := \sqrt{p} + \sqrt{1 - p}$,
    \begin{equation}\label{eq:exponent_identity}\begin{split}
        D\left( \lambda \,\big\|\, p \right) - H(\lambda)
        &= \lambda \log_2 \frac{\lambda}{p} + \left( 1 - \lambda \right) \log_2 \frac{1 - \lambda}{1 - p}
        + \lambda \log_2 \lambda + \left( 1 - \lambda \right) \log_2 \left( 1 - \lambda \right)\\
        &= 2 \left( \lambda \log_2 \frac{\lambda}{\sqrt{p}} + \left( 1 - \lambda \right) \log_2 \frac{1 - \lambda}{\sqrt{1 - p}} \right)\\
        &= 2 D\left( \lambda \,\big\|\, \lambda^* \right) - 2 \log_2 c
        = 2 D\left( \lambda \,\big\|\, \lambda^* \right) - \log_2 \left( 1 + 2 \sqrt{p \left( 1 - p \right)} \right) ,
    \end{split}\end{equation}
    where the last step uses $c^2 = 1 + 2 \sqrt{p \left( 1 - p \right)}$.

    Both branches of Eq.~\eqref{eq:exponent_two_branches} are now a KL divergence $D\left( \lambda \,\big\|\, q \right)$ plus a $\lambda$-independent constant, with $q = \lambda^*$ on the branch for $\lambda \leq 1/2$ by Eq.~\eqref{eq:exponent_identity} and $q = p$ on the branch for $\lambda \geq 1/2$, so their monotonicity follows from the derivative of the KL divergence in its first argument.
    For any $q \in (0,1)$, $\partial_\lambda D\left( \lambda \,\big\|\, q \right) = \log_2 \frac{\lambda \left( 1 - q \right)}{\left( 1 - \lambda \right) q}$, which is negative for $\lambda < q$ and positive for $\lambda > q$ since $\lambda / \left( 1 - \lambda \right)$ is increasing in $\lambda$.
    On the branch for $\lambda \leq 1/2$ this makes $G$ decrease on $[0, \lambda^*]$ and increase on $[\lambda^*, 1/2]$, where $\lambda^* \leq 1/2$ because $p \leq 1 - p$, and on the branch for $\lambda \geq 1/2$, where $\lambda \geq 1/2 \geq p$, it makes $G$ nondecreasing on $[1/2, 1]$.

    Combining the two branches, $G$ decreases on $[0, \lambda^*]$ and is nondecreasing on $[\lambda^*, 1]$, so its minimum over $[\lambda_1, \lambda_2]$ is attained at the point of that interval closest to $\lambda^*$, which is the $\lambda_{\min}$ of Eq.~\eqref{eq:exponent_minimizer}, and the minimum value is $G(\lambda_{\min})$.
    For the case $\lambda_{\min} = \lambda^*$, Eq.~\eqref{eq:exponent_identity} applies because $\lambda^* \leq 1/2$, and $D\left( \lambda^* \,\big\|\, \lambda^* \right) = 0$ gives $G(\lambda^*) = - \log_2 \left( 1 + 2 \sqrt{p \left( 1 - p \right)} \right)$, the third case of Eq.~\eqref{eq:exponent_minimization}.

    It remains to state the positions of $\lambda^*$ as conditions on $p$.
    The map $f(\lambda) := \lambda^2 \big/ \left( \lambda^2 + \left( 1 - \lambda \right)^2 \right)$ equals $1 \big/ \left( 1 + \left( 1/\lambda - 1 \right)^2 \right)$ on $(0, 1]$, where $1/\lambda - 1$ is nonnegative and strictly decreasing, so $f$ strictly increases on $(0, 1]$ and, since $f(0) = 0$, on all of $[0, 1]$.
    It sends $\lambda^*$ to $p$, because $\lambda^{*2} = p/c^2$ and $\left( 1 - \lambda^* \right)^2 = \left( 1 - p \right)/c^2$, so for every $\lambda \in [0, 1]$, $\lambda^* \leq \lambda$ is equivalent to $p \leq f(\lambda)$; and $\lambda^* \geq \lambda$ is equivalent to $p \geq f(\lambda)$.
    At $\lambda = \lambda_1$ and $\lambda = \lambda_2$ these equivalences are the three conditions of Eq.~\eqref{eq:exponent_minimization}.
\end{proof}

The threshold of Sec.~\ref{SM_sec:iid_error_bound} is fixed by the behavior of this minimum as a function of $p$.

\begin{lemma}[Monotonicity of the minimum in $p$]\label{lem:exponent_monotone}
    Let $0 < \lambda_1 < \lambda_2 \leq 1$ and write
    \begin{equation}\label{eq:exponent_monotone_E}
        E(p) := \min_{\lambda \in \left[ \lambda_1, \lambda_2 \right]} \left( D\left( \lambda \,\big\|\, p \right) - \bar{H}(\lambda) \right)
    \end{equation}
    for the minimum of Lemma~\ref{lem:exponent_minimization}.
    \begin{enumerate}[label=(\roman*)]
        \item On the range $0 < p \leq \min\left\{ \lambda_2,\, 1/2 \right\}$, the function $E$ is continuous and strictly decreasing, and $E(p) \to \infty$ as $p \to 0^{+}$.
        \item Let $c \geq 0$ and $t > 0$ be constants.
        If $\lambda_2 \leq 1/2$, or if $\lambda_2 = 1$, $\lambda_1 \leq 1/2$ and $t > c - 1$, then there is a unique $p^* \in \left( 0, \min\left\{ \lambda_2,\, 1/2 \right\} \right)$ with $\min\left\{ c + E(p^*),\; D\left( \lambda_2 \,\big\|\, p^* \right) \right\} = t$.
        \item For this $p^*$, a given $p \in \left( 0, \min\left\{ \lambda_2,\, 1/2 \right\} \right]$ satisfies $\min\left\{ c + E(p),\; D\left( \lambda_2 \,\big\|\, p \right) \right\} > t$ if and only if $p < p^*$.
    \end{enumerate}
\end{lemma}
\begin{proof}
    Throughout, $p$ ranges over $\left( 0,\, \min\left\{ \lambda_2,\, 1/2 \right\} \right]$, so $p \leq 1/2$ and Lemma~\ref{lem:exponent_minimization} applies.

    For the continuity in item (i), that lemma attains the minimum at the point $\lambda_{\min}$ of Eq.~\eqref{eq:exponent_minimizer}, so $E(p) = D\left( \lambda_{\min} \,\big\|\, p \right) - \bar{H}(\lambda_{\min})$.
    Here $\lambda^*$ is continuous in $p$ and taking the maximum and the minimum with the constants $\lambda_1$ and $\lambda_2$ preserves continuity, so $\lambda_{\min}$ is continuous in $p$, and $D\left( \lambda \,\big\|\, p \right) - \bar{H}(\lambda)$ is continuous on $[0,1] \times (0,1)$, so $E$ is continuous.

    For the strict decrease in item (i), the three cases of Eq.~\eqref{eq:exponent_minimization} compare $p$ with $\lambda^2 \big/ \left( \lambda^2 + \left( 1 - \lambda \right)^2 \right)$ at $\lambda = \lambda_1$ and $\lambda = \lambda_2$, and this quantity equals $1 \big/ \left( 1 + \left( 1/\lambda - 1 \right)^2 \right)$, which increases in $\lambda$, so the three cases occupy three consecutive subintervals of the range.
    On the subinterval where $p \leq \lambda_1^2 \big/ \left( \lambda_1^2 + \left( 1 - \lambda_1 \right)^2 \right)$, $E(p) = D\left( \lambda_1 \,\big\|\, p \right) - \bar{H}(\lambda_1)$ and $p \leq \lambda_1$, because $\lambda_1^2 \big/ \left( \lambda_1^2 + \left( 1 - \lambda_1 \right)^2 \right) \leq \lambda_1$ rearranges to $\left( 1 - \lambda_1 \right) \left( 1 - 2 \lambda_1 \right) \geq 0$ when $\lambda_1 \leq 1/2$, while $p \leq 1/2 < \lambda_1$ when $\lambda_1 > 1/2$, so $E$ is strictly decreasing there by Lemma~\ref{lem:divergence_monotone}(i) with $a = \lambda_1$.
    On the subinterval where $p \geq \lambda_2^2 \big/ \left( \lambda_2^2 + \left( 1 - \lambda_2 \right)^2 \right)$, $E(p) = D\left( \lambda_2 \,\big\|\, p \right) - \bar{H}(\lambda_2)$ with $p \leq \lambda_2$, so Lemma~\ref{lem:divergence_monotone}(i) with $a = \lambda_2$ applies again.
    On the remaining subinterval, which lies between these two, $E(p) = - \log_2\left( 1 + 2 \sqrt{p \left( 1 - p \right)} \right)$, which is strictly decreasing because $p \left( 1 - p \right)$ is strictly increasing on $\left( 0,\, 1/2 \right]$.
    The three subintervals cover the whole range and $E$ is continuous, so $E$ is strictly decreasing there.

    For the limit in item (i), expand the KL divergence as $D\left( \lambda \,\big\|\, p \right) = - \lambda \log_2 p - \left( 1 - \lambda \right) \log_2 \left( 1 - p \right) - H(\lambda)$.
    For every $\lambda \in [\lambda_1, \lambda_2]$ the first term is at least $\lambda_1 \log_2 \frac{1}{p}$, the second term is nonnegative, and $H(\lambda) \leq 1$ and $\bar{H}(\lambda) \leq 1$, so $D\left( \lambda \,\big\|\, p \right) - \bar{H}(\lambda) \geq \lambda_1 \log_2 \frac{1}{p} - 2$.
    Taking the minimum over $\lambda$ gives $E(p) \geq \lambda_1 \log_2 \frac{1}{p} - 2$, which tends to $\infty$ as $p \to 0^{+}$.

    For items (ii) and (iii), write $\bar{p} := \min\left\{ \lambda_2,\, 1/2 \right\}$ for the right end of the range of item (i), so that $\bar{p} = \lambda_2$ when $\lambda_2 \leq 1/2$ and $\bar{p} = 1/2$ when $\lambda_2 = 1$.
    On this range, both entries of the minimum $\min\left\{ c + E(p),\; D\left( \lambda_2 \,\big\|\, p \right) \right\}$ are continuous and strictly decreasing in $p$, the first by item (i) and the second by Lemma~\ref{lem:divergence_monotone}(i) with $a = \lambda_2$, so this minimum is continuous and strictly decreasing on $\left( 0, \bar{p} \right]$.
    As $p \to 0^{+}$, the first entry tends to $\infty$ by item (i) and the second by Lemma~\ref{lem:divergence_monotone}(ii), so the minimum exceeds $t$ at every sufficiently small $p$.
    At $p = \bar{p}$ the minimum is below $t$.
    If $\lambda_2 \leq 1/2$, it is at most $D\left( \lambda_2 \,\big\|\, \lambda_2 \right) = 0 < t$ by Lemma~\ref{lem:divergence_monotone}(ii).
    If $\lambda_2 = 1$, then $\bar{p} = 1/2$ and $E(1/2) = -1$ by the definition Eq.~\eqref{eq:exponent_monotone_E}, because $D\left( \lambda \,\big\|\, 1/2 \right) - \bar{H}(\lambda) = 1 - H(\lambda) - \bar{H}(\lambda) \geq -1$ for every $\lambda \in [0,1]$, with equality at $\lambda = 1/2$, which lies in $\left[ \lambda_1, 1 \right]$ because $\lambda_1 \leq 1/2$, so the first entry of the minimum is $c + E(1/2) = c - 1 < t$.
    By the intermediate value theorem the minimum equals $t$ at some $p^* \in \left( 0, \bar{p} \right)$, and the strict decrease makes this $p^*$ unique, which is item (ii), and gives that a $p \in \left( 0, \bar{p} \right]$ satisfies $\min\left\{ c + E(p),\; D\left( \lambda_2 \,\big\|\, p \right) \right\} > t$ if and only if $p < p^*$, which is item (iii).
\end{proof}

\clearpage
\section{Simple Query State Constructions}\label{SM_sec:query_states}

We first prove the two reformulations of the EOC used in the main text, as uniform marginal distributions of the seed state (Lemma~\ref{lem:uniform_marginals}) and, for permutation-symmetric seed states, as the Krawtchouk moment conditions on the weight distribution (Lemma~\ref{lem:krawtchouk_moments}).
We then verify the conditions and the matched query powers for Constructions 1 and 2, and analyze Construction 4.
The near-optimal Construction 3 is considerably more involved and is deferred to Sec.~\ref{SM_sec:construction3}.
Throughout, the seed states take the permutation-symmetric form $\ket{\Theta_s} = \sum_{w=0}^n \sqrt{p_w}\, \ket{D_w^n}$ of Eq.~\eqref{eq:seed_state_dicke}.

\subsection{Error orthogonality as uniform marginals}\label{SM_sec:uniform_marginals}
The EOC Eq.~\eqref{SM_eq:error_orthogonality} involve only the diagonal operators $Z^a Z^b$, so they can be restated as a classical property of the computational-basis distribution of the seed state.
\begin{lemma}[Error orthogonality as uniform marginals]\label{lem:uniform_marginals}
    The seed state $\ket{\Theta_s}$ satisfies the EOC Eq.~\eqref{SM_eq:error_orthogonality} if and only if its computational-basis distribution $p(z) = |\braket{z|\Theta_s}|^2$ has uniform marginal distributions on every subset of at most $2r$ bits.
\end{lemma}
\begin{proof}
    Since $Z^a Z^b = Z^{a\oplus b}$ and every bitstring of weight at most $2r$ splits as $a \oplus b$ with $|a|,|b| \leq r$, the EOC hold if and only if $\bra{\Theta_s} Z^c \ket{\Theta_s} = 0$ for every $c$ with $0 < |c| \leq 2r$.
    Each $Z^c$ is diagonal, so $\bra{\Theta_s} Z^c \ket{\Theta_s} = \sum_z (-1)^{c\cdot z} p(z)$ is the Fourier coefficient of $p$ at $c$.
    For a subset $S$ of at most $2r$ bits and $s \in \{0,1\}^{|S|}$, the marginal distribution of $p$ on $S$ is the expectation of the projector onto $\ket{s}$ on the qubits in $S$.
    Expanding each factor as $\ket{s_i}\bra{s_i} = (I + (-1)^{s_i} Z_i)/2$ writes this marginal as
    \begin{equation}
        p^S(s) = 2^{-|S|} \sum_{\operatorname{supp}(c) \subseteq S} (-1)^{c|_S \cdot s}\, \bra{\Theta_s} Z^c \ket{\Theta_s},
    \end{equation}
    where $c|_S \in \{0,1\}^{|S|}$ denotes the restriction of $c$ to the bits in $S$.
    If all Fourier coefficients with $0 < |c| \leq 2r$ vanish, only the identity term survives and $p^S(s) = 2^{-|S|}$ for every $s$, so the marginals are uniform.
    Conversely, if all marginals on at most $2r$ bits are uniform, then for any $c$ with $0<|c|\leq 2r$, taking $S = \operatorname{supp}(c)$ gives $\bra{\Theta_s} Z^c \ket{\Theta_s} = \sum_{s} (-1)^{c|_S \cdot s} p^S(s) = 0$.
\end{proof}

\subsection{Error orthogonality as Krawtchouk moments}\label{SM_sec:krawtchouk_moments}
For permutation-symmetric seed states, the uniform-marginal characterization reduces further, to moment conditions on the weight distribution.
The conditions are written in terms of the binary Krawtchouk polynomials
\begin{equation}\label{eq:def_K_SM}
    K_k(w) := \sum_{j=0}^{k} (-1)^j \binom{w}{j}\binom{n-w}{k-j},
\end{equation}
where $\deg K_k = k$ and $K_k(0) = \binom{n}{k}$.

\begin{lemma}[Error orthogonality as Krawtchouk moments]\label{lem:krawtchouk_moments}
A permutation-symmetric state $\ket{\psi}$ on $n$ qubits (possibly entangled with additional registers) satisfies the EOC Eq.~\eqref{SM_eq:error_orthogonality} if and only if
\begin{equation}\label{eq:krawtchouk_moments_SM}
    \sum_{w=0}^n p_w\, K_k(w) = \delta_{k,0}, \quad k = 0, 1, \ldots, 2r,
\end{equation}
where $p_w = \bra{\psi} \Pi_w \ket{\psi}$ is the weight distribution and $\Pi_w$ projects onto the Hamming weight-$w$ subspace.
\end{lemma}

\begin{proof}
Since $Z^{b} Z^{b'} = Z^{b\oplus b'}$ and every bitstring $a$ of weight at most $2r$ splits as $a = b \oplus b'$ with $|b|,|b'| \leq r$, the EOC are equivalent to $\bra{\psi} Z^a \ket{\psi} = \delta_{|a|,0}$ for all $|a| \leq 2r$.
We compute the expectation of a single $Z$-string on a Dicke state, through the symmetrized $Z$-observables
\begin{equation}
    A_k = \sum_{\substack{a \in \{0,1\}^n \\ |a| = k}} Z^a.
\end{equation}
$A_k$ is diagonal in the computational basis, so we compute its action on a basis state $\ket{z}$ with $|z| = w$.
Using $Z^a \ket{z} = (-1)^{a \cdot z}\ket{z}$ and grouping the strings by $j = a \cdot z$, of which there are $\binom{w}{j}\binom{n-w}{k-j}$ for each $j$,
\begin{equation}
    A_k \ket{z} = \left(\sum_{j=0}^{k} (-1)^j \binom{w}{j}\binom{n-w}{k-j}\right) \ket{z} = K_k(w)\,\ket{z}.
\end{equation}
Hence $A_k$ acts as the scalar $K_k(w)$ on the entire weight-$w$ subspace, and in particular $A_k \ket{D_w^n} = K_k(w)\,\ket{D_w^n}$.
Since $\ket{D_w^n}$ is permutationally invariant, $\bra{D_w^n} Z^a \ket{D_w^n}$ depends only on $|a| = k$, so summing over the $\binom{n}{k}$ strings of weight $k$ gives
\begin{equation}
    \bra{D_w^n} Z^a \ket{D_w^n} = \frac{1}{\binom{n}{k}} \bra{D_w^n} A_k \ket{D_w^n} = \frac{K_k(w)}{\binom{n}{k}}.
\end{equation}
Since $\ket{\psi}$ is symmetric, it decomposes into Dicke components, and for any $a$ with $|a| = k$,
\begin{equation}\label{eq:symmetric_Z_expectation}
    \bra{\psi} Z^a \ket{\psi} = \sum_w p_w\, \bra{D_w^n} Z^a \ket{D_w^n} = \frac{1}{\binom{n}{k}} \sum_w p_w\, K_k(w).
\end{equation}
The right-hand side does not depend on $a$, so $\bra{\psi} Z^a \ket{\psi} = 0$ for all $a$ with $|a| = k$ if and only if $\sum_w p_w\, K_k(w) = 0$.
Applying this for $k = 1, \ldots, 2r$ and noting that $k = 0$ reduces to normalization gives the equivalence.
\end{proof}

Constructing a valid seed state is thus the purely classical problem of finding a weight distribution that satisfies Eq.~\eqref{eq:krawtchouk_moments_SM}.

\subsection{Construction 1}\label{SM_sec:construction1}
We restate the seed states for both parities of $n$, then verify that the seed states satisfy the EOC for single-qubit $Z$ errors.
For odd $n \ge 3$, the seed state is Eq.~\eqref{eq:Theta-0-odd} of the main text,
\begin{equation}
    \ket{\Theta_s} = \frac{1}{\sqrt{n+1}} \ket{0^n} + \sqrt{\frac{n}{n+1}} \ket{D_{\frac{n+1}{2}}^n}.
\end{equation}
For even $n \ge 2$, no single Dicke state satisfies both moment conditions together with $\ket{0^n}$, since that requires the weight $(n+1)/2$, so the seed state superposes two Dicke states, of which the one at weight $n/2$ has $K_1(n/2) = 0$ and enters only the $k = 2$ condition,
\begin{equation}\label{eq:Theta-0-even}
    \ket{\Theta_s} = \frac{1}{\sqrt{n+2}} \ket{0^n} + \frac{1}{\sqrt{2}} \ket{D_{\frac{n}{2}}^n} + \sqrt{\frac{n}{2(n+2)}} \ket{D_{\frac{n}{2}+1}^n}.
\end{equation}

For $r = 1$, the nontrivial conditions in Eq.~\eqref{eq:krawtchouk_moments_SM} are those with $k = 1, 2$, with $K_1(w) = n - 2w$ and $K_2(w) = \frac{(n-2w)^2 - n}{2}$.
For odd $n$, both $K_1(\frac{n+1}{2}) = -1$ and $K_2(\frac{n+1}{2}) = \frac{1-n}{2}$ equal $-K_k(0)/n$, so both moment conditions reduce to $p_{\frac{n+1}{2}} = n\, p_0$, which the amplitudes satisfy.
For even $n$, substituting $p_0 = \frac{1}{n+2}$, $p_{\frac{n}{2}} = \frac{1}{2}$, and $p_{\frac{n}{2}+1} = \frac{n}{2(n+2)}$ gives
\begin{align}
    \sum_w p_w K_1(w) &= \frac{n}{n+2} + \frac{1}{2} \cdot 0 - \frac{n}{n+2} = 0, \\
    \sum_w p_w K_2(w) &= \frac{n(n-1)}{2(n+2)} - \frac{n}{4} + \frac{n(4-n)}{4(n+2)} = \frac{n \left[ 2(n-1) - (n+2) + (4-n) \right]}{4(n+2)} = 0.
\end{align}
The matched query power is $\eta = p_0$, which equals $\frac{1}{n+1}$ for odd $n$ and $\frac{1}{n+2}$ for even $n$.

\subsection{Construction 2}\label{SM_sec:construction2}
We verify that $\rho$ defined in~\eqref{eq:encoding2} is a valid density matrix, that its seed state satisfies the EOC for $r$ errors, and that it achieves the matched query power claimed in the main text.
Throughout this subsection we write $\lambda := M_{2r}^{-1}$ and use the equivalent mixture form of Eq.~\eqref{eq:encoding2},
\begin{equation}\label{eq:encoding2_mixture}
    \rho = \lambda \left(\ket{0^n}\!\bra{0^n} - \frac{1}{2^n}\sum_{\substack{a\in\{0,1\}^n \\ 1 \leq |a| \leq 2r}} Z^a\right) + \left(1-\lambda\right) \frac{I}{2^n},
\end{equation}
which follows by expanding $\ket{0^n}\!\bra{0^n}$ over Pauli-$Z$ strings.

We first verify that $\rho$ is a valid density matrix.
Since $\rho$ is a real linear combination of Hermitian operators, it is Hermitian, and since every nontrivial Pauli-$Z$ string is traceless, its trace is $\lambda + (1-\lambda) = 1$.
For positive semidefiniteness, note that $\rho$ is a linear combination of Pauli-$Z$ strings and is therefore diagonal in the computational basis.
Each diagonal entry of $Z^a$ is $\pm 1$ and the projector $\ket{0^n}\!\bra{0^n}$ has nonnegative diagonal, so for any bitstring $x$,
\begin{equation}
    \rho_{xx} \geq -\frac{\lambda}{2^n}\sum_{\substack{a\in\{0,1\}^n \\ 1 \leq |a| \leq 2r}} 1 + \frac{1-\lambda}{2^n} = \frac{-\lambda(M_{2r}-1) + (1-\lambda)}{2^n} = 0,
\end{equation}
where the last equality uses $\lambda = M_{2r}^{-1}$.

We next verify the EOC and the matched query power.
Recall from the main text that the seed state carries the diagonal of $\rho$ as its squared amplitudes,
\begin{equation}
    \ket{\Theta_s} = \sum_{z \in \{0,1\}^n} \sqrt{\bra{z} \rho \ket{z}}\, \ket{z}.
\end{equation}
Every $Z$ string is diagonal in the computational basis, so its expectation value in $\ket{\Theta_s}$ depends only on these squared amplitudes,
\begin{equation}\label{SM_eq:seed_diagonal_expectation}
    \bra{\Theta_s} Z^c \ket{\Theta_s} = \sum_{z} (-1)^{c \cdot z}\, |\!\braket{z|\Theta_s}\!|^2 = \sum_{z} (-1)^{c \cdot z} \bra{z} \rho \ket{z} = \Tr[Z^c \rho].
\end{equation}
For $a = b$, the EOC Eq.~\eqref{SM_eq:error_orthogonality} hold by normalization since $Z^a Z^a = I$.
For $a \neq b$ with $|a|, |b| \leq r$, the product $Z^a Z^b = Z^{a \oplus b}$ is a nontrivial $Z$ string of weight at most $2r$, so the conditions reduce to $\Tr[Z^c \rho] = 0$ for all $c$ with $1 \leq |c| \leq 2r$.
Substituting the mixture form Eq.~\eqref{eq:encoding2_mixture} gives
\begin{equation}
    \Tr[Z^c \rho]
    = \lambda \bra{0^n} Z^c \ket{0^n} - \frac{\lambda}{2^n} \sum_{1 \leq |a| \leq 2r} \Tr[Z^c Z^a] + \frac{1-\lambda}{2^n} \Tr[Z^c]
    = \lambda - \lambda + 0 = 0.
\end{equation}
The first term equals $\lambda$ since $Z$ strings act on $\ket{0^n}$ as the identity.
The sum contributes $-\lambda$ since the term $a = c$ gives trace $2^n$ and all others vanish by Pauli orthogonality.
The last term vanishes since $Z^c$ is a nontrivial Pauli string and is traceless.

For the matched query power, the seed state definition gives $\eta = |\!\braket{0^n|\Theta_s}\!|^2 = \bra{0^n} \rho \ket{0^n}$.
Every $Z$ string acts on $\ket{0^n}$ as the identity, so substituting the mixture form gives
\begin{equation}\begin{split}
    \eta = \bra{0^n}\rho\ket{0^n}
    &= \lambda - \frac{\lambda}{2^n}\sum_{i=1}^{2r}\binom{n}{i} + \frac{1-\lambda}{2^n}
    = \lambda - \frac{1-\lambda}{2^n} + \frac{1-\lambda}{2^n} = \lambda = M_{2r}^{-1},
\end{split}\end{equation}
where $\lambda \sum_{i=1}^{2r}\binom{n}{i} = 1 - \lambda$ follows from the definition of $\lambda$.
This confirms the matched query power claimed in the main text.

We finally show that the states $Z^e \ket{\Theta_s}$ remain approximately orthogonal even when the error weight exceeds $r$.
For any $a \neq b$, the overlap between $Z^a \ket{\Theta_s}$ and $Z^b \ket{\Theta_s}$ is $\bra{\Theta_s} Z^{a \oplus b} \ket{\Theta_s}$, which vanishes for $1 \leq |a \oplus b| \leq 2r$ by the EOC verification above.
The following lemma shows that the overlap remains small in the remaining case $|a \oplus b| > 2r$.
\begin{lemma}[Expectation of high-weight $Z$ strings under Construction 2]\label{lem:construction2_high_weight}
    For every $e$ with $|e| > 2r$, the seed state of Construction 2 satisfies
    \begin{equation}
        \bra{\Theta_s} Z^e \ket{\Theta_s} = M_{2r}^{-1}.
    \end{equation}
\end{lemma}
\begin{proof}
    Since the seed state carries the diagonal of $\rho$, Eq.~\eqref{SM_eq:seed_diagonal_expectation} gives $\bra{\Theta_s} Z^e \ket{\Theta_s} = \Tr[Z^e \rho]$.
    Substituting the form Eq.~\eqref{eq:encoding2} gives
    \begin{equation}
        \Tr[Z^e \rho] = \frac{1}{2^n} \Tr[Z^e] + \frac{1}{2^n M_{2r}} \sum_{|a| > 2r} \Tr[Z^e Z^a] = M_{2r}^{-1}.
    \end{equation}
    The first term vanishes since $Z^e$ is a nontrivial Pauli string and is traceless.
    In the sum, $\Tr[Z^e Z^a] = 2^n \delta_{a,e}$ by Pauli orthogonality, and the term $a = e$ is present since $|e| > 2r$, so the sum contributes $2^n$.
\end{proof}

\subsection{Construction 4}\label{SM_sec:construction4}
We analyze Construction 4 as follows.
We first show that the LP Eq.~\eqref{eq:construction4_LP} computes the exact optimum of the matched query power over all seed states, and record how we solve it numerically.
We then derive the guarantee Eq.~\eqref{eq:peled_lower_bound} from the result of Ref.~\cite{Peled2011}.
Finally, we combine the guarantee with the asymptotic scaling of $M_r$ (Lemma~\ref{lem:Mr_asymptotics}) to obtain the three rows for Construction 4 in Table~\ref{tab:constructions}.

\subsubsection{Symmetrization}
By Lemma~\ref{lem:uniform_marginals}, a seed state satisfies the EOC if and only if its computational-basis distribution $p$ has uniform marginal distributions on every subset of at most $2r$ bits, and its matched query power is $\eta = p(0^n)$.
The optimal matched query power is therefore the maximum of $p(0^n)$ over all such distributions.
We verify that this maximum is attained by a permutation-symmetric distribution, so that it equals the optimum of the LP Eq.~\eqref{eq:construction4_LP}.
Given any feasible $p$, define its permutation average
\begin{equation}
    \bar{p}(z) := \frac{1}{n!} \sum_{\pi \in S_n} p(\pi(z)),
\end{equation}
where $\pi(z)$ denotes the bitstring $z$ with its bits permuted by $\pi$.
For each $\pi$, the marginal of the distribution $z \mapsto p(\pi(z))$ on a subset $S$ equals the marginal of $p$ on the subset $\pi(S)$ up to reordering the bits, and $|\pi(S)| = |S| \leq 2r$, so it is uniform.
The marginal of $\bar{p}$ on $S$ is the average of these uniform marginals, so $\bar{p}$ is again feasible.
Since $\pi(0^n) = 0^n$ for every $\pi$, the objective is unchanged, $\bar{p}(0^n) = p(0^n)$.
Every feasible $p$ can thus be replaced by a permutation-symmetric feasible $\bar{p}$ with the same objective, so the maximum is attained by a permutation-symmetric distribution.

\subsubsection{Linear program over weight distributions}
A permutation-symmetric distribution is determined by its weight distribution $p_w$, the uniform-marginal conditions become the Krawtchouk moment conditions Eq.~\eqref{eq:krawtchouk_moments_SM} (Lemma~\ref{lem:krawtchouk_moments}), and the objective becomes $p_0$, which is exactly the LP Eq.~\eqref{eq:construction4_LP}.
The seed state Eq.~\eqref{eq:seed_state_dicke} built from the LP optimizer then achieves the optimal matched query power.

For the numerical results in Fig.~\ref{fig:construction4_LP_numerics}, we solve every instance with $n = 20, 30, 40, 50$ and $1 \leq r \leq n/2$.
The LP has the $n+1$ variables $p_0, \dots, p_n$ and the $2r+1$ equality constraints of Eq.~\eqref{eq:krawtchouk_moments_SM}, of which the $k = 0$ condition is the normalization.
We solve it with a two-phase tableau simplex method under Bland's pivoting rule, which guarantees termination, implemented in Julia in exact rational arithmetic over arbitrary-precision integers.
The Krawtchouk coefficients $K_k(w)$ are also evaluated exactly, so the computation involves no rounding and returns the optimum $\eta$ as an exact rational number.
The plotted quantities $\ln \eta$ and $\eta/\eta^*$ are evaluated from these exact optima, so the observed bound $\eta/\eta^* \geq 0.55$ is a statement about the exact LP optima rather than about floating-point approximations.

\subsubsection{Analytic guarantee}
We now derive an analytic guarantee on the LP optimum of Construction 4 from the result of Ref.~\cite{Peled2011}.

The LP optimum coincides with a quantity studied in Ref.~\cite{Peled2011}, the maximal probability $M(n,k,q)$ of the all-ones bitstring under a distribution on $n$ bits in which every bit equals $1$ with probability $q$ and every $k$ bits are mutually independent.
We write $q$ for the bit probability, denoted $p$ in Ref.~\cite{Peled2011}, since $p$ here is the computational-basis distribution of the seed state.
For $q = 1/2$, this constraint is the same as uniform marginal distributions on every subset of at most $k$ bits, and flipping all bits preserves it, so $M(n,2r,1/2)$ is also the maximal probability assigned to $0^n$.
By the symmetrization argument above, this maximal probability equals the LP optimum, $\eta = M(n,2r,1/2)$.

Theorem 1.1 of Ref.~\cite{Peled2011} states that there exist constants $c_1, c_2, c_3 > 0$ such that for even $k \leq c_1 \left( nq(1-q) - 1 \right)$,
\begin{equation}\label{eq:peled_theorem}
    M(n,k,q) \;\geq\; \frac{c_3}{k}\, \exp\!\left( - \frac{c_2\, k}{V\!\left(\frac{nq(1-q)-1}{k}\right)} \right) \tilde{M}(n,k,q),
    \qquad
    \tilde{M}(n,k,q) := \frac{q^n}{\Pr\!\left[ \mathrm{Bin}(n, 1-q) \leq k/2 \right]},
\end{equation}
where $V(x) = \exp\left(\sqrt{\log(x)\log\log(x)}\right)$.
For $q = 1/2$ and $k = 2r$, we have $\Pr[\mathrm{Bin}(n,1/2) \leq r] = M_r\, 2^{-n}$, so $\tilde{M}(n,2r,1/2) = M_r^{-1}$.
Taking logarithms of Eq.~\eqref{eq:peled_theorem} gives
\begin{equation}
    \ln\frac{1}{\eta} - \ln M_r \;\leq\; \ln\frac{2r}{c_3} + \frac{2 c_2\, r}{V\!\left(\frac{n/4-1}{2r}\right)}.
\end{equation}
Since $\frac{n/4-1}{2r} = \Theta(n/r)$ and rescaling the argument of $V$ by a constant factor changes $V$ by at most a bounded factor, the right-hand side simplifies to the guarantee
\begin{equation}\label{eq:peled_lower_bound}
    \ln \frac{1}{\eta} - \ln M_r \leq O(\ln r) + O\!\left(\frac{r}{V(n/r)}\right).
\end{equation}
The requirement on $k$ becomes $r \leq c n$ for a constant $c > 0$ that Ref.~\cite{Peled2011} does not make explicit.

\subsubsection{Asymptotic analysis}
The three regimes in Table~\ref{tab:constructions} rest on the following asymptotics of $M_r$.
\begin{lemma}[Asymptotics of $M_r$]\label{lem:Mr_asymptotics}
    Let $M_r = \sum_{j=0}^{r}\binom{n}{j}$ with $1 \leq r \leq n/2$.
    Then
    \begin{equation}\label{eq:Mr_sandwich}
        \binom{n}{r} \;\leq\; M_r \;\leq\; \frac{n-r+1}{n-2r+1}\binom{n}{r},
    \end{equation}
    and consequently
    \begin{equation}\label{eq:Mr_asymptotics}
        M_r =
        \begin{cases}
            \Theta\!\left(n^r\right), & r \text{ a constant},\\[4pt]
            \left(\dfrac{n}{r}\right)^{(1+o(1))\,r}, & r = \omega(1) \text{ and } r = o(n),\\[8pt]
            \Theta\!\left(\dfrac{2^{nH(\alpha)}}{\sqrt{n}}\right), & r = \lfloor \alpha n \rfloor \text{ with constant } \alpha \in (0,1/2),
        \end{cases}
    \end{equation}
    where the implicit constants depend only on $r$ in the first case and only on $\alpha$ in the third.
\end{lemma}

\begin{proof}
    We first prove the two-sided bound Eq.~\eqref{eq:Mr_sandwich} and then read off the three regimes.

    The lower bound holds because $\binom{n}{r}$ is one term of the sum.
    For the upper bound, consecutive terms satisfy, for $0 \leq j \leq r-1$,
    \begin{equation}
        \frac{\binom{n}{j}}{\binom{n}{j+1}} = \frac{j+1}{n-j} \leq \frac{r}{n-r+1} =: \zeta,
    \end{equation}
    since the middle ratio is increasing in $j$.
    Iterating gives $\binom{n}{j} \leq \zeta^{\,r-j} \binom{n}{r}$, and $r \leq n/2$ ensures $\zeta < 1$, so the geometric series bounds
    \begin{equation}
        M_r \leq \binom{n}{r} \sum_{k=0}^{r} \zeta^k < \binom{n}{r}\,\frac{1}{1-\zeta} = \frac{n-r+1}{n-2r+1}\binom{n}{r}.
    \end{equation}

    For constant $r$, the prefactor is $\frac{n-r+1}{n-2r+1} = 1 + O(1/n)$, and expanding the binomial coefficient gives $\binom{n}{r} = \frac{n^r}{r!}\prod_{i=0}^{r-1}\left(1 - \frac{i}{n}\right) = \frac{n^r}{r!}\left(1 + O(1/n)\right)$.
    Hence $M_r = \frac{n^r}{r!}\left(1 + O(1/n)\right) = \Theta(n^r)$.

    For $r = \omega(1)$ with $r = o(n)$, we have $\zeta = r/(n-r+1) = o(1)$, so Eq.~\eqref{eq:Mr_sandwich} sharpens to $M_r = \left(1+o(1)\right)\binom{n}{r}$.
    The standard bounds $\left(\frac{n}{r}\right)^r \leq \binom{n}{r} \leq \left(\frac{en}{r}\right)^r$ give
    \begin{equation}
        r \ln\frac{n}{r} \;\leq\; \ln \binom{n}{r} \;\leq\; r\left(\ln\frac{n}{r} + 1\right),
    \end{equation}
    and $r = o(n)$ forces $\ln(n/r) \to \infty$, so the additive $r$ is $o\!\left(r\ln(n/r)\right)$.
    Therefore $\ln M_r = \left(1+o(1)\right) r \ln(n/r)$, which is the claimed $M_r = (n/r)^{(1+o(1))r}$.

    For $r = \lfloor \alpha n \rfloor$ with constant $\alpha \in (0,1/2)$ and $n \geq 2/\alpha$, the prefactor in Eq.~\eqref{eq:Mr_sandwich} is $1 + \frac{r}{n-2r+1} \leq \frac{1-\alpha}{1-2\alpha}$, a constant, so $M_r = \Theta\!\left(\binom{n}{r}\right)$.
    Robbins' form of Stirling's approximation~\cite{Robbins1955} bounds every factorial as
    \begin{equation}\label{eq:stirling_robbins}
        k! = \sqrt{2\pi k}\,\left(\frac{k}{e}\right)^{k} e^{\varepsilon_k},
        \qquad \frac{1}{12k+1} \leq \varepsilon_k \leq \frac{1}{12k}.
    \end{equation}
    Applying it to the three factorials of $\binom{n}{r} = \frac{n!}{r!\,(n-r)!}$ yields
    \begin{equation}
        \binom{n}{r} = \frac{2^{\,nH(r/n)}}{\sqrt{2\pi\,\frac{r}{n}\left(1-\frac{r}{n}\right)\,n}}\; e^{\varepsilon_n - \varepsilon_r - \varepsilon_{n-r}},
    \end{equation}
    where the exponential factor lies between two absolute constants.
    It remains to replace $r/n$ by $\alpha$.
    Since $r > \alpha n - 1$ and $n \geq 2/\alpha$, we have $\alpha/2 \leq r/n \leq \alpha$, so $\frac{r}{n}\left(1-\frac{r}{n}\right)$ lies between $\frac{\alpha}{2}\left(1 - \frac{\alpha}{2}\right)$ and $\alpha(1-\alpha)$.
    Eq.~\eqref{eq:entropy_increment} in Lemma~\ref{lem:entropy_floor} with $x = r/n$ and $y = \alpha$, together with $\alpha - r/n < 1/n$ and $\log_2 \frac{1 - r/n}{r/n} \leq \log_2 \frac{n}{r} \leq \log_2 \frac{2}{\alpha}$, gives $0 \leq H(\alpha) - H(r/n) \leq \frac{1}{n} \log_2 \frac{2}{\alpha}$, so $2^{nH(r/n)}$ lies between $\frac{\alpha}{2}\, 2^{nH(\alpha)}$ and $2^{nH(\alpha)}$.
    Combining these bounds gives $M_r = \Theta\!\left(2^{nH(\alpha)}/\sqrt{n}\right)$, with constants depending only on $\alpha$.
\end{proof}

We now derive the three rows for Construction 4 in Table~\ref{tab:constructions} by combining the guarantee Eq.~\eqref{eq:peled_lower_bound} with Lemma~\ref{lem:Mr_asymptotics}.
The guarantee bounds $\ln(1/\eta)$ by $\ln M_r$ plus the two correction terms $O(\ln r)$ and $O(r/V(n/r))$, so in each regime it remains to bound the correction terms.

For constant $r$, the first term is a constant, and the second term vanishes as $n \to \infty$ because $V(n/r) \to \infty$.
Hence $\ln(1/\eta) \leq \ln M_r + O(1)$, so the LP optimum satisfies $\eta = \Omega(M_r^{-1})$, and the upper bound Eq.~\eqref{eq:eta_upper_bound} sharpens this to $\eta = \Theta(M_r^{-1})$.

For $r = \omega(1)$ with $r = o(n)$, Lemma~\ref{lem:Mr_asymptotics} gives $\ln M_r = (1+o(1))\, r \ln(n/r)$ with $\ln(n/r) \to \infty$.
The first term is $O(\ln r) = o(r)$, and the second term is $o(r)$ because $V(n/r) \to \infty$, so both are $o(\ln M_r)$.
Hence $\ln(1/\eta) \leq (1+o(1)) \ln M_r$, which is the claimed $\eta \geq M_r^{-(1+o(1))}$.

For $r = \lfloor \alpha n \rfloor$ with constant $\alpha$ below the constant $c$ of Eq.~\eqref{eq:peled_lower_bound}, the argument of $V$ is $\Theta(1/\alpha)$, so the second term is $O(r/V(1/\alpha))$ and grows linearly in $n$, while the first term $O(\ln r) = O(\ln n)$ is negligible compared to it.
Hence $\ln(1/\eta) \leq \ln M_r + O(r/V(1/\alpha))$, which is the claimed $\eta \geq e^{-O(r/V(1/\alpha))}\, M_r^{-1}$.
In rate form, Lemma~\ref{lem:Mr_asymptotics} gives $\log_2 M_r = nH(\alpha) - \Theta(\log n)$, so $\frac{1}{n}\log_2\frac{1}{\eta} \leq H(\alpha) + O\!\left(\frac{\alpha}{V(1/\alpha)}\right)$.

\clearpage
\section{Near-Optimal Query State Construction}\label{SM_sec:construction3}
This section gives the full details of Construction 3 from Sec.~\ref{subsec:query_state_constructions}, whose guarantee is summarized in the following theorem.
\begin{theorem}[Near-optimal query state construction]\label{thm:near_optimal_construction}
    For every constant $0 < \alpha \leq 0.16$ and all sufficiently large $n$, there exists an $n$-qubit seed state $\ket{\Theta_s}$ that satisfies the EOC Eq.~\eqref{SM_eq:error_orthogonality} for $r=\lfloor \alpha n \rfloor$ errors and has matched query power
    \begin{equation}
        \eta = \xi(n)\,2^{-n(H(\alpha)+2\alpha)},
    \end{equation}
    where $\xi(n)$ is any function with $\xi(n) \geq 1$ and $\log\xi(n) = o(n)$.
\end{theorem}

As computed below Eq.~\eqref{eq:construction3_eta}, this matched query power satisfies $\eta^{-1} = M_r^{1+2\alpha/H(\alpha)+o(1)}$, which approaches the optimal scaling $\eta^* = M_r^{-1}$ as $\alpha \to 0$ and is the sense in which the construction is near-optimal.

Following the sketch in Sec.~\ref{subsec:query_state_constructions}, the proof perturbs the state $\ket{+}^{\otimes n}$, which satisfies the EOC for all weights but has matched query power only $2^{-n}$.
Section~\ref{SM_sec:reduction} applies Lemma~\ref{lem:krawtchouk_moments} to reduce the construction to finding a weight distribution that satisfies the Krawtchouk moment conditions, and introduces the reference distribution and its orthonormal Krawtchouk basis.
Section~\ref{SM_sec:ansatz} introduces the ansatz, which raises the mass at $w=0$ to $\eta$ and restores the moments with a low-degree perturbation on a truncated window, and reduces the moment conditions to a linear system for the perturbation coefficients.
Section~\ref{SM_sec:proof} states three estimates for this system and derives the theorem from them.
Section~\ref{SM_sec:lemma_proofs} proves the estimates.
Section~\ref{SM_sec:high_weight} gives the expectation of $Z$ strings of weight between $2r$ and $n - 2r$ in the constructed seed state.

\subsection{Orthonormal Krawtchouk basis}\label{SM_sec:reduction}
We take the seed state to be permutation symmetric with real and non-negative coefficients, which loses no generality by the permutation-averaging argument of Sec.~\ref{subsec:query_state_constructions}.
Hence, we write $\ket{\Theta_s}$ in the Dicke basis, as in Eq.~\eqref{eq:seed_state_dicke},
\begin{equation}\label{eq:seed_dicke}
    \ket{\Theta_s} = \sum_{w=0}^n \sqrt{p_w}\, \ket{D^n_w},
\end{equation}
where $p_w$ is a probability distribution over Hamming weights.
By Lemma~\ref{lem:krawtchouk_moments}, the EOC for $r$ errors are equivalent to the Krawtchouk moment conditions Eq.~\eqref{eq:krawtchouk_moments_SM} on $p_w$, so constructing a valid seed state becomes a purely classical problem, that of finding a distribution $p_w$ satisfying Eq.~\eqref{eq:krawtchouk_moments_SM}.

The reference distribution for this problem is the weight distribution of $\ket{+}^{\otimes n}$, the symmetric binomial distribution
\begin{equation}
    \pi(w) := 2^{-n} \binom{n}{w}.
\end{equation}
The Krawtchouk polynomials Eq.~\eqref{eq:def_K_SM} are orthogonal under $\pi$ (Lemma~\ref{lem:orthonormal} below), and it is convenient to work with the normalized polynomials
\begin{equation}\label{eq:def_R}
    R_k(w) := \frac{K_k(w)}{\sqrt{\binom{n}{k}}}.
\end{equation}

\begin{lemma}[Orthonormal Krawtchouk basis]\label{lem:orthonormal}
For all $j, k \in \{0, 1, \ldots, n\}$,
\begin{equation}\label{eq:orthonormal}
    \sum_{w=0}^n \pi(w)\,R_j(w)\,R_k(w) = \delta_{j,k}.
\end{equation}
\end{lemma}
\begin{proof}
    This is the standard orthogonality relation of the binary Krawtchouk polynomials~\cite{Coleman2011, Levenshtein1995}, normalized by Eq.~\eqref{eq:def_R}.
\end{proof}

Since $K_0 = 1$, the orthogonality relation gives $\sum_w \pi(w) K_k(w) = 0$ for all $k \geq 1$, so $\pi$ satisfies the Krawtchouk moment conditions Eq.~\eqref{eq:krawtchouk_moments_SM} for every weight $r$.
By Lemma~\ref{lem:krawtchouk_moments}, this confirms that $\ket{+}^{\otimes n}$ satisfies the EOC for all weights, reflecting the fact that every $Z$ error maps each $\ket{+}$ to the orthogonal state $\ket{-}$ and is hence detectable by $X$ measurement.
However, the mass at $w = 0$ is the matched query power of the seed state, and $\pi$ places only $2^{-n}$ there.
The ansatz of Sec.~\ref{SM_sec:ansatz} raises this mass to $\eta$ while preserving the moment conditions up to $k = 2r$.

\subsection{Ansatz and linear system}\label{SM_sec:ansatz}

\subsubsection{Truncated window and low-degree ansatz}
The seed state places mass $\eta$ at weight zero and the remaining $1-\eta$ according to a distribution $\{q_w\}_{w=1}^n$ over the nonzero weights,
\begin{equation}\label{eq:ansatz_dicke}
    \ket{\Theta_s} = \sqrt{\eta} \ket{D^n_0} + \sqrt{1-\eta} \sum_{w=1}^n \sqrt{q_w} \ket{D^n_w},
\end{equation}
so its matched query power is $\eta$ by construction.
The ansatz builds $q_w$ from the reference distribution $\pi$ in two steps, truncation to a central window and a low-degree multiplicative correction.
We restrict the support to a central window
\begin{equation}
    \mathcal{W} := \{w\in \{1,2,\dots,n\} : |w-n/2|<\beta n\},
\end{equation}
where $\beta\in (0,1/2)$ controls the width, and write the truncated reference distribution
\begin{equation}
    \pi_{\mathcal{W}}(w) := \frac{\pi(w) \mathbf{1}_{w \in \mathcal{W}}}{Z_{\mathcal{W}}}, \qquad Z_{\mathcal{W}} := \sum_{w\in \mathcal{W}} \pi(w).
\end{equation}
The discarded tail mass is exponentially small.
Its decay rate, in bits, is
\begin{equation}\label{eq:def_I}
    I(\beta) := 1 - H\!\left(\tfrac{1}{2}-\beta\right),
\end{equation}
where $H$ is the binary entropy.
Indeed, the entropy bound on partial binomial sums (Lemma~\ref{lem:entropy_binomial}), applied with $\lambda = \tfrac{1}{2}-\beta$, bounds the lower tail of $\pi$ by
\begin{equation}
    \sum_{w \leq (1/2-\beta)n} \pi(w) \leq 2^{-n}\, 2^{nH\left(\frac{1}{2}-\beta\right)} = 2^{-nI(\beta)}.
\end{equation}
The upper tail obeys the same bound by the symmetry of $\pi$ about $n/2$, so
\begin{equation}\label{eq:tail_Z}
    1-Z_{\mathcal{W}} \leq 2^{\,1-n I(\beta)}.
\end{equation}
On the window, the second step applies a low-degree multiplicative correction to $\pi_{\mathcal{W}}$,
\begin{equation}\label{eq:ansatz_q}
    q_w := \pi_{\mathcal{W}}(w)\bigl(1+u(w)\bigr), \qquad u(w) := \sum_{j=0}^{2r} c_j R_j(w).
\end{equation}
The free parameters are the window width $\beta$ and the coefficient vector $\vec{c} = (c_0, c_1, \dots, c_{2r})^T$, which will be determined by the moment conditions.
For the ansatz to be a valid probability distribution, we need $q_w \geq 0$ for every $w\in \mathcal{W}$, or equivalently $u(w) \geq -1$.

\subsubsection{Linear system for the ansatz coefficients}
We now impose the moment conditions Eq.~\eqref{eq:krawtchouk_moments_SM} on the full weight distribution, $p_0 = \eta$ and $p_w = (1-\eta)\,q_w$ for $w \geq 1$.
Dividing the $k$-th condition by $(1-\eta)\sqrt{\binom{n}{k}}$ and using that $q_w$ is supported on $\mathcal{W}$ turns the conditions into
\begin{equation}\label{eq:moment_W}
    \sum_{w\in \mathcal{W}} q_w\, R_k(w) = \frac{\delta_{k,0} - \eta R_k(0)}{1-\eta},\quad k=0,1,\dots,2r.
\end{equation}
Expanding $q_w = \pi_{\mathcal{W}}(w)(1+u(w))$, the left-hand side becomes
\begin{equation}
    \sum_{w \in \mathcal{W}} q_w\,R_k(w) = \frac{1}{Z_{\mathcal{W}}} \left( \sum_{w \in \mathcal{W}} \pi(w)\,R_k(w) + \sum_{w \in \mathcal{W}} \pi(w)\,R_k(w)\,u(w) \right).
\end{equation}
For the first term, we complete the sum over the full support using Lemma~\ref{lem:orthonormal},
\begin{equation}
    \sum_{w \in \mathcal{W}} \pi(w)\,R_k(w) = \delta_{k,0} - \sum_{w \notin \mathcal{W}} \pi(w)\,R_k(w).
\end{equation}
For the second term, substituting $u(w) = \sum_j c_j R_j(w)$ and exchanging the order of summation gives
\begin{equation}
    \sum_{w \in \mathcal{W}} \pi(w)\,R_k(w)\,u(w) = \sum_{j=0}^{2r} c_j \sum_{w \in \mathcal{W}} \pi(w)\,R_k(w)\,R_j(w) = (\mathbf{H}\vec{c})_k,
\end{equation}
where $\mathbf{H}$ is the $(2r+1) \times (2r+1)$ truncated Gram matrix
\begin{equation}\label{eq:definition_H}
    \mathbf{H}_{kj} := \sum_{w \in \mathcal{W}} \pi(w)\,R_k(w)\,R_j(w), \quad 0 \leq k, j \leq 2r.
\end{equation}
Combining the two terms and equating with the right-hand side of Eq.~\eqref{eq:moment_W} gives
\begin{equation}\label{eq:linear_system}
    \mathbf{H}\vec{c} = \vec{b},
\end{equation}
where the source vector $\vec{b} \in \mathbb{R}^{2r+1}$ splits as $b_k = e_k + s_k$ into a tail vector $\vec{e} := (e_0, \ldots, e_{2r})$ and a signal vector $\vec{s} := (s_0, \ldots, s_{2r})$, with
\begin{equation}\label{eq:definition_e_s}
    e_k := \sum_{w \notin \mathcal{W}} \pi(w)\,R_k(w), \qquad s_k := -\frac{\eta\,Z_{\mathcal{W}}}{1-\eta}\,R_k(0), \qquad k = 1, \ldots, n,
\end{equation}
and $e_0 = s_0 = 0$.
The tail vector captures the moment error from restricting $\pi$ to $\mathcal{W}$, and the signal vector captures the moment shift from raising the matched query power to $\eta$.
The $k=0$ row of the system, $b_0 = 0$, is the normalization constraint on $q$.
Since $R_0 = 1$, it requires $\sum_{w \in \mathcal{W}} \pi_{\mathcal{W}}(w)\,u(w) = 0$, so the multiplicative correction preserves the total mass.
The components with $k > 2r$ do not enter the linear system, but they obey a similar identity, which will be used in Sec.~\ref{SM_sec:high_weight}.
Fix $k$ with $2r < k \leq n$.
Since $k \geq 1$, Lemma~\ref{lem:orthonormal} gives $\sum_{w=0}^{n} \pi(w)\,R_k(w) = 0$, so $\sum_{w \in \mathcal{W}} \pi(w)\,R_k(w) = -e_k$ by the definition Eq.~\eqref{eq:definition_e_s}, and expanding $q_w = \pi_{\mathcal{W}}(w)(1+u(w))$ gives
\begin{equation}\label{eq:high_weight_identity}
    \sum_{w \in \mathcal{W}} q_w\,R_k(w)
    = \frac{1}{Z_{\mathcal{W}}} \left( \sum_{w \in \mathcal{W}} \pi(w)\,R_k(w) + \sum_{w \in \mathcal{W}} \pi(w)\,R_k(w)\,u(w) \right)
    = \frac{1}{Z_{\mathcal{W}}} \left( -e_k + \sum_{w \in \mathcal{W}} \pi(w)\,R_k(w)\,u(w) \right).
\end{equation}

\subsection{Proof of Theorem~\ref{thm:near_optimal_construction}}\label{SM_sec:proof}
By Lemma~\ref{lem:krawtchouk_moments} and the reduction of Sec.~\ref{SM_sec:ansatz}, it suffices to find a solution $\vec{c}$ of the linear system Eq.~\eqref{eq:linear_system} for which the multiplicative correction $u = \sum_{j} c_j R_j$ satisfies $|u(w)| < 1$ for all $w \in \mathcal{W}$.
Indeed, such a solution makes every $q_w = \pi_{\mathcal{W}}(w)(1+u(w))$ positive, and the linear system enforces normalization and the Krawtchouk moment conditions Eq.~\eqref{eq:krawtchouk_moments_SM}, so the seed state Eq.~\eqref{eq:ansatz_dicke} satisfies the EOC for $r = \lfloor \alpha n \rfloor$ errors with matched query power $\eta$.
The proof rests on three estimates and an elementary entropy inequality, all proven in Sec.~\ref{SM_sec:lemma_proofs}.

The first lemma shows that $\mathbf{H}$ is close to the identity, so the system is solvable with $\|\vec{c}\|_2 \leq 2\|\vec{b}\|_2$.

\begin{lemma}[Invertibility of $\mathbf{H}$]\label{lem:bound_H}
    If $I(\beta) > 6\alpha$, then for sufficiently large $n$ the matrix $\mathbf{H}$ is invertible with $\|\mathbf{H}^{-1}\|_{\mathrm{op}}\leq 2$.
\end{lemma}

The second bounds the two components of the source vector.

\begin{lemma}[Bounds on the source vector]\label{lem:bound_b}
    Every tail sum with $1 \leq k \leq n$ satisfies
    \begin{equation}\label{eq:bound_e_component}
        |e_k| \leq \sqrt{2}\, 2^{-nI(\beta)/2},
    \end{equation}
    so the tail vector satisfies
    \begin{equation}
        \left\| \vec{e} \right\|_2 \leq 2\sqrt{\alpha n}\, 2^{-nI(\beta)/2},
    \end{equation}
    and, if $\alpha < 1/4$ and $\eta \leq 1/2$, the signal vector satisfies
    \begin{equation}
        \left\| \vec{s} \right\|_2 \leq 2\eta \cdot 2^{nH(2\alpha)/2}.
    \end{equation}
\end{lemma}

The third shows that the pointwise fluctuation of a low-degree polynomial on the central window is bounded.
The relevant quantity is the diagonal reproducing kernel $\kappa_{d}(w) := \sum_{i=0}^{d} R_i(w)^2$, since the Cauchy--Schwarz inequality bounds any degree-$d$ polynomial $\sum_{i} c_i R_i$ pointwise by $\|\vec{c}\|_2\,\sqrt{\kappa_{d}(w)}$.

\begin{lemma}[Reproducing-kernel bound on the window]\label{lem:kernel_window}
    For all $w\in\mathcal{W}$,
    \begin{equation}
        \kappa_{2r}(w) \leq (n+1)^{2/3}\, 2^{\,n\left(\frac{2I(\beta)}{3} + 2\alpha\right)}.
    \end{equation}
\end{lemma}

The last lemma is an elementary inequality between binary entropies.

\begin{lemma}[Entropy inequality]\label{lem:entropy_inequality}
    For all $\alpha \in (0, 1/2)$,
    \begin{equation}\label{eq:entropy_inequality}
        H(\alpha) - \alpha - \frac{1}{2}H(2\alpha) > \frac{\alpha^2}{2}.
    \end{equation}
\end{lemma}

Using the four lemmas, we can now prove Theorem~\ref{thm:near_optimal_construction}.
\begin{proof}[Proof of Theorem~\ref{thm:near_optimal_construction}]
Fix any $\xi(n) \geq 1$ with $\log\xi(n)=o(n)$, and set
\begin{equation}\label{eq:parameter_choice}
    \beta = I^{-1}\left(6\alpha + \alpha^2\right) ,\quad
    \eta = \xi(n)\,2^{-n \left(H(\alpha) + 2\alpha \right)}.
\end{equation}
We first check that $\beta$ is well defined and that the conditions of Lemmas~\ref{lem:bound_H} and~\ref{lem:bound_b} hold.
By the assumption of the Theorem, $\alpha \leq 0.16$, so $6\alpha + \alpha^2 \leq 0.9856 < 1$.
Since $I$ is continuous and strictly increasing on $[0, 1/2)$ with $I(0) = 0$ and $I(1/2^-) = 1$, there is a unique $\beta \in (0, 1/2)$ satisfying $I(\beta) = 6\alpha + \alpha^2$.
Since $\alpha^2 > 0$, this choice satisfies $I(\beta) > 6\alpha$, the condition of Lemma~\ref{lem:bound_H}.
The conditions of Lemma~\ref{lem:bound_b} hold as well, since $\alpha \leq 0.16 < 1/4$ and, for sufficiently large $n$, $\eta \leq 1/2$.

By Lemma~\ref{lem:bound_H}, the linear system admits the unique solution $\vec{c} = \mathbf{H}^{-1}\vec{b}$.
Combining $\|\vec{c}\|_2 \leq \|\mathbf{H}^{-1}\|_{\mathrm{op}}\, \|\vec{b}\|_2$ with the bound $\|\mathbf{H}^{-1}\|_{\mathrm{op}} \leq 2$ of Lemma~\ref{lem:bound_H}, the triangle inequality $\|\vec{b}\|_2 \leq \|\vec{e}\|_2 + \|\vec{s}\|_2$, and the bounds of Lemma~\ref{lem:bound_b} gives
\begin{equation}\label{eq:bound_c}
    \|\vec{c}\|_2 \leq 2 \left( \|\vec{e}\|_2 + \|\vec{s}\|_2 \right) \leq 4\sqrt{\alpha n}\, 2^{-nI(\beta)/2} + 4\eta\, 2^{nH(2\alpha)/2}.
\end{equation}
It remains to show $|u(w)| < 1$ on $\mathcal{W}$.
Applying the Cauchy--Schwarz inequality to the definition of $u$ in Eq.~\eqref{eq:ansatz_q} gives $|u(w)| \leq \|\vec{c}\|_2 \sqrt{\kappa_{2r}(w)}$, and substituting Eq.~\eqref{eq:bound_c} and Lemma~\ref{lem:kernel_window} gives a tail contribution and a signal contribution,
\begin{equation}
    |u(w)| \leq \underbrace{4\sqrt{\alpha n}\,(n+1)^{1/3}\,2^{-n\left(\frac{I(\beta)}{6} - \alpha\right)}}_{\text{tail}} + \underbrace{4(n+1)^{1/3}\eta\,2^{\,n\left(\frac{I(\beta)}{3} + \alpha + \frac{1}{2} H(2\alpha)\right)}}_{\text{signal}}.
\end{equation}
We show that each contribution decays exponentially in $n$.

For the tail contribution, substituting $I(\beta) = 6\alpha + \alpha^2$ gives $I(\beta)/6 - \alpha = \alpha^2/6 > 0$, so the factor $2^{-n\alpha^2/6}$ decays exponentially and dominates the polynomial prefactor $\sqrt{\alpha n}\,(n+1)^{1/3}$.

For the signal contribution, substituting $\eta = \xi(n)\,2^{-n(H(\alpha) + 2\alpha)}$ gives $4\,\xi(n)\,(n+1)^{1/3}\,2^{n g(\alpha)}$ with exponent
\begin{equation}
    g(\alpha) := \frac{I(\beta)}{3} + \alpha + \frac{1}{2}H(2\alpha) - \bigl(H(\alpha) + 2\alpha\bigr)
    = \frac{1}{3}\alpha^2 - \left( H(\alpha) - \alpha - \frac{1}{2}H(2\alpha) \right).
\end{equation}
By Lemma~\ref{lem:entropy_inequality}, $g(\alpha) < 0$.
Hence $2^{n g(\alpha)}$ decays exponentially and dominates the subexponential prefactor $\xi(n)\,(n+1)^{1/3}$, so the signal contribution vanishes as well.
For sufficiently large $n$, the sum of the two contributions drops below $1$, giving $|u(w)| < 1$ on $\mathcal{W}$ and completing the proof.
\end{proof}

\subsection{Proofs of Lemmas~\ref{lem:bound_H}--\ref{lem:entropy_inequality}}\label{SM_sec:lemma_proofs}
We first bound the sum $\sum_{w \in S} \pi(w) f(w)^2$ for a polynomial $f$ of low degree and unit norm under $\pi$ in terms of the probability of the set $S$ under $\pi$, the tool behind the proofs of Lemmas~\ref{lem:bound_H} and~\ref{lem:kernel_window}.
We then prove Lemmas~\ref{lem:bound_H}, \ref{lem:bound_b}, \ref{lem:kernel_window}, and~\ref{lem:entropy_inequality} in the order they were stated.

\subsubsection{Hypercontractive mass bound}
\begin{lemma}[Hypercontractive mass bound]\label{lem:small_set}
    Let $f: \{0,\ldots,n\} \to \mathbb{R}$ be any polynomial of degree at most $d$, normalized so that $\sum_{w=0}^n \pi(w)\, f(w)^2 = 1$.
    Then for any subset $S \subseteq \{0, \ldots, n\}$,
    \begin{equation}\label{eq:small_set}
        \sum_{w \in S} \pi(w)\, f(w)^2 \leq 2^d \left(\sum_{w \in S} \pi(w)\right)^{1/3}.
    \end{equation}
\end{lemma}

\begin{proof}
The proof lifts $f$ to the Boolean cube, bounds its third moment by hypercontractivity, and localizes the mass to $S$ with H\"older's inequality.

Let $Z_1,\ldots,Z_n$ be i.i.d.\ uniform on $\{-1,+1\}$, so that the Hamming weight $W = \sum_{j=1}^n (1-Z_j)/2$ follows the reference distribution $\pi$.
The lifted function
\begin{equation}
    g(z_1,\ldots,z_n) = f\!\left(\frac{n - \sum_{j=1}^n z_j}{2}\right)
\end{equation}
is multilinear of degree at most $d$ after reducing powers using $z_j^2 = 1$.
Since $g(\vec{z}) = f(W)$ and $W$ follows $\pi$, the moments of $g$ under the uniform measure equal those of $f$ under $\pi$.
Since $g$ is multilinear of degree at most $d$, the Bonami--Beckner hypercontractivity inequality~\cite{ODonnell2014} with exponent $3$ gives $\mathbb{E}[|g|^3] \leq 2^{3d/2}\, \mathbb{E}[g^2]^{3/2}$ under the uniform measure.
Transferring to $f$ and using the normalization of $f$,
\begin{equation}
    \sum_{w=0}^{n} \pi(w)\, |f(w)|^3 \leq 2^{3d/2}.
\end{equation}
Writing $\pi(w)\, f(w)^2 = \pi(w)^{1/3} \cdot \pi(w)^{2/3} f(w)^2$, we apply H\"older's inequality in the form $\sum_{w \in S} a_w b_w \leq \bigl(\sum_{w \in S} a_w^3\bigr)^{1/3} \bigl(\sum_{w \in S} b_w^{3/2}\bigr)^{2/3}$ with $a_w = \pi(w)^{1/3}$ and $b_w = \pi(w)^{2/3} f(w)^2$, and extend the second sum to all $w$,
\begin{equation}
    \sum_{w \in S} \pi(w)\, f(w)^2
    \leq \left( \sum_{w \in S} \pi(w) \right)^{1/3} \left( \sum_{w=0}^{n} \pi(w)\, |f(w)|^3 \right)^{2/3}
    \leq 2^d \left( \sum_{w \in S} \pi(w) \right)^{1/3}.
\end{equation}
\end{proof}

\subsubsection{Proof of Lemma~\ref{lem:bound_H} (invertibility of $\mathbf{H}$)}

\begin{proof}[Proof of Lemma~\ref{lem:bound_H}]
We write $\mathbf{H} = I - E$, bound every entry of the error matrix $E$ with the hypercontractive mass bound, and conclude with the Neumann series.

The orthonormality relation of Lemma~\ref{lem:orthonormal} states $\sum_{w=0}^{n} \pi(w)\, R_i(w)\, R_j(w) = \delta_{i,j}$, so subtracting the definition of $\mathbf{H}$ (Eq.~\eqref{eq:definition_H}) leaves the part of this sum outside the window,
\begin{equation}
    E_{ij} = \delta_{i,j} - \mathbf{H}_{ij} = \sum_{w \notin \mathcal{W}} \pi(w)\, R_i(w)\, R_j(w), \quad 0 \leq i, j \leq 2r.
\end{equation}
The Cauchy--Schwarz inequality bounds the magnitude of this tail,
\begin{equation}
    |E_{ij}| \leq \left( \sum_{w \notin \mathcal{W}} \pi(w)\, R_i(w)^2 \right)^{1/2} \left( \sum_{w \notin \mathcal{W}} \pi(w)\, R_j(w)^2 \right)^{1/2}.
\end{equation}
Each $R_i$ with $i \leq 2r$ is a polynomial of degree at most $2r$ with unit norm under $\pi$ (Lemma~\ref{lem:orthonormal}), so the hypercontractive mass bound (Lemma~\ref{lem:small_set}) with $S = \{w : w \notin \mathcal{W}\}$, followed by the tail bound Eq.~\eqref{eq:tail_Z}, gives
\begin{equation}
    \sum_{w \notin \mathcal{W}} \pi(w)\, R_i(w)^2 \leq 2^{2r} \left(1 - Z_{\mathcal{W}}\right)^{1/3} \leq 2^{1/3}\, 2^{\,2r - \frac{n I(\beta)}{3}}.
\end{equation}
Since $2r \leq 2\alpha n$, every entry of $E$ satisfies
\begin{equation}
    |E_{ij}| \leq 2^{1/3}\, 2^{-n \left( \frac{I(\beta)}{3} - 2\alpha \right)}.
\end{equation}

With the entrywise bound in hand, we now control the operator norm of $E$.
Since $E$ is a $(2r+1) \times (2r+1)$ symmetric matrix, its operator norm equals its largest eigenvalue in absolute value, which is bounded by the induced infinity norm,
\begin{equation}
    \|E\|_{\mathrm{op}} \leq \|E\|_{\infty} = \max_{0\leq i \leq 2r} \sum_{j=0}^{2r} |E_{ij}| \leq (2r+1)\cdot 2^{1/3}\, 2^{-n \left( \frac{I(\beta)}{3} - 2\alpha \right)}.
\end{equation}
The lemma assumes $I(\beta) > 6\alpha$, so the exponential decay dominates the polynomial prefactor $2r+1 \leq 2\alpha n + 1$ and $\|E\|_{\mathrm{op}} < 1/2$ for sufficiently large $n$.
Since $\mathbf{H}$ is real and symmetric, its eigenvalues are real, and each eigenvalue $\lambda$ of $\mathbf{H} = I - E$ satisfies $|\lambda - 1| \leq \|E\|_{\mathrm{op}} < 1/2$.
Every eigenvalue therefore exceeds $1/2$, so $\mathbf{H}$ is invertible.
The Neumann series gives a quantitative bound on the inverse,
\begin{equation}
    \|\mathbf{H}^{-1}\|_{\mathrm{op}} = \left\|(I-E)^{-1}\right\|_{\mathrm{op}} \leq \sum_{k=0}^{\infty} \|E\|_{\mathrm{op}}^k = \frac{1}{1-\|E\|_{\mathrm{op}}} \leq 2.
\end{equation}
\end{proof}

\subsubsection{Proof of Lemma~\ref{lem:bound_b} (source vector $\vec{b}$)}

\begin{proof}[Proof of Lemma~\ref{lem:bound_b}]
We bound the tail sums first.
Fix any $k$ with $1 \leq k \leq n$.
By Cauchy--Schwarz,
\begin{equation}
    e_k^2
    = \left( \sum_{w=0}^{n} \pi(w)\,R_k(w)\,\mathbf{1}_{w \notin \mathcal{W}} \right)^2
    \leq \left( \sum_{w=0}^{n} \pi(w)\,R_k(w)^2 \right) \left( \sum_{w \notin \mathcal{W}} \pi(w) \right)
    = 1 \cdot (1 - Z_{\mathcal{W}}),
\end{equation}
where the first factor equals $1$ by Lemma~\ref{lem:orthonormal} and the second is the tail mass $1 - Z_{\mathcal{W}}$.
Equation~\eqref{eq:tail_Z} bounds the tail mass, giving $e_k^2 \leq 2^{\,1-nI(\beta)}$, which is Eq.~\eqref{eq:bound_e_component}.
Since $e_0 = 0$ by the definition Eq.~\eqref{eq:definition_e_s}, summing Eq.~\eqref{eq:bound_e_component} over the $2r \leq 2\alpha n$ nonzero components of $\vec{e}$ gives
\begin{equation}
    \|\vec{e}\|_2 = \left(\sum_{k=1}^{2r} e_k^2\right)^{1/2} \leq 2\sqrt{\alpha n}\, 2^{-nI(\beta)/2}.
\end{equation}

We now bound the signal vector.
Since $s_0 = 0$, only the components with $k \geq 1$ contribute, and by the definition Eq.~\eqref{eq:definition_e_s},
\begin{equation}
    \|\vec{s}\|_2 = \frac{\eta\,Z_{\mathcal{W}}}{1-\eta}\left(\sum_{k=1}^{2r} R_k(0)^2\right)^{1/2}.
\end{equation}
By Eq.~\eqref{eq:def_K_SM}, $K_k(0) = \binom{n}{k}$, and by Eq.~\eqref{eq:def_R}, $R_k(0) = \sqrt{\binom{n}{k}}$, so
\begin{equation}
    \sum_{k=1}^{2r} R_k(0)^2 = \sum_{k=1}^{2r} \binom{n}{k} \leq \sum_{k=0}^{2r} \binom{n}{k} \leq 2^{nH(2\alpha)},
\end{equation}
where the last inequality is the entropy bound Eq.~\eqref{eq:entropy_binomial} of Lemma~\ref{lem:entropy_binomial} applied with $\lambda = 2\alpha$, using $2r \leq 2\alpha n$ and the assumption $\alpha < 1/4$.
Using $Z_{\mathcal{W}} \leq 1$ and $\eta/(1-\eta) \leq 2\eta$, valid under the assumption $\eta \leq 1/2$, we obtain
\begin{equation}
    \|\vec{s}\|_2 \leq 2\eta \cdot 2^{nH(2\alpha)/2}.
\end{equation}
\end{proof}

\subsubsection{Proof of Lemma~\ref{lem:kernel_window} (reproducing kernel $\kappa_{2r}$)}

\begin{proof}[Proof of Lemma~\ref{lem:kernel_window}]
The proof has three steps.
We first bound the kernel at a single weight by applying the hypercontractive mass bound to a single point, then lower bound $\pi(w)$ on the window, and finally combine the two.

Fix a weight $w \in \mathcal{W}$ and define the degree-$2r$ polynomial $f^*(x) := \kappa_{2r}(w)^{-1/2}\sum_{i=0}^{2r} R_i(w)\,R_i(x)$.
Since $\{R_i\}$ are orthonormal under $\pi$ (Lemma~\ref{lem:orthonormal}),
\begin{equation}
    \sum_{x=0}^n \pi(x)\, f^*(x)^2 = \kappa_{2r}(w)^{-1}\sum_{i=0}^{2r} R_i(w)^2 = 1, \qquad
    f^*(w) = \kappa_{2r}(w)^{-1/2}\sum_{i=0}^{2r} R_i(w)^2 = \kappa_{2r}(w)^{1/2},
\end{equation}
so $f^*$ has unit norm under $\pi$.
The hypercontractive mass bound (Lemma~\ref{lem:small_set}) applied to the single point $S = \{w\}$ gives
\begin{equation}
    \pi(w)\,\kappa_{2r}(w) = \pi(w)\, f^*(w)^2 \leq 2^{2r}\, \pi(w)^{1/3},
\end{equation}
or equivalently
\begin{equation}\label{eq:kernel_point}
    \kappa_{2r}(w) \leq 2^{2r}\, \pi(w)^{-2/3}.
\end{equation}

Next we lower bound $\pi$ pointwise on the window.
By the entropy lower bound on binomial coefficients, Eq.~\eqref{eq:binom_entropy_lower} of Lemma~\ref{lem:entropy_binomial}, $\binom{n}{w} \geq \frac{1}{n+1}\,2^{nH(w/n)}$.
Since $H$ is symmetric about $1/2$ and increasing on $[0,1/2]$, the bound $|w/n - 1/2| < \beta$ gives $H(w/n) \geq H(1/2-\beta) = 1 - I(\beta)$ by Eq.~\eqref{eq:def_I}.
Multiplying by $2^{-n}$ then gives
\begin{equation}\label{eq:pi_lower}
    \pi(w) \geq \frac{1}{n+1}\,2^{-n(1 - H(w/n))} \geq \frac{1}{n+1}\,2^{-nI(\beta)}.
\end{equation}

Finally, we combine the two steps.
Substituting Eq.~\eqref{eq:pi_lower} into Eq.~\eqref{eq:kernel_point} and using $2r \leq 2\alpha n$ gives
\begin{equation}
    \kappa_{2r}(w) \leq 2^{2r}\, \pi(w)^{-2/3} \leq 2^{2\alpha n}\,(n+1)^{2/3}\, 2^{2nI(\beta)/3} = (n+1)^{2/3}\, 2^{\,n\left(\frac{2I(\beta)}{3} + 2\alpha\right)}.
\end{equation}
\end{proof}

\subsubsection{Proof of Lemma~\ref{lem:entropy_inequality} (entropy inequality)}

\begin{proof}[Proof of Lemma~\ref{lem:entropy_inequality}]
Write $G(\alpha) := H(\alpha) - \alpha - \frac{1}{2}H(2\alpha)$, which vanishes at $\alpha = 0$ since $H(0) = 0$.
We show that $G'(0) = 0$ and that $G''(\alpha) > 1$ on $(0, 1/2)$, and then obtain the claim by integrating twice.
Differentiating $H(x) = -x\log_2 x - (1-x)\log_2(1-x)$ gives $H'(x) = \log_2\frac{1-x}{x}$, so
\begin{equation}
    G'(\alpha) = H'(\alpha) - 1 - H'(2\alpha) = \log_2\frac{1-\alpha}{\alpha} - 1 - \log_2\frac{1-2\alpha}{2\alpha} = \log_2\frac{2(1-\alpha)}{1-2\alpha} - 1 = \log_2\frac{1-\alpha}{1-2\alpha},
\end{equation}
which vanishes at $\alpha = 0$.
Differentiating $G'(\alpha) = \log_2(1-\alpha) - \log_2(1-2\alpha)$ once more gives
\begin{equation}
    G''(\alpha) = \frac{1}{\ln 2}\left( \frac{2}{1-2\alpha} - \frac{1}{1-\alpha} \right) = \frac{1}{\ln 2}\cdot\frac{2(1-\alpha) - (1-2\alpha)}{(1-\alpha)(1-2\alpha)} = \frac{1}{\ln 2}\cdot\frac{1}{(1-\alpha)(1-2\alpha)}.
\end{equation}
For $\alpha \in (0, 1/2)$ both factors $1-\alpha$ and $1-2\alpha$ lie in $(0,1)$, so $G''(\alpha) > \frac{1}{\ln 2} > 1$.
Since $G'(0) = 0$, the fundamental theorem of calculus gives $G'(t) = \int_0^t G''(s)\, \mathrm{d}s$, and since $G(0) = 0$, integrating once more gives
\begin{equation}
    G(\alpha) = \int_0^{\alpha} G'(t)\, \mathrm{d}t = \int_0^{\alpha} \int_0^{t} G''(s)\, \mathrm{d}s\, \mathrm{d}t = \int_0^{\alpha} (\alpha - s)\, G''(s)\, \mathrm{d}s,
\end{equation}
where the last equality swaps the order of integration over the region $0 \leq s \leq t \leq \alpha$, so that the inner integral over $t \in [s, \alpha]$ produces the factor $\alpha - s$.
Combined with $G''(s) > 1$, this gives
\begin{equation}
    G(\alpha) > \int_0^{\alpha} (\alpha - s)\, \mathrm{d}s = \frac{\alpha^2}{2}.
\end{equation}
\end{proof}

\subsection{Expectation of high-weight $Z$ strings}\label{SM_sec:high_weight}
We finally show that the states $Z^e \ket{\Theta_s}$ remain approximately orthogonal even when the error weight exceeds $r$.
For any $a \neq b$, the overlap between $Z^a \ket{\Theta_s}$ and $Z^b \ket{\Theta_s}$ is $\bra{\Theta_s} Z^{a \oplus b} \ket{\Theta_s}$, which vanishes for $1 \leq |a \oplus b| \leq 2r$ since the seed state satisfies the EOC (Theorem~\ref{thm:near_optimal_construction}).
The following lemma bounds the overlap for $2r < |a \oplus b| \leq n - 2r$.
\begin{lemma}[Expectation of high-weight $Z$ strings under Construction 3]\label{lem:construction3_high_weight}
    For any constant $\alpha \in (0, 0.16]$, any function $\xi(n) \geq 1$ with $\log\xi(n) = o(n)$, and all sufficiently large $n$, the seed state of Theorem~\ref{thm:near_optimal_construction} satisfies, for every $e$ with $2r < |e| \leq n - 2r$,
    \begin{equation}\label{eq:construction3_high_weight}
        \left| \bra{\Theta_s} Z^e \ket{\Theta_s} \right| \leq \left[ \xi(n) + \frac{1-\alpha}{\alpha} \sqrt{n+1} \left( 2\sqrt{2} + 8\sqrt{\alpha n} + 8\xi(n) \right) \right] 2^{-n \left( 3\alpha + \frac{\alpha^2}{2} + \frac{H(2\alpha)}{2} \right)}.
    \end{equation}
\end{lemma}
\begin{proof}
Write $k := |e|$, so $2r < k \leq n - 2r$.
We first express the expectation through the weight distribution of the seed state and the moment identity Eq.~\eqref{eq:high_weight_identity}, then bound the two resulting terms, and finally compare each term of the resulting bound with the claimed rate, using the entropy bound on $\binom{n}{k}$ and the parameter choice Eq.~\eqref{eq:parameter_choice}.

By Eq.~\eqref{eq:symmetric_Z_expectation}, the expectation depends only on the weight distribution, $\bra{\Theta_s} Z^e \ket{\Theta_s} = \binom{n}{k}^{-1} \sum_{w} p_w\, K_k(w)$.
The constructed weight distribution is $p_0 = \eta$ and $p_w = (1-\eta)\, q_w$ for $w \geq 1$, with $q_w$ supported on $\mathcal{W}$ (Eq.~\eqref{eq:ansatz_q}).
Substituting $K_k(0) = \binom{n}{k}$ and the normalization $K_k = \sqrt{\binom{n}{k}}\, R_k$ (Eq.~\eqref{eq:def_R}), and then the moment identity Eq.~\eqref{eq:high_weight_identity}, valid since $k > 2r$, gives
\begin{equation}\label{eq:high_weight_split}
    \bra{\Theta_s} Z^e \ket{\Theta_s}
    = \eta + \frac{1-\eta}{\sqrt{\binom{n}{k}}} \sum_{w \in \mathcal{W}} q_w\, R_k(w)
    = \eta + \frac{1-\eta}{\sqrt{\binom{n}{k}}\, Z_{\mathcal{W}}} \left( -e_k + \sum_{w \in \mathcal{W}} \pi(w)\, R_k(w)\, u(w) \right).
\end{equation}

We next bound the two terms in the parentheses of Eq.~\eqref{eq:high_weight_split} separately.
Lemma~\ref{lem:bound_b} bounds the tail sum by $|e_k| \leq \sqrt{2}\, 2^{-nI(\beta)/2}$.
For the second term, since $k > 2r$, Lemma~\ref{lem:orthonormal} makes $R_k$ orthogonal under $\pi$ to every $R_j$ appearing in $u$, so it equals $-\sum_{w \notin \mathcal{W}} \pi(w)\, R_k(w)\, u(w)$, and Cauchy--Schwarz with the unit norm of $R_k$ (Lemma~\ref{lem:orthonormal}) bounds its magnitude by
\begin{equation}
    \left|\sum_{w \notin \mathcal{W}} \pi(w) R_k(w) u(w) \right|
    \leq \left(\sum_{w \notin \mathcal{W}} \pi(w) R_k(w)^2\right)^{1/2} \left(\sum_{w \notin \mathcal{W}} \pi(w)\, u(w)^2\right)^{1/2}
    \leq \left(\sum_{w \notin \mathcal{W}} \pi(w)\, u(w)^2\right)^{1/2}.
\end{equation}
Extending the sum to the full support and using $\sum_{w} \pi(w)\, u(w)^2 = \|\vec{c}\|_2^2$, which holds by Lemma~\ref{lem:orthonormal} since $u = \sum_j c_j R_j$, bounds the second term by $\|\vec{c}\|_2$, and Eq.~\eqref{eq:bound_c} bounds $\|\vec{c}\|_2 \leq 4\sqrt{\alpha n}\, 2^{-nI(\beta)/2} + 4\eta\, 2^{nH(2\alpha)/2}$.

We now combine the two bounds.
For sufficiently large $n$, the tail bound Eq.~\eqref{eq:tail_Z} gives $1 - Z_{\mathcal{W}} \leq 1/2$, so $Z_{\mathcal{W}}^{-1} \leq 2$, and with $1 - \eta \leq 1$,
\begin{equation}\label{eq:high_weight_combined}
    \left| \bra{\Theta_s} Z^e \ket{\Theta_s} \right|
    \leq \eta + \frac{2}{\sqrt{\binom{n}{k}}} \left( |e_k| + \|\vec{c}\|_2 \right)
    \leq \eta + \frac{1}{\sqrt{\binom{n}{k}}} \left( \left( 2\sqrt{2} + 8\sqrt{\alpha n} \right) 2^{-nI(\beta)/2} + 8\eta\, 2^{nH(2\alpha)/2} \right).
\end{equation}

Then we lower bound the binomial coefficient in Eq.~\eqref{eq:high_weight_combined}.
The condition $2r < k \leq n - 2r$ places $k/n$ between $2r/n$ and $1 - 2r/n$, and $\alpha \leq 0.16 < 1/4$ gives $2r/n \leq 2\alpha < 1/2$, so the monotonicity of $H$ on $[0,1/2]$ together with the symmetry $H(1-x) = H(x)$ gives $H(k/n) \geq H(2r/n)$.
Since $n \geq 2/\alpha$ for sufficiently large $n$, Lemmas~\ref{lem:entropy_binomial} and~\ref{lem:entropy_floor} then give
\begin{equation}\label{eq:high_weight_binomial}
    \frac{1}{\sqrt{\binom{n}{k}}} \leq \sqrt{n+1}\, 2^{-nH(k/n)/2} \leq \sqrt{n+1}\, 2^{-nH(2r/n)/2} \leq \frac{1-\alpha}{\alpha} \sqrt{n+1}\, 2^{-nH(2\alpha)/2}.
\end{equation}

Finally we compare the decay rate of each of the three terms of Eq.~\eqref{eq:high_weight_combined} with the decay rate claimed in Eq.~\eqref{eq:construction3_high_weight}.
Every bound obtained above carries its exponent through $I(\beta)$ and $H(2\alpha)$, so we write the claimed decay rate in the same two quantities, using the parameter choice $I(\beta) = 6\alpha + \alpha^2$ of Eq.~\eqref{eq:parameter_choice},
\begin{equation}
    3\alpha + \frac{\alpha^2}{2} + \frac{H(2\alpha)}{2} = \frac{I(\beta) + H(2\alpha)}{2}.
\end{equation}
The first term of Eq.~\eqref{eq:high_weight_combined} is the matched query power $\eta = \xi(n)\, 2^{-n(H(\alpha)+2\alpha)}$ of Eq.~\eqref{eq:parameter_choice}, whose decay rate exceeds the claimed one, since
\begin{equation}
    H(\alpha) + 2\alpha - \frac{I(\beta) + H(2\alpha)}{2}
    = \left( H(\alpha) - \alpha - \frac{H(2\alpha)}{2} \right) - \frac{\alpha^2}{2} > 0
\end{equation}
by Lemma~\ref{lem:entropy_inequality}, so $\eta \leq \xi(n)\, 2^{-n(I(\beta) + H(2\alpha))/2}$.
The second term carries the factor $2^{-nI(\beta)/2}$ and the binomial prefactor Eq.~\eqref{eq:high_weight_binomial}, whose exponential factor is $2^{-nH(2\alpha)/2}$, so its decay rate is exactly the claimed one and the term is at most $\frac{1-\alpha}{\alpha} \sqrt{n+1} \left( 2\sqrt{2} + 8\sqrt{\alpha n} \right) 2^{-n(I(\beta)+H(2\alpha))/2}$.
In the third term the factor $2^{nH(2\alpha)/2}$ cancels the exponential factor of the same binomial prefactor and leaves $8\eta\, \frac{1-\alpha}{\alpha} \sqrt{n+1}$, which the bound on $\eta$ above makes at most $8\xi(n)\, \frac{1-\alpha}{\alpha} \sqrt{n+1}\, 2^{-n(I(\beta)+H(2\alpha))/2}$.
Adding the three bounds gives Eq.~\eqref{eq:construction3_high_weight}.
\end{proof}

\clearpage
\section{Optimality of Boolean Oracle Distillation}\label{SM_sec:od_optimality}

In this section we prove the two lower bounds stated in the main text, Theorems~\ref{thm:OD_grover_lower_bound} and~\ref{thm:OD_label_indep}, on the query complexity of OD protocols.
Both theorems apply to protocols that distill a family of ideal Boolean oracles into itself and whose pre-query subspaces satisfy the KLC for $Z$ errors of bounded weight, the class defined in Sec.~\ref{sec:optimality}.
The motivation for this class and the outline of the two proofs are given in Sec.~\ref{sec:optimality}, and this section supplies the full arguments.

We first recall the notation of Sec.~\ref{sec:optimality}.
The projector $P_{t,f}$ projects onto the pre-query subspace at the $t$-th query of the run with the ideal oracle $\mathcal{O}_f$, and may depend on both the query round $t$ and the oracle label $f$.
Each $P_{t,f}$ satisfies the KLC for all $Z$ errors of weight at most $r$, restated from Eq.~\eqref{eq:KLC_prequery},
\begin{equation}\label{eq:KLC}
    P_{t,f} Z^a Z^b P_{t,f} = C_{t,f}^{a,b} P_{t,f},\quad a,b \in \{0,1\}^n,\ |a|,|b|\leq r
\end{equation}
where the numbers $C_{t,f}^{a,b}$ form the KL matrix $C_{t,f}$, with rows and columns labeled by the $Z$ strings $a$ and $b$.
We call Eq.~\eqref{eq:KLC} the KLC for $r$ errors, and drop the indices $t,f$ when they are not needed.
Since $Z$ errors on the index register commute with the ideal oracle, the subspace just after each query, with projector $O_f P_{t,f} O_f^{\dagger}$, inherits the KLC~\eqref{eq:KLC} with the same KL matrix, so imposing the KLC just before each query loses no generality.
Since $Z^a Z^a = I$, every diagonal entry of the KL matrix equals one, so
\begin{equation}\label{eq:trace_C}
    \Tr C_{t,f} = M_r, \quad M_r := \sum_{j=0}^r \binom{n}{j}
\end{equation}
with $M_r$ as defined in Eq.~\eqref{eq:def_Mr} of the main text.

The section is organized as follows.
Section~\ref{SM_sec:query_power} proves that the KLC bounds the query power that any state in the pre-query subspace places on each bitstring by the Christoffel function of the KL matrix, and that the sum of these bounds over all bitstrings is at most $N/M_r$.
Section~\ref{SM_sec:grover_lb} combines these bounds with the pair test, averaged over pairs of Grover oracles, to prove Theorem~\ref{thm:OD_grover_lower_bound}, in which the KL matrices may depend on both the query round and the oracle label.
Section~\ref{SM_sec:label_indep_lb} proves Theorem~\ref{thm:OD_label_indep}, which assumes label-independent KL matrices and in exchange applies to protocols that distill an arbitrary family of ideal Boolean oracles.
Section~\ref{SM_sec:useful_precision} justifies the claim of Sec.~\ref{sec:optimality} that an application with constant success probability never benefits from a target precision below $\epsilon = \Theta(1/N)$.

\subsection{Bounds on the query power}\label{SM_sec:query_power}
In this subsection we show that the KLC forces the query power of any codespace state to be small.
As in the main text, the query power of a state on a bitstring $z$ is the population that the state places on $\ket{z}$ on the $n$ qubits carrying the $Z$ errors.
We first show that the KL matrix inherits a special structure from the group structure of $Z$ errors.
We then prove two bounds.
The first bounds the query power on each single bitstring, and the second bounds its total over all bitstrings.
Both bounds are stated through the Christoffel function of the KL matrix, which we introduce below.

We use the Boolean characters $\chi^a\colon \{0,1\}^n \rightarrow \{+1,-1\}$, labeled by $a \in \{0,1\}^n$,
\begin{equation}
    \chi^a(z) := (-1)^{\sum_{i=1}^n a_i z_i} = \bra{z} Z^a \ket{z}.
\end{equation}
Since each $Z^a$ is diagonal in the computational basis, $Z^a = \sum_z \chi^a(z) \Pi_z$, where $\Pi_z$ projects onto $\ket{z}$ on the $n$ designated qubits and acts as the identity on all other registers.

For each $z \in \{0,1\}^n$, we collect the character values into an $M_r$-dimensional vector $v(z)$, whose entries are $v(z)_a = \chi^a(z)$ for $|a| \leq r$, indexed in the same order as the rows of the KL matrix.
Since $|\chi^a(z)| = 1$, we have $v(z)^{\top} v(z) = M_r$.
The characters are also orthogonal,
\begin{equation}
    \sum_z \chi^a(z)\chi^b(z) = \sum_z \bra{z} Z^{a\oplus b} \ket{z} = N\delta_{a,b},
\end{equation}
which in vector form reads
\begin{equation}\label{eq:v_resolution}
    \sum_z v(z) v(z)^{\top} = NI.
\end{equation}

Because all the $Z^a$ are simultaneously diagonal in the computational basis, the KL matrix can only take a special form.
\begin{lemma}[KL matrices are structured]\label{lem:KL_structured}
    Let $P$ satisfy the KLC for $r$ errors with KL matrix $C$.
    Then for every normalized vector $\ket{\psi}$ in the range of $P$,
    \begin{equation}
        C = \sum_{z \in \{0,1\}^n} p_{\psi}(z)\, v(z) v(z)^{\top},
        \qquad
        p_{\psi}(z) = \bra{\psi} \Pi_z \ket{\psi}.
    \end{equation}
    In particular, $C$ is positive semidefinite, and the right-hand side is the same for every choice of $\ket{\psi}$ in the codespace.
\end{lemma}
\begin{proof}
    Since $Z^a$ and $Z^b$ are both diagonal in the computational basis, $Z^a Z^b = \sum_z \chi^a(z) \chi^b(z) \Pi_z$.
    Using $P\ket{\psi} = \ket{\psi}$, the KLC gives
    \begin{equation}
        C^{a,b}
        = \bra{\psi} P Z^a Z^b P \ket{\psi}
        = \sum_{z \in \{0,1\}^n} p_{\psi}(z)\, \chi^a(z) \chi^b(z),
    \end{equation}
    which is the $(a,b)$ entry of the claimed decomposition.
\end{proof}

For a KL matrix $C$, we define the Christoffel function
\begin{equation}\label{eq:christoffel_variational}
    \Lambda_C(z) := \min \left\{u^{\top} C u : u \in \mathbb{R}^{M_r},\ u^{\top} v(z) = 1\right\}.
\end{equation}
We call a vector $u \in \mathbb{R}^{M_r}$ feasible for $z$ if it satisfies the constraint $u^{\top} v(z) = 1$.
Since $C$ is positive semidefinite, $\Lambda_C(z)\geq 0$.
The next lemma shows that $\Lambda_C(z)$ bounds the query power that any codespace state can place on the bitstring $z$.
\begin{lemma}[Pointwise bound on query power]\label{lem:pointwise_qp}
    Let $P$ satisfy the KLC for $r$ errors with KL matrix $C$.
    Then for every $z \in \{0,1\}^n$,
    \begin{equation}
        P \Pi_z P \preceq \Lambda_C(z) P.
    \end{equation}
\end{lemma}
\begin{proof}
    Fix a real vector $u$ with $u^{\top} v(z)=1$ and let $Q := \sum_{|a|\leq r} u_a Z^a$.
    The KLC gives
    \begin{equation}
        PQ^2 P = \sum_{|a|,|b|\leq r} u_a u_b P Z^a Z^b P = (u^{\top} C u) P.
    \end{equation}
    On the other hand, since $Q$ is a diagonal operator, it admits a decomposition in the computational basis
    \begin{equation}
        Q^2 = \sum_{z'} (u^{\top} v(z'))^2 \Pi_{z'}
        \succeq
        (u^{\top} v(z))^2 \Pi_z = \Pi_z.
    \end{equation}
    Sandwiching with $P$ gives $(u^{\top} C u) P \succeq P \Pi_z P$.
    Minimizing over $u$ finishes proving the lemma.
\end{proof}

The minimum defining $\Lambda_C(z)$ can be evaluated in closed form, which turns the lemma into a concrete bound on the query power.
If $v(z)$ lies in the range of $C$, then $\Lambda_C(z) = 1/\big(v(z)^{\top} C^{+} v(z)\big)$, where $C^{+}$ is the pseudoinverse of $C$.
Indeed, writing $v(z) = C^{1/2} C^{+1/2} v(z)$, the Cauchy--Schwarz inequality gives $1 = (u^{\top} v(z))^2 \leq (u^{\top} C u)\big(v(z)^{\top} C^{+} v(z)\big)$ for every feasible $u$, and $u = C^{+} v(z) / \big(v(z)^{\top} C^{+} v(z)\big)$ attains the bound.
Otherwise the component $v_K$ of $v(z)$ in the kernel of $C$ is nonzero, and the feasible vector $u = v_K / (v_K^{\top} v_K)$ gives $\Lambda_C(z) = 0$.

The pointwise bound already constrains a single query state.
\begin{corof}{lem:pointwise_qp}{1}[Query power of a single query state]\label{cor:seed_qp_bound}
    Every state $\ket{\Theta}$ that satisfies the EOC Eq.~\eqref{SM_eq:error_orthogonality} for $r$ errors obeys, for every $z \in \{0,1\}^n$,
    \begin{equation}\label{eq:seed_qp_bound}
        \bra{\Theta} \Pi_z \ket{\Theta} \leq \frac{1}{M_r}.
    \end{equation}
    In particular, any family of query states satisfying the EOC has matched query power $\eta \leq M_r^{-1}$.
\end{corof}
\begin{proof}
    The rank-one projector $P = \ket{\Theta}\bra{\Theta}$ satisfies $P Z^a Z^b P = \bra{\Theta} Z^a Z^b \ket{\Theta} P$, so the EOC state that $P$ satisfies the KLC for $r$ errors with KL matrix $C = I$ (Sec.~\ref{SM_sec:prelim_klc}).
    Since $I$ has full rank, $\Lambda_I(z) = 1/\big(v(z)^{\top} v(z)\big) = 1/M_r$, and Lemma~\ref{lem:pointwise_qp} gives $\bra{\Theta} \Pi_z \ket{\Theta} \leq 1/M_r$.
    Taking $z = x$ for the query state $\ket{\Theta(x)}$ bounds $\eta_x$, and minimizing over $x$ bounds $\eta$.
\end{proof}

We now prove the second bound.
The pointwise bound still allows the query power to be large on a few bitstrings.
The next lemma shows that this cannot happen on many bitstrings at once, since the Christoffel functions have a bounded sum.
\begin{lemma}[Total bound on query power]\label{lem:total_qp}
    Let $P$ satisfy the KLC for $r$ errors with KL matrix $C$. Then
    \begin{equation}
        \sum_{z \in \{0,1\}^n} \Lambda_C(z) \leq \frac{N}{M_r}.
    \end{equation}
\end{lemma}
\begin{proof}
    For each $z$, the vector $u(z) = v(z)/M_r$ is feasible, since $u(z)^{\top} v(z) = v(z)^{\top} v(z)/M_r = 1$.
    Hence $\Lambda_C(z) \leq u(z)^{\top} C u(z) = v(z)^{\top} C v(z)/M_r^2$.
    Summing over $z$, and using the cyclic property of the trace together with Eq.~\eqref{eq:v_resolution} and Eq.~\eqref{eq:trace_C},
    \begin{equation}
        \sum_{z \in \{0,1\}^n} \Lambda_C(z)
        \leq \frac{1}{M_r^2} \sum_{z} \Tr\left[C\, v(z) v(z)^{\top}\right]
        = \frac{N \Tr C}{M_r^2}
        = \frac{N}{M_r}.
    \end{equation}
\end{proof}

\subsection{Lower bound for distilling the Grover family}\label{SM_sec:grover_lb}
In this subsection we prove a general lower bound on the query complexity of oracle distillation.
We only assume that the pre-query subspaces $P_{t,f}$ have a code structure, satisfying the KLC for $r$ errors with KL matrices $C_{t,f}$ that may vary from query round to query round and from oracle label to oracle label.

To prove the lower bound, we split the argument into two steps.
First, we consider a simple pair test, in which the same OD protocol is run on an identical, specifically chosen initial state but queries two different Grover oracles.
We show that if the protocol distills with precision $\epsilon$, the two final states must be almost perfectly distinguishable.
Second, we study the average behavior of this pair test over pairs of Grover oracles, and show that the average distinguishability can only grow slowly with each query, so the query complexity must be high.

We begin with the pair test.
In a pair test between two different oracles $O_f$ and $O_g$, we choose a bitstring $x_0$ with $f(x_0) \neq g(x_0)$, and execute the same OD protocol on the same initial state $\psi^0 = \ket{x_0}\ket{+}^{\otimes m}$, with the response register in $\ket{+}^{\otimes m}$, querying $O_f$ in one execution and $O_g$ in the other.
A pair test is thus specified by the ordered pair $(f,g)$ together with the choice of $x_0$.
Writing the protocol in the form of Eq.~\eqref{eq:od_protocol} with a purified ancilla state $\ket{A}$, denote by
\begin{equation}\label{eq:pair_test_states}
    \ket{\psi^t_{f\mid(f,g)}} := U_{t-1}\, O_f\, U_{t-2} \cdots O_f\, U_0 \left(\ket{x_0}\ket{+}^{\otimes m}\otimes\ket{A}\right),
    \quad t = 1, \dots, T_{\mathrm{OD}}+1,
\end{equation}
the purified states of the execution querying $O_f$, and similarly $\ket{\psi^t_{g\mid(f,g)}}$ for the execution querying $O_g$.
The subscript records which oracle the branch queries and which pair test it belongs to, while the dependence on $x_0$ is left implicit.
For $t \leq T_{\mathrm{OD}}$, the state $\ket{\psi^t_{f\mid(f,g)}}$ is the pre-query state of the $t$-th query, so it lies in the range of $P_{t,f}$.
For $t = T_{\mathrm{OD}}+1$, the state $\ket{\psi^{T_{\mathrm{OD}}+1}_{f\mid(f,g)}}$ is the final state of the protocol, and Eq.~\eqref{eq:od_protocol} gives $\widehat{\mathcal{O}}_f(\psi^0) = \Tr_A[\psi^{T_{\mathrm{OD}}+1}_{f\mid(f,g)}]$.
In particular, $\ket{\psi^1_{f\mid(f,g)}} = \ket{\psi^1_{g\mid(f,g)}} = U_0 \left(\ket{x_0}\ket{+}^{\otimes m}\otimes\ket{A}\right)$ contains no query.
\begin{lemma}[Successful distillation forces distinguishability]\label{lem:pair_test_final}
    Suppose an OD protocol uses $T_{\mathrm{OD}}$ queries to distill each of two different Boolean oracles $\mathcal{O}_f$ and $\mathcal{O}_g$ into an $\epsilon$-approximate version of itself, with $0 \leq \epsilon < 1$.
    Then, for any initial state $\ket{x_0}\ket{+}^{\otimes m}$ with $f(x_0) \neq g(x_0)$, the pair test between $\mathcal{O}_f$ and $\mathcal{O}_g$ ends with nearly orthogonal final states,
    \begin{equation}
        \left| \braket{\psi^{T_{\mathrm{OD}}+1}_{f\mid(f,g)} | \psi^{T_{\mathrm{OD}}+1}_{g\mid(f,g)}}\right| \leq \sqrt{2\epsilon - \epsilon^2}.
    \end{equation}
\end{lemma}
\begin{proof}
    The ideal oracles map the initial state to $O_f \ket{x_0}\ket{+}^{\otimes m} = \ket{x_0}\, Z^{f(x_0)}\ket{+}^{\otimes m}$ and $O_g \ket{x_0}\ket{+}^{\otimes m} = \ket{x_0}\, Z^{g(x_0)}\ket{+}^{\otimes m}$.
    Since $Z\ket{+} = \ket{-}$, the two response registers are products of $\ket{\pm}$ states that differ on every qubit $j$ with $f(x_0)_j \neq g(x_0)_j$, of which there is at least one, so the two outputs are orthogonal,
    \begin{equation}
        \left\|\mathcal{O}_f(\psi^0) - \mathcal{O}_g(\psi^0)\right\|_1 = 2.
    \end{equation}

    We now compare with the outputs of the distilled oracles.
    By the triangle inequality,
    \begin{equation}
        2 = \left\|\mathcal{O}_{f}(\psi^0) - \mathcal{O}_{g}(\psi^0) \right\|_1
        \leq \left\|\mathcal{O}_{f}(\psi^0) - \widehat{\mathcal{O}}_{f}(\psi^0)\right\|_1 + \left\|\widehat{\mathcal{O}}_{f}(\psi^0) - \widehat{\mathcal{O}}_{g}(\psi^0)\right\|_1 + \left\|\widehat{\mathcal{O}}_{g}(\psi^0) - \mathcal{O}_{g}(\psi^0)\right\|_1.
    \end{equation}
    The first and the third terms are each at most $\epsilon$, since the trace distance on any input state is bounded by the diamond norm of the difference of the two channels (Lemma~\ref{lem:channel_norm_facts}), and the protocol distills $\epsilon$-approximate oracles.
    For the middle term, $\widehat{\mathcal{O}}_{f}(\psi^0) = \Tr_A[\psi^{T_{\mathrm{OD}}+1}_{f\mid(f,g)}]$ and $\widehat{\mathcal{O}}_{g}(\psi^0) = \Tr_A[\psi^{T_{\mathrm{OD}}+1}_{g\mid(f,g)}]$, so the data processing inequality for the partial trace (Lemma~\ref{lem:channel_norm_facts}) gives
    \begin{equation}
        \left\|\widehat{\mathcal{O}}_{f}(\psi^0) - \widehat{\mathcal{O}}_{g}(\psi^0)\right\|_1
        = \left\|\Tr_A\left[\psi^{T_{\mathrm{OD}}+1}_{f\mid(f,g)}\right] - \Tr_A\left[\psi^{T_{\mathrm{OD}}+1}_{g\mid(f,g)}\right]\right\|_1
        \leq \left\|\psi^{T_{\mathrm{OD}}+1}_{f\mid(f,g)} - \psi^{T_{\mathrm{OD}}+1}_{g\mid(f,g)}\right\|_1.
    \end{equation}
    Combining the three bounds gives $2 \leq 2\epsilon + \left\|\psi^{T_{\mathrm{OD}}+1}_{f\mid(f,g)} - \psi^{T_{\mathrm{OD}}+1}_{g\mid(f,g)}\right\|_1$.
    The final states are pure, so $\left\|\psi^{T_{\mathrm{OD}}+1}_{f\mid(f,g)} - \psi^{T_{\mathrm{OD}}+1}_{g\mid(f,g)}\right\|_1 = 2\sqrt{1 - |\braket{\psi^{T_{\mathrm{OD}}+1}_{f\mid(f,g)}|\psi^{T_{\mathrm{OD}}+1}_{g\mid(f,g)}}|^2}$ (Lemma~\ref{lem:pure_state_trace_dist}).
    Rearranging and squaring, which is valid since $\epsilon < 1$, gives $|\braket{\psi^{T_{\mathrm{OD}}+1}_{f\mid(f,g)}|\psi^{T_{\mathrm{OD}}+1}_{g\mid(f,g)}}|^2 \leq 2\epsilon - \epsilon^2$.
\end{proof}

During a pair test, the interleaved unitaries are the same in both executions and preserve the inner product, so only the queries can change the overlap.
The next lemma bounds the change caused by a single query, in terms of the Christoffel functions at the inputs where the two oracles disagree.
\begin{lemma}[Change in overlap from a single query]\label{lem:single_query}
    Let $P_1$ and $P_2$ be projectors onto two codespaces on a system containing the $n$ designated qubits and the $m$-qubit response register, each satisfying the KLC for $r$ errors, with KL matrices $C_1$ and $C_2$ respectively.
    Then, for any $\ket{\psi_1}$ in the range of $P_1$, any $\ket{\psi_2}$ in the range of $P_2$, and any Boolean oracles $O_f$ and $O_g$,
    \begin{equation}
        \left|\braket{\psi_1 | \psi_2} - \bra{\psi_1} O_f^\dagger O_g \ket{\psi_2}\right| \leq 2 \sum_{x:\, f(x)\neq g(x)} \sqrt{\Lambda_{C_1}(x)\, \Lambda_{C_2}(x)}.
    \end{equation}
\end{lemma}
\begin{proof}
    Writing $O_f = \sum_x \Pi_x \otimes Z^{f(x)}$, we have $O_f^\dagger O_g = \sum_x \Pi_x \otimes Z^{f(x) \oplus g(x)}$, hence
    \begin{equation}
        I - O_f^\dagger O_g = 2\sum_{x:\, f(x)\neq g(x)} \Pi_x \otimes \Pi^{\mathrm{odd}}_{f(x)\oplus g(x)},
    \end{equation}
    where $\Pi^{\mathrm{odd}}_{s} := \sum_{y:\, s \cdot y = 1} \ket{y}\bra{y}$ projects the response register onto the computational basis states $y$ with $s \cdot y := \bigoplus_{j=1}^m s_j y_j$ equal to $1$.
    Indeed, $Z^{s}\ket{y} = (-1)^{s \cdot y}\ket{y}$ gives $I - Z^{s} = 2\,\Pi^{\mathrm{odd}}_{s}$, and $\Pi^{\mathrm{odd}}_{0^m} = 0$, so the inputs with $f(x) = g(x)$ drop out of the sum.
    Taking the matrix element between $\bra{\psi_1}$ and $\ket{\psi_2}$ and applying the triangle inequality,
    \begin{equation}
        \left|\braket{\psi_1 | \psi_2} - \bra{\psi_1} O_f^\dagger O_g \ket{\psi_2}\right|
        \leq 2 \sum_{x:\, f(x)\neq g(x)} \left|\bra{\psi_1} \Pi_x \otimes \Pi^{\mathrm{odd}}_{f(x)\oplus g(x)} \ket{\psi_2}\right|.
    \end{equation}
    Since each $\Pi_x \otimes \Pi^{\mathrm{odd}}_{f(x)\oplus g(x)}$ is a projector, the Cauchy--Schwarz inequality gives
    \begin{equation}
        \left|\bra{\psi_1} \Pi_x \otimes \Pi^{\mathrm{odd}}_{f(x)\oplus g(x)} \ket{\psi_2}\right|
        \leq \sqrt{\bra{\psi_1} \Pi_x \otimes \Pi^{\mathrm{odd}}_{f(x)\oplus g(x)} \ket{\psi_1}\, \bra{\psi_2} \Pi_x \otimes \Pi^{\mathrm{odd}}_{f(x)\oplus g(x)} \ket{\psi_2}}.
    \end{equation}
    Since $\Pi_x \otimes \Pi^{\mathrm{odd}}_{f(x)\oplus g(x)} \preceq \Pi_x$ and $\ket{\psi_1} = P_1 \ket{\psi_1}$, Lemma~\ref{lem:pointwise_qp} gives $\bra{\psi_1} \Pi_x \otimes \Pi^{\mathrm{odd}}_{f(x)\oplus g(x)} \ket{\psi_1} \leq \bra{\psi_1} P_1 \Pi_x P_1 \ket{\psi_1} \leq \Lambda_{C_1}(x)$, and likewise $\bra{\psi_2} \Pi_x \otimes \Pi^{\mathrm{odd}}_{f(x)\oplus g(x)} \ket{\psi_2} \leq \Lambda_{C_2}(x)$.
    Substituting into the sum proves the lemma.
\end{proof}

We now study the average behavior of the pair test over all pairs of Grover oracles.
For $x \in \{0,1\}^n$, let $f_x$ denote the Grover function marked at $x$, given by $f_x(z) = \delta_{x,z}$.
For each ordered pair $(f_x, f_y)$ with $x \neq y$, we run the pair test between $O_{f_x}$ and $O_{f_y}$ with the choice $x_0 = x$, on which $f_x(x) = 1$ and $f_y(x) = 0$, and track its progress by
\begin{equation}
    W_{(f_x,f_y)}(t) := \Re \braket{\psi^t_{f_x\mid(f_x,f_y)} | \psi^t_{f_y\mid(f_x,f_y)}},
\end{equation}
with the states defined in Eq.~\eqref{eq:pair_test_states}.
Before the first query the two branches coincide, so $W_{(f_x,f_y)}(1) = 1$; after all $T_{\mathrm{OD}}$ queries, the real part of the inner product is at most its absolute value, so Lemma~\ref{lem:pair_test_final} gives $W_{(f_x,f_y)}(T_{\mathrm{OD}}+1) \leq \sqrt{2\epsilon - \epsilon^2}$.
We then define the average progress
\begin{equation}
    \bar{W}(t) := \frac{1}{N(N-1)} \sum_{x \neq y} W_{(f_x,f_y)}(t).
\end{equation}
Hence, $\bar{W}(1) = 1$ and $\bar{W}(T_{\mathrm{OD}}+1) \leq \sqrt{2\epsilon - \epsilon^2}$.

The next lemma bounds the change in the average progress caused by a single query.
Although different pair tests start from different initial states, the bound depends only on the pre-query subspaces $P_{t,f_x}$, which are fixed by the protocol and do not depend on the initial state.
\begin{lemma}[Bound on change in average progress]\label{lem:avg_progress_change}
    For any $1 \leq t \leq T_{\mathrm{OD}}$, the change in the average progress caused by the $t$-th query satisfies
    \begin{equation}\label{eq:Dt_def}
        \left|\bar{W}(t+1) - \bar{W}(t) \right| \leq 4\sqrt{\frac{D_t}{M_r(N-1)}},
        \quad\text{where}\quad
        D_t := \sum_{x \in \{0,1\}^n} \Lambda_{C_{t,f_x}}(x).
    \end{equation}
\end{lemma}
\begin{proof}
    By the triangle inequality,
    \begin{equation}\label{eq:avg_triangle}
        \left|\bar{W}(t+1) - \bar{W}(t) \right| \leq \frac{1}{N(N-1)} \sum_{x\neq y} \left| W_{(f_x,f_y)}(t+1) - W_{(f_x,f_y)}(t) \right|.
    \end{equation}
    We bound each summand by Lemma~\ref{lem:single_query}.
    The interleaved unitary preserves the inner product, so $\braket{\psi^{t+1}_{f_x\mid(f_x,f_y)} | \psi^{t+1}_{f_y\mid(f_x,f_y)}} = \bra{\psi^t_{f_x\mid(f_x,f_y)}} O_{f_x}^\dagger O_{f_y} \ket{\psi^t_{f_y\mid(f_x,f_y)}}$, and the real parts differ by at most the difference of the inner products.
    The two pre-query states lie in the ranges of $P_{t,f_x}$ and $P_{t,f_y}$, and $f_x$ and $f_y$ disagree exactly at the two inputs $x$ and $y$, so Lemma~\ref{lem:single_query} gives
    \begin{equation}\label{eq:per_pair_progress}
        \left| W_{(f_x,f_y)}(t+1) - W_{(f_x,f_y)}(t) \right| \leq 2\left(\sqrt{\Lambda_{C_{t,f_x}}(x) \Lambda_{C_{t,f_y}}(x)} + \sqrt{\Lambda_{C_{t,f_x}}(y) \Lambda_{C_{t,f_y}}(y)}\right).
    \end{equation}
    We sum the two terms on the right over all pairs separately.
    For the first term, the Cauchy--Schwarz inequality gives
    \begin{equation}
        \sum_{x\neq y} \sqrt{\Lambda_{C_{t,f_x}}(x) \Lambda_{C_{t,f_y}}(x)} \leq \sqrt{\sum_{x\neq y} \Lambda_{C_{t,f_x}}(x)}\sqrt{\sum_{x\neq y} \Lambda_{C_{t,f_y}}(x)}.
    \end{equation}
    In the first factor the summand does not depend on $y$, so $\sum_{x\neq y} \Lambda_{C_{t,f_x}}(x) = (N-1)D_t$.
    In the second factor, since $\Lambda_C(z) \geq 0$, adding the $x = y$ terms and applying Lemma~\ref{lem:total_qp} for each $y$ gives
    \begin{equation}
        \sum_{x\neq y} \Lambda_{C_{t,f_y}}(x) \leq \sum_{y} \sum_{x} \Lambda_{C_{t,f_y}}(x) \leq \frac{N^2}{M_r}.
    \end{equation}
    The second term in Eq.~\eqref{eq:per_pair_progress} satisfies the same bound with the roles of $x$ and $y$ exchanged.
    Plugging the two bounds back into Eq.~\eqref{eq:avg_triangle}, we get
    \begin{equation}
        \left| \bar{W}(t+1) - \bar{W}(t) \right| \leq \frac{4}{N(N-1)} \sqrt{(N-1)D_t} \sqrt{\frac{N^2}{M_r}} = 4\sqrt{\frac{D_t}{M_r(N-1)}}. \qedhere
    \end{equation}
\end{proof}

It remains to bound $D_t$.
Before the first query, the protocol has not yet acted with any oracle, so the pre-query subspaces $P_{1,f_x}$ coincide for all $x$, and Lemma~\ref{lem:total_qp} gives $D_1 \leq N/M_r$.
As the number of queries grows, the codespaces of different branches can deviate from one another; the branch querying $O_{f_x}$ can concentrate weight on the marked string $x$, driving $\Lambda_{C_{t,f_x}}(x)$ toward $1$ and $D_t$ toward $N$.
This deviation, however, is built one query at a time, so $D_t$ cannot grow quickly.
The following lemma makes this concrete.
\begin{lemma}[Slow growth of $D_t$]\label{lem:Dt_growth}
    For any $1\leq t\leq T_{\mathrm{OD}}$,
    \begin{equation}
        D_t \leq \min\left\{N, \left(\sqrt{N/M_r} + 2(t-1)\right)^2\right\}.
    \end{equation}
\end{lemma}
\begin{proof}
    We first show $D_t\leq N$.
    The vector $u=(1,0,\dots,0)^\top$ satisfies $u^{\top}v(x) = \chi^{0}(x) = 1$, so it is feasible for every $x$, and it gives $\Lambda_C(x) \leq u^{\top}Cu = C^{0,0} = 1$ for any KL matrix $C$.
    Since $D_t$ is a sum of $N$ Christoffel functions, $D_t \leq N$.

    We then show the remaining bound $D_t \leq \left(\sqrt{N/M_r} + 2(t-1)\right)^2$, in three steps.
    First, we fix an initial state and define one branch for each Grover oracle, in which the protocol queries that oracle, together with a reference branch in which every query is replaced by the identity.
    Second, we bound each Christoffel function in $D_t$ by a norm of the corresponding oracle-branch state.
    Third, we split the bound into two terms, one depending only on the reference branch and the other measuring the difference between each oracle branch and the reference branch, and bound them separately.

    Consider a fixed initial state $\ket{\varphi_0} := \ket{\psi^0} \otimes \ket{A}$, where $\ket{\psi^0}$ is an arbitrary input state and $\ket{A}$ is the purified ancilla state of the protocol.
    For each Grover oracle $O_{f_x}$, define the branch $\ket{\varphi_{t,f_x}} := U_{t-1}\, O_{f_x}\, U_{t-2} \cdots O_{f_x}\, U_0 \ket{\varphi_0}$, the state just before the $t$-th query when the protocol queries $O_{f_x}$, which is normalized and lies in the range of $P_{t,f_x}$.
    We also define the reference branch $\ket{\varphi_{t,I}} := U_{t-1}\, U_{t-2} \cdots U_0 \ket{\varphi_0}$, in which every query is replaced by the identity.

    Second, we bound each Christoffel function in $D_t$.
    Since $\ket{\varphi_{t,f_x}}$ is a normalized state in the range of $P_{t,f_x}$, Lemma~\ref{lem:KL_structured} gives $C_{t,f_x} = \sum_{z} p_{\varphi_{t,f_x}}(z) v(z)v(z)^{\top}$.
    The vector $u = v(x)/M_r$ satisfies $u^{\top}v(x) = v(x)^{\top}v(x)/M_r = 1$, so it is feasible in Eq.~\eqref{eq:christoffel_variational}, and
    \begin{equation}
        \Lambda_{C_{t,f_x}}(x) \leq u^{\top}C_{t,f_x}u = \frac{1}{M_r^2}\sum_{z} p_{\varphi_{t,f_x}}(z) (v(x)^{\top}v(z))^2.
    \end{equation}
    Defining the positive semidefinite operator
    \begin{equation}
        H_x := \sum_{z} \frac{(v(x)^{\top} v(z))^2}{M_r^2} \Pi_z,
    \end{equation}
    the right-hand side equals $\bra{\varphi_{t,f_x}} H_x \ket{\varphi_{t,f_x}}$, so
    \begin{equation}
        \Lambda_{C_{t,f_x}}(x) \leq \left\|H_x^{1/2} \ket{\varphi_{t,f_x}}\right\|_2^2.
    \end{equation}

    Finally, we bound the sum of these norms.
    Summing the bound above over $x$ and taking square roots, Eq.~\eqref{eq:Dt_def} gives
    \begin{equation}
        \sqrt{D_t} \leq \sqrt{\sum_x \left\|H_x^{1/2} \ket{\varphi_{t,f_x}}\right\|_2^2 }.
    \end{equation}
    For each $x$, we decompose
    \begin{equation}
        H_x^{1/2}\ket{\varphi_{t,f_x}} = H_x^{1/2}\ket{\varphi_{t,I}} + H_x^{1/2}\left(\ket{\varphi_{t,f_x}} - \ket{\varphi_{t,I}}\right),
    \end{equation}
    and abbreviate the norms of the two terms on the right as $a_x := \|H_x^{1/2}\ket{\varphi_{t,I}}\|_2$ and $b_x := \|H_x^{1/2}(\ket{\varphi_{t,f_x}} - \ket{\varphi_{t,I}})\|_2$.
    The triangle inequality gives $\|H_x^{1/2}\ket{\varphi_{t,f_x}}\|_2 \leq a_x + b_x$ for each $x$; expanding $\sum_x (a_x + b_x)^2$ and applying the Cauchy--Schwarz inequality $\sum_x a_x b_x \leq \sqrt{\sum_x a_x^2}\sqrt{\sum_x b_x^2}$ to the cross term, we get $\sum_x (a_x + b_x)^2 \leq \left(\sqrt{\sum_x a_x^2} + \sqrt{\sum_x b_x^2}\right)^2$.
    Hence
    \begin{equation}
        \sum_x \left\|H_x^{1/2} \ket{\varphi_{t,f_x}}\right\|_2^2
        \leq \left( \sqrt{\sum_x \left\|H_x^{1/2} \ket{\varphi_{t,I}}\right\|_2^2 } + \sqrt{\sum_x \left\|H_x^{1/2} \left(\ket{\varphi_{t,f_x}} - \ket{\varphi_{t,I}}\right)\right\|_2^2} \right)^2.
    \end{equation}
    We bound the two sums under the square roots separately.

    For the first sum, the operators $H_x$ sum to a multiple of the identity,
    \begin{equation}
        \sum_x H_x
        = \sum_{z} \frac{\sum_x \left(v(x)^{\top} v(z)\right)^2}{M_r^2} \Pi_z
        = \sum_{z} \frac{v(z)^{\top}\left(\sum_x v(x)v(x)^{\top}\right)v(z)}{M_r^2} \Pi_z
        = \frac{N}{M_r} \sum_z \Pi_z
        = \frac{N}{M_r}\, I,
    \end{equation}
    where we used $\sum_x v(x)v(x)^{\top} = NI$ (Eq.~\eqref{eq:v_resolution}), $v(z)^{\top} v(z)=M_r$, and $\sum_z \Pi_z = I$.
    Hence, since $\ket{\varphi_{t,I}}$ is normalized,
    \begin{equation}\label{eq:Dt_first_term}
        \sum_x \left\|H_x^{1/2} \ket{\varphi_{t,I}}\right\|_2^2
        = \bra{\varphi_{t,I}} \Big(\sum_x H_x\Big) \ket{\varphi_{t,I}}
        = \frac{N}{M_r}.
    \end{equation}

    For the second sum, we use the hybrid argument underlying the optimality of Grover search~\cite{Bennett1997}.
    Since $(v(x)^{\top} v(z))^2 \leq M_r^2$, we have $H_x \preceq I$, hence
    \begin{equation}
        \sum_x \left\|H_x^{1/2} \left(\ket{\varphi_{t,f_x}} - \ket{\varphi_{t,I}}\right)\right\|_2^2
        \leq  \sum_x \left\|\ket{\varphi_{t,f_x}} - \ket{\varphi_{t,I}}\right\|_2^2.
    \end{equation}
    The two branches differ only in the first $t-1$ queries.
    Replacing the queries by the identity one at a time, the difference becomes a sum of $t-1$ terms, the $s$-th of which changes only the $s$-th query; everything applied after that query preserves the norm, so the triangle inequality gives
    \begin{equation}
        \left\|\ket{\varphi_{t,f_x}} - \ket{\varphi_{t,I}}\right\|_2
        \leq \sum_{s=1}^{t-1} \left\| (O_{f_x}-I) \ket{\varphi_{s,I}}\right\|_2
        \leq \sqrt{(t-1) \sum_{s=1}^{t-1} \left\| (O_{f_x}-I) \ket{\varphi_{s,I}}\right\|_2^2},
    \end{equation}
    where the second inequality is the Cauchy--Schwarz inequality.
    Since $O_{f_x} - I = -2\,\Pi_x \otimes \Pi_1$ and $\sum_x \Pi_x\otimes\Pi_1 \preceq I$,
    \begin{equation}
        \sum_{x} \left\| (O_{f_x}-I) \ket{\varphi_{s,I}}\right\|_2^2
        = 4 \sum_x \bra{\varphi_{s,I}} \Pi_x\otimes \Pi_1 \ket{\varphi_{s,I}} \leq 4.
    \end{equation}
    Hence
    \begin{equation}\label{eq:Dt_second_term}\begin{split}
        \sum_x \left\|\ket{\varphi_{t,f_x}} - \ket{\varphi_{t,I}}\right\|_2^2
        \leq (t-1)\sum_{s=1}^{t-1}\sum_x \left\| (O_{f_x}-I) \ket{\varphi_{s,I}}\right\|_2^2 \leq 4(t-1)^2.
    \end{split}\end{equation}

    Combining Eq.~\eqref{eq:Dt_first_term} and Eq.~\eqref{eq:Dt_second_term}, $\sqrt{D_t} \leq \sqrt{N/M_r} + 2(t-1)$, which is the claimed bound.
\end{proof}

Combining Lemmas~\ref{lem:pair_test_final}, \ref{lem:avg_progress_change}, and \ref{lem:Dt_growth}, we now prove a lower bound on the query complexity of OD protocols for Grover oracles.
We restate the theorem from the main text.
\odgroverlb*
\begin{proof}[Proof of Theorem~\ref{thm:OD_grover_lower_bound}]
    The endpoints of the average progress satisfy $\bar{W}(1) = 1$ and $\bar{W}(T_{\mathrm{OD}}+1) \leq \sqrt{2\epsilon - \epsilon^2}$.
    By the triangle inequality and Lemma~\ref{lem:avg_progress_change},
    \begin{equation}
        1 - \sqrt{2\epsilon - \epsilon^2}
        \leq \left| \bar{W}(1) - \bar{W}(T_{\mathrm{OD}}+1)\right|
        \leq \sum_{t=1}^{T_{\mathrm{OD}}} \left| \bar{W}(t+1) - \bar{W}(t)\right|
        \leq \frac{4}{\sqrt{M_r(N-1)}} \sum_{t=1}^{T_{\mathrm{OD}}} \sqrt{D_t}.
    \end{equation}
    Lemma~\ref{lem:Dt_growth} gives $\sqrt{D_t} \leq \sqrt{N/M_r} + 2(t-1)$, so
    \begin{equation}
        \sum_{t=1}^{T_{\mathrm{OD}}} \sqrt{D_t}
        \leq T_{\mathrm{OD}}\sqrt{\frac{N}{M_r}} + T_{\mathrm{OD}}(T_{\mathrm{OD}}-1)
        \leq T_{\mathrm{OD}}\sqrt{\frac{N}{M_r}} + T_{\mathrm{OD}}^2.
    \end{equation}
    Substituting the second bound into the first,
    \begin{equation}
        1 - \sqrt{2\epsilon - \epsilon^2}
        \leq \frac{4T_{\mathrm{OD}}}{M_r}\sqrt{\frac{N}{N-1}} + \frac{4T_{\mathrm{OD}}^2}{\sqrt{M_r(N-1)}}.
    \end{equation}
    At least one of the two terms on the right-hand side is at least $\left(1 - \sqrt{2\epsilon - \epsilon^2}\right)/2$, and solving for $T_{\mathrm{OD}}$ in each case gives
    \begin{equation}\label{eq:OD_grover_explicit}
        T_{\mathrm{OD}} \geq \min\left\{\frac{(1-\sqrt{2\epsilon - \epsilon^2})\, M_r}{8}\sqrt{\frac{N-1}{N}},\ \sqrt{\frac{1-\sqrt{2\epsilon - \epsilon^2}}{8}} \left[M_r(N-1)\right]^{1/4}\right\}.
    \end{equation}
    Since $\epsilon \leq 1/2$, the prefactor satisfies $1-\sqrt{2\epsilon - \epsilon^2} \geq 1-\sqrt{3}/2$, so Eq.~\eqref{eq:OD_grover_explicit} implies Eq.~\eqref{eq:OD_grover_lower_bound}.
\end{proof}

\begin{corof}{thm:OD_grover_lower_bound}{1}[Query complexity lower bound for distilling Grover oracles while correcting $r = \lfloor \alpha n \rfloor$ errors]\label{cor:OD_grover_entropy_rate}
    Any OD protocol as in Theorem~\ref{thm:OD_grover_lower_bound}, with error weight $r = \lfloor \alpha n \rfloor$ for a constant $\alpha \in (0, 0.061]$, must satisfy
    \begin{equation}\label{eq:OD_grover_entropy_rate}
        T_{\mathrm{OD}} = \tilde{\Omega}\left(N^{H(\alpha)}\right).
    \end{equation}
\end{corof}
\begin{proof}
    For $r = \lfloor \alpha n \rfloor$ with constant $\alpha \in (0, 1/2)$, the third case of Eq.~\eqref{eq:Mr_asymptotics} in Lemma~\ref{lem:Mr_asymptotics} gives $M_r = \Theta\!\left(2^{nH(\alpha)}/\sqrt{n}\right) = \Theta\!\left(N^{H(\alpha)}/\sqrt{\log_2 N}\right)$.
    The first branch of Eq.~\eqref{eq:OD_grover_lower_bound} is then $\tilde{\Omega}(N^{H(\alpha)})$, and the second branch is $\tilde{\Omega}(N^{(1+H(\alpha))/4})$.
    Since $\alpha \leq 0.061$ implies $H(\alpha) \leq 1/3$, which is equivalent to $H(\alpha) \leq (1+H(\alpha))/4$, the minimum in Eq.~\eqref{eq:OD_grover_lower_bound} is attained by the first branch, which gives the claim.
\end{proof}

\subsection{Lower bound for distilling arbitrary families of Boolean oracles}\label{SM_sec:label_indep_lb}
In this subsection we prove the second lower bound, for OD protocols that distill an arbitrary family $\mathfrak{O}_{F_0} = \{\mathcal{O}_f\}_{f\in F_0}$, $F_0 \subseteq F$, of ideal Boolean oracles into itself, and whose pre-query subspaces satisfy the KLC~\eqref{eq:KLC} with a label-independent KL matrix, $C_{t,f} = C_t$ for all $f \in F_0$.

The proof follows the same strategy as the previous subsection.
We again study the average behavior of pair tests, but now weight each ordered pair $(f,g)$ of distinct functions $f, g \in F_0$ by $\mu_{(f,g)}$, and define the weighted average progress
\begin{equation}
    \bar{W}_{\mu}(t) := \sum_{(f,g)} \mu_{(f,g)} W_{(f,g)}(t),
\end{equation}
where $W_{(f,g)}(t) := \Re \braket{\psi^t_{f\mid(f,g)} | \psi^t_{g\mid(f,g)}}$ is the progress of the pair test between $O_f$ and $O_g$, with the states defined in Eq.~\eqref{eq:pair_test_states} and an arbitrary fixed choice of the bitstring $x_0$ with $f(x_0) \neq g(x_0)$.
The weighted average over pairs is reminiscent of the weighted adversary method \cite{Ambainis2002, Hoyer2007}, but our argument is not an instance of it, since each pair test runs the protocol on its own initial state $\ket{x_0}\ket{+}^{\otimes m}$ chosen for that pair, whereas the adversary method compares executions of one algorithm on a single fixed initial state.

We restate the theorem from the main text.
\odlabelindep*
\begin{proof}
    The proof has two steps.
    We first pin down the weighted average progress at the start and at the end.
    We then bound its change at each query.

    Before the first query the two executions of every pair test coincide, so $W_{(f,g)}(1) = 1$ for every pair, and the normalization of $\mu$ gives $\bar{W}_\mu(1) = 1$.
    After all $T_{\mathrm{OD}}$ queries, Lemma~\ref{lem:pair_test_final} gives $W_{(f,g)}(T_{\mathrm{OD}}+1) \leq \sqrt{2\epsilon - \epsilon^2}$ for every pair, hence $\bar{W}_\mu(T_{\mathrm{OD}}+1) \leq \sqrt{2\epsilon - \epsilon^2}$.

    We now bound the change caused by the $t$-th query.
    The interleaved unitaries preserve each inner product, so only the query itself changes the overlaps, and the triangle inequality gives
    \begin{equation}
        \left|\bar{W}_\mu(t+1) - \bar{W}_\mu(t)\right|
        \leq \sum_{(f,g)} \mu_{(f,g)} \left|W_{(f,g)}(t+1) - W_{(f,g)}(t)\right|.
    \end{equation}
    The two executions of the pair test $(f,g)$ lie in the ranges of $P_{t,f}$ and $P_{t,g}$, whose KL matrices are both $C_t$ by assumption, so Lemma~\ref{lem:single_query} gives
    \begin{equation}
        \left|W_{(f,g)}(t+1) - W_{(f,g)}(t)\right| \leq 2\sum_{z:\, f(z) \neq g(z)} \Lambda_{C_t}(z).
    \end{equation}
    Multiplying by the weights and summing over the pairs, the definition of $l_z$ in Eq.~\eqref{eq:lz_def} gives
    \begin{equation}
    \begin{split}
        \left|\bar{W}_\mu(t+1) - \bar{W}_\mu(t)\right|
        &\leq 2\sum_{(f,g)} \mu_{(f,g)} \sum_{z:\, f(z) \neq g(z)} \Lambda_{C_t}(z)
        = 2\sum_{z} l_z \Lambda_{C_t}(z) \\
        &\leq 2 \left(\max_{z'} l_{z'}\right) \sum_z \Lambda_{C_t}(z)
        \leq \frac{2N}{M_r} \max_{z} l_{z},
    \end{split}
    \end{equation}
    where the last inequality is Lemma~\ref{lem:total_qp}.
    Combining the two endpoints with the per-query bound through the triangle inequality,
    \begin{equation}
        1- \sqrt{2\epsilon - \epsilon^2}
        \leq \left|\bar{W}_\mu(T_{\mathrm{OD}}+1) - \bar{W}_\mu(1)\right|
        \leq \sum_{t=1}^{T_{\mathrm{OD}}} \left|\bar{W}_\mu(t+1) - \bar{W}_\mu(t)\right|
        \leq \frac{2NT_{\mathrm{OD}}}{M_r}\max_{z} l_{z},
    \end{equation}
    which rearranges to $T_{\mathrm{OD}} \geq \left(1-\sqrt{2\epsilon - \epsilon^2}\right) M_r / (2N \max_z l_z)$.
    For $\epsilon \leq 1/2$ the prefactor is at least $1-\sqrt{3}/2$, which gives Eq.~\eqref{eq:OD_label_indep_bound}.
\end{proof}

Theorem~\ref{thm:OD_label_indep} holds for every choice of the weights $\mu$, and we now instantiate it on three families, the Grover oracles, the fixed-period Simon oracles, and the balanced oracles.
The first corollary formalizes the instance stated after Theorem~\ref{thm:OD_label_indep} in the main text.
The other two show that the bound is not tied to the Grover family, covering a family with a structured promise and a family of size $2^{\Theta(N)}$.
In each case we choose $\mu$ uniform over pairs that disagree on at most four inputs, spread so that $l_z = O(1/N)$ for every input $z$, and Eq.~\eqref{eq:OD_label_indep_bound} then gives $T_{\mathrm{OD}} = \Omega(M_r)$ at every error weight $r$.
The Grover and balanced oracles have a single response qubit, and the corresponding corollaries also apply to protocols whose oracles have $m > 1$ response qubits, because a Boolean oracle can be viewed as an oracle with $m$ response qubits that acts trivially on the last $m - 1$ of them.

\begin{corof}{thm:OD_label_indep}{1}[Query complexity lower bound for distilling Grover oracles with label-independent KL matrices]\label{cor:OD_label_indep_grover}
    Any OD protocol as in Theorem~\ref{thm:OD_label_indep} that distills the family of ideal Grover oracles $\mathfrak{O}_{F_G}$ into itself has query complexity $T_{\mathrm{OD}} = \Omega(M_r)$.
\end{corof}
\begin{proof}
    Let $\mu$ be uniform over the $N(N-1)$ ordered pairs of distinct functions in $F_G$.
    Two Grover functions with marked elements $x \neq y$ disagree exactly on the two inputs $x$ and $y$.
    For a fixed input $z$, the pairs that disagree on $z$ are those with $x = z$ or $y = z$, and there are $2(N-1)$ of them, so $l_z = 2/N$ for every $z$.
    Eq.~\eqref{eq:OD_label_indep_bound} then gives $T_{\mathrm{OD}} = \Omega(M_r)$.
\end{proof}

\begin{corof}{thm:OD_label_indep}{2}[Query complexity lower bound for distilling fixed-period Simon oracles with label-independent KL matrices]\label{cor:OD_label_indep_simon}
    For any nonzero $s \in \{0,1\}^n$, let $F^{\mathrm{Sim}}_s$ be the set of functions $f: \{0,1\}^n \rightarrow \{0,1\}^n$ that satisfy the promise of Simon's problem \cite{Simon1997} with period $s$, namely $f(x) = f(y)$ if and only if $x \oplus y \in \{0^n, s\}$.
    Any OD protocol as in Theorem~\ref{thm:OD_label_indep} that distills $\mathfrak{O}_{F^{\mathrm{Sim}}_s}$ into itself, with oracles carrying $m = n$ response qubits, has query complexity $T_{\mathrm{OD}} = \Omega(M_r)$.
\end{corof}
\begin{proof}
    Each $f \in F^{\mathrm{Sim}}_s$ is constant on the $N/2$ cosets of the subgroup $\{0^n, s\}$ and takes distinct values on distinct cosets, so its image contains exactly $N/2$ strings.
    We define the weights $\mu$ by drawing $f$ uniformly from $F^{\mathrm{Sim}}_s$ and then drawing $g$ from a distribution conditioned on $f$.
    Namely, we draw an unordered pair $\{a, b\}$ of distinct strings in the image of $f$ uniformly, and set $g := \tau_{ab} \circ f$, where $\tau_{ab}$ swaps the outputs $a$ and $b$.
    The function $g$ again lies in $F^{\mathrm{Sim}}_s$, and $f$ and $g$ disagree exactly on the four inputs of the two cosets $f^{-1}(a)$ and $f^{-1}(b)$.
    For a fixed input $z$, the pair $(f, g)$ disagrees on $z$ if and only if $f(z) \in \{a, b\}$, which has probability $2/(N/2) = 4/N$.
    Hence $l_z = 4/N$ for every $z$, and Eq.~\eqref{eq:OD_label_indep_bound} gives $T_{\mathrm{OD}} = \Omega(M_r)$.
\end{proof}

\begin{corof}{thm:OD_label_indep}{3}[Query complexity lower bound for distilling balanced oracles with label-independent KL matrices]\label{cor:OD_label_indep_balanced}
    Let $F_B \subset F$ be the set of balanced functions, those with $|f^{-1}(1)| = N/2$.
    Any OD protocol as in Theorem~\ref{thm:OD_label_indep} that distills $\mathfrak{O}_{F_B}$ into itself has query complexity $T_{\mathrm{OD}} = \Omega(M_r)$.
\end{corof}
\begin{proof}
    We define the weights $\mu$ by drawing $f$ uniformly from $F_B$ and then drawing $g$ from a distribution conditioned on $f$.
    Namely, we draw an input $x$ uniformly from $f^{-1}(1)$ and an input $y$ uniformly from $f^{-1}(0)$, and let $g$ agree with $f$ everywhere except that $g(x) = 0$ and $g(y) = 1$.
    The function $g$ is again balanced, and $f$ and $g$ disagree exactly on the two inputs $x$ and $y$.
    For a fixed input $z$, conditioned on $f$, the input $x$ equals $z$ with probability $2/N$ if $f(z) = 1$ and zero otherwise, and $f(z) = 1$ with probability $1/2$ under the uniform $f$, so $\Pr[x = z] = 1/N$, and likewise $\Pr[y = z] = 1/N$.
    Hence $l_z = 2/N$ for every $z$, and Eq.~\eqref{eq:OD_label_indep_bound} gives $T_{\mathrm{OD}} = \Omega(M_r)$.
\end{proof}

\subsection{Smallest useful target precision}\label{SM_sec:useful_precision}
In this subsection we justify the claim of Sec.~\ref{sec:optimality} that, for any number $m$ of response qubits, an application with constant success probability never benefits from a target precision below $\epsilon = \Theta(1/N)$.
The reason is a simple algorithm that learns the label of the noisy oracle and then synthesizes the ideal oracle exactly, so any application demanding such a precision is served more cheaply by learning the label than by distilling to that precision.

The algorithm queries the distilled oracle once on each input $x \in \{0,1\}^n$ and reads out all $m$ response bits of $x$ in that single query.
Suppose we distill each query to precision $\epsilon_1 = \delta/N$ for an arbitrary constant $\delta \in (0,1)$, so each query implements a channel $\widehat{\mathcal{O}}_f$ with $\|\widehat{\mathcal{O}}_f - \mathcal{O}_f\|_\diamond \leq \epsilon_1$.
For each $x \in \{0,1\}^n$, we apply one distilled query to the state $\ket{x}\ket{+}^{\otimes m}$ and measure the $m$ response qubits in the $X$ basis.
The ideal oracle maps
\begin{equation}
    O_f \ket{x}\ket{+}^{\otimes m} = \ket{x}\bigotimes_{j=1}^m
    \begin{cases}
        \ket{+}, & f_j(x)=0,\\
        \ket{-}, & f_j(x)=1,
    \end{cases}
\end{equation}
so under the ideal oracle the measurement returns the response bits $f_1(x), \ldots, f_m(x)$ with certainty.
The run consists of $N$ distilled queries and records one $m$-bit string for each input $x$.
Each query is $\epsilon_1$-close in diamond norm, so the hybrid argument (Lemma~\ref{lem:hybrid_argument}) bounds the deviation of the joint distribution of these $N$ strings from that of the ideal run by $N \epsilon_1 = \delta$ in total variation.
This accounting only adds the diamond norm errors of the queries, so it needs no independence between the queries and holds for adversarial noise.
Hence with probability at least $1 - \delta$ the run returns the full truth table of $f$, which identifies the label.
Once the label is known, the ideal oracle $O_f$ is a known unitary and can be synthesized exactly, with no further queries to the noisy oracle.
The application then runs on the oracle synthesized from the recorded strings, so its output distribution is a function of the joint distribution of these strings and is within total variation distance $\delta$ of the output distribution of the application run on ideal queries.

We now compare the two routes, distilling every query of the application directly, and learning the label as above, both within total variation distance $\delta$ of the application run on ideal queries, so both lose at most $\delta$ in success probability, which stays constant for a small enough constant $\delta$.
Under adversarial noise by Theorem~\ref{thm:boolean_OD_adv}, and under i.i.d.\ depolarizing noise by Theorem~\ref{thm:iid_threshold_at_exponent}, one distilled query at precision $\epsilon$ costs $T_{\mathrm{OD}}(\epsilon) = 2 \lceil A \ln (B/\epsilon) \rceil$ noisy queries, where $A > 0$ and $B \geq 4$ do not depend on $\epsilon$, so $T_{\mathrm{OD}}(\epsilon)$ is positive and never decreases when $\epsilon$ decreases.
Since the diamond norm errors of the distilled queries add up over the run, an application that makes $T_Q$ distilled queries stays within total variation distance $\delta$ with precision $\epsilon = \delta/T_Q$ per query, so the direct route costs $T_Q\, T_{\mathrm{OD}}(\delta/T_Q)$ noisy queries.
The learning route makes $N$ distilled queries at precision $\epsilon_1 = \delta/N$ and serves every later query by synthesis, so it costs $N\, T_{\mathrm{OD}}(\delta/N)$ noisy queries regardless of $T_Q$.
If $T_Q \leq N$, the direct route already runs at a precision $\delta/T_Q \geq \delta/N$.
If $T_Q > N$, then $\delta/T_Q < \delta/N$ gives $T_{\mathrm{OD}}(\delta/T_Q) \geq T_{\mathrm{OD}}(\delta/N) > 0$, so
\begin{equation}
    N\, T_{\mathrm{OD}}(\delta/N) < T_Q\, T_{\mathrm{OD}}(\delta/T_Q) ,
\end{equation}
and the learning route is strictly cheaper than the direct route.
In both cases no precision below $\delta/N$ is needed, so for constant $\delta$ a useful target precision always satisfies $\epsilon = \Omega(1/N)$, and $\ln \epsilon^{-1} \leq n \ln 2 + \ln \delta^{-1} = O(n)$.

\clearpage
\section{A Threshold Theorem for Boolean Oracle Problems}\label{SM_sec:threshold}

This section proves the threshold theorem for Boolean oracle problems (Theorem~\ref{thm:threshold_formal} of the main text).
We first define Boolean oracle problems and review their quantum and classical query complexities.
We then prove Lemma~\ref{lem:distilled_simulation}, which bounds the query complexity of running any quantum algorithm on distilled queries with bounded failure probability, and derive Theorem~\ref{thm:threshold_formal} from it.
The remaining subsections supplement the theorem.
Section~\ref{SM_sec:noisy_classical} shows that oracle noise can only increase the classical query complexity, so the comparison against noiseless classical queries in Theorem~\ref{thm:threshold_formal} is conservative.
Section~\ref{SM_sec:multi_oracle} packs problems specified by several functions, such as $k$-forrelation and claw finding, into the single-oracle framework.
Section~\ref{SM_sec:forrelation_threshold} proves the claim of Sec.~\ref{subsec:forrelation_example} that the threshold of the $k$-forrelation problem tends to $3/4$ as $k \to \infty$.
Section~\ref{SM_sec:simon_lsn} proves the claim of Sec.~\ref{subsec:simon_example} that when Simon's algorithm is run with the i.i.d.-depolarizing oracle, the measured bitstrings are distributed exactly as in the learning Simon with noise (LSN) problem.

\subsection{Boolean oracle problems}\label{SM_sec:boolean_oracle_problems}

We work in the standard quantum query model~\cite{Beals2001, Buhrman2002, Ambainis2002}, which we now state in the form used throughout this section.
A Boolean oracle problem is specified, for each input size $n$, by a set $\mathcal{X}_n$ of instances, a set $\mathcal{Y}_n$ of allowed outputs, and a relation $\mathcal{R}_n \subset \mathcal{X}_n \times \mathcal{Y}_n$ that collects the correct outputs for each instance.
An instance is a Boolean function $f: \{0,1\}^n \rightarrow \{0,1\}^m$ promised to lie in $\mathcal{X}_n$, where $m$ may depend on $n$.

An algorithm accesses the instance only through queries to the oracle.
Quantum query access to $f$ means access to the Boolean oracle $O_f\ket{x}\ket{y} = \ket{x}Z^{f(x)}\ket{y}$, whose unitary channel is $\mathcal{O}_f$.
Classical query access to $f$ means access to $\Delta \circ \mathcal{O}_f \circ \Delta$, where $\Delta$ denotes the completely dephasing channel acting in the computational basis on the index register and in the $X$ basis on the response register.
An algorithm interleaves its queries with arbitrary unitaries that do not depend on $f$, and measures at the end to produce an output $y \in \mathcal{Y}_n$.

The algorithm solves the problem with success probability $p_s$ if, for every $f \in \mathcal{X}_n$, its output satisfies $(f, y) \in \mathcal{R}_n$ with probability at least $p_s$, and solves the problem if it does so for some constant $p_s > 1/2$.
The two access types enter our results asymmetrically, so we define $T_C$ and $T_Q$ differently.
The classical query complexity $T_C$ of the problem at success probability $p^s_C$ is the minimum number of queries made by any algorithm with classical query access that solves it with success probability $p^s_C$.
The minimum ranges over algorithms with arbitrarily many ancilla qubits and arbitrary oracle-independent gates, so solving the problem with success probability $p^s_C$ using fewer than $T_C$ queries beats every such classical algorithm.
For quantum query access, we fix one quantum algorithm that solves the problem with success probability $p^s_Q$, and let $T_Q$ denote its number of queries.
Whenever $p^s_Q > p^s_C$, the loss $\delta$ in Theorem~\ref{thm:threshold_formal} can be chosen below $p^s_Q - p^s_C$, so the noisy quantum algorithm still succeeds with probability at least $p^s_C$ and the comparison with $T_C$ is made at a success probability no lower than the classical one.

The complexities $T_Q$ and $T_C$ above refer to noiseless oracles, and the theorem below shows that when $T_Q$ is far enough below $T_C$, the quantum advantage survives even when every quantum query suffers i.i.d.\ depolarizing noise.
As in the main text, the \textit{advantage budget} of the problem is the ratio $T_C/T_Q$, and its \textit{budget exponent} is
\begin{equation}\label{eq:budget_exponent}
    c := \liminf_{n\rightarrow \infty} \frac{1}{n} \log_2 \frac{T_C}{T_Q},
\end{equation}
and we write $q := \limsup_{n\rightarrow\infty} \frac{1}{n} \log_2 T_Q$ for the exponent of the quantum query complexity.

\subsection{Solving oracle problems with distilled queries}\label{SM_sec:solving_distilled}

The following lemma uses oracle distillation to convert any quantum algorithm querying the ideal oracle into one querying the i.i.d.-depolarizing oracle, at the cost of a larger query complexity and a small loss in success probability.

\begin{lemma}[Running a quantum algorithm on distilled queries]\label{lem:distilled_simulation}
    Consider a Boolean oracle problem with $\log_2 m = o(n)$, and a quantum algorithm that solves it with success probability $p^s_Q$ using $T_Q$ queries with finite exponent $q$.
    If the i.i.d.-depolarizing oracle has a constant per-qubit error rate $p$ below the distillation threshold $p_{\mathrm{th}}(\gamma, q)$ of Eq.~\eqref{eq:iid_pth_at_exponent} for some constant $\gamma \in (0, 1)$, then for every constant $\delta \in (0, p^s_Q)$ and all sufficiently large $n$ the problem can be solved with success probability $p^s_Q - \delta$ using
    \begin{equation}\label{eq:distilled_simulation_cost}
        \tilde{T}_Q = 2\, T_Q \left\lceil \frac{16 \left( 3 - 2 p_t \right)}{\left( 1 - 2 p_t \right)^2} \, N^{\gamma} \, \ln \frac{12\, m T_Q}{\delta} \right\rceil ,
        \qquad p_t = 2p/3 ,
    \end{equation}
    queries to this oracle.
\end{lemma}
\begin{proof}
    We run the given algorithm on distilled queries of precision $\epsilon := \delta/T_Q$, that is, with every ideal query replaced by a distilled query.
    We first bound the cost of one distilled query using Theorem~\ref{thm:iid_threshold_at_exponent}.
    The precision lies in $(0,1)$ because $\delta < p^s_Q \leq 1 \leq T_Q$, and its exponent is $\nu = \limsup_{n\to\infty} \frac{1}{n}\log_2\frac{1}{\epsilon} = q$ because $\delta$ is a constant.
    Together with the assumptions $\log_2 m = o(n)$ and $p < p_{\mathrm{th}}(\gamma, q)$ of the lemma, the theorem applies, and substituting $\epsilon = \delta/T_Q$ into Eq.~\eqref{eq:T_OD_at_exponent} gives, for all sufficiently large $n$, the query complexity
    \begin{equation}
        T_{\mathrm{OD}} = 2 \left\lceil \frac{16 \left( 3 - 2 p_t \right)}{\left( 1 - 2 p_t \right)^2} \, N^{\gamma} \, \ln \frac{12\, m T_Q}{\delta} \right\rceil
    \end{equation}
    of one distilled query.

    It remains to count the noisy queries and bound the loss in success probability.
    The algorithm run on distilled queries makes $T_Q$ distilled queries, each built from $T_{\mathrm{OD}}$ noisy queries, so it makes $\tilde{T}_Q = T_Q T_{\mathrm{OD}}$ noisy queries in total, which is Eq.~\eqref{eq:distilled_simulation_cost}.
    Each distilled query is $\epsilon$-close in diamond norm to the ideal query, so the hybrid argument (Lemma~\ref{lem:hybrid_argument}) bounds the total variation distance between the output distributions of the algorithm run on distilled queries and the algorithm run on ideal queries by $T_Q \epsilon = \delta$.
    For every $f \in \mathcal{X}_n$, the algorithm run on ideal queries outputs a $y$ with $(f,y) \in \mathcal{R}_n$ with probability at least $p^s_Q$, so the algorithm run on distilled queries does so with probability at least $p^s_Q - \delta$.
\end{proof}

We now compare the algorithm run on distilled queries against every classical algorithm.
It has smaller query complexity than every classical algorithm as long as the error rate $p$ is below a constant threshold, so quantum queries beat noiseless classical queries even when they are noisy.
The threshold depends on the problem only through the budget exponent $c$ and the exponent $q$ of $T_Q$.
We first record a technical property of the threshold $p_{\mathrm{th}}$ in Lemma~\ref{lem:pth_smaller_exponent}, and then prove Theorem~\ref{thm:threshold_formal}, which states the threshold as $p_{\mathrm{th}}(c, q)$.

\begin{lemma}[Constant error rate stays below the threshold at a smaller overhead exponent]\label{lem:pth_smaller_exponent}
    For any constant overhead exponent $\gamma_0 \in (0, 1)$, any decay rate $\nu \in [0, \infty)$, and any constant error rate $p < p_{\mathrm{th}}(\gamma_0, \nu)$, there is a constant overhead exponent $\gamma \in (0, \gamma_0)$ with $p < p_{\mathrm{th}}(\gamma, \nu)$.
\end{lemma}
\begin{proof}
If $p = 0$, any $\gamma \in (0, \gamma_0)$ works because $p_{\mathrm{th}}(\gamma, \nu)$ is positive, so let $p > 0$.
Since $p < p_{\mathrm{th}}(\gamma_0, \nu)$, the error rate lies below at least one of the two entries of the maximum in Eq.~\eqref{eq:iid_pth_at_exponent} at $\gamma_0$, and we fix the index $i \in \{3, 2\}$ of that entry for the rest of the proof.
For this $i$ we write $I$ for the range of $\alpha$ in the definition of $\alpha^{(i)}$, $h(\alpha)$ for the exponent that this definition bounds by $\gamma$, $\alpha_{\mathrm{seq}}(\alpha)$ for the second argument of $p_{\mathrm{th}}^{(i)}$ in Eq.~\eqref{eq:iid_pth_at_exponent}, and $B(\alpha)$ for the constant $B$ of the corollary that defines $p_{\mathrm{th}}^{(i)}$.
That is, for $i = 3$ we have $I = (0, 0.16]$, $h(\alpha) = H(\alpha) + 2\alpha$, $\alpha_{\mathrm{seq}}(\alpha) = \frac{1}{2} - \alpha$ and $B(\alpha) = 6\alpha + \alpha^2 + H(2\alpha)$, and for $i = 2$ we have $I = (0, 1/4)$, $h(\alpha) = H(2\alpha)$, $\alpha_{\mathrm{seq}}(\alpha) = 1$ and $B(\alpha) = 2 H(2\alpha)$.
In both cases $h$ is positive, continuous and strictly increasing on $I$, $B$ is continuous on $I$, $\alpha_{\mathrm{seq}}$ is continuous and nonincreasing on $I$, and for every $\alpha \in I$ the constants $\alpha$, $\alpha_{\mathrm{seq}}(\alpha)$, $h(\alpha)$, $B(\alpha)$ and $\nu$ satisfy condition (i) of Lemma~\ref{lem:threshold_equation} for $i = 3$ and condition (ii) for $i = 2$, the latter because $\alpha < 1/4 \leq 1/2$ and $h(\alpha) + \nu - \left( B(\alpha) - 1 \right) = 1 - H(2\alpha) + \nu > 0$.
By definition, $\alpha^{(i)}$ is the largest $\alpha \in I$ with $h(\alpha) \leq \gamma$, and entry $i$ of Eq.~\eqref{eq:iid_pth_at_exponent} at $\gamma$ is $p_{\mathrm{th}}^{(i)}\left( \alpha^{(i)}, \alpha_{\mathrm{seq}}(\alpha^{(i)}), \nu \right)$.
Writing $\alpha_0$ for $\alpha^{(i)}$ at $\gamma_0$, the choice of $i$ gives
\begin{equation}\label{eq:pth_entry_at_alpha0}
    p < p_{\mathrm{th}}^{(i)}\left( \alpha_0, \alpha_{\mathrm{seq}}(\alpha_0), \nu \right)
    \qquad\text{and}\qquad
    h(\alpha_0) \leq \gamma_0 .
\end{equation}

We now show that the first inequality in Eq.~\eqref{eq:pth_entry_at_alpha0} still holds when $\alpha_0$ is replaced by a slightly smaller $\alpha \in I$.
Write $p_{\mathrm{eff}}$ for the effective index error rate Eq.~\eqref{eq:iid_peff} at $p$, which lies in $(0, 1/2]$ because $p > 0$ and $p_{\mathrm{eff}} = 2 p_t (1 - p_t) \leq 1/2$, and for $\alpha \in I$ write $E_0(\alpha)$ for the exponent Eq.~\eqref{eq:p_fail_E0} at this $p_{\mathrm{eff}}$ over the interval $\left[ \alpha, \alpha_{\mathrm{seq}}(\alpha) \right]$.
The threshold $p_{\mathrm{th}}^{(i)}\left( \alpha_0, \alpha_{\mathrm{seq}}(\alpha_0), \nu \right)$ is the threshold of Lemma~\ref{lem:threshold_equation} at $\alpha_0$, $\alpha_{\mathrm{seq}}(\alpha_0)$, $h(\alpha_0)$, $B(\alpha_0)$ and $\nu$, so by the characterization Eq.~\eqref{eq:iid_pth_characterization} of that lemma, the first inequality in Eq.~\eqref{eq:pth_entry_at_alpha0} says that $p_{\mathrm{eff}} \leq \alpha_{\mathrm{seq}}(\alpha_0)$ and
\begin{equation}\label{eq:threshold_slack}
    \min\left\{ B(\alpha_0) + E_0(\alpha_0),\; D\left( \alpha_{\mathrm{seq}}(\alpha_0) \,\big\|\, p_{\mathrm{eff}} \right) \right\} > h(\alpha_0) + \nu .
\end{equation}
We fix $\alpha \in I$ with $\alpha < \alpha_0$ close enough to $\alpha_0$ that Eq.~\eqref{eq:threshold_slack} remains true with $\alpha_0$ replaced by $\alpha$, which is possible because every quantity in it is continuous in $\alpha$ on $I$, where for $E_0(\alpha)$ this holds because it is the minimum of a continuous function over the interval $\left[ \alpha, \alpha_{\mathrm{seq}}(\alpha) \right]$ whose endpoints are continuous in $\alpha$.
Since $\alpha_{\mathrm{seq}}$ is nonincreasing, also $p_{\mathrm{eff}} \leq \alpha_{\mathrm{seq}}(\alpha_0) \leq \alpha_{\mathrm{seq}}(\alpha)$, so both conditions of the characterization of Lemma~\ref{lem:threshold_equation} at $\alpha$, $\alpha_{\mathrm{seq}}(\alpha)$, $h(\alpha)$, $B(\alpha)$ and $\nu$ hold, and that characterization gives $p < p_{\mathrm{th}}^{(i)}\left( \alpha, \alpha_{\mathrm{seq}}(\alpha), \nu \right)$.

It remains to identify this entry with the threshold at a smaller overhead exponent.
Set $\gamma := h(\alpha)$, which lies in $(0, \gamma_0)$ because $h$ is positive and $\gamma = h(\alpha) < h(\alpha_0) \leq \gamma_0$ by the strict increase of $h$ and Eq.~\eqref{eq:pth_entry_at_alpha0}.
Again because $h$ is strictly increasing, $\alpha$ is the largest element of $I$ at which $h$ is at most $\gamma$, so $\alpha^{(i)}$ at overhead exponent $\gamma$ is $\alpha$ and entry $i$ of Eq.~\eqref{eq:iid_pth_at_exponent} at $\gamma$ is $p_{\mathrm{th}}^{(i)}\left( \alpha, \alpha_{\mathrm{seq}}(\alpha), \nu \right)$.
Since $p_{\mathrm{th}}(\gamma, \nu)$ is the maximum of the two entries, $p_{\mathrm{th}}(\gamma, \nu) \geq p_{\mathrm{th}}^{(i)}\left( \alpha, \alpha_{\mathrm{seq}}(\alpha), \nu \right) > p$.
\end{proof}

\odthresholdformal*
\begin{proof}
We first check that $p^* = p_{\mathrm{th}}(c, q)$ is well defined.
Making one classical query to each of the $N$ inputs learns $f$ completely and therefore solves any oracle problem, so $T_C \leq N$, and $c > 0$ then gives $T_Q < T_C \leq N$ for all sufficiently large $n$.
Hence $q \leq 1 < \infty$, which together with the hypothesis $0 < c < 1$ shows that $\gamma = c$ and $\nu = q$ satisfy the conditions $\gamma \in (0, 1)$ and $\nu < \infty$ of Theorem~\ref{thm:iid_threshold_at_exponent}, and the theorem guarantees that $p^* = p_{\mathrm{th}}(c, q)$ exists and is a positive constant.

Now fix a constant $p < p^*$.
Lemma~\ref{lem:pth_smaller_exponent} at $\gamma_0 = c$ and $\nu = q$ gives a constant $\gamma \in (0, c)$ with $p < p_{\mathrm{th}}(\gamma, q)$.
Lemma~\ref{lem:distilled_simulation} applies at this $\gamma$, because the problem has $\log_2 m = o(n)$ and the fixed quantum algorithm has the finite exponent $q$, and it solves the problem with success probability $p^s_Q - \delta$ for all sufficiently large $n$ using the query complexity $\tilde{T}_Q$ of Eq.~\eqref{eq:distilled_simulation_cost}.

It remains to verify $\tilde{T}_Q < T_C$, which we do by showing that the exponent of the ratio $\tilde{T}_Q / T_C$ is negative.
The ceiling in Eq.~\eqref{eq:distilled_simulation_cost} encloses a quantity at least one, so dropping it at most doubles the bound, and dividing by $T_C$ and taking $\frac{1}{n}\log_2$ gives
    \begin{equation}
        \frac{1}{n}\log_2 \frac{\tilde{T}_Q}{T_C}
        \leq \gamma
        + \frac{1}{n}\log_2\left( \frac{64 \left( 3 - 2 p_t \right)}{\left( 1 - 2 p_t \right)^2} \ln\frac{12\, m T_Q}{\delta} \right)
        - \frac{1}{n}\log_2\frac{T_C}{T_Q}.
    \end{equation}
The middle term is $o(1)$, because its prefactor is a constant while $T_Q \leq N = 2^n$ and $\log_2 m = o(n)$ give $\ln(12\, m T_Q/\delta) = O(n)$, so the term is at most $\frac{1}{n}\log_2 (C n)$ for a constant $C$, which is $o(1)$.
The last term is at most $-c + o(1)$, because the definition~\eqref{eq:budget_exponent} of $c$ as a liminf gives $\frac{1}{n}\log_2\frac{T_C}{T_Q} \geq c - o(1)$.
The right-hand side is therefore at most $\gamma - c + o(1)$, which is negative for all sufficiently large $n$ because $\gamma < c$ is a constant, so $\tilde{T}_Q < T_C$.
\end{proof}

\subsection{Noisy classical query complexity}\label{SM_sec:noisy_classical}

Theorem~\ref{thm:threshold_formal} counts noisy quantum queries against noiseless classical queries.
This comparison is conservative, because oracle noise can only increase the classical query complexity, as we now show.
A fair comparison requires the noisy classical query to carry the same noise as the noisy quantum query.
We therefore define the noisy classical query complexity $\tilde{T}_C$ as the quantum query complexity of the dephased noisy oracle $\Delta \circ \tilde{\mathcal{O}}_f^{\mathrm{iid}} \circ \Delta$, so the classical query inherits exactly the noise channel of the quantum query.

The argument rests on one fact, that the single-qubit depolarizing channel $\mathcal{D}_p$ commutes with the single-qubit completely dephasing channel in any basis.
The channel $\Delta$ is the tensor product of single-qubit completely dephasing channels, in the computational basis on the index qubits and in the $X$ basis on the response qubits, so $\mathcal{D}_p$ commutes with each of its factors.
Applied to each of the $n+m$ qubits, the noise channel therefore moves past the first dephasing layer,
$\Delta \circ \tilde{\mathcal{O}}_f^{\mathrm{iid}} \circ \Delta = \mathcal{D}_p^{\otimes(n+m)} \circ \Delta \circ \mathcal{O}_f \circ \Delta$,
so a noisy classical query is a noiseless classical query followed by the channel $\mathcal{D}_p^{\otimes(n+m)}$.
This channel does not depend on $f$, so an algorithm with noiseless classical query access can simulate each noisy classical query by one noiseless classical query followed by applying $\mathcal{D}_p^{\otimes(n+m)}$ itself.
Any algorithm solving the problem with $\tilde{T}_C$ noisy classical queries therefore yields one solving it with $\tilde{T}_C$ noiseless classical queries, so $\tilde{T}_C \geq T_C$, and any lower bound on $T_C$ is automatically a lower bound on $\tilde{T}_C$.
The following lemma supplies the commutation fact used above and completes the argument.

\begin{lemma}[Commutation properties of the depolarizing channel]\label{lem:depol_facts}
    The single-qubit depolarizing channel $\mathcal{D}_p$ of Eq.~\eqref{iidnoise} satisfies the following properties.
    \begin{enumerate}
        \item $\mathcal{D}_p$ is unitarily covariant, $\mathcal{D}_p(U \rho U^{\dagger}) = U \mathcal{D}_p(\rho) U^{\dagger}$ for every single-qubit unitary $U$.
        \item $\mathcal{D}_p$ commutes with the single-qubit completely dephasing channel in any basis.
    \end{enumerate}
\end{lemma}
\begin{proof}
Both properties follow from rewriting the depolarizing channel as $\mathcal{D}_p(\rho) = \left(1 - \tfrac{4p}{3}\right) \rho + \tfrac{2p}{3} \Tr[\rho]\, I$.
This form follows from Eq.~\eqref{iidnoise} by substituting the identity $X\rho X + Y\rho Y + Z\rho Z = 2\Tr[\rho]\, I - \rho$, valid for every $2 \times 2$ matrix $\rho$, which gives $\mathcal{D}_p(\rho) = (1-p)\rho + \tfrac{p}{3}\left(2\Tr[\rho]\, I - \rho\right) = \left(1 - \tfrac{4p}{3}\right) \rho + \tfrac{2p}{3} \Tr[\rho]\, I$.
Unitary covariance follows because the trace is invariant under unitary conjugation, so $\mathcal{D}_p(U \rho U^{\dagger}) = \left(1 - \tfrac{4p}{3}\right) U \rho U^{\dagger} + \tfrac{2p}{3} \Tr[\rho]\, I = U \mathcal{D}_p(\rho) U^{\dagger}$, where the last step uses $U I U^{\dagger} = I$.
    For the commutation with dephasing, write $\Lambda$ for the single-qubit completely dephasing channel in an arbitrary basis.
Since $\Lambda$ preserves trace and satisfies $\Lambda(I) = I$, both $\Lambda \circ \mathcal{D}_p$ and $\mathcal{D}_p \circ \Lambda$ map $\rho$ to $\left(1 - \tfrac{4p}{3}\right) \Lambda(\rho) + \tfrac{2p}{3} \Tr[\rho]\, I$, so the two compositions are equal.
\end{proof}

\subsection{Problems with multiple oracle functions}\label{SM_sec:multi_oracle}

Forrelation, $k$-forrelation, and claw finding are specified by several Boolean functions rather than one, while our framework gives query access to a single function.
These problems fit the framework by packing the $k$ functions into the single function $F(j, x) := f_j(x)$ on $n' = n + \lceil \log_2 k \rceil$ bits, where the extra index bits select the function.
One query to $F$ addresses one function, possibly in superposition over $j$, so quantum algorithms and classical lower bounds for the multi-function problem carry over to the packed oracle with the same query complexities.
For constant $k$ we have $n' = O(n)$, so the exponents $q$ and $c$ are unchanged, and the noisy oracle in Theorem~\ref{thm:threshold_formal} is understood as the i.i.d.-depolarizing packed oracle on all $n' + m$ qubits.

\subsection{$k$-forrelation}\label{SM_sec:forrelation_threshold}

The main text (Sec.~\ref{subsec:forrelation_example}) claims that the threshold of the $k$-forrelation problem tends to $3/4$ as $k \to \infty$.
The problem has budget exponent $c = 1 - 1/k$ and quantum query exponent $q = 0$, so its threshold is $p_{\mathrm{th}}\left( 1 - 1/k, 0 \right)$.
The following lemma, applied with $\gamma = 1 - 1/k$, proves the claim.

\begin{lemma}[Distillation threshold at zero precision exponent]\label{lem:pth_nu_zero}
    The distillation threshold Eq.~\eqref{eq:iid_pth_at_exponent} at $\nu = 0$ satisfies $p_{\mathrm{th}}\left( \gamma, 0 \right) \to 3/4$ as $\gamma \to 1$.
\end{lemma}
\begin{proof}
    Fix $\gamma \in (0, 1)$, and set
    \begin{equation}\label{eq:pth_nu_zero_pstar}
        p^* := \frac{3}{4} \left( 1 - \left( 1 - \left( 2^{\gamma} - 1 \right)^2 \right)^{1/4} \right) .
    \end{equation}
    Since $2^{\gamma} - 1 \in (0, 1)$, we have $p^* \in (0, 3/4)$.
    We claim that every $p \in \left( 0, p^* \right)$ satisfies $p < p_{\mathrm{th}}\left( \gamma, 0 \right)$.
    The claim gives $p_{\mathrm{th}}\left( \gamma, 0 \right) \geq p^*$, and $p^* \to 3/4$ as $\gamma \to 1$ because then $2^{\gamma} - 1 \to 1$, so together with $p_{\mathrm{th}}\left( \gamma, 0 \right) < 3/4$ from Lemma~\ref{lem:threshold_equation} the claim proves the lemma.
    Throughout, $p_{\mathrm{eff}} = 2 p_t \left( 1 - p_t \right)$ with $p_t = 2p/3$ is the effective index error rate Eq.~\eqref{eq:iid_peff} of $p$, and $0 < p < 3/4$ gives $0 < p_t < 1/2$ and hence $0 < p_{\mathrm{eff}} < 1/2$.

    To prove the claim, we first rewrite the inequality $p < p_{\mathrm{th}}\left( \gamma, 0 \right)$ as an explicit condition on $p$, and then verify this condition for every $p \in \left( 0, p^* \right)$.
    Since $p_{\mathrm{th}}\left( \gamma, 0 \right) \geq p_{\mathrm{th}}^{(2)}\left( \alpha^{(2)},\, 1,\, 0 \right)$ by Eq.~\eqref{eq:iid_pth_at_exponent}, it suffices that $p < p_{\mathrm{th}}^{(2)}\left( \alpha^{(2)},\, 1,\, 0 \right)$.
    The largest $\alpha \in (0, 1/4)$ with $H(2\alpha) \leq \gamma$ satisfies $H\left( 2 \alpha^{(2)} \right) = \gamma$, because $H(2\alpha)$ is continuous and strictly increasing on $(0, 1/4)$ with limits $0$ and $1$ at the two ends and $\gamma < 1$.
    The constants of Corollary~\ref{cor:iid_boolean_query_complexity_C2} at $\alpha = \alpha^{(2)}$ are therefore $h = \gamma$ and $B = 2\gamma$, with $\alpha_{\mathrm{seq}} = 1$ and $\nu = 0$, so the characterization Eq.~\eqref{eq:iid_pth_characterization} of Lemma~\ref{lem:threshold_equation} states that $p \in (0, 3/4)$ satisfies $p < p_{\mathrm{th}}^{(2)}\left( \alpha^{(2)},\, 1,\, 0 \right)$ exactly when $p_{\mathrm{eff}} \leq 1$ and
    \begin{equation}\label{eq:pth_nu_zero_condition}
        \min\left\{ 2 \gamma + E_0 ,\; D\left( 1 \,\big\|\, p_{\mathrm{eff}} \right) \right\} > \gamma ,
    \end{equation}
    where $E_0$ is the exponent Eq.~\eqref{eq:p_fail_E0} at $p_{\mathrm{eff}}$ over the interval $\left[ \alpha^{(2)}, 1 \right]$.
    Neither the condition $p_{\mathrm{eff}} \leq 1$ nor the second term of the minimum is ever binding, because $p_{\mathrm{eff}} < 1/2$ gives $p_{\mathrm{eff}} \leq 1$ and $D\left( 1 \,\big\|\, p_{\mathrm{eff}} \right) = \log_2 \frac{1}{p_{\mathrm{eff}}} > 1 > \gamma$.

    Hence it remains to verify $2 \gamma + E_0 > \gamma$ for every $p \in \left( 0, p^* \right)$, and for this we bound $E_0$ from below.
    $E_0$ is the minimum of $D\left( \lambda \,\big\|\, p_{\mathrm{eff}} \right) - \bar{H}(\lambda)$ over $\lambda \in \left[ \alpha^{(2)}, 1 \right]$, hence at least its minimum over $\lambda \in [0, 1]$.
    Lemma~\ref{lem:exponent_minimization} with $\lambda_1 = 0$ and $\lambda_2 = 1$ evaluates the latter minimum to $- \log_2 \left( 1 + 2 \sqrt{p_{\mathrm{eff}} \left( 1 - p_{\mathrm{eff}} \right)} \right)$, since its first two cases would require $p_{\mathrm{eff}} \leq 0$ or $p_{\mathrm{eff}} \geq 1$.
    Now let $p \in \left( 0, p^* \right)$.
    Then $1 - 4p/3 > 1 - 4p^*/3 = \left( 1 - \left( 2^{\gamma} - 1 \right)^2 \right)^{1/4} > 0$, so $\left( 1 - 4p/3 \right)^4 > 1 - \left( 2^{\gamma} - 1 \right)^2$.
    Since $1 - 2 p_{\mathrm{eff}} = 1 - 4 p_t + 4 p_t^2 = \left( 1 - 2 p_t \right)^2 = \left( 1 - 4p/3 \right)^2$, we have
    \begin{equation}\label{eq:pth_nu_zero_peff_bound}
        4 p_{\mathrm{eff}} \left( 1 - p_{\mathrm{eff}} \right) = 1 - \left( 1 - 2 p_{\mathrm{eff}} \right)^2 = 1 - \left( 1 - 4p/3 \right)^4 < \left( 2^{\gamma} - 1 \right)^2 ,
    \end{equation}
    so $2 \sqrt{p_{\mathrm{eff}} \left( 1 - p_{\mathrm{eff}} \right)} < 2^{\gamma} - 1$ and $E_0 > - \log_2 2^{\gamma} = - \gamma$.
    Therefore $2 \gamma + E_0 > \gamma$, which proves the claim.
\end{proof}

\subsection{Simon's problem}\label{SM_sec:simon_lsn}

The main text (Sec.~\ref{subsec:simon_example}) claims that plugging the i.i.d.-depolarizing oracle directly into Simon's algorithm produces measured bitstrings distributed exactly as in the LSN problem.
The following lemma proves this claim.

\begin{lemma}[Noisy Simon sampling reproduces the LSN problem]\label{lem:noisy_simon_lsn}
    Simon's algorithm, run with the i.i.d.-depolarizing Simon oracle $\tilde{\mathcal{O}}_f = \mathcal{D}_p^{\otimes 2n} \circ \mathcal{O}_f$ for an $f$ satisfying the promise of Simon's problem with hidden string $s$, samples bitstrings distributed exactly as in the LSN problem~\cite{May2021} with parameters $n$ and $\tau = \frac{1}{2}\big(1 - (1 - \frac{4p}{3})^{|s|}\big)$.
    Explicitly, the bitstring $y$ measured in each round is distributed as
    \begin{equation}\label{eq:lsn_single_round}
        \Pr[y] = \frac{2}{N} \times
        \begin{cases}
            1-\tau, & y \cdot s = 0 \bmod 2\\
            \tau, & y \cdot s = 1 \bmod 2,
        \end{cases}
    \end{equation}
    that is, conditioned on a bit $b$ that equals one with probability $\tau$, the bitstring $y$ is uniform on $\{y : y \cdot s = b \bmod 2\}$, and bitstrings measured in different rounds are independent.
\end{lemma}
\begin{proof}
    Label the index register $A$ and the response register $B$, each of $n$ qubits.
    One round of Simon's algorithm prepares $\ket{0^n}_A\ket{+}^{\otimes n}_B$, applies $H^{\otimes n}$ on $A$, applies $\tilde{\mathcal{O}}_f$ once, applies $H^{\otimes n}$ on $A$ again, and measures $A$ in the computational basis to obtain $y$.
    The noise factors over the registers as $\mathcal{D}_p^{\otimes 2n} = \mathcal{N}_A \otimes \mathcal{N}_B$ with $\mathcal{N}_A = \mathcal{N}_B = \mathcal{D}_p^{\otimes n}$.
    Any channel acting on $B$ alone, including a measurement, commutes with the operations on $A$ and disappears under the partial trace over $B$, so it leaves the distribution of $y$ unchanged.
    We may therefore drop the noise $\mathcal{N}_B$ and measure $B$ in the $X$ basis right after the query.
    The depolarizing channel is unitarily covariant (Lemma~\ref{lem:depol_facts}), so $\mathcal{N}_A$ commutes qubit by qubit with the final Hadamard transform on $A$.
    One noisy round is thus equivalent to one noiseless round followed by $\mathcal{N}_A$ applied just before the measurement of $A$.

    The noiseless round is standard, and we write $y_0$ for its measurement outcome.
    After the query the state is $\frac{1}{\sqrt{N}}\sum_x \ket{x}_A\ket{(-1)^{f(x)}}_B$, where $\ket{(-1)^{f(x)}} := \ket{(-1)^{f(x)_1}} \otimes \cdots \otimes \ket{(-1)^{f(x)_n}}$ records $f(x)$ in the $X$ basis.
    By the promise, $f(x) = f(y)$ holds if and only if $x \oplus y \in \{0^n, s\}$, so the $X$-basis measurement of $B$ selects a uniformly random coset $\{x, x\oplus s\}$ and collapses $A$ onto $\frac{\ket{x} + \ket{x \oplus s}}{\sqrt{2}}$.
    The final Hadamard transform maps this state to $\frac{1}{\sqrt{2N}} \sum_{y_0} (-1)^{x \cdot y_0}\big(1 + (-1)^{s \cdot y_0}\big) \ket{y_0}$.
    Whichever coset the measurement of $B$ selects, the amplitude of $\ket{y_0}$ vanishes when $y_0 \cdot s = 1 \bmod 2$ and has modulus $\sqrt{2/N}$ when $y_0 \cdot s = 0 \bmod 2$, so $y_0$ is uniform on $\{y_0 : y_0 \cdot s = 0 \bmod 2\}$.

    It remains to analyze how the noise $\mathcal{N}_A$, now applied just before the measurement, affects the outcome.
    The measurement reads out only the diagonal matrix elements in the computational basis, and on these each single-qubit factor of $\mathcal{N}_A$ acts as a random classical bit flip.
    On each qubit, every state $\rho$ satisfies $\bra{y} Z \rho Z \ket{y} = \bra{y} \rho \ket{y}$ and $\bra{y} X \rho X \ket{y} = \bra{y} Y \rho Y \ket{y} = \bra{\bar{y}} \rho \ket{\bar{y}}$, where $\bar{y}$ flips the bit of $y$ on that qubit.
Applying these identities to the state of $A$ qubit by qubit, the $I$ and $Z$ branches of $\mathcal{N}_A$ leave the outcome on that qubit unchanged, while the $X$ and $Y$ branches, each of probability $\frac{p}{3}$, flip it.
The measured bitstring is therefore $y = y_0 \oplus a$, where the flip pattern $a \in \{0,1\}^n$ is independent of $y_0$ and has independent bits, each equal to one with probability $\frac{p}{3} + \frac{p}{3} = \frac{2p}{3}$.

    Finally we identify the distribution of $y = y_0 \oplus a$.
    For fixed $a$, the map $y_0 \mapsto y_0 \oplus a$ is a bijection from $\{y_0 : y_0 \cdot s = 0 \bmod 2\}$ onto $\{y : y \cdot s = a \cdot s \bmod 2\}$, so conditioned on $a$ the bitstring $y$ is uniform on this set.
    The conditional distribution depends on $a$ only through the bit $b := a \cdot s$, so conditioned on $b$ the bitstring $y$ is uniform on $\{y : y \cdot s = b \bmod 2\}$.
The bit $b$ is the parity of the $|s|$ independent bits $a_i$ with $i \in \mathrm{supp}(s)$, so $\mathbb{E}[(-1)^b] = \prod_{i \in \mathrm{supp}(s)} \mathbb{E}[(-1)^{a_i}] = \left(1 - \frac{4p}{3}\right)^{|s|}$, using $\mathbb{E}[(-1)^{a_i}] = 1 - 2 \cdot \frac{2p}{3}$.
Hence $\Pr[b = 1] = \frac{1}{2}\left(1 - \mathbb{E}[(-1)^b]\right) = \frac{1}{2}\left(1 - \left(1 - \frac{4p}{3}\right)^{|s|}\right) = \tau$, and combining this with the conditional uniformity gives Eq.~\eqref{eq:lsn_single_round}.
    Each round applies this procedure to a fresh copy of the initial state with an independent realization of the oracle noise, so the joint process over all rounds is a product channel and the samples are independent.
\end{proof}

\clearpage
\section{Distillation of Fractional and Continuous-time Boolean Oracles}\label{SM_sec:frac_ct_distillation}
In this section, we consider distilling the family of fractional and continuous-time Boolean oracles into ideal Boolean oracles.
Throughout this section we fix an integer $J \geq 1$, and the ideal fractional oracles are defined by
\begin{equation}\label{eq:frac_oracle_def}
    O_f(\theta) := e^{-i \theta H_f}, \qquad \theta := \frac{\pi}{J}, \qquad H_f := \sum_{x \in \{0,1\}^n} \ket{x}\bra{x} \otimes \sum_{j=1}^{m} f(x)_j \ket{1}\bra{1}_j ,
\end{equation}
where $\ket{1}\bra{1}_j$ acts on the $j$-th response qubit.
The Hamiltonian $H_f$ is diagonal with eigenvalue $f(x) \cdot y = \sum_{j} f(x)_j y_j \in \{0, \ldots, m\}$ on $\ket{x}\ket{y}$, so $\|H_f\|_{\mathrm{op}} \leq m$ and $O_f(\theta)$ is the diagonal unitary with entries $e^{-i\theta\, f(x) \cdot y}$.
At $\theta = \pi$ the entry is $(-1)^{f(x) \cdot y}$, so $O_f(\pi) \ket{x}\ket{y} = \ket{x} Z^{f(x)} \ket{y}$ is the Boolean oracle $O_f$ of Sec.~\ref{sec:grover_distillation}, and $O_f(\theta)^J = O_f(J\theta) = O_f(\pi)$, so $J$ fractional oracle queries compose to one Boolean oracle query.
A query to $O_f(\theta)$ runs the oracle Hamiltonian $H_f$ for time $\theta$, and a Boolean oracle query runs it for time $\pi$.
We therefore measure the cost of a protocol in this section by its \textit{total evolution time}, the query complexity times the evolution time $\theta$ of one query, and compare it with $\pi\, T_{\mathrm{OD}}$ for the Boolean oracle.

We consider two families of noisy oracles, both modeling a query that runs the oracle for a fraction of the full angle $\pi$, with noise that shrinks in proportion to the fraction.
The \textit{family of two-sided i.i.d.-depolarizing fractional oracles} is
\begin{equation}\label{eq:frac_family}
    \tilde{\mathfrak{O}}_F^{\mathrm{frac}}(\theta) := \left\{ \tilde{\mathcal{O}}_f(\theta) \right\}_{f \in F},
    \qquad
    \tilde{\mathcal{O}}_f(\theta) := \mathcal{D}_{p_\theta}^{\otimes (n+m)} \circ \mathcal{O}_f(\theta) \circ \mathcal{D}_{p_\theta}^{\otimes (n+m)} ,
\end{equation}
where $\mathcal{O}_f(\theta)(\cdot) := O_f(\theta) (\cdot) O_f(\theta)^{\dagger}$ and the per-qubit error rate of each of the two depolarizing layers is
\begin{equation}\label{eq:p_theta_def}
    p_\theta := \frac{\theta}{2\pi}\, p = \frac{p}{2J} ,
\end{equation}
proportional to the angle, so that the error rates of the $2J$ depolarizing layers accompanying one Boolean oracle query sum to $p$.
The \textit{family of continuous-time i.i.d.-depolarizing oracles} is
\begin{equation}\label{eq:ct_family}
    \tilde{\mathfrak{O}}_F^{\mathrm{ct}}(\theta) := \left\{ e^{\theta \mathcal{L}_f} \right\}_{f \in F},
    \qquad
    \mathcal{L}_f(\rho) := -i[H_f, \rho] + \frac{\Gamma}{3} \sum_{j=1}^{n+m} \sum_{P \in \{X, Y, Z\}} \left( P_j \rho P_j - \rho \right) ,
\end{equation}
where $\mathcal{L}_f$ is the Lindbladian that generates the dissipative quantum dynamics, describing evolution under the oracle Hamiltonian $H_f$ with i.i.d.\ depolarizing noise at rate $\Gamma$ on every qubit, so that $e^{\theta \mathcal{L}_f}$ is the evolution for time $\theta$.

In the previous sections, noise acts only after the ideal oracle, and the stage 1 gadget converts it into phase noise.
Both families here are more realistic, with noise acting before and after the oracle or throughout the evolution, and an error before the oracle propagates through the oracle before the stage 1 gadget can act on it.
Distillation still works because in a query of angle $\theta$ an error before the oracle occurs with probability $O(\theta)$, and such an error commutes with $O_f(\theta)$ up to $O(\theta)$, so each query differs from one with noise only after the oracle by $O(\theta^2)$.
Composing $J = \pi/\theta$ queries into one Boolean oracle query then leaves an error of $O(\theta)$, which vanishes as $\theta \to 0$.

The distillation protocol for both families shares the same three steps.
First, we apply the stage 1 gadget to each noisy query, a fractional oracle query in Sec.~\ref{SM_subsec:frac_distillation} or a continuous-time oracle query of evolution time $\theta$ in Sec.~\ref{SM_subsec:ct_distillation}, and show that its output is an i.i.d.-dephasing fractional oracle $(\mathcal{D}^Z_q)^{\otimes (n+m)} \circ \mathcal{O}_f(\theta)$ up to an error of $O(\theta^2)$.
This is the only step that differs between the two families.
Second, we compose $J$ such outputs, which gives an i.i.d.-dephasing Boolean oracle up to $J$ times that error, with a per-qubit error rate bounded by a constant independent of $J$.
Third, we run the stage 2 protocol of Theorem~\ref{thm:iid_threshold_at_exponent} on the composed queries.
The resulting protocols, Theorems~\ref{thm:fractional_OD_two_sided} and~\ref{thm:continuous_OD_two_sided}, have total evolution time $\pi\, T_{\mathrm{OD}}$, where $T_{\mathrm{OD}}$ is the query complexity Eq.~\eqref{eq:T_OD_at_exponent} of the Boolean oracle at precision $\epsilon/2$, so in evolution time the fractional and continuous-time oracles cost the same as the Boolean oracle, while the number of queries grows as $1/\theta$.
The price is that the angle must satisfy $\theta \leq \theta_{\mathrm{th}}$ for a threshold $\theta_{\mathrm{th}}$ that is positive for every $n$ but decreases with $n$ and with the precision $\epsilon$.
A small angle keeps $O_f(\theta)$ close to the identity, so an error before the oracle is changed only by $O(\theta)$ when it passes through the oracle, and the stage 1 gadget handles it almost as if it had occurred after the oracle.

The following lemma makes the second step precise.
\begin{lemma}[Composing $J$ i.i.d.-dephasing fractional oracle queries]\label{lem:compose_J_steps}
    Let $\Lambda$ be a channel on the $n+m$ oracle qubits that is within diamond distance $\delta$ of an i.i.d.-dephasing fractional oracle,
    \begin{equation}\label{eq:compose_J_hypothesis}
        \left\| \Lambda - \left( \mathcal{D}^Z_{q} \right)^{\otimes (n+m)} \circ \mathcal{O}_f(\theta) \right\|_{\diamond} \leq \delta ,
    \end{equation}
    for some $q \in [0, 1/2]$ and $\delta \geq 0$.
    Then the $J$-fold composition of $\Lambda$ is within diamond distance $J\delta$ of the i.i.d.-dephasing Boolean oracle with per-qubit error rate $q_J$,
    \begin{equation}\label{eq:compose_J_conclusion}
        \left\| \Lambda^{\circ J} - \left( \mathcal{D}^Z_{q_J} \right)^{\otimes (n+m)} \circ \mathcal{O}_f \right\|_{\diamond} \leq J \delta ,
        \qquad
        1 - 2 q_J := (1 - 2q)^J .
    \end{equation}
\end{lemma}
\begin{proof}
    Write $\Psi := \left( \mathcal{D}^Z_{q} \right)^{\otimes (n+m)} \circ \mathcal{O}_f(\theta)$ for the i.i.d.-dephasing fractional oracle in Eq.~\eqref{eq:compose_J_hypothesis}.
    We first show that $\Psi^{\circ J}$ is the i.i.d.-dephasing Boolean oracle in Eq.~\eqref{eq:compose_J_conclusion}, and then bound the distance between $\Lambda^{\circ J}$ and $\Psi^{\circ J}$.

    Both $O_f(\theta)$ and every Pauli $Z$ string are diagonal in the computational basis, so $\left( \mathcal{D}^Z_{q} \right)^{\otimes (n+m)}$ and $\mathcal{O}_f(\theta)$ commute, and
    \begin{equation}\label{eq:compose_J_ideal}
        \Psi^{\circ J}
        = \left( \left( \mathcal{D}^Z_{q} \right)^{\circ J} \right)^{\otimes (n+m)} \circ \mathcal{O}_f(\theta)^{\circ J}
        = \left( \left( \mathcal{D}^Z_{q} \right)^{\circ J} \right)^{\otimes (n+m)} \circ \mathcal{O}_f ,
    \end{equation}
    using $O_f(\theta)^J = O_f$.
    On a single qubit, $\left( \mathcal{D}^Z_{q} \right)^{\circ J}$ applies $Z^{s_1 \oplus \cdots \oplus s_J}$ where $s_1, \ldots, s_J$ are independent bits, each equal to one with probability $q$, so it is the dephasing channel with rate $u := \Pr[s_1 \oplus \cdots \oplus s_J = 1]$.
    Since $\mathbb{E}\left[ (-1)^{s_1 \oplus \cdots \oplus s_J} \right] = \prod_{j=1}^{J} \mathbb{E}\left[ (-1)^{s_j} \right] = (1 - 2q)^J$ and this expectation equals $1 - 2u$, we get $u = q_J$.
    Hence $\Psi^{\circ J} = \left( \mathcal{D}^Z_{q_J} \right)^{\otimes (n+m)} \circ \mathcal{O}_f$.

    It remains to show $\left\| \Lambda^{\circ J} - \Psi^{\circ J} \right\|_{\diamond} \leq J\delta$.
    Replacing the $J$ copies of $\Lambda$ by $\Psi$ one at a time gives the telescoping sum
    \begin{equation}\label{eq:compose_J_telescope}
        \Lambda^{\circ J} - \Psi^{\circ J} = \sum_{j=1}^{J} \Psi^{\circ (j-1)} \circ \left( \Lambda - \Psi \right) \circ \Lambda^{\circ (J-j)} .
    \end{equation}
    By the triangle inequality and Lemma~\ref{lem:channel_norm_facts}(vi), each term has diamond norm at most $\left\| \Lambda - \Psi \right\|_{\diamond} \leq \delta$, and summing the $J$ terms gives the claim.
\end{proof}

\subsection{Distillation of fractional oracles}\label{SM_subsec:frac_distillation}
We first show that the stage 1 gadget converts the noisy fractional oracle $\tilde{\mathcal{O}}_f(\theta)$ into an i.i.d.-dephasing fractional oracle up to a small error.
Post-oracle Pauli errors and pre-oracle $Z$ errors are handled exactly as in Corollary~\ref{cor:stage1_depol}, since $O_f(\theta)$ is diagonal.
A pre-oracle $X$ error is not, since moving it past the oracle conjugates $O_f(\theta)$ by an $X$ string, which changes each phase by at most $m\theta$.
Such an error occurs with probability $O((n+m) p_\theta)$ per query, so the error that the stage 1 gadget leaves is of order $m (n+m)\, p_\theta\, \theta$, second order in $\theta$ since $p_\theta$ is proportional to $\theta$.

\begin{lemma}[Stage 1 on one two-sided i.i.d.-depolarizing fractional oracle query]\label{lem:stage1_two_sided_frac}
    For every $\theta \geq 0$ and $p_\theta \in [0, 1]$, applying stage 1 to the two-sided i.i.d.-depolarizing fractional oracle $\tilde{\mathcal{O}}_f(\theta)$ of Eq.~\eqref{eq:frac_family} leaves an i.i.d.-dephasing fractional oracle up to error $\delta_{\mathrm{frac}}$,
    \begin{equation}\label{eq:stage1_two_sided_frac}
        \left\| \mathsf{Rec}_{\mathrm{rep}}^{\otimes (n+m)} \circ \left( \tilde{\mathcal{O}}_f(\theta) \otimes \mathcal{I} \right) \circ \mathsf{Enc}_{\mathrm{rep}}^{\otimes (n+m)} - \left( \mathcal{D}^Z_{q} \right)^{\otimes (n+m)} \circ \mathcal{O}_f(\theta) \right\|_{\diamond}
        \leq \delta_{\mathrm{frac}} ,
    \end{equation}
    where
    \begin{equation}\label{eq:stage1_two_sided_frac_params}
        \delta_{\mathrm{frac}} := \frac{4 m (n+m)\, p_\theta\, \theta}{3} ,
        \qquad
        q := \frac{4 p_\theta}{3} \left( 1 - \frac{2 p_\theta}{3} \right) ,
    \end{equation}
    and $\tilde{\mathcal{O}}_f(\theta) \otimes \mathcal{I}$ acts as $\tilde{\mathcal{O}}_f(\theta)$ on the third qubit of each code block and identity on the rest.
\end{lemma}
\begin{proof}
    We expand the two depolarizing layers into Pauli strings, move the pre-oracle string past the oracle at the price of conjugating the oracle by an $X$ string, apply Lemma~\ref{lem:stage1_rewrite} to each term, and finally bound the distance between the conjugated oracle and the original one.

    We first write $\tilde{\mathcal{O}}_f(\theta)$ as a mixture of Pauli strings after the oracle.
    Each depolarizing layer is a mixture of Pauli channels, the post-oracle one of $\mathcal{X}^{a} \mathcal{Z}^{b}$ with weight $\mu(a, b)$ and the pre-oracle one of $\mathcal{X}^{c} \mathcal{Z}^{d}$ with weight $\mu(c, d)$, where $a, b, c, d \in \{0,1\}^{n+m}$.
    $Z^{d}$ commutes with the diagonal unitary $O_f(\theta)$ and merges with $Z^{b}$, while $X^{c}$ passes the oracle by conjugation, $O_f(\theta) X^{c} = X^{c} O_f^{(c)}(\theta)$ with $O_f^{(c)}(\theta) := X^{c} O_f(\theta) X^{c}$, and merges with $X^{a}$.
    Hence
    \begin{equation}\label{eq:two_sided_frac_expansion}
        \tilde{\mathcal{O}}_f(\theta) = \sum_{a, b, c, d} \mu(a, b)\, \mu(c, d)\; \mathcal{X}^{a \oplus c} \mathcal{Z}^{b \oplus d} \circ \mathcal{O}_f^{(c)}(\theta) ,
    \end{equation}
    where $\mathcal{O}_f^{(c)}(\theta)$ is the unitary channel of $O_f^{(c)}(\theta)$.
    Phases and signs of the Pauli operators drop out of the channels.
    Since $X^{c}$ permutes computational basis states, $O_f^{(c)}(\theta)$ is again diagonal.

    We now apply stage 1 to each term of Eq.~\eqref{eq:two_sided_frac_expansion}.
    Since $O_f^{(c)}(\theta)$ is diagonal, it commutes with the encoding and can be pulled out to act before $\mathsf{Enc}_{\mathrm{rep}}^{\otimes (n+m)}$, and Lemma~\ref{lem:stage1_rewrite} gives $\mathsf{Rec}_{\mathrm{rep}}^{\otimes (n+m)} \circ \left( \mathcal{X}^{a \oplus c} \mathcal{Z}^{b \oplus d} \otimes \mathcal{I} \right) \circ \mathsf{Enc}_{\mathrm{rep}}^{\otimes (n+m)} = \mathcal{Z}^{b \oplus d}$, where $\mathcal{X}^{a \oplus c} \mathcal{Z}^{b \oplus d} \otimes \mathcal{I}$ acts on the third qubit of each code block.
    Hence
    \begin{equation}\label{eq:two_sided_frac_after_stage1}
        \mathsf{Rec}_{\mathrm{rep}}^{\otimes (n+m)} \circ \left( \tilde{\mathcal{O}}_f(\theta) \otimes \mathcal{I} \right) \circ \mathsf{Enc}_{\mathrm{rep}}^{\otimes (n+m)}
        = \sum_{a, b, c, d} \mu(a, b)\, \mu(c, d)\; \mathcal{Z}^{b \oplus d} \circ \mathcal{O}_f^{(c)}(\theta) .
    \end{equation}

    We compare Eq.~\eqref{eq:two_sided_frac_after_stage1} term by term with the target $\left( \mathcal{D}^Z_{q} \right)^{\otimes (n+m)} \circ \mathcal{O}_f(\theta)$, which we also write as a mixture of the Pauli channels $\mathcal{Z}^{b \oplus d}$ with the same weights $\mu(a, b)\, \mu(c, d)$.
    Under the marginal $\mu_Z$ of $\mu$ on the $Z$ string each bit equals one with probability $2 p_\theta / 3$ independently, and $b$ and $d$ come from different layers and are independent, so each bit of $b \oplus d$ equals one with probability $2 \cdot \frac{2 p_\theta}{3} \left( 1 - \frac{2 p_\theta}{3} \right) = q$ independently, and hence
    \begin{equation}\label{eq:two_sided_frac_target_expansion}
        \left( \mathcal{D}^Z_{q} \right)^{\otimes (n+m)} \circ \mathcal{O}_f(\theta)
        = \sum_{a, b, c, d} \mu(a, b)\, \mu(c, d)\; \mathcal{Z}^{b \oplus d} \circ \mathcal{O}_f(\theta) .
    \end{equation}
    The two sums differ only in the terms with $c \neq 0$, where $\mathcal{O}_f^{(c)}(\theta)$ replaces $\mathcal{O}_f(\theta)$.
    By the triangle inequality for the diamond norm and Lemma~\ref{lem:channel_norm_facts}(vi), the latter to drop the unitary channel $\mathcal{Z}^{b \oplus d}$,
    \begin{equation}\label{eq:two_sided_frac_replacement_cost}
    \begin{split}
        &\left\| \mathsf{Rec}_{\mathrm{rep}}^{\otimes (n+m)} \circ \left( \tilde{\mathcal{O}}_f(\theta) \otimes \mathcal{I} \right) \circ \mathsf{Enc}_{\mathrm{rep}}^{\otimes (n+m)} - \left( \mathcal{D}^Z_{q} \right)^{\otimes (n+m)} \circ \mathcal{O}_f(\theta) \right\|_{\diamond} \\
        \leq &\sum_{a, b, c, d} \mu(a, b)\mu(c, d) \left\| \mathcal{O}_f^{(c)}(\theta) - \mathcal{O}_f(\theta) \right\|_{\diamond} \\
        = &\sum_{c} \mu_X(c) \left\| \mathcal{O}_f^{(c)}(\theta) - \mathcal{O}_f(\theta) \right\|_{\diamond} .
    \end{split}
    \end{equation}
    For $c \neq 0$, every diagonal entry of $O_f^{(c)}(\theta) - O_f(\theta)$ has the form $e^{-i \theta k} - e^{-i \theta k'}$ with $k, k' \in \{0, \ldots, m\}$, and $\left| e^{-i \theta k} - e^{-i \theta k'} \right| = 2 \left| \sin\left( \theta (k - k') / 2 \right) \right| \leq \theta |k - k'| \leq m \theta$.
    Using $\left\| \mathcal{U}_A - \mathcal{U}_B \right\|_{\diamond} \leq 2 \left\| A - B \right\|_{\mathrm{op}}$ for unitary channels~\cite{Haah2023},
    \begin{equation}\label{eq:conjugated_oracle_distance}
        \left\| \mathcal{O}_f^{(c)}(\theta) - \mathcal{O}_f(\theta) \right\|_{\diamond}
        \leq 2 \left\| O_f^{(c)}(\theta) - O_f(\theta) \right\|_{\mathrm{op}}
        \leq 2 m \theta .
    \end{equation}
    Finally, under the marginal $\mu_X$ of $\mu$ on the $X$ string each bit equals one with probability $2 p_\theta / 3$, so $\Pr_{\mu_X}[c \neq 0] \leq (n+m) \cdot \frac{2 p_\theta}{3}$ by a union bound over the $n+m$ bits of $c$, and the sum is at most $(n+m) \cdot \frac{2 p_\theta}{3} \cdot 2 m \theta = \delta_{\mathrm{frac}}$.
\end{proof}

We now combine Lemma~\ref{lem:stage1_two_sided_frac} with Lemma~\ref{lem:compose_J_steps} and reduce the distillation of fractional oracles to Theorem~\ref{thm:iid_threshold_at_exponent}.
The per-qubit error rate of the composed query depends on $J$, whereas the theorem runs at a constant error rate, so we pad the composed query with one more dephasing channel that raises its rate to the constant $2p/3$.

\begin{formalthm}[a]{theorem}{twosided}[Distillation of fractional Boolean oracles under two-sided i.i.d.\ depolarizing noise]\label{thm:fractional_OD_two_sided}
    For any constant overhead exponent $\gamma \in (0, 1)$, any precision $\epsilon = \epsilon(n) \in (0,1)$ with finite decay rate $\nu := \limsup_{n\to\infty} \frac{1}{n} \log_2 \frac{1}{\epsilon}$, any $m$ with $\log_2 m = o(n)$, any constant per-qubit error rate $p < p_{\mathrm{th}}\left( \gamma, \nu \right)$ of Eq.~\eqref{eq:iid_pth_at_exponent}, any angle $\theta = \pi/J \leq \theta_{\mathrm{th}}$ with integer $J$ and
    \begin{equation}\label{eq:frac_theta_condition}
        \theta_{\mathrm{th}} := \frac{3 \epsilon}{8 m \left( n + m \right) p\, L} ,
        \qquad
        L := \left\lceil \frac{48 \left( 9 - 4 p \right)}{\left( 3 - 4 p \right)^2} \, N^{\gamma} \, \ln \frac{24 m}{\epsilon} \right\rceil ,
    \end{equation}
    and all sufficiently large $n$, there exists a protocol that distills the family of two-sided i.i.d.-depolarizing fractional oracles $\tilde{\mathfrak{O}}_F^{\mathrm{frac}}(\theta)$ into the family of ideal Boolean oracles $\mathfrak{O}_F$ with precision $\epsilon$ and total evolution time
    \begin{equation}\label{eq:frac_query_complexity}
        \theta\, T_{\mathrm{OD}}^{\mathrm{frac}} = 2 \pi L ,
    \end{equation}
    that is, with query complexity $T_{\mathrm{OD}}^{\mathrm{frac}} = 2 \pi L / \theta$.
\end{formalthm}
\begin{proof}
    We run the stage 2 protocol of Theorem~\ref{thm:iid_threshold_at_exponent} at precision $\epsilon/2$, and replace each raw oracle query it makes by the composition of $J$ fractional oracle queries, each passed through the stage 1 gadget.
    The total error then has two sources, the distance between this composed query and an i.i.d.-dephasing Boolean oracle, and the error of the stage 2 protocol itself, and we show that the two sum to at most $\epsilon$.

    We first apply stage 1 to one fractional oracle query and write $\Lambda := \mathsf{Rec}_{\mathrm{rep}}^{\otimes (n+m)} \circ \left( \tilde{\mathcal{O}}_f(\theta) \otimes \mathcal{I} \right) \circ \mathsf{Enc}_{\mathrm{rep}}^{\otimes (n+m)}$ for its output.
    Lemma~\ref{lem:stage1_two_sided_frac} with $\theta = \pi/J$ and $p_\theta = p/(2J)$ bounds the distance between $\Lambda$ and the i.i.d.-dephasing fractional oracle with per-qubit error rate $q = \frac{2p}{3J} \left( 1 - \frac{p}{3J} \right)$ by
    \begin{equation}\label{eq:frac_q_and_delta}
        \left\| \Lambda - \left( \mathcal{D}^Z_{q} \right)^{\otimes (n+m)} \circ \mathcal{O}_f(\theta) \right\|_{\diamond}
        \leq \delta_{\mathrm{frac}} = \frac{4 m \left( n + m \right)}{3} \cdot \frac{p}{2J} \cdot \frac{\pi}{J} = \frac{2 \pi\, m \left( n + m \right) p}{3 J^2} ,
    \end{equation}
    and this value of $q$ satisfies $1 - 2q = \left( 1 - \frac{2p}{3J} \right)^2$.

    We next compose $J$ such queries.
    Lemma~\ref{lem:compose_J_steps} applied to Eq.~\eqref{eq:frac_q_and_delta} bounds the distance between $\Lambda^{\circ J}$ and the i.i.d.-dephasing Boolean oracle with per-qubit error rate $q_J$ given by $1 - 2 q_J = (1 - 2q)^J = \left( 1 - \frac{2p}{3J} \right)^{2J}$ by
    \begin{equation}\label{eq:frac_composed}
        \left\| \Lambda^{\circ J} - \left( \mathcal{D}^Z_{q_J} \right)^{\otimes (n+m)} \circ \mathcal{O}_f \right\|_{\diamond} \leq J \delta_{\mathrm{frac}} = \frac{2 \pi\, m \left( n + m \right) p}{3 J} .
    \end{equation}
    Since $(1 - x)^k \geq 1 - kx$ for $x \in [0,1]$ and $k \geq 1$, we have $1 - 2 q_J \geq 1 - \frac{4p}{3}$, that is, $q_J \leq 2p/3$.

    We now run the stage 2 protocol of Theorem~\ref{thm:iid_threshold_at_exponent} at precision $\epsilon/2$ on the composed query.
    The theorem requires the raw oracle to have constant error rate, whereas the rate $q_J$ in Eq.~\eqref{eq:frac_composed} depends on $n$ through $J$.
    To overcome this, we apply one more layer of dephasing channel at a rate $s$ to the composed query, such that the two channels together dephase at rate exactly $2p/3$.
    Since $0 \leq q_J \leq 2p/3 < 1/2$, where $p < 3/4$ because the hypothesis $p < p_{\mathrm{th}}\left( \gamma, \nu \right)$ and Lemma~\ref{lem:threshold_equation} give $p < p_{\mathrm{th}}\left( \gamma, \nu \right) < 3/4$, the value of $s$ can be defined by
    \begin{equation}\label{eq:frac_padding_rate}
        1 - 2s := \frac{1 - 4p/3}{1 - 2 q_J} ,
    \end{equation}
    so $s$ lies in $[0, 1/2]$.
    On a single qubit, $\mathcal{D}^Z_{s} \circ \mathcal{D}^Z_{q_J}$ applies $Z^{s_1 \oplus s_2}$ for independent bits $s_1, s_2$ equal to one with probabilities $s$ and $q_J$, and $\mathbb{E}\left[ (-1)^{s_1 \oplus s_2} \right] = (1 - 2s)(1 - 2 q_J) = 1 - 4p/3$, so $\mathcal{D}^Z_{s} \circ \mathcal{D}^Z_{q_J} = \mathcal{D}^Z_{2p/3}$.
    Appending $\left( \mathcal{D}^Z_{s} \right)^{\otimes (n+m)}$ after both channels in Eq.~\eqref{eq:frac_composed} does not increase their diamond distance by Lemma~\ref{lem:channel_norm_facts}(vi), so
    \begin{equation}\label{eq:frac_padded_target}
        \left\| \left( \mathcal{D}^Z_{s} \right)^{\otimes (n+m)} \circ \Lambda^{\circ J} - \left( \mathcal{D}^Z_{2p/3} \right)^{\otimes (n+m)} \circ \mathcal{O}_f \right\|_{\diamond} \leq \frac{2 \pi\, m \left( n + m \right) p}{3 J} .
    \end{equation}
    Theorem~\ref{thm:iid_threshold_at_exponent} applies at precision $\epsilon/2$ for all sufficiently large $n$ because $\epsilon/2$ has the same decay rate $\nu$ as $\epsilon$.
    Its stage 2 protocol makes $2L$ queries to $\left( \mathcal{D}^Z_{2p/3} \right)^{\otimes (n+m)} \circ \mathcal{O}_f$ with $L$ as in Eq.~\eqref{eq:frac_theta_condition}, and outputs a channel within diamond distance $\epsilon/2$ of $\mathcal{O}_f$.
    Our distilled oracle is this protocol with each of the $2L$ queries replaced by the padded composed query on the left-hand side of Eq.~\eqref{eq:frac_padded_target}.
    Each padded composed query makes $J$ fractional oracle queries, so the distilled oracle makes $2JL = 2 \pi L / \theta$ fractional oracle queries, which is $T_{\mathrm{OD}}^{\mathrm{frac}}$ of Eq.~\eqref{eq:frac_query_complexity}.
    Replacing the $2L$ queries one at a time writes the difference between the distilled oracle and the output of the protocol as a telescoping sum of $2L$ terms, each the difference of two channels that differ in one query only.
    By Lemma~\ref{lem:channel_norm_facts}(vi) and Eq.~\eqref{eq:frac_padded_target}, each term has diamond norm at most $J \delta_{\mathrm{frac}}$, so by the triangle inequality the diamond distance between the distilled oracle and the output of the protocol is at most $2 L \cdot J \delta_{\mathrm{frac}} = 2 L \cdot \frac{2 \pi\, m \left( n + m \right) p}{3 J} = \frac{4 m \left( n + m \right) p\, L}{3} \, \theta \leq \frac{\epsilon}{2}$ by $\theta = \pi / J$ and $\theta \leq \theta_{\mathrm{th}}$ of Eq.~\eqref{eq:frac_theta_condition}.
    By the triangle inequality with the error $\epsilon/2$ of the protocol itself, the distilled oracle is within diamond distance $\epsilon$ of $\mathcal{O}_f$.
\end{proof}

\subsection{Distillation of continuous-time oracles}\label{SM_subsec:ct_distillation}
In the continuous-time oracle $e^{\theta \mathcal{L}_f}$ the noise acts throughout the evolution, at the same time as the oracle Hamiltonian, rather than in two layers around the oracle.
We handle this by splitting the generator in Eq.~\eqref{eq:ct_family} into its oracle part and its noise part,
\begin{equation}\label{eq:ct_generator_split}
    \mathcal{L}_f = \mathcal{L}_H + \mathcal{L}_N ,
    \qquad
    \mathcal{L}_H(\rho) := -i [H_f, \rho] ,
    \qquad
    \mathcal{L}_N(\rho) := \frac{\Gamma}{3} \sum_{j=1}^{n+m} \sum_{P \in \{X, Y, Z\}} \left( P_j \rho P_j - \rho \right) .
\end{equation}
The oracle part alone generates the fractional oracle, $e^{\theta \mathcal{L}_H} = \mathcal{O}_f(\theta)$ by Eq.~\eqref{eq:frac_oracle_def}, and the noise part alone generates an i.i.d.\ depolarizing channel with per-qubit error rate $O(\Gamma \theta)$.
We show that for a short evolution time $\theta$ the continuous-time oracle is close to the fractional oracle followed by a single layer of i.i.d.\ depolarizing noise, by bounding the error of the first-order product formula that separates the two parts of the generator.
This error is quadratic in $\theta$, because in time $\theta$ a Pauli error occurs with probability $O(\Gamma \theta)$, and moving it past the oracle changes the phase angle by at most $m \theta$.
Once the noise acts after the oracle, stage 1 converts it into dephasing by Corollary~\ref{cor:stage1_depol}, which applies because $O_f(\theta)$ is diagonal.

\begin{lemma}[Product formula for the continuous-time i.i.d.-depolarizing oracle]\label{lem:ct_product_formula}
    For every $\theta \geq 0$ and $\Gamma \geq 0$, the continuous-time i.i.d.-depolarizing oracle $e^{\theta \mathcal{L}_f}$ of Eq.~\eqref{eq:ct_family} is within diamond distance $\delta_{\mathrm{ct}}$ of the fractional oracle followed by one layer of i.i.d.\ depolarizing noise,
    \begin{equation}\label{eq:ct_split_bound}
        \left\| e^{\theta \mathcal{L}_f} - \mathcal{D}_{p(\theta)}^{\otimes (n+m)} \circ \mathcal{O}_f(\theta) \right\|_{\diamond} \leq \delta_{\mathrm{ct}} ,
    \end{equation}
    where
    \begin{equation}\label{eq:ct_split_params}
        \delta_{\mathrm{ct}} := \frac{2 m (n+m)\, \Gamma\, \theta^2}{3} ,
        \qquad
        p(\theta) := \frac{3}{4} \left( 1 - e^{-4 \Gamma \theta / 3} \right) .
    \end{equation}
\end{lemma}
\begin{proof}
    We first compute the channel generated by the noise part, and then bound the error of the first-order product formula that separates it from the oracle part.

    Write $\mathcal{L}_N = \sum_{j=1}^{n+m} \mathcal{L}_N^{(j)}$ with $\mathcal{L}_N^{(j)}(\rho) := \frac{\Gamma}{3} \sum_{P \in \{X, Y, Z\}} \left( P_j \rho P_j - \rho \right)$.
    The terms act on different qubits and commute, so $e^{\theta \mathcal{L}_N} = \bigotimes_{j=1}^{n+m} e^{\theta \mathcal{L}_N^{(j)}}$.
    On qubit $j$, $\mathcal{L}_N^{(j)}$ maps the identity to zero and maps each non-identity Pauli $Q$ to $-\frac{4 \Gamma}{3} Q$, because $Q$ commutes with one of $X, Y, Z$ and anticommutes with the other two, so that $\sum_{P \in \{X, Y, Z\}} P Q P = -Q$.
    Hence $e^{\theta \mathcal{L}_N^{(j)}}$ fixes the identity and multiplies each non-identity Pauli by $e^{-4 \Gamma \theta / 3}$, while the depolarizing channel $\mathcal{D}_p$ of Eq.~\eqref{iidnoise} fixes the identity and multiplies each non-identity Pauli by $1 - \frac{4p}{3}$, since $\mathcal{D}_p(Q) = (1-p) Q + \frac{p}{3} \sum_{P \in \{X, Y, Z\}} P Q P = \left( 1 - \frac{4p}{3} \right) Q$.
    The four Paulis span the single-qubit operators, so the two channels agree exactly when $1 - \frac{4p}{3} = e^{-4 \Gamma \theta / 3}$, that is, at $p = p(\theta)$ of Eq.~\eqref{eq:ct_split_params}, and hence
    \begin{equation}\label{eq:ct_noise_exponential}
        e^{\theta \mathcal{L}_N} = \mathcal{D}_{p(\theta)}^{\otimes (n+m)} .
    \end{equation}

    Since $\mathcal{L}_H$, $\mathcal{L}_N$ and $\mathcal{L}_f$ are Lindbladians, the first-order product formula bound for Lindbladians (Theorem~5 of Ref.~\cite{Wang2026}) and the triangle inequality give
    \begin{equation}\label{eq:ct_product_formula}
        \left\| e^{\theta \mathcal{L}_f} - e^{\theta \mathcal{L}_N} \circ e^{\theta \mathcal{L}_H} \right\|_{\diamond}
        \leq \frac{\theta^2}{2} \left\| \left[ \mathcal{L}_N, \mathcal{L}_H \right] \right\|_{\diamond}
        \leq \frac{\theta^2}{2} \sum_{j=1}^{n+m} \left\| \left[ \mathcal{L}_N^{(j)}, \mathcal{L}_H \right] \right\|_{\diamond} .
    \end{equation}
    To bound the commutator for one qubit $j$, write $\Phi_{j,P}(\rho) := P_j \rho P_j$ for the unitary channel of $P_j$ and $\mathcal{L}_K(\rho) := -i [K, \rho]$ for any Hermitian $K$.
    The $-\rho$ terms of $\mathcal{L}_N^{(j)}$ are a multiple of the identity channel and commute with $\mathcal{L}_H$, so $\left[ \mathcal{L}_N^{(j)}, \mathcal{L}_H \right] = \frac{\Gamma}{3} \sum_{P \in \{X, Y, Z\}} \left[ \Phi_{j,P}, \mathcal{L}_H \right]$, and moving the conjugation by $P_j$ inside the commutator gives
    \begin{equation}\label{eq:ct_commutator_single}
    \begin{split}
        \left[ \Phi_{j,P}, \mathcal{L}_H \right](\rho)
        &= -i P_j [H_f, \rho] P_j + i \left[ H_f, P_j \rho P_j \right] \\
        &= -i \left[ P_j H_f P_j, P_j \rho P_j \right] + i \left[ H_f, P_j \rho P_j \right]
        = -i \left[ P_j H_f P_j - H_f,\; P_j \rho P_j \right] ,
    \end{split}
    \end{equation}
    that is, $\left[ \Phi_{j,P}, \mathcal{L}_H \right] = \mathcal{L}_{K_{j,P}} \circ \Phi_{j,P}$ with $K_{j,P} := P_j H_f P_j - H_f$.
    Since $H_f$ is diagonal, $Z_j$ commutes with it and $K_{j,Z} = 0$, while $Y_j H_f Y_j = X_j Z_j H_f Z_j X_j = X_j H_f X_j$ gives $K_{j,Y} = K_{j,X}$.
    Conjugation by $X_j$ permutes the diagonal entries of $H_f$, so $X_j H_f X_j$ is diagonal and each diagonal entry of $K_{j,X}$ is the difference of two eigenvalues of $H_f$, which lie in $\{0, \ldots, m\}$ by Eq.~\eqref{eq:frac_oracle_def}, so $\left\| K_{j,X} \right\|_{\mathrm{op}} \leq m$.
    For any Hermitian $K$, $\left\| \mathcal{L}_K \right\|_{\diamond} \leq 2 \left\| K \right\|_{\mathrm{op}}$, because $\left\| \left( \mathcal{L}_K \otimes \mathcal{I} \right)(A) \right\|_1 \leq \left\| (K \otimes I) A \right\|_1 + \left\| A (K \otimes I) \right\|_1 \leq 2 \left\| K \right\|_{\mathrm{op}} \left\| A \right\|_1$ for every operator $A$ on the system and a reference.
    Hence $\left\| \left[ \Phi_{j,P}, \mathcal{L}_H \right] \right\|_{\diamond} = \left\| \mathcal{L}_{K_{j,P}} \circ \Phi_{j,P} \right\|_{\diamond} \leq \left\| \mathcal{L}_{K_{j,P}} \right\|_{\diamond} \leq 2 \left\| K_{j,P} \right\|_{\mathrm{op}}$, which is at most $2m$ for $P \in \{X, Y\}$ and zero for $P = Z$, and hence $\left\| \left[ \mathcal{L}_N^{(j)}, \mathcal{L}_H \right] \right\|_{\diamond} \leq \frac{\Gamma}{3} \left( 2m + 2m + 0 \right) = \frac{4 m \Gamma}{3}$.
    Inserting this into the right-hand side of Eq.~\eqref{eq:ct_product_formula}, and replacing its left-hand side using $e^{\theta \mathcal{L}_H} = \mathcal{O}_f(\theta)$ and Eq.~\eqref{eq:ct_noise_exponential}, gives
    \begin{equation*}
        \left\| e^{\theta \mathcal{L}_f} - \mathcal{D}_{p(\theta)}^{\otimes (n+m)} \circ \mathcal{O}_f(\theta) \right\|_{\diamond}
        \leq \frac{\theta^2}{2} \cdot (n+m) \cdot \frac{4 m \Gamma}{3}
        = \delta_{\mathrm{ct}} ,
    \end{equation*}
    and proves Eq.~\eqref{eq:ct_split_bound}.
\end{proof}

We now combine Lemma~\ref{lem:ct_product_formula} and the stage 1 gadget with Lemma~\ref{lem:compose_J_steps} and reduce the distillation of continuous-time oracles to Theorem~\ref{thm:iid_threshold_at_exponent}.

\begin{formalthm}[b]{theorem}{twosided}[Distillation of continuous-time Boolean oracles under i.i.d.\ depolarizing noise]\label{thm:continuous_OD_two_sided}
    For any constant overhead exponent $\gamma \in (0, 1)$, any precision $\epsilon = \epsilon(n) \in (0,1)$ with finite decay rate $\nu := \limsup_{n\to\infty} \frac{1}{n} \log_2 \frac{1}{\epsilon}$, any $m$ with $\log_2 m = o(n)$, any constant noise rate $\Gamma > 0$ satisfying
    \begin{equation}\label{eq:ct_threshold}
        \Gamma < \frac{3}{4 \pi} \ln \frac{3}{3 - 4 p_{\mathrm{th}}\left( \gamma, \nu \right)} ,
    \end{equation}
    with $p_{\mathrm{th}}\left( \gamma, \nu \right)$ of Eq.~\eqref{eq:iid_pth_at_exponent}, any angle $\theta = \pi/J \leq \theta_{\mathrm{th}}$ with integer $J$ and
    \begin{equation}\label{eq:ct_theta_condition}
        \theta_{\mathrm{th}} := \frac{3 \epsilon}{8 \pi m \left( n + m \right) \Gamma\, L} ,
        \qquad
        L := \left\lceil 16 \left( 2 + e^{-4 \pi \Gamma / 3} \right) e^{8 \pi \Gamma / 3} \, N^{\gamma} \, \ln \frac{24 m}{\epsilon} \right\rceil ,
    \end{equation}
    and all sufficiently large $n$, there exists a protocol that distills the family of continuous-time i.i.d.-depolarizing oracles $\tilde{\mathfrak{O}}_F^{\mathrm{ct}}(\theta)$ into the family of ideal Boolean oracles $\mathfrak{O}_F$ with precision $\epsilon$ and total evolution time
    \begin{equation}\label{eq:ct_query_complexity}
        \theta\, T_{\mathrm{OD}}^{\mathrm{ct}} = 2 \pi L ,
    \end{equation}
    that is, with query complexity $T_{\mathrm{OD}}^{\mathrm{ct}} = 2 \pi L / \theta$.
\end{formalthm}
\begin{proof}
    We run the stage 2 protocol of Theorem~\ref{thm:iid_threshold_at_exponent} at precision $\epsilon/2$, and replace each raw oracle query it makes by the composition of $J$ continuous-time oracle queries, each passed through the stage 1 gadget.
    The total error then has two sources, the distance between this composed query and an i.i.d.-dephasing Boolean oracle, and the error of the stage 2 protocol itself, and we show that the two sum to at most $\epsilon$.

    We first apply stage 1 to one continuous-time oracle query and write $\Lambda := \mathsf{Rec}_{\mathrm{rep}}^{\otimes (n+m)} \circ \left( e^{\theta \mathcal{L}_f} \otimes \mathcal{I} \right) \circ \mathsf{Enc}_{\mathrm{rep}}^{\otimes (n+m)}$ for its output, where $e^{\theta \mathcal{L}_f} \otimes \mathcal{I}$ acts as $e^{\theta \mathcal{L}_f}$ on the third qubit of each code block and identity on the rest.
    Lemma~\ref{lem:ct_product_formula} with $\theta = \pi/J$ bounds the distance between $e^{\theta \mathcal{L}_f}$ and $\mathcal{D}_{p(\theta)}^{\otimes (n+m)} \circ \mathcal{O}_f(\theta)$ by $\delta_{\mathrm{ct}}$, and Corollary~\ref{cor:stage1_depol} at the diagonal unitary $U = O_f(\theta)$ of Eq.~\eqref{eq:frac_oracle_def} gives
    \begin{equation}\label{eq:ct_after_stage1}
        \mathsf{Rec}_{\mathrm{rep}}^{\otimes (n+m)} \circ \left( \left( \mathcal{D}_{p(\theta)}^{\otimes (n+m)} \circ \mathcal{O}_f(\theta) \right) \otimes \mathcal{I} \right) \circ \mathsf{Enc}_{\mathrm{rep}}^{\otimes (n+m)}
        = \left( \mathcal{D}^Z_{q} \right)^{\otimes (n+m)} \circ \mathcal{O}_f(\theta) ,
    \end{equation}
    with $q := 2 p(\theta) / 3 = \frac{1}{2} \left( 1 - e^{-4 \Gamma \theta / 3} \right)$.
    By Lemma~\ref{lem:channel_norm_facts}(vii) and (vi), tensoring both channels of Eq.~\eqref{eq:ct_split_bound} with the identity on the first two qubits of each code block and composing them with the encoding and the recovery does not increase their diamond distance, so Eq.~\eqref{eq:ct_after_stage1} gives
    \begin{equation}\label{eq:ct_q_and_delta}
        \left\| \Lambda - \left( \mathcal{D}^Z_{q} \right)^{\otimes (n+m)} \circ \mathcal{O}_f(\theta) \right\|_{\diamond}
        \leq \delta_{\mathrm{ct}} = \frac{2 m \left( n + m \right) \Gamma}{3} \cdot \frac{\pi^2}{J^2} .
    \end{equation}

    Lemma~\ref{lem:compose_J_steps} applied to Eq.~\eqref{eq:ct_q_and_delta} bounds the distance between the composed query $\Lambda^{\circ J}$ and the i.i.d.-dephasing Boolean oracle with per-qubit error rate $q_J$ given by $1 - 2 q_J = (1 - 2q)^J = e^{-4 \Gamma \theta J / 3} = e^{-4 \pi \Gamma / 3}$ by
    \begin{equation}\label{eq:ct_composed}
        \left\| \Lambda^{\circ J} - \left( \mathcal{D}^Z_{q_J} \right)^{\otimes (n+m)} \circ \mathcal{O}_f \right\|_{\diamond} \leq J \delta_{\mathrm{ct}} .
    \end{equation}
    Writing $p_{\Gamma} := \frac{3}{4} \left( 1 - e^{-4 \pi \Gamma / 3} \right)$, we have $q_J = \frac{1}{2} \left( 1 - e^{-4 \pi \Gamma / 3} \right) = 2 p_{\Gamma} / 3$, so the i.i.d.-dephasing Boolean oracle in Eq.~\eqref{eq:ct_composed} is exactly the oracle on which the stage 2 protocol of Theorem~\ref{thm:iid_threshold_at_exponent} runs when its raw oracle is the i.i.d.-depolarizing Boolean oracle with per-qubit error rate $p_{\Gamma}$ (Corollary~\ref{cor:stage1_depol}).
    The composed query is therefore within diamond distance $J \delta_{\mathrm{ct}}$ of the input of that stage 2 protocol.

    We now run the stage 2 protocol of Theorem~\ref{thm:iid_threshold_at_exponent} at precision $\epsilon/2$ and per-qubit error rate $p_{\Gamma}$ on the composed query.
    The theorem requires $p_{\Gamma} < p_{\mathrm{th}}\left( \gamma, \nu \right)$, and since $p_{\mathrm{th}}\left( \gamma, \nu \right) < 3/4$ by Lemma~\ref{lem:threshold_equation}, this is equivalent to $e^{-4 \pi \Gamma / 3} > \frac{3 - 4 p_{\mathrm{th}}\left( \gamma, \nu \right)}{3} > 0$, which gives Eq.~\eqref{eq:ct_threshold} after taking logarithms.
    Theorem~\ref{thm:iid_threshold_at_exponent} applies for all sufficiently large $n$ because $\epsilon/2$ has the same decay rate $\nu$ as $\epsilon$.
    Its query complexity Eq.~\eqref{eq:T_OD_at_exponent} at precision $\epsilon/2$ and $p = p_{\Gamma}$ is $2L$ with $L$ as in Eq.~\eqref{eq:ct_theta_condition}, because $9 - 4 p_{\Gamma} = 3 \left( 2 + e^{-4 \pi \Gamma / 3} \right)$ and $\left( 3 - 4 p_{\Gamma} \right)^2 = 9 e^{-8 \pi \Gamma / 3}$ give $\frac{48 \left( 9 - 4 p_{\Gamma} \right)}{\left( 3 - 4 p_{\Gamma} \right)^2} = 16 \left( 2 + e^{-4 \pi \Gamma / 3} \right) e^{8 \pi \Gamma / 3}$.
    Hence its stage 2 protocol makes $2L$ queries to $\left( \mathcal{D}^Z_{q_J} \right)^{\otimes (n+m)} \circ \mathcal{O}_f$ and outputs a channel within diamond distance $\epsilon/2$ of $\mathcal{O}_f$.
    Our distilled oracle is this protocol with each of the $2L$ queries replaced by the composed query $\Lambda^{\circ J}$.
    Each composed query makes $J$ continuous-time oracle queries of evolution time $\theta$, so the distilled oracle makes $2JL = 2 \pi L / \theta$ continuous-time oracle queries, which is $T_{\mathrm{OD}}^{\mathrm{ct}}$ of Eq.~\eqref{eq:ct_query_complexity}, and its total evolution time is $2 J L \theta = 2 \pi L$.
    Replacing the $2L$ queries one at a time writes the difference between the distilled oracle and the output of the protocol as a telescoping sum of $2L$ terms, each the difference of two channels that differ in one query only.
    By Lemma~\ref{lem:channel_norm_facts}(vi) and Eq.~\eqref{eq:ct_composed}, each term has diamond norm at most $J \delta_{\mathrm{ct}}$, so by the triangle inequality the diamond distance between the distilled oracle and the output of the protocol is at most $2 L \cdot J \delta_{\mathrm{ct}} = \frac{4 \pi^2 m \left( n + m \right) \Gamma L}{3 J} = \frac{4 \pi m \left( n + m \right) \Gamma L}{3} \, \theta \leq \frac{\epsilon}{2}$ by $\theta = \pi / J$ and $\theta \leq \theta_{\mathrm{th}}$ of Eq.~\eqref{eq:ct_theta_condition}.
    By the triangle inequality with the error $\epsilon/2$ of the protocol itself, the distilled oracle is within diamond distance $\epsilon$ of $\mathcal{O}_f$, which completes the proof of Theorem~\ref{thm:continuous_OD_two_sided}.
\end{proof}

\clearpage
\IfFileExists{paper/ref.bib}{\bibliography{paper/ref}}{\bibliography{ref}}

\end{document}